\documentclass[11pt]{article}
\usepackage[margin=1in]{geometry}
\usepackage{amsmath}
\usepackage{graphicx}
\usepackage{enumerate}
\usepackage{enumitem}
\usepackage{natbib}
\usepackage{url} 

\usepackage{algorithmic} 
\usepackage{algorithm} 
\usepackage{setspace} 
\usepackage{rotating} 

\usepackage{amsfonts, amssymb} 
\usepackage{amsthm}
\usepackage[T1]{fontenc}
\usepackage{lmodern}

\usepackage{epstopdf} 
\usepackage{hyperref} 
\usepackage{bm} 
\usepackage{bbm}
\usepackage{multirow} 
\usepackage{latexsym} 
\usepackage{rotating} 
\usepackage{titlesec} 
\usepackage{caption} 
\usepackage{mathtools} 
\usepackage{color} 
\usepackage{diagbox}
\usepackage{pict2e}
\usepackage{etoolbox}
\makeatletter
\@ifundefined{c@lofdepth}{}{\let\c@lofdepth\undefined}
\@ifundefined{c@lotdepth}{}{\let\c@lotdepth\undefined}
\makeatother

\usepackage{tocloft}

\usepackage[english]{babel}
\usepackage{empheq}

\usepackage{cases}
\usepackage{tikz}
\usetikzlibrary{arrows.meta,shapes,shapes.arrows,
shapes.geometric,shapes.multipart,decorations.pathmorphing,
positioning,shapes.swigs,}

\usepackage{lato}

\newcommand{\mymatrixTwo}[3]{ 
\begin{array}{#1}
#2 \\ #3 
\end{array}}
\newcommand{\mymatrixThree}[4]{ 
\begin{array}{#1}
#2 \\ #3 \\ #4
\end{array}}
\newcommand{\mymatrixFour}[5]{ 
\begin{array}{#1}
#2 \\ #3 \\ #4 \\ #5
\end{array}}

\newtheorem{theorem}{Theorem}[section]
\newtheorem{lemma}[theorem]{Lemma}

\theoremstyle{definition}

\definecolor{mycolor}{RGB}{0,200,200} 

\newcommand{\indep}{\rotatebox[origin=c]{90}{$\models$} }

\DeclareMathOperator*{\argmin}{arg\,min}
\DeclareMathOperator*{\median}{median}
\newcommand{\EXP}{E}
\newcommand{\VAR}{\text{Var}}

\newcommand{\AVER}{\mathbbm{P}}
\newcommand{\EMP}{\mathbbm{G}}
\newcommand{\EXPk}{\EXP\LSS}
\newcommand{\VARk}{\text{Var}\LSS}
\newcommand{\AVERk}{\mathbbm{P}_{\mathcal{I}_k}}
\newcommand{\EMPk}{\mathbbm{G}_{\mathcal{I}_k}}

\newcommand{\cond}{\, \big| \,}
\newcommand{\COND}{\, \bigg| \,}
\newcommand{\con}{ ; }
\newcommand{\R}{\mathbbm{R}}
\newcommand{\ind}{\mathbbm{1}}
\newcommand{\T}{^\intercal}
\newcommand{\sT}{^{*,\intercal}}

\newcommand{\bX}{X}
\newcommand{\obX}{\overline{\bX}}
\newcommand{\ubX}{\underline{\bX}}
\newcommand{\bx}{x}
\newcommand{\obx}{\overline{\bx}}
\newcommand{\ubx}{\underline{\bx}}
\newcommand{\oD}{\overline{D}}
\newcommand{\uD}{\underline{D}}

\newcommand{\bZ}{Z}
\newcommand{\bz}{z}
\newcommand{\bW}{W}
\newcommand{\bw}{w}
\newcommand{\bU}{U}
\newcommand{\bu}{u}
\newcommand{\bL}{L}
\newcommand{\obL}{\overline{L}}

\newcommand{\bO}{O}
\newcommand{\bo}{o}

\newcommand{\potY}[1]{Y^{(#1)}}
\newcommand{\potD}[1]{D^{(#1)}}
\newcommand{\potDt}[2]{D_{#2}^{(#1)}}

\newcommand{\bv}{v}

\newcommand{\HL}[1]{\hyperlink{(#1)}{(#1)}}
\newcommand{\HT}[1]{\hypertarget{(#1)}{(#1)}}
\newcommand{\ETA}{^{(\eta)}}
\newcommand{\LSS}{^{(-k)}}
\newcommand{\tLSS}{^{(-k), \intercal}}
\newcommand{\SSS}{^{(k)}}

\newcommand{\M}{\mathcal{M}}
\newcommand{\InfFt}{{\normalfont\texttt{IF}}}
\newcommand{\cInfFt}{{\normalfont\texttt{cIF}}}
\newcommand{\HH}{\mathcal{H}}

\newcommand{\Eh}[1]{\widehat{h}_{#1}\LSS}
\newcommand{\Eq}[1]{\widehat{q}_{#1}\LSS}

\newcommand{\EhOone}{\widehat{h}_{1}\LSS (W,\obX_{\TIME})}
\newcommand{\EhOzero}{\widehat{h}_{0}\LSS (W,X_{0})}
\newcommand{\EhOzone}{\widehat{h}_{0}\LSS (W,X_{0} \con a_Y=1)}
\newcommand{\EhOzzero}{\widehat{h}_{0}\LSS (W,X_{0} \con a_Y=0)}
\newcommand{\EhOtwo}{\widehat{h}_{2}\LSS (W,X_{0})}

\newcommand{\ThOone}{h_{1}^*(W,\obX_{\TIME})}
\newcommand{\ThOzero}{h_{0}^*(W,X_{0})}
\newcommand{\ThOtwo}{h_{2}^*(W,X_{0})}

\newcommand{\EqOzero}{\widehat{q}_{0}\LSS (Z,X_{0})}
\newcommand{\EqOone}{\widehat{q}_{1}\LSS (Z,\obX_{\TIME})}

\newcommand{\TqOzero}{q_{0}^*(Z,X_{0})}
\newcommand{\TqOone}{q_{1}^*(Z,\obX_{\TIME})}

\newcommand{\Th}[1]{{h}_{#1}^*}
\newcommand{\Tq}[1]{{q}_{#1}^*}
\newcommand{\bIF}{\bm{IF}}
\newcommand{\cbIF}{\bm{cIF}}
\newcommand{\RD}{\text{rand}}
\newcommand{\CSE}{\text{CSE}}
\newcommand{\TIME}{T}
\newcommand{\stime}{t}
\newcommand{\SUPP}{Supplement}
\newcommand\numeq{\addtocounter{equation}{1}\tag{\theequation}}
\newcommand{\CBF}{CBF}
\newcommand{\CBFs}{CBFs}

\newcommand{\CIone}{$\bZ  \indep Y \cond (A,\oD_{\TIME+1},\obX_{\TIME},\bU)$}
\newcommand{\CItwo}{$\bZ \indep (\uD_{1},\ubX_{1}) \cond (A, \bX_{0}, \bU)$}
\newcommand{\CIthree}{$\bW \indep (A,\bZ) \cond (\oD_{\TIME+1},\obX_{\TIME},\bU)$}
\newcommand{\CIfour}{$\bW \indep (A,\uD_{1},\ubX_{1},\bZ) \cond (\bX_{0}, \bU)$}
\newcommand{\CIfive}{$\bW \indep \bZ \cond (A,\oD_{\TIME+1},\obX_{\TIME},\bU) $}
\newcommand{\CIsix}{$\bW \indep (\uD_{1},\ubX_{1},\bZ) \cond (A,\bX_{0}, \bU) $}

\newcommand{\AssumptionProxy}{\HL{A6}}
\newcommand{\AssumptionCompleteness}{\HL{A7}}
\newcommand{\AssumptionDismissible}{\HL{A5}}
\newcommand{\AssumptionDismissibleTV}{\HL{A5}}
\newcommand{\AssumptionMixBiasDenom}{\HL{A11}}
\newcommand{\AssumptionMixBiasNumer}{\HL{A10}}
\newcommand{\AssumptionConsistency}{\HL{A9}}
\newcommand{\AssumptionStrongOv}{\HL{A8}}

\newcommand{\NUMER}{{\normalfont\texttt{Num}}}
\newcommand{\DENOM}{{\normalfont\texttt{Den}}}

\hypersetup{
colorlinks=true,
linkcolor=blue,
filecolor=blue,
urlcolor=blue,
citecolor=blue
}

\graphicspath{ {plot/} }

\definecolor{red1}{RGB}{255,64,64}
\definecolor{blue1}{RGB}{128,255,255}
\definecolor{green1}{RGB}{0,205,0}

\definecolor{red1}{RGB}{255,204,204}
\definecolor{blue1}{RGB}{204,204,255}
\definecolor{light-gray}{gray}{0.7}

\allowdisplaybreaks
\begin{document}

\title{Proximal Causal Inference for Conditional Separable Effects}
\author{
Chan Park$^{a}$, Mats J. Stensrud$^{b}$, Eric J. Tchetgen Tchetgen$^{c}$\\[0.25cm]
\makebox[1cm][c]{{\footnotesize $^{a}$Department of Statistics, University of Illinois Urbana-Champaign, Champaign, IL 61820, U.S.A.}}\\
\makebox[1cm][c]{{\footnotesize $^{b}$Department of Mathematics, \'Ecole Polytechnique F\'ed\'erale de Lausanne, 
Lausanne 1015, Switzerland}}\\
\makebox[1cm][c]{{\footnotesize $^{c}$Department of Statistics and Data Science, University of Pennsylvania, Philadelphia, PA 19104, U.S.A.}}
}
 \date{}
  \maketitle
\begin{abstract}
Scientists regularly pose questions about treatment effects on outcomes conditional on a post-treatment event. However, causal inference in such settings requires care, even in perfectly executed randomized experiments. Recently, the conditional separable effect (CSE) was proposed as an interventionist estimand that corresponds to scientifically meaningful questions in these settings. However, existing results for the CSE require no unmeasured confounding between the outcome and post-treatment event, an assumption frequently violated in practice. In this work, we address this concern by developing new identification and estimation results for the CSE that allow for unmeasured confounding. We establish nonparametric identification of the CSE in observational and experimental settings with time-varying confounders, provided that certain proxy variables for hidden common causes of the post-treatment event and outcome are available. For inference, we characterize an influence function for the CSE under a semiparametric model where nuisance functions are a priori unrestricted. Using modern machine learning methods, we construct nonparametric nuisance function estimators and establish convergence rates that improve upon existing results. Moreover, we develop a consistent, asymptotically linear, and locally semiparametric efficient estimator of the CSE. We illustrate our framework with simulation studies and a real-world cancer therapy trial.

\end{abstract}
\noindent%
{\it Keywords:}  confounding bridge function, controlled direct effect, mixed-bias property, principal stratum effect, proxy maximum moment restriction, truncation by death

\newpage

\section{Introduction} \label{sec:Introduction}

\subsection{Review: Conditional Separable Effects and Proximal Causal Inference}

In many observational studies and randomized trials, outcomes of interest are defined conditional on a post-treatment event. For example, \citet{Ding2011}, \citet{Wang2017}, \citet{YangDing2018}, and \citet{Stensrud2022CSE} analyzed the Southwest Oncology Group (SWOG) Trial, where male refractory prostate cancer patients were assigned to one of two chemotherapies. Their objective was to estimate the causal effect of cancer chemotherapies on quality of life in a subsequent follow-up period. Quality of life can only be assessed for patients who have survived until the follow-up period, and the causal effect of interest is only relevant to survivors. As another example, \citet{Jemiai2007} analyzed an HIV vaccine trial to study vaccine effects on viral load. Since uninfected individuals have a viral load of essentially zero, their question focused on the causal effects among individuals who experience the post-randomization event of HIV infection.


Several causal estimands can be considered in the presence of post-treatment events. \citet{RG1992} and \citet{Pearl2001} introduced the concepts of controlled direct effects and pure (natural) direct effects. The controlled direct effects refer to the effect of treatment on the outcome in a hypothetical scenario where a particular post-treatment event can be prevented. In contrast, pure (natural) direct effects capture the effect of treatment on the outcome when the post-treatment event is allowed to take the level it would naturally attain under a given treatment condition. \citet{FR2002} defined a so-called principal stratum effect, which is a causal effect in the subset of individuals who would experience a meaningful post-treatment event status regardless of the treatment they received. 
The principal stratum effect was also described earlier in  \citet{Robins1986} in the context of truncation by death. In the SWOG trial, the controlled direct effect is defined as the effect of cancer treatment on quality of life had, contrary to fact, every patient been kept alive via an external intervention. In contrast, an instance of a principal stratum effect is the effect of cancer treatment on quality of life among ``always survivors,'' i.e., patients who would survive regardless of the cancer treatments they received; this principal stratum effect is also known as the survivor average causal effect.

However, these causal estimands have some limitations. First, the controlled direct effect is defined under an intervention where everyone experiences a post-treatment event, which may be unrealistic and infeasible. For instance, assessing the causal effect of cancer therapy on quality of life in a hypothetical scenario where everyone survives may not be of substantive interest, given that such a hypothetical intervention is difficult, if not impossible, to conceive of. Second, pure (natural) direct effects are not defined with respect to interventions that can be implemented in real-life experiments, and their classical identification relies on untestable cross-world independence assumptions; see \citet{AndrewsDidelez2021} for a detailed discussion of cross-world assumptions. Third, although principal stratum effects have been advocated in various settings \citep{Rubin2006, Vanderweele2011, Ding2016}, this effect is defined in a particular subgroup that cannot be identified from real-world data without strong assumptions. Moreover, this subgroup may represent an atypical (possibly non-existent) subset of the population, limiting its practical utility. For more in-depth discussion on this topic, see \citet{Robins1986}, \citet{Robins2007Discussion}, \citet{Joffe2011}, \citet{RobinsRichardson2011}, \citet{DawidDidelez2012}, \citet{RRS2022}, \citet{Stensrud2022}, and \citet{Stensrud2022CSE}.

Instead of considering the aforementioned estimands, \citet{Stensrud2022CSE} proposed the conditional separable effect (CSE), which is applicable when the primary outcome is only of interest conditional on a post-treatment event. The definition of the CSE is inspired by the seminal treatment composition idea of \citet{RobinsRichardson2011}. Considering the CSE requires that the treatment in the current study can be conceived as a composite of two well-defined components, say $A_Y$- and $A_D$-components, where the $A_Y$-component directly affects the outcome but does not affect the post-treatment event and the $A_D$-component directly affects the post-treatment event but may or may not directly affect the outcome. The CSE is then defined as the causal effect of the $A_Y$-component on the outcome within a subgroup of the population experiencing a post-treatment event, provided that the $A_D$-component is set at a certain level. This implies that the CSE can be interpreted as the direct effect of treatment on the outcome through the $A_Y$-component in a specific subgroup of the population. Unlike principal stratum effects, the two key conditions can in principle be falsified in future experiments provided that $A_D$ and $A_Y$ are well-defined separable components of the treatment. In other words, unlike the subgroup used to define the principal stratum effects, such as always survivors, the subgroup used to define the CSE is potentially experimentally observable. In addition, unlike the controlled direct effect, the CSE does not require conceiving of hypothetical interventions on the post-treatment event.  

For identification and estimation of the CSE, \citet{Stensrud2022CSE} relied on the assumption that there is no unmeasured common cause of the outcome and the post-treatment event. This assumption is often justified by the belief that the measured covariates are rich enough to account for confounding between the outcome and the post-treatment event. However, in many real-world applications, collecting all confounding variables is a daunting task. In the examples mentioned above, unmeasured variables such as medical history, chronic illnesses, lifestyle elements (e.g., smoking, drinking), and sociodemographic status can affect both the outcome and post-treatment event, casting doubt on the no unmeasured confounding assumption, even in randomized experiments.

Recent works \citep{Lipsitch2010, Kuroki2014, Miao2018, Miao2020, Shi2020, Shi2020_Epi, Kallus2021, PMMR2021, Ghassami2022, Dukes2023_ProxMed, Ying2023, Bennett2024, Cui2023, LuoLiMiaoHe2024, TT2024_Proximal} have focused on mitigating bias due to unmeasured confounding by carefully leveraging proxies for the latter. This constitutes the so-called proximal causal inference framework, which relies on the key assumption that one has access to two types of proxies: (i) treatment-confounding proxies, which may or may not be causally related to the treatment of interest, and are related to the outcome only through an association with unmeasured confounders of the treatment and outcome; (ii) outcome-confounding proxies, which are potential causes of the outcome and are related with the treatment only through an association with unmeasured confounders. 

Using treatment- and outcome-confounding proxies appropriately, one can potentially identify and estimate causal effects. Below, we summarize some previous works discussing such methodologies. \citet{Miao2018} established sufficient conditions for proxy-based nonparametric identification of the average treatment effect in the point treatment setting. Building upon this work, \citet{TT2024_Proximal} extended the result to time-varying treatments using proximal g-computation, a generalization of Robins’ g-computation algorithm \citep{Robins1986}. \citet{Cui2023} and \citet{Deaner2018} provided alternative identification conditions for proximal causal inference while the former developed a doubly robust semiparametric locally efficient estimator of the average treatment effect using proxies. \citet{Kallus2021}, \citet{PMMR2021}, and \citet{Ghassami2022} discussed estimation of the average treatment effect using nonparametric methods. While these previous works in proximal causal inference aimed to infer causal estimands that do not depend on a post-treatment event, some recent extensions have considered such settings. \citet{Dukes2023_ProxMed} studied semiparametric proximal causal inference in mediation settings, focusing on identification and estimation of natural direct and indirect effects. \citet{Ghassami2024} considered proximal causal inference in mediation settings where the mediator of interest is unobserved. \citet{LuoLiMiaoHe2024} established identification and estimation of principal stratum effects using proxies when the treatment and principal strata are confounded by unobserved variables. While similar to our work in proposing a proximal inference approach to the truncation by death problem, they focused on a different estimand.

\subsection{Contributions} \label{sec:Contribution}

The main goals of this paper concern identification and estimation of the CSE in both observational and experimental settings, allowing for the possibility of unmeasured confounding. Therefore, the paper complements \citet{Stensrud2022CSE}, which a priori ruled out the presence of such confounding. The main contributions of the paper are as follows:
\begin{itemize}[leftmargin=0.0cm]

\item[1.] We extend the proximal causal inference framework by establishing the formal identification of the CSE. This is achieved by introducing new confounding bridge functions (CBFs; \citealp{Miao2020}), which capture the impact of hidden common causes of the outcome and post-treatment events. Incorporating these \CBFs\ is nontrivial, as existing approaches in the proximal causal inference framework are not directly applicable to the CSE. This is because the CSE is defined conditional on a post-treatment event, a scenario not addressed by the prior methods, and fundamentally different from mediation effects and principal stratum effects studied by \citet{Dukes2023_ProxMed}, \citet{Ghassami2024}, and \citet{LuoLiMiaoHe2024}. Notably, these studies rely on a cross-world independence assumption \citep{AndrewsDidelez2021}, and \citet{Dukes2023_ProxMed} conceives an intervention on the mediator to define causal effects; both of which we bypass.

\item[2.] We explicitly account for time-varying confounders in our framework in order to broaden the applicability of our approach to longitudinal settings. This is a significant distinction from most proximal causal inference approaches, which typically do not address time-varying confounders, with some exceptions such as \citet{Ying2023}, \citet{Bennett2024}, and \citet{TT2024_Proximal}. Thus, we expand the scope of the proximal causal inference methodology.

\item[3.] Using semiparametric efficiency theory, we characterize an influence function for the CSE within a semiparametric model of the observed data law that allows for unrestricted \CBFs. The influence function is distinct from those in prior papers, such as \citet{Cui2023} and \citet{Dukes2023_ProxMed}, and is defined in terms of new \CBFs.  This distinction arises from the difference between the CSE and other estimands, as well as the inclusion of time-varying confounders. Consequently, the influence function-based methods from previous works cannot be readily applied for inference, necessitating the development of a new approach.

\item[4.] We use modern machine learning methods to nonparametrically estimate \CBFs\ without the need for specifying restrictive parametric models. This approach sets our work apart from prior works in proximal causal inference that rely on parametric methods. Specifically, we build upon the proxy maximum moment restriction (PMMR) framework, originally introduced by \citet{PMMR2021}. We show that our estimator achieves a faster convergence rate than the original PMMR method, which provides, to the best of our knowledge, the sharpest available bounds for this class of estimators. Furthermore, we extend the PMMR framework to estimate nested CBFs, an area that has remained largely unexplored in the existing literature.

\item[5.] Based on the influence function in 3 and the nonparametric \CBF\ estimators in 4, we develop a consistent and asymptotically normal estimator for the CSE, which attains the semiparametric local efficiency bound under certain conditions. In addition, we show that the estimator has the so-called mixed-bias property \citep{Rotnitzky2020}, meaning that it is asymptotically normal even if some nuisance components are estimated at a slow rate, provided that other components can be estimated at a faster rate to compensate.

\end{itemize}

The replication code is publicly available at \url{https://github.com/qkrcks0218/PCSE}.

\section{Setup} \label{sec:setup}

Let $N$ be the number of observed study units in the data, indexed by the subscript $i \in \{ 1,\ldots,N\}$. We will suppress the subscript unless necessary. Let $\stime \in \{0,\ldots,\TIME+1\}$ index discrete time intervals. For a variable $V$, $\overline{V}_{\stime} = (V_0,\ldots,V_{\stime})$ denotes the history of $V$ up to time $t$, and $\underline{V}_{\stime+1} = (V_{\stime+1},\ldots,V_{\TIME+1})$ denotes its future after time $\stime$. At the final time $\TIME+1$, we observe independent and identically distributed (i.i.d.) data for each study unit, given by $\bO = (Y, \oD_{\TIME+1} ,A, \overline{\bL}_{\TIME})$; here, $Y$ is the outcome of interest at time $\TIME+1$, and $\oD_{\TIME+1} = (D_{1},\ldots,D_{\TIME+1})$ is the history of $D$, where $D_{\stime} \in \{0,1\}$ is a binary post-treatment event at time $\stime \in \{1,\ldots,\TIME+1\}$ which may occur prior to $Y$. Without loss of generality, we encode $D_{\stime}=0$ to indicate that a unit experiences the post-treatment event at time $t$ for which $Y$ becomes relevant; under this definition, $D_{\stime}$ is nondecreasing in $t$. Next, $A \in \{0,1\}$ denotes the treatment status at baseline where $A=0$ and $A=1$ indicate that a study unit is assigned to the control and treatment arms in the current two-arm study, respectively. The vector $\overline{\bL}_{\TIME} = (\bL_{0}, \ldots, \bL_{\TIME})$ consists of observed baseline covariates $\bL_{0}$ and time-varying covariates $\bL_{\stime}$ at time $t\in \{1,\ldots,\TIME\}$. We use the temporal convention $(\bL_{0}, A, \bL_{1}, D_{1},\ldots,D_{\TIME},\bL_{\TIME},D_{\TIME+1},Y)$. In the SWOG trial, $Y$ is defined as change in quality of life between baseline and 12-month follow-up, $D_{\stime}$ is the indicator of whether a patient died at time $\stime$ within the studied period (i.e., truncation by death), $A$ encodes the cancer therapy assigned to a patient. Pre-treatment covariates, $\bL_{0}$, include age and race, while time-varying covariates, $\bL_{\stime}$, capture disease progression at time $t$. Clearly, quality of life at follow up is only relevant, and in fact only well-defined for a person who is alive at follow-up; see Section \ref{sec:Application} for further details.

Throughout the paper, we suppose that the binary treatment can reasonably be conceptualized as a composition of two binary components, say $A_Y \in \{0,1\}$ and $A_D \in \{0,1\}$. Specifically, $A_Y$- and $A_D$-components are defined as modified treatment, which,  e.g., could be components of $A$, that only directly affect the outcome $Y$ and the post-treatment event $D_{\stime}$, respectively; see \HL{A4} and related discussions for details. In particular, this decomposition is valid when the treatment in the current study comprises two physically distinct components. In the SWOG trial, each cancer therapy consisted of two drugs: a pain reliever and an inhibitor of cancer cell proliferation, which can be reasonably seen as $A_Y$- and $A_D$-components, respectively; see Section \ref{sec:Application} for more discussions. More generally, it is not required for one to be able to physically decompose the original treatment. Rather, we require the conceptualization of modified treatments $A_Y$ and $A_D$ that, when set to the same value $a$, leads to exactly the same values of $Y$ and $D_{\stime}$ as assigning $A$ to $a$ \citep[Section 5]{Stensrud2022}. 


In the current two-arm study, the treatment $A$ and its $A_Y$- and $A_D$-components satisfy a deterministic relationship. Specifically, $A$ is defined as a specific combination of $(A_Y,A_D)$ in the sense that $A=a$ if and only if $A_Y=A_D=a$ for $a \in \{ 0,1 \}$. On the other hand, consider a four-arm trial that can potentially be conducted in the future, in which each study unit is assigned to one of the following four treatment levels $(A_Y,A_D) \in \{ (0,0),(1,0),(0,1),(1,1) \}$. According to the deterministic relationship, the two treatment combinations $(A_Y,A_D) \in \{ (0,0) , (1,1) \}$ are assigned in the two-arm study, whereas the other two treatment combinations $(A_Y,A_D) \in \{ (1,0) , (0,1) \}$ are not. Notably, an analogous hypothetical trial has been considered in the context of mediation \citep{RobinsRichardson2011, RRS2022} and competing risk \citep{Stensrud2021}, and the CSE \citep{Stensrud2022CSE}. 

Let $\bL_{\stime}^{(a_Y,a_D)}$, $\potDt{a_Y,a_D}{\stime} \in \{0,1\}$, and $\potY{a_Y,a_D} \in \R$ be the potential time-varying confounders at time $\stime \in \{0,\ldots,\TIME\}$, post-treatment event at time $\stime \in \{1,\ldots,\TIME+1\}$, and outcome at the final time $\TIME+1$, respectively, had $(A_Y,A_D)$ been assigned to $(a_Y,a_D) \in \{ (0,0), (1,0), (0,1), (1,1) \}$ in the four-arm trial. The potential post-treatment event and the potential outcome in the two-arm study reduce to $\potDt{a}{t}= \potDt{a_Y=a,a_D=a}{t}$ and $\potY{a}= \potY{a_Y=a,a_D=a}$ for $a \in \{0,1\}$. In addition, following the notation in \citet{Stensrud2022CSE}, let $G$ be the potential future four-arm trial, and let $V(G)$ be a node represented on the causal directed acyclic graph (DAG; \citealp{Pearl2009}). For example, we use $Y(G)$ to denote the outcome that would be observed in the four-arm trial. 

Throughout the paper, we allow for the possibility of hidden confounding factors, denoted by $\bU$, which are common causes of $(Y,D_{\stime},A)$. Based on the relationship with $\bU$, we assume that the measured baseline covariate $\bL_{0}$ can be partitioned into three variables $(\bW, \bZ, \bX_{0})$, which are distinguished as follows. First, $\bW$ and $\bZ$ are outcome- and treatment-confounding proxies, respectively; see \HL{A6} for formal conditions that $W$ and $Z$ must satisfy as valid outcome and treatment-confounding proxies. Second, $\bX_{0}$ is a collection of measured baseline confounders that may causally impact $(Y,D_{\stime},A)$. We formalize the relationships between these variables in the next section. The time-varying covariates, $\bL_{\stime}$ for $t \in \{1,\ldots,\TIME\}$, are treated as confounders, and for notational consistency, we adaptively denote them as $\bX_{\stime}$. 

We introduce additional notation used throughout. The abbreviation ``a.s.'' stands for ``almost surely.'' Let $\mathcal{L}_{2}(V)$ be a Hilbert space of functions of a random variable $V$ with finite variance equipped with the inner product $\langle f_1, f_2 \rangle = \EXP \{ f_1(V) f_2(V) \}$. For a function $f (\bO)$, let $\big\| f \big\|_{\infty} = \sup_{\bO} \big| f(\bO) \big|$ be the supremum norm of $f$ and $\big\| f \big\|_{P,2} = \big[ \EXP \big\{ f^2(\bO) \big\} \big]^{1/2}$ be the $L_2(P)$-norm of $f$. For a sequence of random variables $ V_N $ and a sequence of numbers $ r_N $, let $V_N=O_P(r_N)$ and $V_N=o_P(r_N)$ denote that $V_N/r_N$ is stochastically bounded and converges to zero in probability as $N \rightarrow \infty$, respectively. Let $V_N \stackrel{D}{\rightarrow} V$ mean that $V_N$ converges weakly to a random variable $V$ as $N \rightarrow \infty$. Lastly, the symbol $\indep$ denotes statistical independence.

\section{Assumptions} \label{sec:assumption}

We begin by introducing key assumptions. 
\begin{itemize}
\item[\HT{A1}] (\textit{Consistency}) $Y = \potY{a=A}$, $\oD_{\TIME+1} = \oD_{\TIME+1}^{(a=A)}$, $ \obX_{\TIME} = \obX_{\TIME}^{(a=A)}$ a.s.

\item[\HT{A2}] (\textit{Latent Ignorability}) For $a \in \{0,1\}$, $(\potY{a} , \uD_{1}^{(a)}, \underline{\bX}_{1}^{(a)} ) \indep A \cond (\bX_{0} , \bU)$.

\item[\HT{A3}] (\textit{Latent Positivity}) For $a \in \{0,1\}$, \\
(i) $\Pr(D_{\TIME+1}=0, \ubX_{1} \cond A=a,\bX_{0},\bU) > 0$ a.s. and (ii) $\Pr(A=a \cond \bX_{0},\bU)>0$ a.s.
\end{itemize}  
\HL{A1}-\HL{A3} are standard consistency, ignorability, and positivity conditions. Note that the probability notation $\Pr$ in \HL{A3}-(i) can be interpreted as a density when $\ubX_{1}$ includes continuous components. Both \HL{A2} and \HL{A3} are fairly mild conditions as $U$ needs not be observed. Importantly, we do not impose a conditional independence assumption considered by \citet{Stensrud2022CSE}, that is, $(\potY{a}, \uD_{1}^{(a)}, \underline{\bX}_{1}^{(a)}) \indep A \cond \bX_{0}$. This assumption is more restrictive than \HL{A2}, thereby highlighting a critical distinction between our setting and theirs. In addition, \HL{A2} also allows for unmeasured confounding of $D_{t}$ and $Y$. Therefore, this condition is less restrictive than the principal ignorability condition \citep{JoStuart2009, Ding2016, Feller2017, Laura2018}, which posits the absence of unmeasured confounding between $D_{t}$ and $Y$. In experimental settings, $A$ is unconfounded by virtue of randomization. However, unmeasured confounders may still exist for the association between $D_{t}$ and $Y$; see {\SUPP} \ref{sec: exp setting main} for details.

Next, following \citet{Stensrud2022} and \citet{Stensrud2022CSE}, we formalize the required conditions for the $A_Y$-component in a hypothetical four-arm trial $G$:

\begin{itemize}
\item[\HT{A4}] (\textit{\hypertarget{$A_Y$-Partial Isolation}{$A_Y$-Partial Isolation}}) For $a_D \in \{0,1\}$, $\potDt{a_Y=1,a_D}{t} (G) = \potDt{a_Y=0,a_D}{t} (G)$ a.s. for all $t \in \{ 1,\ldots,\TIME+1 \}$.
\end{itemize}
\HL{A4} states that, in a hypothetical future four-arm trial, the post-treatment event would not be affected by the $A_Y$-component. This assumption could in principle be falsified from a future four-arm trial in which $A_Y$ and $A_D$ are randomized. 

It is useful to consider the following condition on the $A_D$-component, even though we do not formally introduce it as an assumption:
\begin{itemize}
\item[] (\textit{\hypertarget{$A_D$-Partial Isolation}{$A_D$-Partial Isolation}}) For $a_Y \in \{0,1\}$,  $\potY{a_Y,a_D=1}(G) = \potY{a_Y,a_D=0}(G)$ a.s. \\
if $\potDt{a_Y,a_D=1}{\TIME+1}(G) = \potDt{a_Y,a_D=0}{\TIME+1}(G) = 0$. 
\end{itemize}
Similar to \HL{A4}, \hyperlink{$A_D$-Partial Isolation}{($A_D$-Partial Isolation)} states that the $A_D$-component does not affect the outcome, provided the unit does not experience the post-treatment event irrespective of the intervention on the $A_D$-component. While this assumption is not required for identification or estimation, the interpretation of the CSE depends on \hyperlink{$A_D$-Partial Isolation}{($A_D$-Partial Isolation)}; see the next paragraph for an example. However, this assumption is unlikely to hold in the presence of time-varying confounders due to the causal pathway $A_D \rightarrow \bX_{t} \rightarrow Y$. 


As discussed in Section \ref{sec:Introduction}, we focus on the CSE as the estimand of interest, which is formally defined as $\tau_{\CSE}^*
=
\psi^*(1) - \psi^*(0)$ where $
\psi^*(a_Y)
=
\EXP \big\{ \potY{a_Y,a_D=0} \cond \potDt{a_Y,a_D=0}{\TIME+1} = 0 \big\}$. 
Without loss of generality, we focus on the $a_D=0$ case, but the results established below naturally extend to the $a_D=1$ case. Under \HL{A4} (i.e., \HL{$A_Y$-Partial Isolation}), the CSE reduces to $\tau_{\CSE}^* = \EXP \big\{ \potY{a_Y=1,a_D=0} - \potY{a_Y=0,a_D=0} \cond \potDt{a_D=0}{\TIME+1} = 0 \big\}$, which is interpreted as the average causal effect of $A_Y$ on $Y$ when the other treatment component is assigned to $A_D = 0$ among the subset of study units who do not experience the post-treatment event under $A_D=0$. In the SWOG trial, the CSE can be viewed as the average causal effect of pain relievers on change in quality of life among patients who would survive had they received a specific cancer cell proliferation inhibitor; see Section \ref{sec:Application} for details. Without \hyperlink{$A_D$-Partial Isolation}{($A_D$-Partial Isolation)}, there can be causal paths from $A_D$ to $Y$ that are not mediated by $D$, such as $A_D \rightarrow \bX_{t} \rightarrow Y$, indicating that $\tau_{\CSE}^*$ captures the direct effect, but not necessarily all effects of $A$ on $Y$ relative to the post-treatment event. On the other hand, under \HL{A4} and \hyperlink{$A_D$-Partial Isolation}{($A_D$-Partial Isolation)}, i.e., full isolation, all causal paths from $A_D$ to $Y$ are mediated by $D_{\TIME+1}$, and all causal paths from $A_Y$ to $Y$ are not mediated by $D_{\TIME+1}$, implying that the CSE can be viewed as the total effect of $A$ on $Y$ not through survival.

In order to establish identification of $\tau_{\CSE}^*$, we make the following condition.
\begin{itemize}
\item[\HT{A5}] (\textit{Latent Dismissible Condition}) \begin{itemize}[leftmargin=0.25cm]
\item[(i)]  $Y(G) \indep A_D(G) \cond  ( A_Y(G), D_{\TIME+1}(G) = 0, \overline{\bX}_{\TIME}(G) , \bU(G) ) $; 
\item[(ii)]  $D_{\stime+1}(G) \indep A_Y(G) \cond ( A_D(G), D_{\stime}(G)=0, \overline{\bX}_{\stime} (G), \bU(G) )$ for all $t \in \{ 0,\ldots,\TIME\}$; 
\item[(iii)]  $\bX_{\stime+1}(G) \indep A_Y(G) \cond ( A_D(G), D_{\stime+1}(G)=0, \overline{\bX}_{\stime} (G), \bU(G) )$ for all $t \in \{ 0,\ldots,\TIME\}$. 
\end{itemize} 
\end{itemize}
\HL{A5}-(i) states that, in a hypothetical future four-arm trial, the outcome would be independent of the $A_D$-component conditional on $(A_Y,\obX_{\TIME},\bU)$ among units experiencing the post-treatment event. Likewise, \HL{A5}-(ii) states that the post-treatment event at time $\stime+1$ would be independent of the $A_Y$-component conditional on $( A_D,\obX_{\stime},\bU)$. A similar interpretation applies to \HL{A5}-(iii). \HL{A5} can be interpreted as conditions on causal mechanisms between $(A_Y,A_D)$ and $(Y,D_{\stime})$. Specifically, suppose that a causal DAG for variables $(A_Y(G),A_D(G),Y(G), 
D_{\stime}(G),U(G))$ represents a Finest Fully Randomized Causally Interpreted Structural Tree Graph model \citep{Robins1986, SWIG2013} under $G$ so that a missing arrow on the DAG encodes the assumption that an individual-level causal effect is absent for every individual in the study population. Then, \HL{A5} implies \HL{A4}, as established by \citet{Stensrud2022CSE}.

Under \HL{A1}-\HL{A5}, we obtain $\tau_{\CSE}^* = \{ \psi_{\NUMER}^*(1) -\psi_{\NUMER}^*(0) \} / \psi_{\DENOM}^*$ where
\begin{align}
&  
\psi_{\NUMER}^*(1)
=
\EXP \bigg\{
\EXP \big( Y \cond A = 1 , D_{\TIME+1} = 0 , \obX_{\TIME} , \bU  \big)
\frac{\Pr ( A=0, D_{\TIME+1} = 0 \, | \, \obX_{\TIME}, \bU ) }{
\Pr ( A=0 \, | \, \bX_{0}, \bU )}
\bigg\} 
\ ,
\nonumber
\\ 
&
\psi_{\NUMER}^*(0)
=
\EXP \big\{
\EXP \big( Y \cond A = 0 , D_{\TIME+1} = 0 , \bX_{0} , \bU  \big)
\Pr ( D_{\TIME+1} = 0 \cond A=0, \bX_{0}, \bU ) 
\big\} 
\ ,  
\label{eq-psi} 
\\ 
&
\psi_{\DENOM}^* 
=
\EXP \big\{
\Pr \big( D_{\TIME+1} = 0 \cond A=0, \bX_{0}, \bU \big)
\big\}  
\ . 
\nonumber
\end{align}
See {\SUPP} \ref{sec:supp:psi} for details on establishing \eqref{eq-psi}. However, \eqref{eq-psi} cannot serve as an identification formula as $\bU$ is unmeasured. In order to establish identification, one can use a treatment-confounding proxy $\bZ$ and an outcome-confounding proxy $\bW$, which are formally defined as follows:
\begin{itemize}
\item[\HT{A6}] (\textit{Proxies}) For $a \in \{0,1\}$, 
\begin{align*}
\begin{array}{llll}
\text{(i)} 
&  
\bZ \indep \potY{a}  \cond  (A,\oD_{\TIME+1}^{(a)}, \obX_\TIME^{(a)}, \bU) \ ;
\hspace*{1cm}
&
\text{(ii)}
&
\bZ \indep (\ubX_{1}^{(a)},\uD_{1}^{(a)}) \cond  (A, \bX_{0}, \bU)
\ ;
\\
\text{(iii)} 
&
\bW \indep (\uD_{1}^{(a)} , \ubX_{1}^{(a)} )  \cond (\bX_{0}, \bU)
\ ;
\hspace*{1cm}
&
\text{(iv)}
& 
\bW \indep (A,\bZ)  \cond  (\oD_{\TIME+1}^{(a)}, \overline{\bX}_{\TIME}^{(a)}, \bU) \ .
\end{array}
\end{align*}
\end{itemize}
\HL{A6} states that, $(\potY{a},\uD_{1}^{(a)},\ubX_{1}^{(a)})$ are conditionally independent of $\bZ$ given $(\bX_0,\bU)$, and $(A,\uD_{1}^{(a)})$ are likewise conditionally independent of $\bW$ given $(\bX_{0},\bU)$. Additionally, $\bW$ and $\bZ$ are associated only to the extent that they are associated with $\bU$. These conditions extend analogous proxy conditions in \citet{Miao2018}, \citet{Dukes2023_ProxMed}, and \citet{TT2024_Proximal} to the current setting. Figure \ref{fig:SWIG} illustrates a single world intervention graph \citep{SWIG2013} compatible with  \HL{A4}-\HL{A6}.

\begin{figure}[!htp]
\centering 
\vspace*{-0.1cm}
\scalebox{0.575}{
\begin{tikzpicture}
\tikzset{line width=1pt, outer sep=0.5pt,
ell/.style={draw,fill=white, inner sep=3pt,
line width=1pt},
swig vsplit={gap=2.5pt, 
inner line width right=0.5pt,
line width right=1.5pt}};
\node[name=Ay,shape=swig vsplit] at (0,-1.5*0.65){  \nodepart{left}{$A_Y$} \nodepart{right}{$a_Y$} };
\node[name=Ad, shape=swig vsplit] at (0,0) { \nodepart{left}{$A_D$} \nodepart{right}{$a_D$} };
\node[name=A,ell,  shape=ellipse] at (-4,0) {$A$}  ;
\node[name=D1,ell,  shape=ellipse] at (4,0) {$D_1^{(a_D)}$}  ;
\node[name=D2,ell,  shape=ellipse] at (8,0) {$D_2^{(a_D)}$}  ;
\node[name=Y,ell,  shape=ellipse] (Y) at (12,0) {$Y^{(a_Y,a_D)}$};
\node[name=U,ell,  shape=ellipse] at (4,5.5*0.75) {$U$};
\node[name=X0,ell,  shape=ellipse] at (1.75,2.25*0.75) {$X_0$};
\node[name=X1,ell,  shape=ellipse] at (5.5,3.2*0.75) {$X_1^{(a_D)}$};
\node[name=Z,ell,  shape=ellipse] at (-4,2.5*0.75) {$Z$};
\node[name=W,ell,  shape=ellipse] at (12,2.5*0.75) {$W$};
\begin{scope}[>={Stealth[black]},
every edge/.style={draw=black,line width=0.5pt}]
\path [->] (Ad) edge (D1);
\path [->] (Ad) edge (X1);
\path [->] (Ad) edge[bend right=17] (D2);
\path [->] (Ay) edge[bend right=7] (Y);
\path [->] (W) edge (Y);    
\path [->] (D1) edge (D2);
\path [->] (D1) edge (X1);
\path [->] (D2) edge (Y);
\path [->] (U) edge (A);
\path [->] (U) edge (Y);
\path [->] (U) edge (D1);
\path [->] (U) edge (Z);
\path [->] (U) edge (X0);
\path [->] (U) edge (X1);
\path [->] (X0) edge (X1);
\path [->] (X0) edge (D1);
\path [->] (X0) edge (A);
\path [->] (X0) edge (Y);
\path [->] (X1) edge (Y);
\path [->] (X1) edge (D2);
\path [->] (X0) edge (Z);
\path [->] (X0) edge (W);
\path [->] (Z) edge (A);
\path [->] (A) edge (Ad);
\path [->] (A) edge (Ay);
\path [->] (U) edge (W);
\end{scope}

\end{tikzpicture} 
}
\caption{\footnotesize A Single World Intervention Graph compatible with  \protect\HL{A4}-\protect\HL{A6} when $\TIME=1$.}
\label{fig:SWIG}
\vspace*{-0.3cm}
\end{figure}

For identification, we also require certain completeness conditions:
\begin{itemize}
\item[\HT{A7}] (\textit{Completeness}) Let $g(\bU)$ be an arbitrary square-integrable function. Then, we have:
\begin{itemize}[leftmargin=0.0cm]
\item[(i)] $\EXP\big\{ g(\bU) \cond \bZ,A=1,D_{\TIME+1}=0,\obX_{\TIME} \big\} = 0$ for all $(\bZ,\obX_{\TIME})$ a.s. implies $g(\bU)=0$ a.s.; 
\item[(ii)] $\EXP\big\{ g(\bU) \cond \bZ,A=0,\bX_{0} \big\} = 0$ for all $(\bZ,\bX_{0})$ a.s. implies $g(\bU)=0$ a.s.; 
\item[(iii)] $\EXP\big\{ g(\bU) \cond \bW,A=1,D_{\TIME+1}=0,\obX_{\TIME} \big\} = 0$ for all $(\bW,\obX_{\TIME})$ a.s. implies $g(\bU)=0$ a.s.; 
\item[(iv)] $\EXP\big\{ g(\bU) \cond \bW,A=0,\bX_{0} \big\} = 0$ for all $(\bW,\bX_{0})$ a.s. implies $g(\bU)=0$ a.s..
\end{itemize}
\end{itemize}
Completeness has routinely been used for identification in the context of nonparametric instrumental variable regression \citep{Newey2003}, measurement error model \citep{Hu2008}, longitudinal data \citep{Freyberger2017}, nonadditive, endogenous models \citep{Chen2014}, and proximal causal inference \citep{TT2024_Proximal}. In words, \HL{A7}-(i) and (ii) imply that any variation in $\bU$ is associated with some variation in $\bZ$ conditional on $(A=0,D_{\TIME+1}=0,\obX_{\TIME})$ and $(A=1,\bX_{0})$, respectively; \HL{A7}-(iii) and (iv) are interpreted in a similar manner. Therefore, \HL{A7} implies that $\bZ$ and $\bW$ are sufficiently relevant for $U$ so that variation in the latter induces variation in the former. To illustrate, suppose that $(\bZ,\bW,\bU)$ are categorical variables with number of categories $(d_Z,d_W,d_U)$, respectively; \HL{A7} then can be seen as rank conditions on $(\bZ,\bW,\bU)$ and requires $d_Z \geq d_U$ and $d_W \geq d_U$, i.e., $\bZ$ and $\bW$ must have at least as many categories as $\bU$. \HL{A7}-(i), (ii) will be used to establish identification via outcome \CBFs, and \HL{A7}-(iii), (iv) will be used to identify treatment \CBFs; see Theorem \ref{thm:identification} for details.

\section{Nonparametric Proximal Identification} \label{sec:obs setting}

In this section, we establish nonparametric identification of the CSE in observational settings using certain key \CBFs\ which we introduce below. We assume that these \CBFs\ exist; see {\SUPP}  \ref{sec:supp:exist bridge ft}  for sufficient conditions for their existence. To begin, we define outcome \CBFs. Let $h_0^*$, $h_1^*$, and $h_2^*$ be solutions to the following Fredholm integral equations of the first kind:
\begin{align}
&
\EXP \big\{ Y  -  h_1^* (\bW,\obX_{\TIME}) \cond \bZ, A=1, D_{\TIME+1}=0 , \obX_{\TIME} \big\}
=
0 \text{ a.s.}  
\label{eq-defh1}
\\
&
\EXP \big\{ (1-D_{\TIME+1}) h_1^* (\bW,\obX_{\TIME}) - h_0^* (\bW,\bX_{0} \con a_Y=1) \cond \bZ, A=0 , \bX_{0} \big\}
=
0 \text{ a.s.}  
\label{eq-defh0}
\\
&
\EXP \big\{ (1-D_{\TIME+1}) Y - h_0^* (\bW,\bX_{0} \con a_Y=0) \cond \bZ, A=0 , \bX_{0} \big\}
=
0 \text{ a.s.}  
\label{eq-defh00}
\\
&  
\EXP \big\{ (1-D_{\TIME+1}) - h_2^* (\bW,\bX_{0}) \cond \bZ , A=0 , \bX_{0} \big\} 
=
0 \text{ a.s.} 
\label{eq-defh2}
\end{align}

\noindent In words, $h_1^* (\bW,\obX_{\TIME})$ can be viewed as a well-calibrated forecast for $Y$ with respect to $Z$ in the sense that the resulting residual $Y - h_1^*(\bW,\obX_{\TIME})$ is mean zero conditional on $(\bZ, A=1 , D_{\TIME+1}=0 , \obX_{\TIME})$.  Likewise, $h_0^* (\bW,\bX_{0} \con a_Y=1)$, $h_0^* (\bW,\bX_{0} \con a_Y=0)$, and $h_2^* (\bW,\bX_{0})$ can be viewed as well-calibrated forecasts relative to $Z$ of $(1-D_{\TIME+1}) h_1^*(\bW,\obX_{\TIME}) $, $(1-D_{\TIME+1})Y $, and $(1-D_{\TIME+1})$, respectively, in the sense that the corresponding residuals are mean zero conditional on $(\bZ, A=0 , \bX_{0})$. Therefore, $h_0^*(\, \cdot \, \con a_Y=1)$ is indirectly related to $Y$ via the nested relationship between $h_1^*$ and $h_0^*(\, \cdot \, \con a_Y=1)$, i.e., equation \eqref{eq-defh0}.

We now define treatment \CBFs. Suppose that there exist functions $q_0^*$ and $q_1^*$ satisfying the following Fredholm integral equations of the first kind:
\begin{align}
&
\EXP \big\{ 1 - (1-A) q_0^* (\bZ,\bX_{0}) \cond \bW, \bX_{0} \big\}
=
0
\text{ a.s.} 
\label{eq-defq0}
\\
&
\EXP \big\{ (1-A) q_0^* (\bZ,\bX_{0}) -  A q_1^* (\bZ,\obX_{\TIME}) \cond \bW, D_{\TIME+1}=0 , \obX_{\TIME} \big\}
=
0 
\text{ a.s.} 
\label{eq-defq1}
\end{align}
An equivalent form for equation (6) is $1/ \Pr(A = 0
\cond \bW, \bX_{0} ) = \EXP \big\{ q_0^*(\bZ,\bX_{0}) \cond W,A=0,\bX_{0}\big\}$. Therefore, $q_0^*  $ is a well-calibrated forecast relative to $\bW$ of the inverse probability $1/\Pr(A=0 \cond \bW,\bX_{0})$, and the difference between these two functions has mean zero conditional on $(\bW,A=0,\bX_{0})$. 
The other function $q_1^* $ can also be interpreted as a probability weighting term involving both $A$ and $D_{\TIME+1}$. Notably, $q_1^*$ corresponds to the weighting term associated with the mediator and treatment in the efficient influence function of the mediation functional in the absence of $U$ \citep{TTS2012}. Like $h_0^*(\,\cdot\, \con a_Y=1)$ in \eqref{eq-defh0}, $q_1^*$ is defined through a nested relationship with $q_0^*$.

Analogous to other proximal causal inference settings in the literature, equations \eqref{eq-defh1}-\eqref{eq-defq1} do not need to admit a unique solution. In fact, all solutions to these equations uniquely identify $\tau_{\CSE}^*$, as formalized in the following Theorem.

\begin{theorem} \label{thm:identification}
Suppose that Assumptions \HL{A1}-\HL{A6} are satisfied.
\begin{itemize} 
\item[(i)] Further suppose that Assumption \HL{A7}-(i),(ii) is satisfied, and that there exist \CBFs\ $h_0^*$, $h_1^*$, and $h_2^*$ satisfying \eqref{eq-defh1}-\eqref{eq-defh2}. Then, we have
\begin{align*}
\tau_{\CSE}^*
=
\frac{ \EXP \big\{ h_0^*(\bW,\bX_{0} \con a_Y=1) - h_0^*(\bW,\bX_{0} \con a_Y=0) \big\} }
{ \EXP \big\{ h_2^*(\bW,\bX_{0}) \big\} } \ .
\end{align*}

\item[(ii)] Further suppose that Assumption \HL{A7}-(i),(iv) is satisfied, and that there exist \CBFs\ $h_1^*$ and $q_0^*$ satisfying \eqref{eq-defh1} and \eqref{eq-defq0}. Then, we have 
\begin{align*}
\tau_{\CSE}^*
=
\frac{ \EXP \big[ 
(1-A) (1-D_{\TIME+1}) q_0^*(\bZ,\bX_{0}) \big\{ h_1^*(\bW,\obX_{\TIME}) - Y \big\}
\big] }
{ \EXP \big\{ (1-A) (1-D_{\TIME+1}) q_0^*(\bZ,\bX_{0})  \big\} } \ .
\end{align*}

\item[(iii)] Further suppose that Assumption \HL{A7}-(iii),(iv) is satisfied, and that there exist \CBFs\ $q_0^*$ and $q_1^*$ satisfying \eqref{eq-defq0} and \eqref{eq-defq1}. Then, we have 
\begin{align*}
\tau_{\CSE}^*
=
\frac{
\EXP \big[ (1-D_{\TIME+1}) Y \big\{ A q_1^*(\bZ,\obX_{\TIME}) - (1-A) q_0^*(\bZ,\bX_{0})
\big\}
\big] 
}{
\EXP \big\{ 
A (1-D_{\TIME+1}) q_1^*(\bZ,\obX_{\TIME}) 
\big\}
} \ .
\end{align*}
\end{itemize}
\end{theorem}
Theorem \ref{thm:identification} provides three separate identification results which rely on the existence of different \CBFs. 
These identification results complement analogous results in proximal causal inference in other settings and establish a formal approach for leveraging confounding proxies in the context of separable effects. To the best of our knowledge, these results offer a new contribution to the growing literature on proximal causal inference in the challenging setting where the outcome of primary interest is truncated by death. 


\section{Semiparametric Inference}	\label{sec:inference}

\subsection{An Influence Function for Conditional Separable Effects} \label{sec:SEB}

In this section, we derive an influence function for $\tau_{\CSE}^*$ under a certain semiparametric model for the observed data. Consider the semiparametric model $\M$ for the observed data law that solely assumes the existence of a solution to equations \eqref{eq-defh1}-\eqref{eq-defh2} and is otherwise unrestricted. Formally, $\M=\big\{P \cond\text{there exist $h_0^*$, $h_1^*$, $h_2^*$ satisfying \eqref{eq-defh1}-\eqref{eq-defh2}} \big\}$. 
Importantly, note that $\M$ does not impose uniqueness of the outcome \CBFs. In addition, we can characterize the semiparametric local efficiency bound for $\tau_{\CSE}^*$ under model $\M$ when the true data-generating law belongs to a certain submodel, denoted by $\M_{\text{sub}}$. To this end, we introduce the following condition: 
\begin{itemize}
\item[\HT{S1}] Let $\mathcal{T}_0:\mathcal{L}_2(\bW,\bX_{0}) \rightarrow \mathcal{L}_2(\bZ,A=0,\bX_{0})$ and $\mathcal{T}_1:\mathcal{L}_2(\bW,\obX_{\TIME}) \rightarrow \mathcal{L}_2(\bZ,A=1,D_{\TIME+1}=0,\obX_{\TIME})$ be the operators given by $\mathcal{T}_0(g)  =  \EXP \big\{ g(\bW,\bX_{0}) \cond \bZ, A=0,\bX_{0} \big\}$ and $\mathcal{T}_1(g)  =  \EXP \big\{ g(\bW,\obX_{\TIME}) \cond \bZ,A=a_Y,D_{\TIME+1}=0,\obX_{\TIME} \big\}$. Then, $\mathcal{T}_{0}$ and $\mathcal{T}_{1}$ are surjective. 
\end{itemize}
As discussed in \citet{Cui2023}, \citet{Dukes2023_ProxMed}, and \citet{Ying2023}, \HL{S1} states that the Hilbert spaces $\mathcal{L}_2(\bW,\bX_{0})$ and $\mathcal{L}_2(\bW,\obX_{\TIME})$ are sufficiently rich so that any element in $\mathcal{L}_2(\bZ,A=0,\bX_{0})$ and $\mathcal{L}_2(\bZ,A=a_Y,D_{\TIME}=0,\obX_{\TIME})$ can be recovered via the conditional expectation mapping of an element belonging to the former, respectively. Uniqueness of a \CBF\ is ensured under certain completeness conditions; see {\SUPP} \ref{sec:supp:exist bridge ft} for details.

The following Theorem provides an influence function for the CSE under model $\M$ using the \CBFs, which will later serve as a basis of robust inference; it also gives a closed-form characterization of a semiparametric local efficiency bound for $\tau_{\CSE}^*$. 

\begin{theorem} \label{thm:IF} Suppose that Assumptions \HL{A1}-\HL{A7} hold, and that there exist treatment \CBFs\ $q_0^*$ and $q_1^*$ satisfying \eqref{eq-defq0} and \eqref{eq-defq1} at the true data-generating law $P^*$. 
\begin{itemize}[leftmargin=0.25cm]
    \item[(i)]  Then, an influence function for $\tau_{\CSE}^*$ under model $\M$ is given by $\InfFt_{\CSE}^* (\bO)$ where:
\begin{align}	
\InfFt_{\CSE}^* (\bO)
&
=
\frac{ \InfFt_{\NUMER}^*(\bO \con a_Y=1) - \InfFt_{\NUMER}^*(\bO \con a_Y=0) - \tau_{\CSE}^* \InfFt_{\DENOM}^*(\bO) }{ \psi_{\DENOM}^*  } \label{eq-IF CSE}
\ , 
\\
\InfFt_{\NUMER}^*(\bO \con a_Y=1)
&
= A  (1-D_{\TIME+1}) q_1^* (\bZ,\obX_{\TIME} ) \big\{ Y - h_1^* (\bW,\obX_{\TIME}) \big\}+
h
\nonumber_0^* (\bW,\bX_{0} \con a_Y=1) 
\\
&
\quad
+ (1-A) q_0^* (\bZ,\bX_{0} ) 
\big\{  (1-D_{\TIME+1}) h_1^* (\bW,\obX_{\TIME} ) - h_0^* (\bW,\bX_{0}  \con a_Y=1) \big\}
\ , 
\label{eq-IF_N1}
\\
\InfFt_{\NUMER}^* (\bO \con a_Y=0)
&
=
(1-A)  q_0^* (\bZ,\bX_{0}) \big\{ (1-D_{\TIME+1}) Y - h_0^* (\bW,\bX_{0} \con a_Y=0) \big\} 
\nonumber
\\
&
\quad
+
 h_0^* (\bW,\bX_{0} \con a_Y=0)  \ ,
\label{eq-IF_N2}
\\
\InfFt_{\DENOM}^*(\bO)
& 
=
(1-A) q_0^* (\bZ,\bX_{0})  \big\{ (1-D_{\TIME+1}) - h_2^*(\bW,\bX_{0}) \big\}  
+  h_2^* (\bW,\bX_{0})  \ .
\label{eq-IF_D}
\end{align} 

\item[(ii)]  
Further suppose that the true data-generating law $P^*$ belongs to $\M_{\text{sub}}$ where
\begin{align*}
    \M_{\text{sub}} 
=
\big\{
P  \in \M
\cond
\text{\HL{S1} holds, and $h_0^*$, $h_1^*$, $h_2^*$, $q_0^*$, $q_1^*$ satisfying \eqref{eq-defh1}-\eqref{eq-defq1} are unique}
\big\} \ .
\end{align*}
Then, $\InfFt_{\CSE}^*(\bO)$ given in \eqref{eq-IF CSE} is the efficient influence function for $\tau_{\CSE}^*$ under model $\M$ at $P^*$. Therefore, the corresponding semiparametric local efficiency bound for $\tau_{\CSE}^*$ is $\VAR \big\{ \InfFt_{\CSE}^* (\bO) \big\}$. 

\end{itemize} 

\end{theorem}

\noindent 
Theorem \ref{thm:IF}-(i) presents an influence function $\InfFt_{\CSE}^*$, which in fact combines three influence functions. The first two influence functions, $\InfFt_{\NUMER}^*(\,\cdot\, \con a_Y)$, are (uncentered) influence functions for $\psi_{\NUMER}^*(a_Y)$ in \eqref{eq-psi}. We remark that $\InfFt_{\NUMER}^*$ depends on $h_1^*$ and $q_1^*$ when $a_Y = 1$, but not when $a_Y=0$. The third influence function, $\InfFt_{\DENOM}^*$, is an (uncentered) influence function for $\psi_{\DENOM}^* $ in \eqref{eq-psi}. Note that $\InfFt_{\NUMER}^* (\bO \con a_Y=1)$ has a similar form as an influence function of the mediation functional in the proximal causal inference framework \citep{Dukes2023_ProxMed}. Likewise, $\InfFt_{\NUMER}^* (\bO \con a_Y=0)$ and $\InfFt_{\DENOM}^*(\bO)$ have the same form as an influence function of the average treatment effect in the proximal causal inference framework \citep{Cui2023}. The influence function thus bears similarities to these prior works, apparently using some of them as building blocks. Nevertheless, we emphasize that $\InfFt_{\CSE}^*$ in equation \eqref{eq-IF CSE} is novel to proximal causal inference literature and semiparametric theory. Specifically, the estimand $\tau_{\CSE}^*$ and its identifying functional are sufficiently distinct from those in previous works to require a standalone derivation, which is demonstrably nontrivial.  Therefore, the alignment of our results with theirs is not immediately apparent; it only becomes clear upon establishing our key results, necessitating careful consideration due to our more challenging setting. 
Specifically, as the influence function $\InfFt_{\CSE}^*$ explicitly incorporates time-varying confounding, results from \citet{Cui2023} and \citet{Dukes2023_ProxMed} are of limited value given that such confounding is ruled out in these works. Consequently, these prior results are thus technically special cases of our more general results, which one can recover under the additional assumption of no time-varying confounding. Finally, $\InfFt_{\CSE}^*$ differs significantly from the influence function for the CSE assuming no hidden confounding \citep[Appendix C]{Stensrud2022CSE}, as their nuisance functions are standard regression functions, while ours are more challenging to estimate as they correspond to solutions of certain Fredholm integral equations.

Theorem \ref{thm:IF}-(ii) provides a sufficient condition under which the influence function $\InfFt_{\CSE}^*$ becomes the efficient influence function for $\tau_{\CSE}^*$. While the following discussion also appears in \citet{Ying2023}, we reiterate it here for clarity. An important note about \HL{S1} is that it can be quite restrictive. This is because conditional expectation operators are compact under weak conditions, but a compact infinite-dimensional operator on a Banach space cannot be surjective. Therefore, a plausible scenario where \HL{S1} holds, aside from pathological cases, is when all variables $(\bZ,\bW,\obX_{\TIME})$ are categorical (e.g., \citet{Shi2020}), ensuring that the $\mathcal{L}_2$ spaces in the condition are finite-dimensional. This observation further implies that $U$ is categorical under \HL{A7}, thereby narrowing the scope of the method. However, we emphasize that our identification, estimation, and inference procedures will remain valid even if \HL{S1} does not hold. As stated in Theorem \ref{thm:IF}, \HL{S1} is introduced solely for characterizing the efficiency bound and providing its closed-form expression. In the absence of this condition, a closed-form expression for the efficiency bound is typically not available. It is also important to note that this condition only needs to hold at the true data-generating process for the efficiency bound to apply to the broader semiparametric model. Even if it does not hold elsewhere in the model, our estimator will remain consistent and asymptotically normal, though it may be inefficient.

\subsection{A Semiparametric Estimator}	\label{sec:estimator}

We construct a semiparametric estimator of $\tau_{\CSE}^*$ using the influence function from Theorem \ref{thm:IF}. In short, the proposed estimator adopts the cross-fitting approach \citep{Schick1986, Victor2018}, which is implemented as follows. We randomly split $N$ study units, denoted by $\mathcal{I}=\{1,\ldots,N\}$, into $K$ non-overlapping folds, denoted by $\mathcal{I}_1, \ldots,\mathcal{I}_K $. For each $k \in \{1,\ldots,K\}$, we estimate the \CBFs\ using observations in $\mathcal{I}_k^c = \mathcal{I} \setminus \mathcal{I}_k$, and then evaluate the estimated \CBFs\ using observations in $\mathcal{I}_k$ to obtain an estimator of $\tau_{\CSE}^*$. In what follows, we refer to $\mathcal{I}_k^c$ and $\mathcal{I}_k$ as the estimation and evaluation folds, respectively. To use the entire sample, we take the weighted average of the $K$ estimators, where the weights are proportional to the size of the evaluation fold $\mathcal{I}_k$. In the remainder of the section, we provide details on how the estimator is constructed.

\subsubsection{Estimation of Confounding Bridge Functions} \label{sec:PMMR Estimator}

We begin with estimating the \CBFs\ by adopting recently developed nonparametric methods specifically developed to estimate such \CBFs\ in other settings \citep{Singh2019, PMMR2021, Meza2021, Ghassami2022}. In particular, we outline the PMMR approach proposed by \citet{PMMR2021}, with a primary focus on the estimation of $h_1^*$. Details on estimating other \CBFs\ using the PMMR approach, as well as alternative methods, are provided in {\SUPP} \ref{sec:supp:PMMR details} and \ref{sec:supp:minmax}. In what follows, we denote $\mathcal{H} (V)$ be a Reproducing Kernel Hilbert Space (RKHS) associated with a random variable $V$, equipped with the Gaussian kernel, given by $\mathcal{K}(v,v') = \exp \big\{ - \| v - v' \|_2^2 / \kappa \big\}$, where $\kappa \in (0,\infty)$ is a bandwidth parameter.

The definition of $h_1^*$ in \eqref{eq-defh1} implies that the following moment condition is satisfied:
\begin{align}		\label{eq-moment condition}
\EXP \big[
A(1-D_{\TIME+1}) \big\{ Y - h_1^* (\bW,\obX_{\TIME}) \big\} q(\bZ,\obX_{\TIME})
\big] = 0 \ , \quad \forall q \in \mathcal{L}_2(\bZ,\obX_{\TIME}) \ . 
\end{align}
One may use \eqref{eq-moment condition} as a basis to quantify the discrepancy between a candidate $h$ and the true $h_1^*$. Specifically, we consider the following risk function of a candidate $h$:
\begin{align}
R  (h)
=
\max_{ q \in \HH(\bZ, \obX_{\TIME}), \| q \| \leq 1  }
\Big[
\EXP \big[
A(1-D_{\TIME+1}) \big\{ Y - h (\bW,\bX_{\TIME}) \big\} q(\bZ,\obX_{\TIME})
\big]
\Big]^2 \ . 
\label{eq-Risk}
\end{align}
Since the Gaussian kernel is universal and integrally strictly positive definite, we have $R(h) = 0$ if and only if $h$ satisfies \eqref{eq-defh1} \citep{PMMR2021, Zhang2023}; see {\SUPP} \ref{sec:supp:PMMR details} for technical details. Therefore, $R(h)$ admits a unique minimizer if and only if $h_1^*$ solving \eqref{eq-defh1} is uniquely determined, which holds under the completeness condition in {\SUPP} \ref{sec:supp:exist bridge ft}. If \eqref{eq-defh1} has multiple solutions, $h_1^*$ can be chosen as the minimal norm solution, which remains uniquely determined despite the existence of multiple solutions \citep[Theorem 2.5]{Engl2000}. The required convergence condition for the PMMR estimator, discussed in the following section, is then considered relative to this minimal norm solution. We refer readers to \citet{Florens2011} for a similar argument and to {\SUPP} \ref{sec:supp:PMMR details} for technical details. Accordingly, we reasonably may assume throughout that $h_1^*$ is a well-defined and unique function among the set of solutions to \eqref{eq-defh1}.

An estimator of $h_1^*$ can be obtained from the following Tikhonov-regularized empirical analogue of the population-level minimization. 
\begin{align}
&\! 
\widehat{h}_1\LSS (\bW,\obX_{\TIME})
=
\argmin_{h \in \HH(\bW, \obX_{\TIME}) }
\Big\{
\widehat{R} \LSS(h)
+
\lambda  \big\| h \big\|_{\HH}^2
\Big\} \ , 
\label{eq-ERM-mainpaper}
\\
&
\!
\widehat{R}  \LSS (h)
=
\frac{1}{ M_k^2 }
\sum_{i,j \in \mathcal{I}_k^c}
\!
 \big[ A_{i}A_{j}(1-D_{\TIME+1,i})(1-D_{\TIME+1,j})
 \epsilon_{i}(h)
 \epsilon_{j}(h)
 \mathcal{K} \big( (\bZ_i,\obX_{\TIME,i}) , (\bZ_j,\obX_{\TIME,j}) \big)
 \big] 
\ .
\label{eq-risk-mainpaper}
\end{align}
Here, $\lambda \in (0,\infty)$ is a regularization parameter, $\| \cdot \|_{\mathcal{H}}$ is the RKHS norm, $M_k = | \mathcal{I}_k^c |$ is the size of estimation fold, and $\epsilon_{i}(h) = Y_i - h (\bW_i,\obX_{\TIME,i})$. The superscript $^{(-k)}$ indicates that the empirical risk and the corresponding estimator are obtained only based on the estimation fold $\mathcal{I}_k^c$. In words, $\widehat{h}_1\LSS$ is the regularized minimizer of the empirical risk $\widehat{R} \LSS $, which is a kind of V-statistic \citep{Serfling2009}. Based on the representer theorem \citep{KW1970, SHS2001}, the estimated outcome \CBF\ can be written as $\widehat{h}_1\LSS (\bw,\bx)
=
\sum_{i \in \mathcal{I}_k^c}
\widehat{\alpha}_{i} \mathcal{K} \big( (\bW_i, \obX_{\TIME,i}), (\bw,\bx) \big)$ 
where the coefficient $\widehat{\alpha} = (\widehat{\alpha}_{1},\ldots,\widehat{\alpha}_{M_k} )\T$ has a closed-form representation. In {\SUPP} \ref{sec:supp:PMMR details}, we discuss why $\widehat{R} \LSS $ is an empirical analogue of the risk function $R$ and provide a closed-form representation of $\widehat{\alpha}$.

The PMMR approach requires three hyperparameters, the bandwidth parameters of the Gaussian kernel $\mathcal{K}$ associated with RKHSes $\HH(\bZ, \obX_{\TIME})$ and $\HH(\bW, \obX_{\TIME})$, and the regularization parameter $\lambda$. The choice of these hyperparameters can affect the finite-sample performance of the \CBF\ estimator. We select the hyperparameters based on a repeated cross-validation procedure and the median heuristic \citep{Garreau2018}; see {\SUPP} \ref{sec:supp:practical} for details. Briefly, for $\widehat{h}_1 \LSS$ in \eqref{eq-ERM-mainpaper}, the bandwidth parameter for $h \in \HH(\bW, \obX_{\TIME})$ and the regularization parameter $\lambda$ are selected using repeated cross-validation. In each repetition, a different random split of the data is used, and the final parameters are chosen based on their performance across all repetitions. The bandwidth parameter of $\HH(\bZ, \obX_{\TIME})$ (which corresponds to $\mathcal{K} ( (\bZ_i,\obX_{\TIME,i}) , (\bZ_j,\obX_{\TIME,j}) )$ in \eqref{eq-risk-mainpaper}) is selected from the median heuristic. Notably, we recommend fixing the bandwidth of $\HH(\bZ, \obX_{\TIME})$, as it determines the scale of the risk function $R(h)$ in \eqref{eq-Risk}. If it is subject to selection, lower risk values may not necessarily indicate better performance, which could result in the usual cross-validation criterion being invalid. Although no established theory exists, we assess the effectiveness of the proposed cross-validation procedure through a simulation study, which demonstrates its reasonable performance; see {\SUPP} \ref{sec:supp:Simulation CV} for details.
 
Note that $h_1^*$ and $h_0^*(\cdot \con a_Y=1)$ are connected via \eqref{eq-defh0}, and $q_0^*$ and $q_1^*$ are connected via \eqref{eq-defq1}. It would be desirable for the PMMR estimators to also satisfy the finite sample version of the empirical counterparts of these nested relationships within the given dataset. However, unlike parametric estimators, constructing nonparametric estimators that respect these relationships is generally challenging due to the regularization applied during the estimation of each nuisance function. Nevertheless, the limiting \CBFs\ of the nonparametric estimators satisfy relationships \eqref{eq-defh1}-\eqref{eq-defq1} under certain conditions; see {\SUPP} \ref{sec:supp:PMMR details} for details.

\subsubsection{Convergence Rates of the Confounding Bridge Function Estimators} \label{sec:convergence:CBF}

We now establish the convergence rate of $\widehat{h}_1\LSS$ introduced earlier; convergence rates for the other PMMR estimators are presented in {\SUPP} \ref{sec:supp:PMMR details}. In essence, the convergence rate depends on the degree of ill-posedness of the integral equation \eqref{eq-defh1}, which is characterized by the so-called $\beta$-source condition \citep{Bennett2023, Rotnizky2025}. Formally, let $\mathcal{S}_1 : \HH(\bW,\obX_{\TIME}) \rightarrow \HH(\bZ,\obX_{\TIME})$ be the restriction of $\mathcal{T}_1$, the conditional expectation mapping defined in \HL{S1}, to the RKHSes, and let $\mathcal{S}_1^{\star}: \HH(\bZ,\obX_{\TIME}) \rightarrow \HH(\bW, \obX_{\TIME})$ be its adjoint operator. For $\beta>0$, the $\beta$-source condition posits that $h_1^*$, the minimal norm solution to  \eqref{eq-defh1},  lies in the range of the fractional power operator $(\mathcal{S}_1^\star \mathcal{S}_1)^{\beta/2}$, i.e., $h_1^* \in \text{Range} ( (\mathcal{S}_1^\star \mathcal{S}_1)^{\beta/2} )$; see \cite{Plato2025} and references therein for the definition of the fractional powers of the operator. Roughly speaking, the source measures how strongly $h_1^*$ aligns with the well-identified directions of the inverse problem, namely those along which $(\mathcal{S}_1^\star \mathcal{S}_1)$ exhibits the greatest stability. In {\SUPP} \ref{sec:supp:PMMR details}, we further interpret the source condition in the case where $\mathcal{S}_1$ is compact; see also \citet{Carrasco2007}, \citet{Bennett2023}, and \citet{Rotnizky2025} for comprehensive discussions of this condition.

Theorem \ref{thm:convergence:1} establishes the convergence rate of $\widehat{h}_1\LSS$ under the source condition.
\begin{theorem} \label{thm:convergence:1}
Suppose that the observed data $\bO$ have compact support and a uniformly bounded density. Further suppose that there exists $\beta_{h1}>0$ so that $h_1^* \in \text{Range} ( (\mathcal{S}_1^\star \mathcal{S}_1)^{\beta_{h1}/2} )$. If the regularization parameter $\lambda_{h1}$ in \eqref{eq-ERM-mainpaper} is chosen as $\lambda_{h1}  = N^{-\max( 1/(1+\beta_{h1}), 1/3 )}$, the PMMR estimator $\widehat{h}_1\LSS$ achieves a convergence rate of 
\begin{align*}
	\big\| \widehat{h}_1\LSS - h_1^* \big\|_{P,2} 
	=
	\Bigg\{
	\begin{array}{ll}	
	O_P( N^{-\beta_{h1}/(2+2\beta_{h1})} )
	&
	\quad
	\text{ if $\beta_{h1} \in (0,2]$}
	\\
	O_P( N^{-1/3} )
	&
	\quad
	\text{ if $\beta_{h1} \in (2,\infty)$} 
	\end{array}
	\ .
\end{align*}  
\end{theorem}

To the best of our knowledge, the convergence rate established in Theorem \ref{thm:convergence:1} provides the sharpest available bound for the $L_2(P)$-error of the PMMR estimator, a view also supported by \citet{Rotnizky2025}. In contrast, the original PMMR framework of \citet{PMMR2021} derived a convergence bound for $\widehat{h}_1\LSS$ that is at most $O_P(N^{-1/4})$, even in mildly ill-posed settings (i.e., when $\beta_{h1}$ is sufficiently large). This rate is overly conservative, as it fails to guarantee the asymptotic normality of the CSE estimator introduced in the next section, even in mildly ill-posed settings. By comparison, the sharper rate established in Theorem \ref{thm:convergence:1} ensures the asymptotic normality of the CSE estimator in such cases. Hence, our refinement of \citet{PMMR2021} is crucial not only for improving the convergence rate but also for establishing the asymptotic normality of the CSE estimator, the ultimate goal of interest.

We also establish the convergence rates of the PMMR estimators for the nested \CBFs\ (i.e., $h_0^*(\cdot \con a_Y=1)$ in \eqref{eq-defh0} and $q_1^*$ in \eqref{eq-defq1}) under relative source conditions. Compared with the other \CBFs, these estimators present an additional technical challenge due to the nested structure of the estimation procedure. Consequently, very few existing studies investigate convergence rates for nonparametric estimators derived from nested ill-posed integral equations; a notable exception is \citet{Meza2021}. However, their analysis does not rely on the PMMR approach and instead employs the alternative methods introduced in Section \ref{sec:PMMR Estimator}. Thus, our work appears to be the first to consider the nested PMMR estimator and to rigorously characterize its convergence rate. Further details are provided in {\SUPP} \ref{sec:supp:PMMR details}.


\subsubsection{Estimation and Inference of the Conditional Separable Effects}

Once all \CBFs\ are estimated, we evaluate the influence function in Theorem \ref{thm:IF} for the observations over the evaluation fold $\mathcal{I}_k$ for each $k$. Specifically, let $\widehat{\InfFt}_{\NUMER} \LSS$ and $\widehat{\InfFt}_{\DENOM} \LSS$ be the estimated influence function where the true \CBFs\ $\{ h_0^*(\cdot \con a_Y=1),h_0^*(\cdot \con a_Y=0),h_1^*,h_2^*,q_0^*,q_1^*\}$ in \eqref{eq-IF_N1}-\eqref{eq-IF_D} are substituted by their estimated counterparts $\{ \widehat{h}_0\LSS(\cdot \con a_Y=1),\widehat{h}_0\LSS(\cdot \con a_Y=0),\widehat{h}_1\LSS,\widehat{h}_2\LSS,\widehat{q}_0\LSS,\widehat{q}_1\LSS\}$.  
By averaging the estimated influence functions over the evaluation folds, one obtains the estimator of the CSE, given by  
\begin{align*}
&
\widehat{\tau}_{\CSE}
=
\frac{
N^{-1}
\sum_{k=1}^{K}
\sum_{i \in \mathcal{I}_k}
\big\{ \widehat{\InfFt}_{\NUMER} \LSS (\bO_i  \con a_Y=1)
-
\widehat{\InfFt}_{\NUMER} \LSS (\bO_i  \con a_Y=0)
\big\}
}{
N^{-1}
\sum_{k=1}^{K}
\sum_{i \in \mathcal{I}_k}
\big\{ \widehat{\InfFt}_{\DENOM} \LSS (\bO_i)
\big\}
} \ .
\end{align*}

The asymptotic normality of the estimator $\widehat{\tau}_{\CSE}$ follows under additional regularity conditions given below. Consider the following assumptions for the true \CBFs\ and their corresponding estimators:
\begin{itemize}
\item[\HT{A8}] (\textit{Boundedness}) For $a_Y \in \{0,1\}$, there exists a finite constant $C > 0$ such that 
$\big\| \EXP \big( Y^2 \cond \bZ,A=a_Y,D_{\TIME+1}=0,\obX_{\TIME} \big) \big\|_{\infty}$, 
$\big\| h_0^*(\cdot \con a_Y) \big\|_{\infty} $, 
$\big\| h_1^* \big\|_{\infty} $, 
$\big\| h_2^* \big\|_{\infty} $, 
$\big\| q_0^* \big\|_{\infty} $,  
$\big\| q_1^* \big\|_{\infty} $ are bounded above by $C$ a.s.
Also, for all $k \in \{1,\ldots,K\}$ and $a_Y \in \{0,1\}$, 
$\big\| \widehat{h}\LSS_0(\cdot \con a_Y) \big\|_{\infty} $, 
$\big\| \widehat{h}\LSS_1 \big\|_{\infty} $, 
$\big\| \widehat{h}\LSS_2 \big\|_{\infty} $, 
$\big\| \widehat{q}\LSS_0 \big\|_{\infty} $,  
$\big\| \widehat{q}\LSS_1 \big\|_{\infty} $ are bounded above by $C$ a.s.

\item[\HT{A9}] (\textit{Consistency}) For all $k \in \{1,\ldots,K\}$ and $a_Y  \in \{0,1\}$, 
$\big\| \widehat{h}\LSS_0(\cdot \con a_Y) - h_0^*(\cdot \con a_Y) \big\|_{P,2}$,
$\big\| \widehat{h}\LSS_1 - h_1^* \big\|_{P,2}$,
$\big\| \widehat{h}\LSS_2 - h_2^* \big\|_{P,2}$,
$\big\| \widehat{q}\LSS_0 - q_0^* \big\|_{P,2}$,  
$\big\| \widehat{q}\LSS_1 - q_1^* \big\|_{P,2}$ are $o_P(1)$.

\item[\HT{A10}] (\textit{Cross-product Rates for the Numerator}) For all $k \in \{1,\ldots,K\}$ and $a_Y \in \{0,1\}$, 
$\big\| \widehat{h}\LSS_1 - h_1^* \big\|_{P,2}
\times
\big\| \widehat{q}\LSS_1 - q_1^* \big\|_{P,2}$, 
$\big\| \widehat{h}\LSS_1 - h_1^*  \big\|_{P,2}
\times
\big\| \widehat{q}\LSS_0 -  q_0^* \big\|_{P,2}$,  
$\big\| \widehat{h}\LSS_0(\cdot \con a_Y) - h_0^*(\cdot \con a_Y) \big\|_{P,2}
\times
\big\| \widehat{q}\LSS_0 - q_0^* \big\|_{P,2}$ are $o_P(N^{-1/2})$.

\item[\HT{A11}] (\textit{Cross-product Rates for the Denominator}) For all $k \in \{1,\ldots,K\}$,  $\big\| \widehat{h}\LSS_2 - h_2^* \big\|_{P,2}
\times
\big\| \widehat{q}\LSS_0 - q_0^* \big\|_{P,2}$ is $o_P(N^{-1/2})$.

\end{itemize}
\HL{A8} implies that the conditional second moment of the outcome given $(\bZ,A=a_Y,D_{\TIME+1}=0,\obX_{\TIME})$, the true \CBFs, and the estimated \CBFs\ are uniformly bounded. 
\HL{A9} states that the estimated \CBFs\ are consistent for the true \CBFs\ in the $L_2(P)$-norm. 
\HL{A10} means that some, but not necessarily all, pairs of numerator-related \CBFs\ are estimated at sufficiently fast rates. Specifically, \HL{A10} would be satisfied, for instance, if at least one of the following conditions were satisfied: (i) both $h_1^*$ and $q_0^*$ are known; or (ii) both $h_0^*$ and $h_1^*$ are known; or (iii) both $q_0^*$ and $q_1^*$ are known. Therefore, \HL{A10} states the level of precision required in the estimation of (i) the pair of outcome and treatment \CBFs\ $h_1^*$ and $q_0^*$; or (ii) the pair of outcome \CBFs\ $h_1^*$ and $h_0^*$; or (iii) the pair of treatment \CBFs\ $q_0^*$ and $q_1^*$.  When $a_Y$ is equal to $a_D$, which we fix at $0$, \HL{A10} reduces to the last condition, i.e., $\big\| \widehat{h}\LSS_0(\cdot \con 0) - h_0^*(\cdot \con 0) \big\|_{P,2} \times  \big\| \widehat{q}\LSS_0 - q_0^* \big\|_{P,2} = o_P(N^{-1/2})$. \HL{A11} is similar to \HL{A10}, and states the required precision in the estimation of $h_2^*$ and $q_0^*$, a pair of outcome and treatment \CBFs\ that are related to the denominator. We remark that \HL{A10} and \HL{A11} are instances of the mixed-bias property described by \citet{Rotnitzky2020}, and are closely related to the conditions given in \citet{Cui2023} and \citet{Dukes2023_ProxMed}; see {\SUPP} \ref{sec:supp:others} for details. A sufficient condition for \HL{A10} and \HL{A11} is that all error terms listed in \HL{A9} are $o_P(N^{-1/4})$, a convergence rate which is attainable provided the true \CBFs\ are sufficiently smooth and the integral equations are only mildly ill-posed \citep{ChenChristensen2018}. For the PMMR estimators, \HL{A9}-\HL{A11} hold under the mildly ill-posed settings, and we evaluate whether the proposed PMMR approach satisfies these conditions through simulation studies; see {\SUPP}s \ref{sec:supp:PMMR details} and \ref{sec:supp:Simulation CV}, respectively.

Under these regularity conditions, we establish the asymptotic normality of  $\widehat{\tau}_{\CSE} $ and provide a consistent variance estimator; Theorem \ref{thm:AN} formally states the result:
\begin{theorem} \label{thm:AN}
Suppose that there exist \CBFs\ $h_0^*$, $h_1^*$, $h_2^*$, $q_0^*$, $q_1^*$ satisfying \eqref{eq-defh1}-\eqref{eq-defq1}, and that Assumptions \HL{A1}-\HL{A11} hold. Then, we have that $    \sqrt{N}
\big(
\widehat{\tau}_{\CSE} 
-
\tau_{\CSE}^*  
\big)
\stackrel{D}{\rightarrow}
N \big( 0 , \sigma_{\CSE}^{*2}   \big)$
where $\sigma_{\CSE}^{*2} = \VAR \{ \InfFt_{\CSE}^*(\bO) \}$. Moreover, a consistent estimator of the asymptotic variance of $\widehat{\tau}_{\CSE}$ is given by
\begin{align*}
& 
\widehat{\sigma}_{\CSE}^2
=
\frac{1}{N}
\sum_{k=1}^{K}
\sum_{i \in \mathcal{I}_k}
\bigg\{
\frac{
\widehat{\InfFt}_{\NUMER}\LSS(\bO_i \con a_Y=1)
-
\widehat{\InfFt}_{\NUMER}\LSS(\bO_i \con a_Y=0 ) 
-
\widehat{\tau}_{\CSE} 
\widehat{\InfFt}_{\DENOM}\LSS(\bO_i )}{ 
N^{-1}
\sum_{k=1}^{K}
\sum_{i \in \mathcal{I}_k}
\widehat{\InfFt}_{\DENOM} \LSS (\bO_i) 
} 
\bigg\}^2 \ .
\end{align*}
\end{theorem}
\noindent 
Using the variance estimator $\widehat{\sigma}_{\CSE}^2 $, Wald confidence intervals for $\tau_{\CSE}^* $ can be constructed. 
 Alternatively, one may construct confidence intervals using the multiplier bootstrap \citep[Chapter 2.9]{VW1996}; see {\SUPP} \ref{sec:supp:practical} for details.

We conclude the section by briefly discussing the important setting in which the treatment in the two-arm study is randomized. By virtue of randomization, there is no confounding of the association between $A$ and $(Y, D_{\stime})$. However, it is still possible that unmeasured confounders exist in the association between $Y$ and $D_{\stime}$, making the approach in \citet{Stensrud2022CSE} not applicable even in this experimental setting. In {\SUPP} \ref{sec: exp setting main}, we extend the proposed approach to experimental settings to establish identification and estimation of the CSE.

\section{Simulation}		\label{sec:Simulation}

We conducted simulation studies to investigate the finite-sample performance of the proposed estimators for both observational and experimental settings with $\TIME=1$. Specifically, we generated an unmeasured confounder $U \in \R$, proxy variables $\bW \in \R$ and $\bZ \in \R$, pre-treatment confounders $\bX_{0} \in \R^{5}$, and a treatment status $A \in \{0,1\}$. We then generated post-treatment time-varying confounders $\bX_{\TIME} \in \R^{5}$. Next, we generated a post-treatment event $D_{\TIME+1} \in \{0,1\}$ and the outcome variable $Y \in \R$. All variables, except for $A$ and $D_{\TIME+1}$, are continuously distributed. Under the data-generating process, the \CBFs\ are uniquely defined and available in closed-form. The details of the data-generating process can be found in {\SUPP} \ref{sec:supp:Simulation CV}. We focused on the CSE under $a_D=0$, which is $\tau_{\CSE}^*=2$.

We considered the number of study units $N \in \{500,1000,1500,2000\}$. The proposed estimator, $\widehat{\tau}_{\CSE}$, was obtained following the approach described in Section \ref{sec:inference}. We used the Gaussian kernel and selected hyperparameters based on 5-fold cross-validation with 5 repetitions and the median heuristic. For inference, we constructed 95\% confidence intervals using the proposed consistent variance estimator and those obtained via the multiplier bootstrap; implementation details are provided in {\SUPP} \ref{sec:supp:practical}. We evaluated the performance of the proposed estimator and the corresponding confidence intervals based on 1000 repetitions for each value of $N$. As a competing estimator, we considered an estimator based on the efficient influence function under the assumption of no unmeasured confounding where $\bZ$ and $\bW$ are treated as measured confounders. The details of this estimator, denoted by $\widetilde{\tau}_{\text{CSE}}$, are described in {\SUPP} \ref{sec:supp:others}.

\begin{figure}[!htb]
\centering
\vspace*{-0.1cm}
\includegraphics[width=0.95\textwidth]{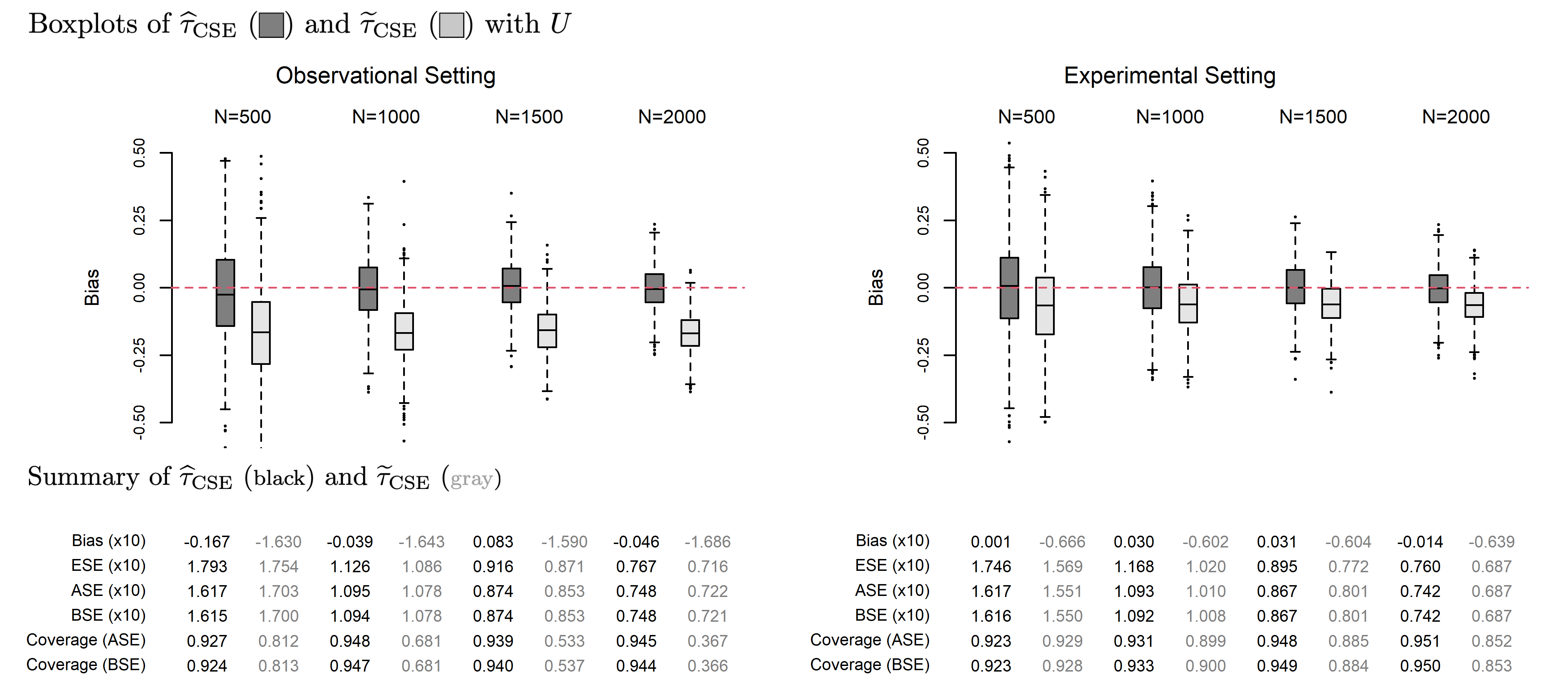}
\vspace*{-0.4cm}
\caption{\footnotesize A Graphical Summary of the Simulation Results. The left and right panels show results for observational and experimental settings, respectively. In the top panel, each column gives boxplots of biases of $\widehat{\tau}_{\CSE}$ and $\widetilde{\tau}_{\CSE}$ for  $N\in \{500,1000,1500,2000\}$, respectively. The bottom panel provides numerical summaries informing the performance of $\widehat{\tau}_{\CSE}$. Each row represents the empirical bias, empirical standard error (ESE), asymptotic standard error based on the proposed variance estimator (ASE), bootstrap standard error (BSE), empirical coverage rates of 95\% confidence intervals based on the ASE and BSE. The bias and standard errors are scaled by factors of 10.}
\vspace*{-0.5cm}
\label{fig:Simulation}
\end{figure}		

The top panel of Figure \ref{fig:Simulation} graphically summarizes the empirical distribution of the estimators. In both settings, we find that the proposed estimator $\widehat{\tau}_{\text{CSE}}$ exhibits negligible bias for all $N$, and its variability decreases as $N$ increases. On the other hand, the competing estimator $\widetilde{\tau}_{\text{CSE}}$ appears to be biased, as it does not account for the possibility of unmeasured confounding. Numerical summaries of $\widehat{\tau}_{\text{CSE}}$ and $\widetilde{\tau}_{\text{CSE}}$ are provided in the bottom panel of Figure \ref{fig:Simulation}. For $\widehat{\tau}_{\text{CSE}}$, we reconfirm that all three standard errors decrease as $N$ increases, and their values are similar. Empirical coverage rates of both proposed confidence intervals appear to attain the nominal coverage of 95\%. On the other hand, $\widetilde{\tau}_{\text{CSE}}$ exhibits significant bias, resulting in a failure to achieve nominal coverage; however, it achieves a smaller standard error compared to $\widehat{\tau}_{\text{CSE}}$. The simulation results suggest that the performance of the proposed estimator $\widehat{\tau}_{\text{CSE}}$ aligns with the asymptotic properties established in Section \ref{sec:estimator} and {\SUPP} \ref{sec: exp setting main}.

In {\SUPP} \ref{sec:supp:Simulation CV}, we present additional simulation results. First, we compare the performance of $\widehat{\tau}_{\CSE}$ and $\widetilde{\tau}_{\CSE}$ under no unmeasured confounding by setting $U=0$ in the data-generating process. While both estimators remain consistent and achieve nominal coverage, $\widetilde{\tau}_{\CSE}$ is more efficient, as expected, given that it is a semiparametric efficient estimator in this setting. Second, we evaluate the estimation error of the six nuisance functions in \eqref{eq-defh1}-\eqref{eq-defq1} to empirically assess the hyperparameter selection procedure proposed in {\SUPP} \ref{sec:supp:practical}. As the training data size increases, the nuisance function estimation errors decrease, with convergence rates appearing to satisfy Assumptions \HL{A9}-\HL{A11}, thus supporting validity of the proposed procedure.

\section{Application: The Southwest Oncology Group Trial}		\label{sec:Application}

We reanalyzed data from the SWOG Trial \citep{SWOG2004}, which has been used to study the effect of chemotherapy on health-related quality of life \citep{Ding2011, Wang2017, YangDing2018, Stensrud2022CSE}. We used data from 487 male patients with refractory prostate cancer who were randomly assigned to one of two chemotherapy regimes: Estramustine and Docetaxel (ED) or Prednisone and Mitoxantrone (PM), which are denoted by $A=\text{ED}$ and $A=\text{PM}$, respectively. Before being assigned to cancer chemotherapy, the following six variables were measured for each patient: type of progression, degree of bone pain, SWOG performance measure, race, age, and quality of life. Cancer progression was measured during a 12-month follow-up period. Quality of life was measured again at the end of this period; however, this follow-up assessment was only available for patients who survived through the study duration. The final time is denoted by $\TIME+1=12$ hereafter. The outcome $Y$ is defined as change in quality of life between the baseline and follow-up period.


Following \citet{Stensrud2022CSE}, we focused on evaluating a CSE of ED versus PM on change in quality of life, with specific details outlined below. First, let $A_Y=\text{E}$ and $A_Y=\text{P}$ indicate receiving only the Estramustine component of ED and the Prednisone component of PM, respectively. As primarily palliative and pain relief medications, there is no clear evidence that Estramustine and Prednisone offer any survival benefit. Therefore, it is reasonable to assume that these chemotherapeutic agents do not have a direct effect on mortality. Estramustine and Prednisone might also induce side effects like nausea, fatigue, and vomiting \citep{SWOG2004}. 
The pain relief properties and the potential for side effects can ultimately influence an individual's overall quality of life.  We use this argument to justify \HL{A4}, $A_Y$-partial isolation, which is not guaranteed by the experimental design. 

Second, let $A_D=\text{D}$ and $A_D=\text{M}$ indicate receiving only the Docetaxel component of ED and the Mitoxantrone component of PM, respectively. This choice was motivated by the fact that Docetaxel and Mitoxantrone are chemotherapeutic agents that can inhibit cancer cell proliferation and reduce the risk of death during the study period. However, it is plausible that these drugs affect quality of life through cancer progression. If this is the case, it implies the existence of a causal relationship between $A_D$ and the outcome $Y$, thus violating \hyperlink{$A_D$-Partial Isolation}{($A_D$-Partial Isolation)}. Under such circumstances, the CSE defined below can be viewed as a direct effect, but it may not capture all effects of $A$ on $Y$. 

Assuming \HL{A4}, \citet{Stensrud2022CSE} focused on the following CSE:
\begin{align*}
\tau_{\CSE}^*(a_D=\text{D})
&
=
\psi^*(a_Y=\text{P}, \, a_D=\text{D})
-
\psi^*(a_Y=\text{E}, \, a_D=\text{D})
\\
&
=
\EXP \big\{ \potY{a_Y=\text{P}, \, a_D=\text{D}} - \potY{a_Y=\text{E}, \, a_D=\text{D}} \cond \potDt{a_D=\text{D}}{\TIME+1} = 0 \big\} \ . 
\end{align*}
In words, $(a_Y=\text{P}, \, a_D=\text{D})$ is a combination of the component of PM that may directly affect quality of life (i.e., Prednisone) and that of ED that affects mortality (i.e., Docetaxel). Therefore, $\tau_{\CSE}^*(\text{D})$ quantifies the causal effect of receiving Prednisone over Estramustine among patients who would survive had they received Docetaxel as part of their treatment. For identification and estimation of $\tau_{\CSE}^*(\text{D})$, \citet{Stensrud2022CSE} relied on the assumption that there is no unmeasured common cause of survival and change in quality of life. However, it is important to acknowledge the potential presence of such unmeasured variables, including medical history, chronic diseases, lifestyle factors (e.g., smoking, drinking), disease progression, and other clinical complications.

In order to address the potential for such unmeasured confounding, we estimated $\tau_{\CSE}^*(\text{D})$ using the proposed approach. Specifically, as chemotherapies were randomly assigned, we implemented the version of the proposed methods tailored for experimental settings, detailed in {\SUPP} \ref{sec: exp setting main}. For proxy variables $\bW$ and $\bZ$, we used baseline quality of life and age, respectively. The remaining four pre-treatment covariates (type of progression, degree of bone pain, SWOG performance measure, race) were used as pre-treatment confounder $\bX_{0}$, while the indicator of cancer progression within the 12-month study period was used as a time-varying confounder $\bX_{\TIME}$. The choice of $\bZ$ was motivated by the hypothesis that age may affect mortality but may not directly affect change in quality of life conditional on $\bX_{0}$ and potential unmeasured confounders $\bU$. Likewise, the choice of $\bW$ was based on the rationale that baseline quality of life likely impacts change in quality of life but is unlikely to impact mortality conditional on $\bX_{0}$ and $\bU$. Test statistics for pairwise partial correlations between $\bW$ and $\bZ$ given $\bX_{0}$ indicated that they are strongly associated, consistent with the presence of $U$; see {\SUPP} \ref{sec:supp:Data} for details. As in the simulation study in Section \ref{sec:Simulation}, we used the Gaussian kernel and selected hyperparameters based on 5-fold cross-validation with 5 repetitions and the median heuristic. To mitigate the impact of a particular random split on the effect estimate, we adopted the median adjustment by repeating cross-fitting 500 times, see {\SUPP} \ref{sec:supp:practical} for details. Lastly, for comparison, we also estimated $\tau_{\CSE}^*(\text{D})$ under the assumption of no unmeasured common causes of death and change in quality of life, following \citet{Stensrud2022CSE}. 

At the 12-month follow-up, some surviving patients had missing quality of life. For these patients, we imputed the missing outcomes of the survived patients by Multivariate Imputation by Chained Equations (MICE) algorithm implemented in \texttt{mice} R-package \citep{MICE2011}. We note that MICE requires the missing at random assumption, and the results presented below remain consistent even when the analysis is restricted to patients who either died during the study period or survived with complete quality of life data at follow-up; see {\SUPP} \ref{sec:supp:Data} for details.

Table \ref{tab:Data} summarizes the results. First, the estimate of $\psi^*(\text{P,D})$, the expected change in quality of life of receiving Prednisone among patients who would survive had they received Docetaxel as part of their treatment, accounting for $\bU$ was $-6.72$, which is larger than the estimate obtained by assuming no $\bU$, $-5.49$. On the other hand, the estimate of ${\psi} ^*(\text{E,D})$, the expected change in quality of life of receiving Estramustine among patients who would survive had they received Docetaxel as part of their treatment, is equal to $-4.81$ in both approaches, because it can be estimated simply by the sample mean due to randomization. Furthermore, the estimate of $\tau_{\CSE}^*(\text{D})$ under the proximal approach, $-1.91$, is larger than that assuming no $\bU$, $-0.68$. At the 5\% level, both approaches conclude that ${\psi}^*(\text{P,D})$ and ${\psi}^*(\text{E,D})$ are significantly different from 0 whereas $\tau_{\CSE}^*(\text{D})$ is not significantly different from 0. Therefore, it appears that there is no significant difference between the effects of Prednisone and Estramustine on change in quality of life in the subgroup of patients who would have survived with Docetaxel treatment, even after accounting for the potential for unmeasured common causes of death and quality of life. This conclusion agrees with the findings presented in \citet{Stensrud2022CSE}.

\begin{table}[!htp]
\renewcommand{\arraystretch}{1.01} 
\fontsize{8}{11}\selectfont
\centering
\setlength{\tabcolsep}{5pt}
\begin{tabular}{|c|c|c|c|c|}
\hline
\multirow{2}{*}{Allow for $\bU$} & \multirow{2}{*}{Statistic} & \multicolumn{3}{c|}{Estimand}                      \\ \cline{3-5} 
& & \multicolumn{1}{c|}{$\psi^*(\text{P,D} )$} & \multicolumn{1}{c|}{$\psi^*(\text{E,D} )$} & $\tau_{\CSE}^* (\text{D} )$ \\ \hline
\multirow{3}{*}{Yes}  &  Estimate  &            -6.72  &           -4.81  &          -1.91 \\ \cline{2-5} 
&       ASE  &             1.88  &            1.37  &           2.20 \\ \cline{2-5} 
&  95\% CI  &  (-10.41, -3.03)  &  (-7.50, -2.12)  &  (-6.22, 2.40)       \\ \hline
\multirow{3}{*}{No}  &  Estimate  &            -5.49  &           -4.81  &          -0.68 \\ \cline{2-5} 
&       ASE  &             1.73  &            1.37  &           1.98 \\ \cline{2-5} 
&  95\% CI  &  (-8.89, -2.09)  &  (-7.50, -2.12)  &  (-4.57, 3.21)       \\ \hline
\end{tabular}
\caption{\footnotesize Summary Statistics of the Estimation of the CSE $\tau_{\CSE}^*(\text{D})$. Statistics in ``Allow for $\bU$=Yes'' and ``Allow for $\bU$=No'' rows are obtained from the proximal approach proposed in {\SUPP} \ref{sec: exp setting main} and from the approach proposed in {\SUPP} \ref{sec:supp:others} developed under unconfoundedness (i.e., without $\bU$), respectively. ASE and 95\% CI stand for the asymptotic standard error obtained from the proposed consistent variance estimator and the corresponding 95\% confidence intervals, respectively.}
\vspace*{-0.5cm}
\label{tab:Data}
\end{table}

\section{Concluding Remarks} \label{sec:Conclusion}

We have proposed a proximal causal inference framework for identifying and estimating the CSE in the presence of unmeasured confounding. Using the influence function for the CSE under model $\mathcal{M}$ given in \eqref{eq-IF CSE}, we constructed a semiparametric estimator of the CSE with all nuisance components estimated nonparametrically. We established that the estimator exhibits a mixed-bias structure in the sense that the estimator is consistent and asymptotically normal if some, but not necessarily all, nuisance components are estimated at sufficiently fast rates. We demonstrated the theoretical properties of the estimator in simulation studies, and then applied our method to a randomized trial to evaluate the effect of cancer chemotherapies on change in quality of life.


While we have focused on the CSE, our results can be extended to the principal stratum effect, such as the survivor average causal effect, under additional assumptions. For instance, if $\potDt{a=1}{\stime} \leq \potDt{a=0}{\stime}$ for all individuals (i.e., monotonicity), 
the CSE is equal to the survivor average causal effect, a result shown by \citet{Stensrud2022CSE}; see {\SUPP} \ref{sec:supp:others} for details. Therefore, our identification and estimation results are applicable for principal stratum effects in the presence of unmeasured confounding, assuming monotonicity and other assumptions, some of which can be tested in a future randomized experiment. This approach sharply contrasts with those relying on assumptions that cannot be verified even in any randomized experiment \citep{RobinsRichardson2011}, like the principal ignorability condition, which rules out the possibility of unmeasured confounding between $D_{\stime}$ and $Y$.

The proposed approach hinges on the plausibility of separable treatments, which, e.g., could be decompositions of the original study treatment, that exert particular causal effects. In some studies, such as the SWOG trial, the treatment actually consists of two distinct components. Then the treatment decomposition seems to be concrete and plausible. However, in cases where the original treatment is not explicitly defined as a composite of two components, a justification for the decomposition is necessary. This justification should be guided by subject-matter expertise, as we generally want the estimand to correspond to practically relevant research questions. Furthermore, the exercise of articulating decompositions can itself be useful for sharpening research questions, enriching the understanding of the treatment mechanisms, and inspiring future treatment strategies \citep{RobinsRichardson2011, RRS2022, Stensrud2021, Stensrud2022CSE, Stensrud2022}. 

From a proximal causal inference perspective, several avenues for future research remain open. Many proximal methods, including ours, rely on multiple \CBFs, which yield distinct identifying formulas for the same functional (e.g., Theorem \ref{thm:identification}). A promising direction for future research is to develop a unified framework to characterize these relationships and design procedures that enable the potential falsification of identifying assumptions. Moreover, while we employed the PMMR approach to estimate the \CBFs\ and derived its convergence rate (see Section \ref{sec:PMMR Estimator}), other approaches may achieve faster convergence rates; see \citet{Rotnizky2025} for a detailed discussion. However, such alternatives typically require tuning two regularization parameters (see {\SUPP} \ref{sec:supp:minmax} for details), whereas PMMR involves only a single regularization parameter, making it simpler to implement. Despite this advantage, several theoretical aspects of PMMR remain unresolved. In particular, it is unclear whether the convergence rate in Theorem \ref{thm:convergence:1} can be improved or whether the proposed hyperparameter tuning procedure admits a formal theoretical justification. However, we note that these open questions are not specific to PMMR; they also arise for other estimators in proximal causal inference, as emphasized by \citet{Rotnizky2025}. We leave these issues for future research.

\section*{Acknowledgment}

The authors appreciate two anonymous reviewers and the editorial board for their valuable comments on this paper. 

\section*{Data Availability}

The replication code for both the simulation studies and the analyses of the Southwest Oncology Group (SWOG) trial dataset is publicly available at \url{https://github.com/qkrcks0218/PCSE}. Since the SWOG dataset is not publicly available, we provide replication code based on a simulated dataset that mirrors the structure of the original SWOG dataset.

\newpage  
\newpage

\appendix

\renewcommand{\theequation}{S.\arabic{equation}}
\setcounter{equation}{0}
 
\section*{Supplementary Material}

This document contains supplementary materials for ``Proximal Causal Inference for  Conditional Separable Effects.'' Section \ref{sec:supp:detail} presents additional results related to the main paper.
Section \ref{sec:supp:proof} presents proofs of the theorems from the main paper and from Section \ref{sec:supp:detail} of this document.


{\small
\begingroup
\makeatletter  
\babel@toc {english}{}\relax 
\contentsline {section}{\numberline {A}Details of the Main Paper}{27}{appendix.A}%
\contentsline {subsection}{\numberline {A.1}Relaxation for All $(a_Y,a_D)$ Cases}{27}{subsection.A.1}%
\contentsline {subsection}{\numberline {A.2}Details of the Representation of $\psi ^* (a_Y,a_D)$}{30}{subsection.A.2}%
\contentsline {subsection}{\numberline {A.3}Existence and Uniqueness of the Confounding Bridge Functions}{32}{subsection.A.3}%
\contentsline {subsection}{\numberline {A.4}Details of the PMMR Estimation}{34}{subsection.A.4}%
\contentsline {subsection}{\numberline {A.5}Alternative Estimation Approach}{40}{subsection.A.5}%
\contentsline {subsection}{\numberline {A.6}Practical Considerations: Hyperparameter Tuning, Median Adjustment, and Multiplier Bootstrap}{41}{subsection.A.6}%
\contentsline {subsection}{\numberline {A.7}Extension: An Experimental Setting}{43}{subsection.A.7}%
\contentsline {subsection}{\numberline {A.8}Comparison to Previous Work}{46}{subsection.A.8}%
\contentsline {subsection}{\numberline {A.9}Details of the Simulation Studies}{50}{subsection.A.9}%
\contentsline {subsubsection}{\numberline {A.9.3}Closed-form Representations of Confounding Bridge Functions}{53}{subsubsection.A.9.3}%
\contentsline {subsection}{\numberline {A.10}Details of the Data Analysis}{58}{subsection.A.10}%
\contentsline {section}{\numberline {B}Proof}{60}{appendix.B}%
\contentsline {subsection}{\numberline {B.1}Proof of Theorem \ref {thm:identification}}{60}{subsection.B.1}%
\contentsline {subsection}{\numberline {B.2}Proof of Theorem \ref {thm:IF}}{67}{subsection.B.2}%
\contentsline {subsection}{\numberline {B.3}Proof of Theorem \ref {thm:AN}}{72}{subsection.B.3}%
\contentsline {subsection}{\numberline {B.4}Proof of Theorem \ref {thm:convergence:1} and Theorem \ref {lemma:convergence:1}}{84}{subsection.B.4}%
\contentsline {subsection}{\numberline {B.5}Proof of Theorem \ref {lemma:convergence:2}}{87}{subsection.B.5}%
\contentsline {subsection}{\numberline {B.6}Proof of Theorem \ref {thm:identifcation rand}}{90}{subsection.B.6}%
\contentsline {subsection}{\numberline {B.7}Proof of Theorem {\ref {thm:IF Rand Supp}}}{91}{subsection.B.7}%
\contentsline {subsection}{\numberline {B.8}Proof of Theorem \ref {thm:AN rand}}{95}{subsection.B.8}%

\makeatother
\endgroup}%

\newpage

{\small
\section{Details of the Main Paper}	\label{sec:supp:detail}

In this Section, we present additional results related to the main paper.

\subsection{Relaxation for All $(a_Y,a_D)$ Cases}

In the Supplementary Material, we present general results by considering all combinations $(a_Y,a_D) \in \{(0,0),(0,1),(1,0),(1,1) \}$. 

\subsubsection{Notation}

In the general $(a_Y,a_D)$ case, we adjust the notations for the causal estimands and confounding bridge functions to explicitly reflect their dependence on $a_D$ as follows:
\begin{itemize}[leftmargin=0.25cm]
    \item Estimands: 
\begin{align*} 
&
\tau_{\CSE}^*(a_D)
=
\psi^*(1, a_D)
-
\psi^*(0,a_D)
\ , 
&&
\psi^*(a_Y,a_D) 
= 
\EXP \big\{ \potY{a_Y,a_D} \cond \potDt{a_Y,a_D}{\TIME+1} = 0 \big\}  \ . 
\end{align*}

\item Under the observational setting:

\begin{itemize}[leftmargin=0.25cm, itemsep=0cm, topsep=0cm, partopsep=0cm,  parsep=0cm]
    \item Under Assumptions \HL{A1}-\HL{A5}, we have $\psi^*(a_Y,a_D) 
=
\psi_{\NUMER}^*(a_Y,a_D)  / \psi_{\DENOM}^*(a_D) $ where
\begin{align} \label{eq-psi in supp}
&
\psi_{\NUMER}^*(a_Y,a_D)
=
\EXP \bigg\{
    \EXP \big( Y \cond A = a_Y, D_{\TIME+1} = 0 , \obX_{\TIME} , \bU  \big)
    \frac{
    \Pr \big( A=a_D, D_{\TIME+1} = 0 \cond \obX_{\TIME}, \bU \big)
    }{
    \Pr \big( A=a_D  \cond \bX_{0}, \bU \big)
    }
\bigg\}
&&
\text{if $a_Y \neq a_D$}
\ , 
\nonumber
\\
&
\psi_{\NUMER}^*(a_D,a_D)
=
\EXP \big\{
    \EXP \big( Y \cond A = a_D, D_{\TIME+1} = 0 , \bX_{0} , \bU  \big)
    \Pr \big( D_{\TIME+1} = 0 \cond A=a_D, \bX_{0}, \bU \big)
\big\}
\ ,
\nonumber
\\
&
\psi_{\DENOM}^*(a_D)
=
\EXP \big\{
\Pr \big( D_{\TIME+1} = 0 \cond A=a_D, \bX_{0}, \bU \big)
\big\}
\ .
&&
\end{align}
See Section \ref{sec:supp:psi obs} for details. 

\item We use the following confounding bridge functions:
\begin{align}
&
\EXP \big\{ Y  -  h_1^* (\bW,\obX_{\TIME} \con a_Y) \cond \bZ, A=a_Y, D_{\TIME+1}=0 , \obX_{\TIME} \big\}
=
0 \text{ a.s.}
\label{eq-bridgeft1 obs}
\\
&
\EXP \bigg\{ 
\begin{array}{l}
(1-D_{\TIME+1}) h_1^* (\bW,\obX_{\TIME} \con a_Y) \\
- h_0^* (\bW,\bX_{0} \con a_Y, a_D)
\end{array} \COND \bZ, A=a_D , \bX_{0} \bigg\}
=
0 \text{ a.s.}  
&&
\text{ if $a_Y \neq a_D$}
\label{eq-bridgeft2 obs}
\\
&
\EXP \bigg\{ 
\begin{array}{l}
(1-D_{\TIME+1}) Y
 \\
- h_0^* (\bW,\bX_{0} \con a_D, a_D)
\end{array} \COND \bZ, A=a_D , \bX_{0} \bigg\}
=
0 \text{ a.s.}  
\label{eq-bridgeft3 obs}
\\
&
\EXP \big\{  
( 1-D_{\TIME+1} ) - h_0^* (\bW,\bX_{0} \con a_D, a_D)
\cond \bZ, A=a_D , \bX_{0} \big\}
=
0 \text{ a.s.}  
\label{eq-bridgeft4 obs}
\\
&
\EXP \big\{ 1 - \ind(A=a_D) q_0^* (\bZ,\bX_{0} \con a_D) \cond \bW, \bX_{0} \big\}
=
0
\text{ a.s.}  
\label{eq-bridgeft5 obs}
\\
&
\EXP \bigg\{ 
\begin{array}{l}
\ind(A=a_D) q_0^* (\bZ,\bX_{0} \con a_D) \\ 
-  \ind(A=a_Y) q_1^* (\bZ,\obX_{\TIME} \con a_Y, a_D)
\end{array} 
\COND \bW, D_{\TIME+1}=0 , \obX_{\TIME} \bigg\}
=
0 
\text{ a.s.} &&
\text{ if $a_Y \neq a_D$}
\label{eq-bridgeft6 obs}
\end{align} 
In addition, we define $q_1^*(Z, \obX_{\TIME} \con a_D,a_D) = q_0^*(Z, \bX_{0} \con a_D)$.

\end{itemize}

\item Under the experimental setting (see Section \ref{sec: exp setting main} for details):

\begin{itemize}[leftmargin=0.25cm, itemsep=0cm, topsep=0cm, partopsep=0cm,  parsep=0cm]
    \item 
Under Assumptions \HL{A1}, \HL{A2'}, \HL{A3}-(i), \HL{A4}, \HL{A5}, $\psi_{\RD}^*(a_Y,a_D) 
=
\psi_{\RD,\NUMER}^*(a_Y,a_D)  / \psi_{\RD,\DENOM}^*(a_D) $ where
\begin{align}
&
\psi_{\RD,\NUMER}^*(a_Y,a_D)
=
\EXP \bigg\{
    \EXP \big( Y \cond A = a_Y, D_{\TIME+1} = 0 , \obX_{\TIME} , \bU  \big)
    \frac{
    \Pr \big( A=a_D, D_{\TIME+1} = 0 \cond \obX_{\TIME}, \bU \big)
    }{
    \Pr \big( A=a_D  \cond \bX_{0},\bU \big)
    }
\bigg\}
&&
\text{if $a_Y \neq a_D$}
\ , 
\nonumber
\\
&
\psi_{\RD,\NUMER}^*(a_D,a_D)
=
\EXP \big\{
    \ind(A=a_D)(1-D_{\TIME+1}) Y
\big\}  
\ ,
\nonumber
\\
&
\psi_{\RD,\DENOM}^*(a_D)
=
\EXP \big\{
\ind(A=a_D)(1-D_{\TIME+1})
\big\}
\ .
\label{eq-psi exp in supp}
\end{align}
See Section \ref{sec:supp:psi exp} for details.

\item We use the following confounding bridge functions:
\begin{align}	
&
\EXP \big\{ Y  - \xi_1^* (\bW,\obX_{\TIME} \con a_Y) \cond \bZ, A=a_Y, D_{\TIME+1}=0 , \obX_{\TIME} \big\} 
=
0
\label{eq-bridgeft1 exp}
\\
&
p_0^*(a_D) = 1/\Pr(A=a_D)
\label{eq-bridgeft4 exp}
\\
&
\EXP \bigg\{ 
\begin{array}{l}
\ind (A=a_D) p_0^*(a_D) \\
 - \ind(A=a_Y) p_1^* (\bZ,\obX_{\TIME} \con a_Y, a_D) 
\end{array} \COND \bW, D_{\TIME+1}=0 , \obX_{\TIME} \bigg\} 
&& 
\text{if $a_Y  \neq a_D$}
\label{eq-bridgeft5 exp}
\end{align}

\end{itemize}

\end{itemize}

\subsubsection{Assumptions}

For readability, we restate Assumptions \HL{A1}–\HL{A6} from the main paper in this Supplementary Material.

\begin{itemize}[leftmargin=0.0cm]
\item[\HL{A1}] (\textit{Consistency}) $Y = \potY{a=A}$, $\oD_{\TIME+1} = \oD_{\TIME+1}^{(a=A)}$, $ \obX_{\TIME} = \obX_{\TIME}^{(a=A)}$ a.s.

\item[\HL{A2}] (\textit{Latent Ignorability}) For $a \in \{ 0,1 \}$, $(\potY{a} , \uD_{1}^{(a)}, \underline{\bX}_{1}^{(a)} ) \indep A \cond (\bX_{0} , \bU)$.

\item[] In the experimental setting in Section \ref{sec: exp setting main}, we assume the following \HL{A2'} instead of \HL{A2}.

\item[\HL{A2'}] (\textit{Randomization})  $A  \indep (\potY{0}, \potY{1},\oD_{\TIME+1}^{(0)}, \oD_{\TIME+1}^{(1)},\obX_{\TIME}^{(0)},\obX_{\TIME}^{(1)},\bW,\bZ,\bU)$ where $\Pr(A=1) \in (0,1)$ is known.

\item[\HL{A3}] (\textit{Latent Positivity}) For $a \in \{ 0,1 \}$, (i) $\Pr(D_{\TIME+1}=0, \ubX_{1} \cond A=a,\bX_{0},\bU) > 0$ a.s. and (ii) $\Pr(A=a \cond \bX_{0},\bU)>0$ a.s.

\item[\HL{A4}] (\textit{$A_Y$-Partial Isolation}) For $a_D \in \{ 0,1 \}$, $\potDt{a_Y=1,a_D}{t+1} (G) = \potDt{a_Y=0,a_D}{t+1} (G)$ for all $t \in \{ 0,\ldots,\TIME \}$.

\item[] Of note, we introduce \hyperlink{$A_D$-Partial Isolation}{($A_D$-Partial Isolation)} below: 
\item \textit{\hyperlink{$A_D$-Partial Isolation}{($A_D$-Partial Isolation)}} \\
For $a_Y \in \{ 0,1 \}$,  $\potY{a_Y,a_D=1}(G) = \potY{a_Y,a_D=0}(G)$ if $\potDt{a_Y,a_D=1}{\TIME+1}(G) = \potDt{a_Y,a_D=0}{\TIME+1}(G) = 0$. 

In the presence of time-varying confounders, \HL{$A_D$-Partial Isolation} is unlikely to hold due to the causal path $A_D \rightarrow \bX_t \rightarrow Y$. 

\item[\HL{A5}] 
(\textit{Latent Dismissible Condition}) 
\begin{itemize} 
	\item[(i)]  $Y(G) \indep A_D(G) \cond  ( A_Y(G), D_{\TIME+1}(G) = 0, \overline{\bX}_{\TIME}(G) , \bU(G) ) $; 
    \item[(ii)]  $D_{\stime+1}(G) \indep A_Y(G) \cond ( A_D(G), D_{\stime}(G)=0, \overline{\bX}_{\stime} (G), \bU(G) )$ for all $t \in \{ 0,\ldots,\TIME \}$; 
    \item[(iii)]  $\bX_{\stime+1}(G) \indep A_Y(G) \cond ( A_D(G), D_{\stime+1}(G)=0, \overline{\bX}_{\stime} (G), \bU(G) )$ for all $t \in \{0,\ldots,\TIME \}$. 
	\end{itemize}

\item[\HL{A6}] (\textit{Proxies}) For $a \in \{0,1\}$, 
\begin{itemize}
	\item[(i)] \makebox[6cm][l]{$\bZ \indep \potY{a}  \cond  (A,\oD_{\TIME+1}^{(a)}, \obX_\TIME^{(a)}, \bU)$;} (ii) $\bZ \indep (\ubX_{1}^{(a)},\uD_{1}^{(a)}) \cond  (A, \bX_{0}, \bU)$; 
	\item[(iii)] \makebox[6cm][l]{$\bW \indep (\uD_{1}^{(a)} , \ubX_{1}^{(a)} )  \cond (\bX_{0}, \bU)$;} (iv) $\bW \indep (A,\bZ)  \cond  (\oD_{\TIME+1}^{(a)}, \overline{\bX}_{\TIME}^{(a)}, \bU)$.
\end{itemize}

\end{itemize}

\subsubsection{Conditional Independence of the Observed Variables}

Based on Assumptions \HL{A1} and \AssumptionProxy, we establish the following conditional independencies, which will be used throughout this document:
\begin{itemize}
\item[\HT{CI1}] \makebox[4.75cm][l]{\CIone;} 
\makebox[5.25cm][l]{\HT{CI2} \CItwo;}
\makebox[5cm][l]{\HT{CI3} \CIthree;} 
\item[\HT{CI4}] \makebox[4.75cm][l]{\CIfour;} 
\makebox[5.25cm][l]{\HT{CI5} \CIfive;}
\makebox[5cm][l]{\HT{CI6} \CIsix.}  
\end{itemize}
To establish the result, we first introduce the graphoid axioms of conditional independence for random variables \citep{Dawid1979}.
\begin{itemize}
\item[\HT{G1}] \makebox[6cm][l]{$S_1 \indep S_2 \cond S_3$ $\Rightarrow$ $S_2 \indep S_1 \cond S_3$;}
\makebox[7.5cm][l]{\HT{G2} $S_1 \indep (S_2,S_3) \cond S_4$ $\Rightarrow$ $S_1 \indep S_2 \cond S_4$ and $S_1 \indep S_3 \cond S_4$;}
\item[\HT{G3}] \makebox[6cm][l]{$S_1 \indep (S_2,S_3) \cond S_4$ $\Rightarrow$ $S_1 \indep S_2 \cond (S_3, S_4)$;}
\makebox[7.5cm][l]{\HT{G4} $S_1 \indep S_2 \cond S_3$ and $S_1 \indep S_4 \cond (S_2,S_3)$ $\Rightarrow$ $S_1 \indep (S_2, S_4) \cond S_3$;} 
\end{itemize}

We now establish conditions \HL{CI1}-\HL{CI6} under \HL{G1}-\HL{G4}. First, \HL{CI1} is established as follows:
\begin{align*}
    & \text{\AssumptionProxy-(i) + \AssumptionProxy-(ii)}
    \\
    & \Rightarrow \quad \bZ \indep (\potY{A},\potDt{A}{\TIME+1}) \cond (A,\oD_{\TIME}^{(A)},\obX_{\TIME},\bU)
    &&
    \Leftarrow \quad \text{\HL{G4}}
    \\
    & \Rightarrow \quad \bZ \indep (Y,D_{\TIME+1}) \cond (A,\oD_{\TIME},\obX_{\TIME}, \bU) && \Leftarrow \quad \text{\HL{A1}}
    \\
    & \Rightarrow \quad \bZ \indep Y \cond (A,\oD_{\TIME+1},\obX_{\TIME},\bU) && \Leftarrow \quad \text{\HL{G3}}
\end{align*} 

Next, \HL{CI2} is trivial from \HL{A1} and \AssumptionProxy-(ii). Lastly, \HL{CI3}-\HL{CI6} are established as follows:
\begin{align*}
    & \text{\AssumptionProxy-(iii) + \AssumptionProxy-(iv)}
    \\
    & \Rightarrow \quad \bW \indep (A,\uD_{1}^{(A)},\ubX_{1}^{(A)},\bZ) \cond (\bX_{0}, \bU)
    && \Leftarrow \quad \text{\HL{G4}}
    \\
    & \Rightarrow \quad \bW \indep (\uD_{1}^{(A)},\ubX_{1}^{(A)},\bZ) \cond (A,\bX_{0}, \bU) && \Leftarrow \quad \text{\HL{G3}}
    \\
    & \Rightarrow \quad \text{\HL{CI6}: } \bW \indep (\uD_{1},\ubX_{1},\bZ) \cond (A,\bX_{0}, \bU) && \Leftarrow \quad \text{\HL{A1}}
    \\
    & \Rightarrow \quad \text{\HL{CI4}: } \bW \indep (A,\uD_{1},\ubX_{1},\bZ) \cond (\bX_{0}, \bU) && \Leftarrow \quad \text{\HL{G4} and $\bW \indep A \cond (\bX_{0}, \bU)$ in the second line}
    \\
    & \Rightarrow \quad \text{\HL{CI3}: } \bW \indep (A,\bZ) \cond (\oD_{\TIME+1},\obX_{\TIME},\bU) && \Leftarrow \quad \text{\HL{G3}}
    \\
    & \Rightarrow \quad \text{\HL{CI5}: } \bW \indep \bZ \cond (A,\oD_{\TIME+1},\obX_{\TIME},\bU)  && \Leftarrow \quad \text{\HL{G3}}
\end{align*}

Next, we establish \HL{CI1}, \HL{CI-3}-\HL{CI-6} under Assumptions \HL{A2'},  and \AssumptionProxy-(i),(iii),(iv). Assumption \HL{A2'} implies that the density of $\bZ \cond (A,\oD_{\TIME+1}^{(a)},\obX_{\TIME}^{(a)},\bU)$ is
\begin{align}
&
f^*(\bZ=\bz, \potY{a}=y \cond A,\oD_{\TIME+1}^{(a)},\obX_{\TIME}^{(a)},\bU)
\nonumber
\\
&
=
\frac{f^*(\bZ=\bz,\potY{a}=y,A,\oD_{\TIME+1}^{(a)},\obX_{\TIME}^{(a)},\bU)}{f^*(A,\oD_{\TIME+1}^{(a)},\obX_{\TIME}^{(a)},\bU)} 
\nonumber
\\
&
=
\frac{f^*(\bZ=\bz,\potY{a}=y, \oD_{\TIME+1}^{(a)},\obX_{\TIME}^{(a)},\bU) f^*(A=a) }{f^*(\oD_{\TIME+1}^{(a)},\obX_{\TIME}^{(a)},\bU) f^*(A=a) }
&&
\Leftarrow \quad \text{\HL{A2'} and evaluate $A=a$}	
\nonumber
\\
&
=
\frac{f^*(\bZ=\bz,\potY{a}=y,A=a,\oD_{\TIME+1}^{(a)},\obX_{\TIME}^{(a)},\bU)}{f^*(A=a,\oD_{\TIME+1}^{(a)},\obX_{\TIME}^{(a)},\bU)}
&&
\Leftarrow \quad \text{\HL{A2'}}	
\nonumber
\\
&
=
\frac{f^*(\bZ=\bz,Y=y,A=a,D_{\TIME+1},\obX_{\TIME},\bU)}{f^*(A=a,D_{\TIME+1},\obX_{\TIME},\bU)}
&&
\Leftarrow \quad \text{\HL{A1}}	
\nonumber
\\
&
=
f^*(\bZ=\bz, Y=y \cond A=a,D_{\TIME+1},\obX_{\TIME},\bU) \ . 
\label{eq-CI11}
\end{align}
Similarly, we establish 
\begin{align}
&
f^*(\bZ=\bz  \cond A,\oD_{\TIME+1}^{(a)},\obX_{\TIME}^{(a)},\bU)
=
f^*(\bZ=\bz  \cond A=a,D_{\TIME+1},\obX_{\TIME},\bU)
\ , 
\label{eq-CI12}
\\
&
f^*(  \potY{a}=y \cond A,\oD_{\TIME+1}^{(a)},\obX_{\TIME}^{(a)},\bU)
=
f^*( Y=y \cond A=a,D_{\TIME+1},\obX_{\TIME},\bU) \ .
\label{eq-CI13}
\end{align}  
Therefore, we obtain	the following result for all $a \in \{0,1\}$:
\begin{align*}
    &
    f^*(\bZ=\bz, Y=y \cond A=a,D_{\TIME+1},\obX_{\TIME},\bU) 
    \\
    &
    =
    f^*(\bZ=\bz, \potY{a}=y \cond A,\oD_{\TIME+1}^{(a)},\obX_{\TIME}^{(a)},\bU)
    &&
    \Leftarrow \quad \eqref{eq-CI11}
    \\
    & 
    =
    f^*(\bZ=\bz \cond A,\oD_{\TIME+1}^{(a)},\obX_{\TIME}^{(a)},\bU)
    f^*(\potY{a}=y \cond A,\oD_{\TIME+1}^{(a)},\obX_{\TIME}^{(a)},\bU)
    &&
    \Leftarrow \quad \text{\AssumptionProxy-(i)}
    \\
    & 
    =
    f^*(\bZ=\bz \cond A=a,D_{\TIME+1},\obX_{\TIME},\bU)
    f^*(Y=y \cond A=a,D_{\TIME+1},\obX_{\TIME},\bU) 
    &&
    \Leftarrow \quad \eqref{eq-CI12}, \eqref{eq-CI13}
\end{align*}
which implies \HL{CI1}: \CIone. Conditions \HL{CI3}-\HL{CI6} can be established under Assumptions \HL{A2'} and \AssumptionProxy-(i),(iii),(iv) based on similar algebra.

\subsection{Details of the Representation of $\psi^* (a_Y,a_D)$}	\label{sec:supp:psi}

We provide details on the representation of the CSE in \eqref{eq-psi} of the main paper (equivalently \eqref{eq-psi in supp}) under Assumptions \HL{A1}-\HL{A5}.

\subsubsection{Under the Observational Setting} \label{sec:supp:psi obs}

First, we find
\begin{align}
\psi^*(a_Y,a_D) 
&
= 
\frac{
\EXP \big[ \potY{a_Y,a_D} \big\{ 1 - \potDt{a_Y,a_D}{\TIME+1} \big\} \big] }{
\Pr \big\{ \potDt{a_Y,a_D}{\TIME+1} = 0 \big\}
}
\nonumber
\\
&
= 
\frac{
\EXP_U \bigg[ 
\displaystyle{\int}
\bigg[ 
\begin{array}{l} 
 \EXP \big\{ \potY{a_Y,a_D} \cond 
\potDt{a_Y,a_D}{\TIME+1} = 0, \obX_{\TIME}^{(a_Y,a_D)}=\obx_{\TIME}, U \big\}
\\
\quad 
\times 
f^* \{ \potDt{a_Y,a_D}{\TIME+1 }= 0 , \obX_{\TIME}^{(a_Y,a_D)} = \obx_{\TIME} \cond   U \big\} 
\end{array}
\bigg]
\, 
d \obx_{\TIME}
\bigg]
}{
\EXP_U \big[
\int  
f^* \{ \potDt{a_Y,a_D}{\TIME+1 }= 0 , \obX_{\TIME}^{(a_Y,a_D)} = \obx_{\TIME}\cond  U \big\} 
\, 
d \obx_{\TIME}
\big] 
} \ .  \label{eq-estimand}
\end{align}

We first establish some results using the dismissible conditions. We find
\begin{align}
&
\EXP \big\{ \potY{a_Y,a_D=1} \cond \potDt{a_Y,a_D=1}{\TIME+1} = 0 , \obX_{\TIME}^{(a_Y,a_D=1)} , \bU  \big\} 
\nonumber
\\
&
=
\EXP \big\{ Y(G) \cond D_{\TIME+1}(G) = 0 , A_Y(G) = a_Y , A_D(G) = 1 , \obX_{\TIME} , \bU (G)  \big\} 
\nonumber
\\
&
=
\EXP \big\{ Y(G) \cond D_{\TIME+1}(G) = 0 , A_Y(G) = a_Y , A_D(G) = 0 , \obX_{\TIME}  , \bU (G)  \big\} 
\nonumber
\\
&
=
\EXP \big\{ \potY{a_Y,a_D=0} \cond \potDt{a_Y,a_D=0}{\TIME+1} = 0 , \obX_{\TIME}^{(a_Y,a_D=0)} , \bU (G) \big\}  \ .
\label{eq-psi numer}
\end{align}
The second line holds from the fact that $A_Y(G)$ and $A_D(G)$ are randomized and the consistency assumption \HL{A1}. The third line holds from \AssumptionDismissibleTV-(i).  The last line can be easily deduced by the same reasons. 

Likewise, one can also establish the following result for $t \in \{0,\ldots,\TIME\}$:
\begin{align}
&
\Pr \big\{ \potDt{a_Y=1,a_D}{\stime+1} = 0 \cond \potDt{a_Y=1,a_D}{\stime}=0, \obX_{\stime}^{(a_Y=1,a_D)}, \bU \big\}
\nonumber
\\
&
=
\Pr \big\{ \potDt{a_Y=0,a_D}{\stime+1} = 0 \cond \potDt{a_Y=0,a_D}{\stime}=0, \obX_{\stime}^{(a_Y=0,a_D)}, \bU \big\}
\ ,
\label{eq-Dism-D}
\\
&
\Pr \big\{ \obX_{\stime+1}^{(a_Y=1,a_D)} \cond \potDt{a_Y=1,a_D}{\stime}=0, \obX_{\stime}^{(a_Y=1,a_D)}, \bU \big\}
\nonumber
\\
&
=
\Pr \big\{ \obX_{\stime+1}^{(a_Y=0,a_D)} \cond \potDt{a_Y=0,a_D}{\stime}=0, \obX_{\stime}^{(a_Y=0,a_D)}, \bU \big\}
\ .
\label{eq-Dism-X}
\end{align}

Next, we find an alternative representation for $f^* \{ \potDt{a_Y,a_D}{\TIME+1 }= 0 , \obX_{\TIME}^{(a_Y,a_D)} = \obx_{\TIME} \cond   U \big\}$. We find 
\begin{align*}
    &
    f^* \{ \potDt{a_Y,a_D}{\TIME+1 }= 0 , \obX_{\TIME}^{(a_Y,a_D)} = \obx_{\TIME} \cond   U \big\} 
    \\
    &
    =
    \left[ 
    \begin{array}{l}
    \Pr\{ \potDt{a_Y,a_D}{\TIME+1 }= 0 \cond \obX_{\TIME}^{(a_Y,a_D)} = \obx_{\TIME}, \potDt{a_Y,a_D}{\TIME }= 0, U \big\}
    \\ 
    \quad 
    \times
    f^* \{ \bX_{\TIME}^{(a_Y,a_D)} = \obx_{\TIME} \cond \obX_{\TIME-1}^{(a_Y,a_D)} = \obx_{\TIME-1}, \potDt{a_Y,a_D}{\TIME }= 0, U \big\}
    \\ 
    \quad \times
    f^* \{ \potDt{a_Y,a_D}{\TIME }= 0 , \obX_{\TIME-1}^{(a_Y,a_D)} = \obx_{\TIME-1} \cond  U \big\} 
    \end{array}
    \right] 
    \\
    &
    =    
    \prod_{\stime=1}^{\TIME} 
    \Bigg[
    \begin{array}{l}    
    \Pr\{ \potDt{a_Y,a_D}{\stime+1 }= 0 \cond \obX_{\stime}^{(a_Y,a_D)}=\obx_{\stime}, \potDt{a_Y,a_D}{\stime}= 0, U \big\}
    \\ \quad \times 
    f^* \{ \bX_{\stime}^{(a_Y,a_D)} = \bx_{\stime} \cond \obX_{\stime-1}^{(a_Y,a_D)}=\obx_{\stime-1}, \potDt{a_Y,a_D}{\stime }= 0, U \big\} 
    \end{array}
    \Bigg] 
    \times 
    \Bigg[
    \begin{array}{l}
    \Pr\{ \potDt{a_Y,a_D}{1}= 0 \cond \bX_{0}=\bx_{0}, U \big\}
         \\
         \quad \times 
    f^*(\bX_{0}=\bx_0 \cond U)   
    \end{array}
    \Bigg]
    \\
    &
    =    
    \prod_{\stime=1}^{\TIME} 
    \Bigg[
    \begin{array}{l}    
    \Pr\{ \potDt{a_D}{\stime+1 }= 0 \cond \obX_{\stime}^{(a_D)}=\obx_{\stime}, \potDt{a_D}{\stime}= 0, U \big\}
    \\ \quad \times 
    f^* \{ \bX_{\stime}^{(a_D)} = \bx_{\stime} \cond \obX_{\stime-1}^{(a_D)}=\obx_{\stime-1}, \potDt{a_D}{\stime }= 0, U \big\} 
    \end{array}
    \Bigg]
    \times 
    \Bigg[
    \begin{array}{l}
    \Pr\{ \potDt{a_D}{1}= 0 \cond \bX_{0}=\bx_{0}, U \big\}
         \\
         \quad \times 
    f^*(\bX_{0}=\bx_0 \cond U)   
    \end{array}
    \Bigg]
    \\
    &
    =    
    \prod_{\stime=1}^{\TIME} 
    \Bigg[
    \begin{array}{l}    
    \Pr \big( D_{\stime+1 }= 0 \cond 
    A=a_D, \obX_{\stime}=\obx_{\stime}, D_{\stime}= 0, U \big)
    \\ \quad \times 
    f^* \big( \bX_{\stime} = \bx_{\stime} \cond A=a_D, \obX_{\stime-1}=\obx_{\stime-1}, D_{\stime }= 0, U \big)
    \end{array}
    \Bigg] 
    \times 
    \Bigg[
    \begin{array}{l}
    \Pr \big( D_1= 0 \cond A=a_D ,  \bX_{0}=\bx_{0}, U \big)
         \\
         \quad \times 
    f^*\big( \bX_{0}=\bx_0 \cond U \big) 
    \end{array}
    \Bigg]
    \\
    &
    =
    f^* \big( D_{\TIME+1}=0, \ubX_{1}=\ubx_{1} \cond A=a_D, \bX_{0}=\bx_{0}, U \big)
    f^* \big( \bX_{0}=\bx_{0} \cond  U \big) 
    \\
    &
    =
    \frac{     
    \Pr( A=a_D, D_{\TIME+1}=0 \cond  \ubX_{1}, \bX_{0}, \bU ) 
    }
    {
    \Pr(A=a_D \cond \bX_{0},U)
    }
    f^*(\ubX_{1}, \bX_{0} \cond U ) 
    \\
    &
    =
    \frac{     
    \Pr( A=a_D, D_{\TIME+1}=0 \cond  \obX_{\TIME}, \bU ) 
    }
    {
    \Pr(A=a_D \cond \bX_{0},U)
    }
    f^*(  \obX_{\TIME} \cond U )  \ .
    \numeq \label{eq-psi numer1}
\end{align*}
The first identity follows from the law of iterated expectation, while the second identity is derived by repeatedly applying the same decomposition used in the previous step. The third identity is from \eqref{eq-Dism-D} and \eqref{eq-Dism-X}. The fourth identity is from \HL{A1} and \HL{A2}. The last two identities are straightforward.  

Similarly, we establish
\begin{align}
&
\EXP \big\{ \potY{a_Y,a_D=0} \cond \potDt{a_Y,a_D=0}{\TIME+1} = 0 , \obX_{\TIME}^{(a_Y,a_D=0)} , \bU  \big\}  
\nonumber
\\
&
=
\EXP \big\{ \potY{a_Y} \cond \potDt{a_Y}{\TIME+1} = 0 , \obX_{\TIME}^{(a_Y)} , \bU  \big\} 
\nonumber
\\
&
=
\EXP \big\{ \potY{a_Y} \cond D_{\TIME+1}^{(a_Y)} = 0 , \ubX_{1}^{(a_Y)}, X_{0} , \bU  \big\} 
\nonumber
\\
&
=
\EXP \big\{ \potY{a_Y} \cond 
A=a_{Y}, D_{\TIME+1}^{(a_Y)} = 0 , \ubX_{1}^{(a_Y)}, X_{0} , \bU  \big\} 
\nonumber
\\  
&
=
\EXP \big( Y \cond 
A=a_{Y}, D_{\TIME+1} = 0 , \obX_{\TIME} , \bU  \big)
\ .
\label{eq-psi numer2}
\end{align}
The second line is from \eqref{eq-psi numer}. 
The third and fourth lines are from Assumption \HL{A2}. 
The last line is from Assumption \HL{A1}. 

Therefore, $\psi^* (a_Y,a_D)$ is represented as
\begin{align*} 
\psi^* (a_Y,a_D) 
&
= 
\frac{
\EXP_U \bigg[ 
\displaystyle{\int}
\bigg[ 
\begin{array}{l} 
 \EXP \big\{ \potY{a_Y,a_D} \cond 
\potDt{a_Y,a_D}{\TIME+1} = 0, \obX_{\TIME}^{(a_Y,a_D)}=\obx_{\TIME}, U \big\}
\\
\quad 
\times 
f^* \{ \potDt{a_Y,a_D}{\TIME+1 }= 0 , \obX_{\TIME}^{(a_Y,a_D)} = \obx_{\TIME} \cond   U \big\} 
\end{array}
\bigg]
\, 
d \obx_{\TIME}
\bigg]
}{
\EXP_U \big[
\int  
f^* \{ \potDt{a_Y,a_D}{\TIME+1 }= 0 , \obX_{\TIME}^{(a_Y,a_D)} = \obx_{\TIME}\cond  U \big\} 
\, 
d \obx_{\TIME}
\big] 
} 
\\
&
= 
\frac{
\displaystyle{
\EXP_U \bigg\{ 
\EXP \big( Y \cond 
A=a_{Y}, D_{\TIME+1} = 0 , \obX_{\TIME}  , \bU  \big)
\frac{     
    \Pr( A=a_D, D_{\TIME+1}=0 \, | \,  \obX_{\TIME}, \bU ) 
    }
    {
    \Pr(A=a_D \, | \, \bX_{0},U)
    }
\bigg\}
}
}{  
\EXP_U \big\{
\Pr \big( D_{\TIME+1}=0 \cond A=a_D, \bX_{0}, U \big)
\big\}
}  
\ . 
\numeq \label{eq-id form with U}
\end{align*}
The second line is from \eqref{eq-estimand}. The last line is from \eqref{eq-psi numer1} and \eqref{eq-psi numer2}. When $\ubX_{1}=\emptyset$, \eqref{eq-id form with U} reduces to
\begin{align*}
&
\psi^* (a_Y,a_D) 
= 
\frac{
\EXP_U \big\{ \EXP \big( Y \cond 
A=a_{Y}, D_{\TIME+1} = 0 , \bX_{0} , \bU  \big)
\Pr \big( D_{\TIME+1}=0 \cond A=a_D, \bX_{0}, U \big)
\big\}
}{
\EXP_U \big\{
\Pr \big( D_{\TIME+1}=0 \cond A=a_D, \bX_{0}, U \big)
\big\}
}  \ .
\end{align*}
If $a_Y=a_D$, \eqref{eq-id form with U} reduces to 
\begin{align*}
\psi^* (a_D,a_D)
=
\frac{
\EXP_U \big\{
\EXP \big( Y \cond 
A=a_{D}, D_{\TIME+1} = 0 , \bX_{0} , \bU  \big)
\Pr\big( D_{\TIME+1}=0 \cond A=a_{D},\bX_{0}, \bU \big)
\big\} 
}{ 
\EXP_U \big\{
\Pr \big( D_{\TIME+1}=0 \cond A=a_D, \bX_{0}, U \big)
\big\}
}   \ .
\end{align*}

\subsubsection{Under the Experimental Setting} \label{sec:supp:psi exp} 
When $a_Y \neq a_D$, the result established under the observational setting in Section \ref{sec:supp:psi obs} remains the same under the experimental setting. Therefore, we establish results under experimental settings when $a_Y = a_D$. Under Assumptions \HL{A1} and \HL{A2'}, it is straightforward to establish that
 \begin{align*}
 & \EXP \big\{ \ind (A=a_D, D_{\TIME+1}=0) p_0^*(a_D) Y \big\}
 \\
& 
=
\EXP \big\{ \ind (A=a_D, \potDt{a_D}{\TIME+1}=0) p_0^*(a_D) \potY{a_D,a_D} \big\}
\\
& 
=
\Pr (A=a_D) p_0^*(a_D)
\EXP \big[ \big\{ 1- \potDt{a_D}{\TIME+1} \big\}  \potY{a_D,a_D} \big]
\\
& 
=
\EXP \big[ \big\{ 1- \potDt{a_D}{\TIME+1} \big\}  \potY{a_D,a_D} \big]
\\
&
=
\psi_{\RD,\NUMER}^*(a_D,a_D) \ .
 \end{align*}
 The first identity is from \HL{A1}.
 The second identity is from \HL{A2'}. The third identity is from $p_0^*(a_D) = 1/\Pr(A=a_D)$. We can establish a similar result for the denominator, i.e., $\EXP \big\{ \ind (A=a_Y, D_{\TIME+1}=0) p_1^* \big\} = \psi_{\RD,\DENOM}^*(a_D)$. Therefore, the target estimand $\psi_{\RD}^*(a_D,a_D)$ simplifies to
\begin{align*}
\psi_{\RD}^* (a_D,a_D)
=
\frac{\psi_{\RD,\NUMER}^*(a_D,a_D)}{\psi_{\RD,\DENOM}^*(a_D)}
=
\frac{ p_0^*(a_D) \EXP\big\{ \ind(A=a_D)(1-D_{\TIME+1}) Y \big\} }{p_0^*(a_D) \EXP\big\{ \ind(A=a_D)(1-D_{\TIME+1}) \big\}} \ .
\end{align*}

 \subsection{Existence and Uniqueness of the Confounding Bridge Functions} \label{sec:supp:exist bridge ft}

We examine the conditions under which the integral equations \eqref{eq-bridgeft1 obs}-\eqref{eq-bridgeft6 obs} admit solutions and ensure their uniqueness. In this discussion, we focus solely on the integral equation \eqref{eq-bridgeft1 obs}, as the results for the remaining equations can be derived in a similar manner.
 
 \subsubsection{Existence}

In brief, we follow the approach in \citet{Miao2018}. The proof relies on Theorem 15.18 of \citet{Kress2014}, which is stated below for completeness.\\[0.25cm]
 \noindent
 \textbf{Theorem 15.18.} \citep{Kress2014}
 Let $A:X \rightarrow Y$ be a compact operator with singular system $\big\{ \mu_n,\phi_n,g_n \big\}_{n=1,2,\ldots}$. The integral equation of the first kind $A\phi = f$ is solvable if and only if the following conditions hold:
 \begin{align*}
 	&
 	f \in \mathcal{N}(A^{\text{ad}})^\perp = \big\{ f \, \big| \, A^{\text{ad}}(f) = 0 \big\}^\perp 
 	&&
 	\text{ and }
 	&&
 	\sum_{n=1}^{\infty} \mu_n^{-2} \big| \langle f,g_n \rangle |^2 < \infty \ .
 \end{align*} 

To apply the Theorem, we introduce some additional notations. For a fixed $\obX_{\TIME}=\bx$, let $\mathcal{L}_{W} = \mathcal{L}_2 (\bW, A=a_Y,D_{\TIME+1}=0,\obX_{\TIME}=\bx)$ and $\mathcal{L}_{Z} = \mathcal{L}_2 (\bZ,A=a_Y,D_{\TIME+1}=0,\obX_{\TIME}=\bx)$ be the spaces of square-integrable functions of $\bW$ and $\bZ$ given $(A=a_Y,D_{\TIME+1}=0,\obX_{\TIME}=\bx)$, respectively, which are equipped with the inner products 
 \begin{align*}
 &
 \langle h, h' \rangle_{W} = \int h(\bw) h'(\bw) \, f_{W|ADX} (\bw \cond a_Y,0,\bx) \, d\bw 
 = \EXP \big\{ h(\bW) h'(\bW) \cond A_Y=a_Y,D_{\TIME+1}=0,\obX_{\TIME}=\bx \big\}
 \ ,
 \\
 &
 \langle g, g' \rangle_{Z} = \int g(\bz) g'(\bz) \, f_{Z|ADX}(\bz \cond a_Y,0,\bx) \, d\bz  
 = \EXP \big\{ g(\bZ) g'(\bZ) \cond A_Y=a_Y,D_{\TIME+1}=0,\obX_{\TIME}=\bx \big\} 
 \ .
 \end{align*} 
 Let $\mathcal{K}_{x}: \mathcal{L}_{W} \rightarrow \mathcal{L}_{Z}$ be the conditional expectation of $h( \bW ) \in \mathcal{L}_{W}$ given $(\bZ,A=a_Y,D_{\TIME+1}=0,\obX_{\TIME}=\bx)$:
 \begin{align*}
 \mathcal{K}_{x} (h) \in \mathcal{L}_{W} 
 \text{ satisfying }
 \big( \mathcal{K}_{x}(h) \big) (\bZ) 
 =
 \EXP \big\{ h(\bW) \cond \bZ,A=a_Y,D_{\TIME+1}=0,\obX_{\TIME}=\bx \big\}
 \text{ for } h \in \mathcal{L}_{W}
 \ .
 \end{align*}
 Let $g_{x}^*(z) := \EXP \big( Y \cond \bZ=z,A=a_Y,D_{\TIME+1}=0,\obX_{\TIME}=x)$.   
  
 Using the notation, the confounding bridge function $h_{1,x}^* \in \mathcal{L}_{W}$ satisfies $\mathcal{K}_x ( h_{1,x}^* ) = g_x^*(z) $, i.e., 
 \begin{align*}
 \int h_{1,x}^* (\bw) 
 f_{W|ZADX}(\bw \cond \bz,a_Y,0,\bx) \, d\bw
  = g_{\bx}^*(\bz) , \ \forall \bz \ .
 \end{align*}

 Now, we assume the following conditions for all $\bx$:
 \begin{itemize}[leftmargin=0cm]
     \item[] \HT{Bridge1} $\iint
 f_{W|ZADX}(\bw \cond \bz,a_Y,0,\bx)
 f_{Z|WADX}(\bz \cond \bw,a_Y,0,\bx)
 \, d\bw \, d \bz
 < \infty$;
 \item[] \HT{Bridge2} For $g \in \mathcal{L}_{Z}$, $\EXP \big\{ g(\bZ) \cond \bW,A=a_Y,D_{\TIME+1}=0,\obX_{\TIME}=\bx \big\} = 0$ implies $g(\bZ)= 0$ almost surely;
 \item[] \HT{Bridge3} $\EXP \big[ \{g_{x}^*(\bZ)\}^2 \cond A=a_Y,D_{\TIME+1}=0,\obX_{\TIME}=\bx \big] < \infty$ almost surely;
 \item[] \HT{Bridge4} $\sum_{n=1}^{\infty} \mu_{n,\bx}^{-2} \big| \langle g_{\bx}^*(\bZ) ,g_{n,\bx} \rangle_Z |^2 < \infty$ where $\big\{ \mu_{n,\bx},\phi_{n,\bx},g_{n,\bx} \big\}_{n}$ is the singular system of $\mathcal{K}_{\bx}$.
 \end{itemize}

 First, we show that $\mathcal{K}$ is a compact operator under \HL{Bridge1}. Let $\mathcal{K}^{\text{ad}} : \mathcal{L}_{Z} \rightarrow \mathcal{L}_{W}$ be the conditional expectation of $g(\bZ) \in \mathcal{L}_{Z}$ given $(\bW,A=a_Y,D_{\TIME+1}=0,\obX_{\TIME}=\bx)$, i.e., 
 \begin{align*}
 \mathcal{K}_{x}^{\text{ad}}  (g) \in \mathcal{L}_{W} 
 \text{ satisfying }
 \big( \mathcal{K}_{x}(g) \big) (\bW) 
 =
 \EXP \big\{ g(\bZ) \cond \bW,A=a_Y,D_{\TIME+1}=0,\obX_{\TIME}=\bx \big\}
 \text{ for } g \in \mathcal{L}_{Z} \ .
 \end{align*}
 Then, $\mathcal{K}_{x}$ and $\mathcal{K}_{x}^{\text{ad}}$ are the adjoint operator of each other as follows:
 \begin{align*}
 &
 \langle \mathcal{K}_{x}(h) , g \rangle_{Z}
 \\
 & 
 =
 \EXP
 \Big[
     \EXP \big\{ h(\bW) \cond \bZ,A=a_Y,D_{\TIME+1}=0,\obX_{\TIME}=\bx \big\}
     g(\bZ)
     \, \Big| \, 
     A=a_Y,D_{\TIME+1}=0,\obX_{\TIME}=\bx
 \Big]	
 \\
 & 
 =
 \EXP
 \Big[
 h(\bW) g(\bZ)
 \, \Big| \, 
     A=a_Y,D_{\TIME+1}=0,\obX_{\TIME}=\bx
 \Big]	
 \\
 &
 =
 \EXP
 \Big[
 h(\bW) \EXP \big\{  g(\bZ) \cond \bW,A=a_Y,D_{\TIME+1}=0,\obX_{\TIME}=\bx \big\}
 \Big]	
 \\
 &
 =
     \langle h , \mathcal{K}_{x}^{\text{ad}}(g) \rangle_{W} \ .
 \end{align*}
 Additionally, as shown in page 5659 of \citet{Carrasco2007}, $\mathcal{K}_{x}$ and $\mathcal{K}_{x}^{\text{ad}}$ are compact operators under \HL{Bridge1}. Moreover, by Theorem 15.16 of \citet{Kress2014}, there exists a singular value decomposition of $\mathcal{K}_{x}$ as  $\big\{ \mu_{n,x},\phi_{n,x},g_{n,x} \big\}_{n}$. 

 Second, we show that $\mathcal{N}(\mathcal{K}_{x}^{\text{ad}})^\perp = \mathcal{L}_{Z}$, which suffices to show $\mathcal{N}(\mathcal{K}_{x}^{\text{ad}}) = \big\{ 0 \big\} \subseteq \mathcal{L}_{Z}$. Under \HL{Bridge2}, we have 
 \begin{align*}
 g \in \mathcal{N}(\mathcal{K}_{x}^{\text{ad}})
 \quad 
 \Rightarrow 
 \quad 
 \EXP \big\{ g (\bZ) \cond \bW=\bw,A=a_Y,D_{\TIME+1}=0,\obX_{\TIME}=\bx \big\}
 =
 0, \ \forall \bw
 \quad \Rightarrow
 \quad
 g(\bZ) = 0  \ ,
 \end{align*}
 where the first arrow is from the definition of the null space $\mathcal{N}$, and the second arrow is from \HL{Bridge2}. Therefore, any $g \in \mathcal{N}(\mathcal{K}_{x}^{\text{ad}})$ must satisfy $g(\bZ) = 0 $ almost surely, i.e., $\mathcal{N}(\mathcal{K}_{x}^{\text{ad}})= \big\{ 0 \big\} \subseteq \mathcal{L}_{Z}$ almost surely. 

 Third, it is trivial that $g_{x}^*(\bZ)  \in \mathcal{L}_{Z} = \mathcal{N}(\mathcal{K}_{x}^{\text{ad}})^\perp  $ under \HL{Bridge3}.

 Combining the three results, we establish that $g_{x}^*(\bZ)$ satisfies the first condition of Theorem 15.18 of \citet{Kress2014}. The second condition of the Theorem is exactly the same as \HL{Bridge4}. Therefore, we establish that the Fredholm integral equation of the first kind $\mathcal{K}_{x} ( h ) = g_{x}^*(\bZ) $ is solvable under \HL{Bridge1}-\HL{Bridge4}, i.e., for each $\bx$, there exists a function $h_{1,x}^*(\bw)$ satisfying
 \begin{align*}
 \int h_{1,x}^* (\bw) 
 f_{W|ZADX}(\bw \cond \bz,a_Y,0,\bx) \, d\bw
  = g_{\bx}^*(\bz) , \ \forall \bz \ .
 \end{align*}
 Therefore, $h_1^*(x,w) := h_{1,x}^*(w)$ satisfies 
 \begin{align*}
     \int h_{1}^* (\bx,\bw) 
 f_{W|ZADX}(\bw \cond \bz,a_Y,0,\bx) \, d\bw
  = g^*(\bz,\bx) , \ \forall (\bz,\bx) \ ,
 \end{align*}
 implying that the confounding bridge function exists.

 \subsubsection{Uniqueness}
 
 Consider the following completeness condition:
 \begin{align*}
 	\EXP \big\{ h(W,\obX_{\TIME}) \cond Z,A=a_Y,D_{\TIME+1}, \obX_{\TIME} \big\} = 0  \text{ for all $Z,\obX_{\TIME}$ a.s. }
 	\quad \text{ if and only if }
 	\quad  h(W,\obX_{\TIME}) = 0 \ .
 \end{align*}
 Then, \citet[Theorem 3.1]{Cui2023} and \citet[Lemma 10]{PMMR2021} established that $h_1^*$ solving \eqref{eq-bridgeft1 obs} is unique.

\subsection{Details of the PMMR Estimation} \label{sec:supp:PMMR details}

We present details of the PMMR estimation, including the closed-form expressions for the empirical risk functions of the confounding bridge functions and the convergence rates of the (nested) PMMR estimators.

\subsubsection{Details of the PMMR Empirical Risk Minimization} \label{sec:supp:risk}

 We provide details on the construction of the empirical risk functions $\widehat{R}_{h1}$, $\widehat{R}_{h0}$, $\widehat{R}_{h2}$, $\widehat{R}_{q0}$, $\widehat{R}_{q0}$. First, we introduce Lemma 2 of \citet{PMMR2021}:\\
 \textbf{Lemma 2} \citep{PMMR2021}: Let $\mathcal{K}(\cdot,\cdot)$ be the kernel of the RKHS of $q$. Suppose that
 \begin{align}	\label{eq-kernel bound}
 \EXP
     \big[
         \ind (A=a_Y,D_{\TIME+1}=0)^2
         \big\{
             Y - h(W,\obX_{\TIME})
         \big\}^2 \mathcal{K} ((Z,\obX_{\TIME}), (Z,\obX_{\TIME}) )
     \big] < \infty \ .
 \end{align}
 Then, we have
 \begin{align*}
 R_{h1}(h)
 &
 =
 \max_{ 
 \substack{ 
 q \in \HH(Z,\obX_{\TIME})
 \\ 
 q: \| q  \| \leq 1} }
     \Big[
     \EXP
     \big[
         q (Z,\obX_{\TIME})
         \ind(A=a_Y,D_{\TIME+1}=0)
         \big\{
             Y - h( W,\obX_{\TIME} )
         \big\}
     \big] \Big]^2
     \\
     &
     =
     \EXP \left[
     \mymatrixTwo{l}{\ind(A=a_Y,D_{\TIME+1}=0)
     \ind(A'=a_Y,D_{\TIME+1}'=0)
     \big\{
             Y - h(W,\obX_{\TIME})
         \big\} \big\{
             Y' - h(W',\obX_{\TIME}')
         \big\}}{\times
         \mathcal{K} ( (Z,\obX_{\TIME}), (Z',\obX_{\TIME}') ) }
     \right] \ ,
 \end{align*}
where $(Y',D_{\TIME+1}',A',W',Z',\obX_{\TIME}')$ denotes an independent copy of $(Y,D_{\TIME+1},A,W,Z,\obX_{\TIME})$.

 The result is also reported in other works, e.g., Theorem 3.3 of \citet{Muandet2020} and Lemma 1 of \citet{Zhang2023}. Condition \eqref{eq-kernel bound} implies that $\EXP \big[ 	\ind (A=a_Y,D_{\TIME+1}=0)
         \big\{
             Y - h(W,\obX_{\TIME})
         \big\} \mathcal{K} (( Z,\obX_{\TIME} ), \cdot ) \big]$ is Bochner integrable \citep[Definition A.5.20]{SVM2008}. One important property of the Bochner integrability is that an integration and a linear operator can be interchanged. Therefore, we find 
             \begin{align*}
 & \max_{\substack{ q \in \HH(Z,\obX_{\TIME}) \\ q: \| q  \| \leq 1}}
     \Big[
     {\EXP}
     \big[
         q (Z,\obX_{\TIME})
         \ind(A=a_Y,D_{\TIME+1}=0)
         \big\{
             Y - h(W,\obX_{\TIME})
         \big\}
     \big] \Big]^2
     \\
     &
     =
     \max_{\substack{ q \in \HH(Z,\obX_{\TIME}) \\ q: \| q  \| \leq 1}}
     \Big[
     {\EXP}
     \big[
         \ind(A=a_Y,D_{\TIME+1}=0)
         \big\{
             Y - h(W,\obX_{\TIME})
         \big\}
         \langle q, \mathcal{K} ( (Z,\obX_{\TIME}), \cdot ) \rangle
     \big] \Big]^2
     \\
     &
     =
     \max_{\substack{ q \in \HH(Z,\obX_{\TIME}) \\ q: \| q  \| \leq 1}}
     \Big[ 
         \Big \langle q, 
         {\EXP}
     \big[ 
         \ind(A=a_Y,D_{\TIME+1}=0)
         \big\{
             Y - h(W,\obX_{\TIME})
         \big\} \mathcal{K} ( (Z,\obX_{\TIME}) , \cdot ) 
         \big]
         \Big \rangle
      \Big]^2
      \\
     &
     =
     \Big\|
         {\EXP}
     \big[ 
         \ind(A=a_Y,D_{\TIME+1}=0)
         \big\{
             Y - h(W,\obX_{\TIME})
         \big\}
             \mathcal{K} ( (Z,\obX_{\TIME}), \cdot ) 
         \big] 
         \Big\|_{\HH(Z,\obX_{\TIME}) }^2
              \\
     &
     =
     \Big\langle
         {\EXP}
     \big[ 
         \ind(A=a_Y,D_{\TIME+1}=0)
         \big\{
             Y - h(W,\obX_{\TIME})
         \big\}
              \mathcal{K} ( (Z,\obX_{\TIME}) , \cdot ) 
         \big] 
         ,
         \\
         & \hspace*{1cm}
         {\EXP}
     \big[ 
         \ind(A=a_Y,D_{\TIME+1}=0)
         \big\{
             Y - h(W,\obX_{\TIME})
         \big\}
             \mathcal{K} ( (Z,\obX_{\TIME}) , \cdot ) 
         \big] 
     \Big\rangle
              \\
     &
     =
     {\EXP}
     \left[ \mymatrixTwo{l}{\Big \langle 
         \ind(A=a_Y,D_{\TIME+1}=0)
         \big\{
             Y - h(W,\obX_{\TIME})
         \big\}
             \mathcal{K} ( (Z,\obX_{\TIME}) , \cdot )  
         , }{\hspace*{1cm}
         {\EXP} 
         \big[
         \ind(A'=a_Y,D_{\TIME+1}'=0)
         \big\{
             Y' - h(\bW',\obX_{\TIME}')
         \big\}
              \mathcal{K} ( (\bZ',\obX_{\TIME}'), \cdot )  
              \big] \Big \rangle} 
     \right] 
              \\
     &
     =
     {\EXP}
     \left[	\mymatrixTwo{l}{
          \Big \langle 
          \ind(A=a_Y,D_{\TIME+1}=0)
         \big\{
             Y - h(W,\obX_{\TIME})
         \big\}
             \mathcal{K} ( (Z,\obX_{\TIME}) , \cdot )  
         ,   }{
         \hspace*{1cm}
         \ind(A'=a_Y,D_{\TIME+1}'=0)
         \big\{
             Y' - h(\bW',\obX_{\TIME}')
         \big\}
              \mathcal{K} ( (\bZ',\obX_{\TIME}'), \cdot )  
             \Big  \rangle} 
     \right] 
     \\
     &
     =
     \EXP \left[ \mymatrixTwo{l}{\ind(A=a_Y,D_{\TIME+1}=0)
     \ind(A'=a_Y,D_{\TIME+1}'=0)
     \big\{
             Y - h(W,\obX_{\TIME})
         \big\} \big\{
             Y' - h(\bW',\obX_{\TIME}')
         \big\}}{ \quad
         \times
         \mathcal{K} ( (\bZ,\obX), (\bZ',\obX_{\TIME}') ) } 
     \right]  \ .
 \end{align*}
 The second line holds from $q \in \HH(Z,\obX_{\TIME})$, implying that $q(Z,\obX_{\TIME})	 = \langle q , \mathcal{K}( (Z,\obX_{\TIME}), \cdot ) \rangle$. 
 The third line holds from the Bochner integrability.
 The fourth line holds from the fact that $\HH(Z,\obX_{\TIME})$ is a vector space, and ${\EXP}
     \big[ 
         \big\{
             Y - h(W,\obX_{\TIME})
         \big\} \mathcal{K} ( (Z,\obX_{\TIME}), \cdot ) 
         \big] \in \HH(Z,\obX_{\TIME})$ from the Bochner integrability. Therefore, by choosing $q \propto {\EXP}
     \big[ 
         \big\{
             Y - h(W,\obX_{\TIME})
         \big\} \mathcal{K} ( (Z,\obX_{\TIME}), \cdot )  \big]$, we obtain the result. 
 The fifth line is trivial from the definition of the norm $\| \cdot \|_{\HH(Z,\obX_{\TIME})}$. 
 The sixth and seventh lines are from the Bochner integrability. 
 The last line is trivial. Therefore, $\widehat{R}_{h1}\LSS(h \con a_Y)$ in Section \ref{sec:estimator} is an empirical analogue of $R_{h1} (h \con a_Y)$.

 The other empirical risk functions can be similarly constructed. Specifically, the population-level risk functions for $h_{1}^*, h_0^*, h_2^*,q_0^* , q_1^*$ are
 \begin{align*}
 &
 R_{h1}(h \con a_Y)  
     =
     \EXP \left[
     \mymatrixTwo{l}{\ind(A=a_Y,D_{\TIME+1}=0)
     \ind(A'=a_Y,D_{\TIME+1}'=0)}{\quad \times
     \big\{
             Y - h(W,\obX_{\TIME})
         \big\} \big\{
             Y' - h(\bW',\obX_{\TIME}')
         \big\}  
         \mathcal{K} ( (\bZ,\obX_{\TIME}), (\bZ',\bX_{\TIME}') ) }
     \right]
     \\
 &
 R_{h0}(h \con a_Y, a_D)
 =
 \EXP 
 \left[
 \mymatrixTwo{l}{\ind(A=a_D) \ind(A'=a_D)  
     \big\{ (1-D_{\TIME+1}) h_1^*(\bW,\obX_{\TIME} \con a_Y)
     -  h (\bW,\bX_{0})  \big\}}{\quad \times 
     \big\{ (1-D_{\TIME+1}') h_1^* (\bW',\obX_{\TIME}' \con a_Y)
     -  h (\bW',\bX_{0}') \big\} 
     \mathcal{K} \big( (\bZ,\bX_{0}) , (\bZ',\bX_{0}') \big)}
     \right] 
     \quad \text{ if $a_Y \neq a_D$}
     \\
         &
 R_{h0}(h \con a_D, a_D)
 =
 \EXP 
 \left[ \mymatrixTwo{l}{\ind(A=a_D) \ind(A'=a_D) 
     \big\{ Y (1-D_{\TIME+1}) 
     -  h (\bW,\bX_{0})  \big\}}{\quad \times 
     \big\{ Y' (1-D_{\TIME+1}') 
     -  h (\bW',\bX_{0}') \big\}
     \mathcal{K} \big( (\bZ,\bX_{0}) , (\bZ',\bX_{0}') \big)}
     \right]  
     \\
     &
     R_{h2}(h \con a_D)
     =
     \EXP 
     \left[ \mymatrixTwo{l}{\ind(A=a_D) \ind(A'=a_D) 
     \big\{ (1-D_{\TIME+1}) 
     -  h (\bW,\bX_{0})  \big\}}{
     \quad \times 
     \big\{ (1-D_{\TIME+1}') 
     -  h (\bW',\bX_{0}') \big\}
     \mathcal{K} \big( (\bZ,\bX_{0}) , (\bZ',\bX_{0}') \big)}
     \right] 
     \\
     &
     R_{q0}(q \con a_D)
     =
     \EXP
     \Big[ \big\{ 1  - \ind(A =a_D)  q(\bZ ,\bX_{0} ) \big\}
     \big\{ 1  -  \ind(A' =a_D) q(\bZ' ,\bX_{0}') \big\}
     \mathcal{K} \big( (\bW ,\bX_{0} ) , (\bW' ,\bX_{0}' ) \big)
     \Big] 
     \\
     &
     R_{q1}(q \con a_Y, a_D)
     =
     \EXP
     \left[ \mymatrixTwo{l}{\ind ( D_{\TIME+1}  = 0 ) \ind ( D_{\TIME+1}' =0 ) 
     \big\{ 
     \ind (A=a_D)
      q_0^* (\bZ ,\bX_{0} )
     -
     \ind (A=a_Y)
     q(\bZ ,\obX_{\TIME} ) \big\} }{\quad \times
     \big\{  
     \ind (A'=a_D)	q_0^* (\bZ' ,\bX_{0}' ) 
     -
     \ind (A'=a_Y)
     q(\bZ' ,\obX_{\TIME}' )  \big\} 
     \mathcal{K} \big( (\bW ,\obX_{\TIME} ) , (\bW' ,\obX_{\TIME}' ) \big)}
     \right] 
     \quad \text{ if $a_Y \neq a_D$}
     \ .
 \end{align*}
 Therefore, the empirical analogues of the population-level risk functions are given by: 
 \begin{align*}
     &
     \widehat{R}_{h1} \LSS (h \con a_Y)
     =
     \frac{1}{ M_k^2 }
     \sum_{i,j \in \mathcal{I}_k^c}
     \left[
     \mymatrixTwo{l}{\ind(A_i=a_Y, A_j=a_Y, D_{\TIME+1,i}=0, D_{\TIME+1,j}=0 )}{\quad \times
     \big\{ Y_i
     -  h (\bW_i,\obX_{\TIME,i})  \big\}
     \big\{ Y_j
     -  h (\bW_j,\obX_{\TIME,j}) \big\} 
     \mathcal{K} \big( (\bZ_i,\obX_{\TIME,i}) , (\bZ_j,\obX_{\TIME,j}) \big)}
     \right] 
     \\
     &
     \widehat{R}_{h0} \LSS (h \con a_Y, a_D)
     =
     \frac{1}{ M_k^2 }
     \sum_{i,j \in \mathcal{I}_k^c}
     \left[
     \mymatrixFour{l}{\ind(A_i=a_D, A_j=a_D ) }
     {\quad \times \big\{ (1-D_{\TIME+1,i}) \widehat{h}_1\LSS(\bW_i,\obX_{\TIME,i} \con a_Y)
     -  h (\bW_i,\bX_{0,i})  \big\}}{\quad \times
     \big\{ (1-D_{\TIME+1,j}) \widehat{h}_1\LSS(\bW_j,\obX_{\TIME,j} \con a_Y)
     -  h (\bW_j,\bX_{0,j}) \big\}}{\quad \times
     \mathcal{K} \big( (\bZ_i,\bX_{0,i}) , (\bZ_j,\bX_{0,j}) \big)}
     \right] 
     \quad \text{ if $a_Y \neq a_D$}
     \\
     &
     \widehat{R}_{h0} \LSS (h \con a_D, a_D)
     =
     \frac{1}{ M_k^2 }
     \sum_{i,j \in \mathcal{I}_k^c}
     \left[
     \mymatrixTwo{l}{\ind(A_i=a_D, A_j=a_D ) 
     \big\{ Y_i (1-D_{\TIME+1,i})
     -  h (\bW_i,\bX_{0,i})  \big\}}{\quad \times
     \big\{ Y_j (1-D_{\TIME+1,j}) 
     -  h (\bW_j,\bX_{0,j}) \big\}
     \mathcal{K} \big( (\bZ_i,\bX_{0,i}) , (\bZ_j,\bX_{0,j}) \big)}
     \right] 
     \\
     &
     \widehat{R}_{h2} \LSS (h \con a_D)
     =
     \frac{1}{ M_k^2 }
     \sum_{i,j \in \mathcal{I}_k^c}
     \left[ 
     \mymatrixTwo{l}{\ind(A_i=a_D, A_j=a_D ) 
     \big\{ (1-D_{\TIME+1,i})  
     -  h (\bW_i,\bX_{0,i})  \big\}}{
     \quad \times \big\{ (1-D_{\TIME+1,j})  
     -  h (\bW_j,\bX_{0,j}) \big\}
     \mathcal{K} \big( (\bZ_i,\bX_{0,i}) , (\bZ_j,\bX_{0,j}) \big)
     }
     \right] 
     \\
     &
     \widehat{R}_{q0} \LSS (q \con a_D)
     =
     \frac{1}{ M_k^2 }
     \sum_{i,j \in \mathcal{I}_k^c}
     \Big[ 
     \big\{ 1 - \ind(A_i = a_D) q(\bZ_i, \bX_{0,i}) \big\} 
     \big\{ 1 - \ind(A_j = a_D)  q(\bZ_j, \bX_{0,j}) \big\}
     \mathcal{K} \big( (\bW_i,\bX_{0,i}) , (\bW_j,\bX_{0,j}) \big)
     \Big]
     \\
     &
     \widehat{R}_{q1} \LSS (q \con a_Y, a_D)
     =
     \frac{1}{ M_k^2 }
     \sum_{i,j \in \mathcal{I}_k^c}
     \left[ 
     \mymatrixFour{l}{\ind ( D_{\TIME+1,i}  = 0 , D_{\TIME+1,j} =0)}{\quad \times 
     \big\{ 
     \ind(A_i = a_D) \widehat{q}_0\LSS (\bZ_i,\bX_{0,i} \con a_D)
     -
     \ind(A_i = a_Y)
     q(\bZ_i ,\obX_{\TIME,i} ) \big\} }{\quad \times
     \big\{ 
     \ind(A_j = a_D) \widehat{q}_0\LSS (\bZ_j,\bX_{0,j} \con a_D)
     -
     \ind(A_j = a_Y)
     q(\bZ_j ,\obX_{\TIME,j} ) \big\} }{\quad \times \mathcal{K} \big( (\bW_i,\obX_{\TIME,i}) , (\bW_j,\obX_{\TIME,j}) \big)}
     \right] 
     \quad \text{ if $a_Y \neq a_D$}
     \ .
 \end{align*} 
 We again note that these empirical risk functions are motivated from relationships \eqref{eq-bridgeft2 obs}-\eqref{eq-bridgeft6 obs}. Due to the nested relationships, the empirical risk functions for $h_0^*$ and $q_1^*$ rely on the estimated confounding bridge functions $\widehat{h}_1\LSS$ and $\widehat{q}_0\LSS$, respectively; we remark that a similar nonparametric approach for estimating nested functionals was considered in \citet{Singh2022}. 

 Note that all regularized empirical risk functions share the same form as
 \begin{align} \label{eq-EmpRisk}
 & 
 \widehat{R}\LSS \big( f (V_f), V_g \big) 
 + \lambda \big\| f \big\|_{\mathcal{H}}^2
 \nonumber
 \\
 &
 =
 \frac{1}{M_k^2}
 \sum_{i,j \in \mathcal{I}_k^c}
  \Big[ 
  \big\{ 
  S_i - T_i \cdot f(V_{f,i})
  \big\}
  \big\{ 
  S_j - T_j \cdot f(V_{f,j})
  \big\}
  \mathcal{K} (V_{g,i},V_{g,j})
  \Big] + \lambda \big\| f \big\|_{\mathcal{H}}^2
  \ .
 \end{align}
 Therefore, the estimated confounding bridge functions share the same closed form based on the representer theorem. Specifically, the empirical risk minimizers of \eqref{eq-EmpRisk} is given by $\widehat{f}\LSS(v_f) = \sum_{i \in \mathcal{I}_k^c}
     \widehat{\alpha}_{i} \mathcal{K} \big( V_{f,i}, v_f \big) $ 
     where $\widehat{\alpha}  =(\widehat{\alpha}_i)_{i \in \mathcal{I}_k^c}$ is equal to $
         \widehat{\alpha}
     =
     \big(
     \mathcal{F} \mathcal{T} \mathcal{G} \mathcal{T} \mathcal{F}
     +    
     M_k^{2}
     \lambda
     \mathcal{F} \big)^{-1}
     \big( 
     \mathcal{F} \mathcal{T} 
     \mathcal{G}   \mathcal{S} \big)$
 where 
 \begin{align*}
 & \mathcal{F}
 =
 \Big[
     \mathcal{K} \big( V_{f,i}, V_{f,j} \big)
 \Big]_{i,j \in \mathcal{I}_k^c}  
   \in \R^{M_k \times M_k}
 \quad , 
 &&
 \mathcal{G}
 =
 \Big[
     \mathcal{K} \big( V_{g,i}, V_{g,j} \big)
 \Big]_{i,j \in \mathcal{I}_k^c}  
   \in \R^{M_k \times M_k}
 \\
 &
 \mathcal{T} = \text{diag} \Big[ 
     T_i
 \Big]_{i \in \mathcal{I}_k^c}  \in \R^{M_k \times M_k}
 \quad , 
 &&
 \mathcal{S} = \Big[ S_i \Big]_{i \in \mathcal{I}_k^c}  \in \R^{M_k} \ .
 \end{align*}
 Therefore, the confounding bridge function estimators are readily available by using $(V_f,V_g,S,T)$ accordingly; see the table below for details:
 \begin{table}[!htp]
     \renewcommand{\arraystretch}{1.4} \centering
     \footnotesize
     \setlength{\tabcolsep}{3pt}
 \begin{tabular}{|c|c|c|c|c|c|}
 \hline
 Setting                        & Nuisance Function                            & $V_f$ & $V_g$ & $S$                                                                                                                                            & $T$                \\ \hline
 \multirow{6}{*}{Observational} & $h_1^*(\cdot \con a_Y)$                      & $W,\obX_{\TIME}$ & $Z,\obX_{\TIME}$ & $\ind(A=a_Y)(1-D_{\TIME+1})Y$                                                                                                                            & $\ind(A=a_Y)(1-D_{\TIME+1})$ \\ \cline{2-6} 
 & $h_0^*(\cdot \con a_Y,a_D)$ $(a_Y \neq a_D)$ & $W,\bX_{0}$ & $Z,\bX_{0}$ & $\ind(A=a_D)(1-D_{\TIME+1})\widehat{h}_1\LSS(W,\obX_{\TIME} \con a_Y)$                                                                                              & $\ind(A=a_D)$      \\ \cline{2-6} 
 & $h_0^*(\cdot \con a_D,a_D)$                  & $W,\bX_{0}$ & $Z,\bX_{0}$ & $\ind(A=a_D)(1-D_{\TIME+1})Y$                                                                                                                            & $\ind(A=a_D)$      \\ \cline{2-6} 
 & $h_2^*(\cdot \con a_D)$                      & $W,\bX_{0}$ & $Z,\bX_{0}$ & $\ind(A=a_D)(1-D_{\TIME+1})$                                                                                                                             & $\ind(A=a_D)$      \\ \cline{2-6} 
 & $q_0^*(\cdot \con a_D)$                      & $Z,\bX_{0}$ & $W,\bX_{0}$ & $1$                                                                                                                                            & $\ind(A=a_D)$      \\ \cline{2-6} 
 & $q_1^*(\cdot \con a_Y,a_D)$ $(a_Y \neq a_D)$ & $Z,\obX_{\TIME}$ & $W,\obX_{\TIME}$ & $\ind(A=a_D)(1-D_{\TIME+1}) \widehat{q}_0\LSS(Z,\bX_{0} \con a_D)$                                                                                             & $\ind(A=a_Y)(1-D_{\TIME+1})$ \\ \hline
 \multirow{2}{*}{Experimental}  & $\xi_1^*(\cdot \con a_Y)$                    & $W,\obX_{\TIME}$ & $Z,\obX_{\TIME}$ & $\ind(A=a_Y)(1-D_{\TIME+1})Y$                                                                                                                            & $\ind(A=a_Y)(1-D)$ \\ \cline{2-6} 
 & $p_1^*(\cdot \con a_Y,a_D)$ $(a_Y \neq a_D)$ & $Z,\obX_{\TIME}$ & $W,\obX_{\TIME}$ & $\ind(A=a_D)(1-D_{\TIME+1}) /\Pr(A=a_D)$ & $\ind(A=a_Y)(1-D_{\TIME+1})$ \\ \hline
 \end{tabular}
 \caption{\footnotesize Variable Specification for Empirical Risk Minimization.}
 \label{tab:variables}
 \end{table} 
 
 These closed-form representations can be obtained as follows. From the representer theorem, we have $\widehat{f}\LSS(V_{f,i}) = (\mathcal{F} \widehat{\alpha})_i$ for $i \in \mathcal{I}_k\LSS$. Then, the regularized empirical risk is evaluated at $\widehat{\alpha}$ is
 \begin{align*}
 \widehat{R} \LSS (\widehat{\alpha})
 + \lambda \big\| \widehat{f}\LSS \big\|_{\mathcal{H}}^2 
 &
 =
 \frac{1}{M_k^2}
 \sum_{i,j \in \mathcal{I}_k^c}
  \Big[ 
  \big\{ 
  S_i - T_i \cdot \widehat{f}\LSS(V_{f,i})
  \big\}
  \big\{ 
  S_j - T_j \cdot \widehat{f}\LSS(V_{f,j})
  \big\}
  \mathcal{K} (V_{g,i},V_{g,j})
  \Big] + \lambda \big\| \widehat{f}\LSS \big\|_{\mathcal{H}}^2
  \\
 &
 =
 \frac{1}{M_k^2}
 \sum_{i,j \in \mathcal{I}_k^c}
  \Big[ 
  \big\{ 
  S_i - T_i \cdot (\mathcal{F} \widehat{\alpha})_i
  \big\}
  \big\{ 
  S_j - T_j \cdot (\mathcal{F} \widehat{\alpha})_j
  \big\}
  \mathcal{K} (V_{g,i},V_{g,j})
  \Big] + \lambda \widehat{\alpha}\T \mathcal{F} \widehat{\alpha}
   \\
 &
 =
 \frac{1}{M_k^2}
 \sum_{i,j \in \mathcal{I}_k^c}
  \Big[ 
  \big\{ 
  S_i - ( \mathcal{T} \cdot \mathcal{F} \widehat{\alpha})_i
  \big\}
  \mathcal{G}_{ij}
  \big\{ 
  S_j - ( \mathcal{T} \cdot \mathcal{F} \widehat{\alpha})_j
  \big\} 
  \Big] + \lambda \widehat{\alpha}\T \mathcal{F} \widehat{\alpha}
    \\
 &
 =
 \frac{ \big(
  \mathcal{S} - \mathcal{T} \mathcal{F} \widehat{\alpha}
  \big)\T
  \mathcal{G}
  \big(
  \mathcal{S} - \mathcal{T} \mathcal{F} \widehat{\alpha}
  \big)
  + M_k^{2} \lambda \widehat{\alpha}\T \mathcal{F} \widehat{\alpha} }{M_k^{2}}
    \ .
 \end{align*}
 Therefore, it is trivial that $\widehat{\alpha}
 =
 \big(
     \mathcal{F} \mathcal{T} \mathcal{G} \mathcal{T} \mathcal{F}
     +    
     M_k^{2}
     \lambda
     \mathcal{F} \big)^{-1}
     \big( 
     \mathcal{F} \mathcal{T} 
     \mathcal{G}   \mathcal{S} \big)$ is the minimizer of the above quadratic function.

\subsubsection{Convergence Rates for the PMMR Estimators} \label{sec:supp:convergence}

We complement the results in Section \ref{sec:convergence:CBF} by providing further discussion and presenting the convergence rate of the nested PMMR estimators. We present results only for $\widehat{h}_1\LSS$ and $\widehat{h}_0\LSS(\cdot \con a_Y)$, as the corresponding results for other confounding bridge function estimators can be established analogously. Following Section \ref{sec:inference} of the main paper, we denote $\mathcal{H} (V)$ be a RKHS associated with a random variable $V$, equipped with the Gaussian kernel, given by $\mathcal{K}(v,v') = \exp \big\{ - \| v - v' \|_2^2 / \kappa \big\}$, where $\kappa \in (0,\infty)$ is a bandwidth parameter. Define the operators $\mathcal{S}_1: \HH(W,  \obX_{\TIME}) \rightarrow \HH (Z, \obX_{\TIME})$ and  $\mathcal{S}_0: \HH(W, \bX_{0}) \rightarrow \HH(Z, \bX_{0})$ by
\begin{align*}
&
\mathcal{S}_1(h) = \EXP \big\{ h(W,\obX_{\TIME}) \cond Z,A=a_Y,D_{\TIME+1}=0,\obX_{\TIME} \big\} \ ,
\\
&
\mathcal{S}_0(h) = \EXP \big\{ h(W,\bX_{0}) \cond Z,A=a_D,\bX_{0} \big\}
\ .
\end{align*}
Let  $\mathcal{S}_1^{\text{ad}}: \HH (Z, \obX_{\TIME})\rightarrow \HH(W,  \obX_{\TIME}) $ and $\mathcal{S}_0^{\text{ad}}: \HH(Z, \bX_{0}) \rightarrow \HH(W, \bX_{0})$ be the adjoint operator of $\mathcal{S}_1$ and $\mathcal{S}_0$, respectively.  Following \citet{PMMR2021}, \citet{Bennett2023}, and and \citet{Rotnizky2025}, we consider the following source conditions:
\begin{itemize}[leftmargin=1.5cm]
	\item[\hypertarget{SC1}{(SC-$\mathcal{S}_1$)}] The minimal norm solution  $h_1^*$ in \eqref{eq-bridgeft1 obs} satisfies $h_1^* \in \text{Range}( (\mathcal{S}_1^{\text{ad}} \mathcal{S}_1)^{\beta/2} )$ some $\beta > 0$.
	\item[\hypertarget{SC2}{(SC-$\mathcal{S}_0$)}] The minimal norm solution  $h_0^*(\cdot \con a_Y)$ in \eqref{eq-bridgeft2 obs} satisfies $h_0^*(\cdot \con a_Y) \in \text{Range}( (\mathcal{S}_0^{\text{ad}} \mathcal{S}_0)^{\beta/2} )$ some $\beta > 0$.
\end{itemize}
See \cite{Plato2025} and references therein for the definition of the fractional powers of the operator. Roughly speaking, condition \hyperlink{SC1}{(SC-$\mathcal{S}_1$)} measures how strongly $h_1^*$ aligns with the well-identified directions of the inverse problem, namely those along which $(\mathcal{S}_1^{\text{ad}} \mathcal{S}_1)$ exhibits the greatest stability. The same interpretation applies to \hyperlink{SC2}{(SC-$\mathcal{S}_0$)}.

In order to enhance interpretation of the source condition, we illustrate under the case when $\mathcal{S}_1$ be a compact operator. Then, it admits a singular value decomposition, denoted by $\big\{ \mu_{n},\phi_{n},g_{n} \big\}_{n=1,2,\ldots}$ where $\mu_1 \geq \mu_2 \geq \ldots \geq 0$ are the singular values, and $\phi_n$ and $g_n$ form an orthonormal basis of $\HH(\bW,\obX_{\TIME})$ and $\HH(\bZ,\obX_{\TIME}) $, respectively \citep{Kress2014}. Then, following \citet{Carrasco2007},  \hyperlink{SC1}{(SC-$\mathcal{S}_1$)} can be written as $h_1^* \in \mathcal{H}_{\beta} (\mathcal{S}_1)$ where 
\begin{align}
    & 
    \mathcal{H}_{\beta} (\mathcal{S}_1)
	=
	\Bigg\{ 
		h_1 \in \mathcal{N}(\mathcal{S}_1)^{\perp}
		\, \Bigg| \, 
		\sum_{n=1}^{\infty} \frac{ \langle h_1 , \phi_{n} \rangle }{ \mu_{n}^{2 \beta}} < \infty 
	\Bigg\} \ , 
 \label{eq-regularity space}
\end{align}
where $\mathcal{N}(\mathcal{S}_1)$ is the null space of $\mathcal{S}_1$. Note that $\mathcal{H}_{\beta}(\mathcal{S}_1)$ is a decreasing space as $\beta$ increases. In particular, as $\beta$ goes to infinity, condition \eqref{eq-regularity space} implies that $h_1^*$ behaves as being supported only on a finite set of basis functions $\phi_n$, and excluding the possibility that $h_1^*$ has substantial components in directions with near-zero singular values of $\mathcal{T}_1$. Therefore, the parameter $\beta$ controls the degree of ill-posedness, where larger $\beta$ implies that \eqref{eq-defh1} is more well-posed.  A similar interpretation applies to $\mathcal{H}_{\beta}(\mathcal{S}_0)$.

We now restate Theorem \ref{thm:convergence:1} from the main paper, which characterizes the convergence rate of $\widehat{h}_1\LSS$. For notational convenience, we write $a \vee b = \max(a, b)$ and $a \wedge b = \min(a, b)$.

\begin{theorem} \label{lemma:convergence:1}
Suppose that the observed data $\bO$ have compact support and a uniformly bounded density. Further suppose that  \hyperlink{SC1}{(SC-$\mathcal{S}_1$)} holds for some $\beta_{h1}>0$. If the regularization parameter $\lambda_{h1}$ for the PMMR estimator $\widehat{h}_1\LSS$ in \eqref{eq-EmpRisk} is chosen as $\lambda_{h1}  = N^{-1/(1+ \beta_{h1} \wedge 2) }$, $\widehat{h}_1\LSS$ achieves a convergence rate of 
\begin{align*}
	\big\| \widehat{h}_1\LSS - h_1^* \big\|_{P,2} 
	=
	O_P \bigg(
	N^{ \text{\scalebox{1.35}{$- \frac{ \beta_{h1}\wedge 2 }{2+2 (\beta_{h1} \wedge 2)}$}} }
	\bigg)
	=
	\left\{
	\begin{array}{ll}	
	O_P( N^{-\beta_{h1}/(2+2\beta_{h1})} )
	&
	\quad
	\text{ if $\beta_{h1} \in (0,2]$}
	\\
	O_P( N^{-1/3} )
	&
	\quad
	\text{ if $\beta_{h1} \in (2,\infty)$} 
	\end{array}
	\right. \ .
\end{align*}  
\end{theorem}

Next, we study the convergence rate of $\widehat{h}_{0}\LSS(\cdot \con a_Y)$. In order to do so, we first define $\mathcal{H}_c = \big\{ f \in \HH (W, \obX_{\TIME}) \cond \|f \|_{\mathcal{H}} \leq c \big\}$, which is a collection of functions in $\HH (W, \obX_{\TIME})$ with a finite RKHS norm. We consider constant $c$ is chosen sufficiently large so that $h_1^* \in \mathcal{H}_c$ and $\widehat{h}_1\LSS \in \mathcal{H}_c$, where we can consider that the PMMR estimator $\widehat{h}_1\LSS$ is estimated over $\mathcal{H}_c$.

 \begin{theorem} \label{lemma:convergence:2}
 Suppose that the conditions in Theorem \ref{lemma:convergence:1} hold. Let $r_{h1}$ be the convergence rate of $\widehat{h}_1\LSS$, i.e., $\big\| \widehat{h}_{1}\LSS - h_1^* \big\|_{P,2} = O_P(r_{h1})$. 
 Further suppose that  \hyperlink{SC2}{(SC-$\mathcal{S}_0$)} holds for some $\beta_{h0}(a_Y)>0$. If the regularity parameter $\lambda_{h0}(a_Y)$ for the PMMR estimator $\widehat{h}_{0}\LSS(\cdot \con a_Y)$ in \eqref{eq-EmpRisk} is chosen as $\lambda_{h0}(a_Y) = r_{h1}^{ 2 / \{ 1+\beta_{h0}(a_Y) \wedge 2 \} }$, $\widehat{h}_{0}\LSS(\cdot \con a_Y)$ achieves a convergence rate of 
\begin{align*}
&
\big\| h_0^*(\cdot \con a_Y) - \widehat{h}_0\LSS(\cdot \con a_Y) \big\|_{P,2}
=
O_P \bigg( 
r_{h1}^{ \text{\scalebox{1.35}{$\frac{ \beta_{h0}(a_Y)\wedge 2 }{1+\{ \beta_{h0}(a_Y) \wedge 2\} }$}} }
\bigg)
=
	 \left\{
	 	\begin{array}{ll}
	 	O_P \big( 
		r_{h1}^{\beta_{h0}/(\beta_{h0}+1) }
		\big)
	 		& \text{ if $\beta_{h0} \in (0,2]$}
	 		\\
	 		O_P \big( 
	 	r_{h1}^{2/3}
	 	\big)
	 		& \text{ if $\beta_{h0} \in (2,\infty)$}
	 	\end{array}
	 \right.  
	  \ .
\end{align*}
\end{theorem}

The proofs of these Theorems can be found in Sections \ref{sec:proof:lemma:convergence:1} and \ref{sec:proof:lemma:convergence:2}. 
The convergence rate established in Theorem \ref{lemma:convergence:1} refines the original PMMR framework of \citet{PMMR2021} derived a convergence bound for $\widehat{h}_1\LSS$ that is at most $O_P(N^{-1/4})$, even in mildly ill-posed settings (i.e., when $\beta_{h1}$ is sufficiently large); see Section \ref{sec:proof:lemma:convergence:1} for the technical details underlying this refinement.  The convergence rate established in Theorem \ref{lemma:convergence:2} is novel, as the original PMMR framework did not consider the nested estimation of ill-posed integral equations.

We use $\beta_{h0}(a_D), \beta_{h2}, \beta_{q0}, \beta_{q1}, \beta_{\xi 1}, \beta_{p1}$ to denote the parameters satisfying the respective source conditions for their corresponding confounding bridge functions $h_0(\cdot \con a_D), h_2, q_0, q_1, \xi_1,$ and $p_1$, analogous to \hyperlink{SC1}{(SC-$\mathcal{S}_1$)} and \hyperlink{SC2}{(SC-$\mathcal{S}_0$)}. Building on similar results from Theorems \ref{lemma:convergence:1} and \ref{lemma:convergence:2}, the following convergence rates are achieved when the regularization parameters are chosen appropriately.
\begin{table}[!htp]
 \renewcommand{\arraystretch}{2.5} \centering
 \footnotesize
 \setlength{\tabcolsep}{2pt}
\begin{tabular}{|c|c|c|c|c|}
\hline
Setting                        & Nuisance Function    & Regularity parameter                        & Rate of $\lambda$                & Estimation error rate                                                         \\ \hline
\multirow{6}{*}{Observational} & $h_1^*(\cdot \con a_Y)$      & $\beta_{h1}$                & $N^{\text{\scalebox{1.35}{$-\frac{1}{1+\beta_{h1} \wedge 2}$}}}$           & $r_{h1} = N^{ \text{\scalebox{1.35}{$- \frac{ \beta_{h1}\wedge 2 }{2+2 (\beta_{h1} \wedge 2)}$}} } $                    \\ \cline{2-5} 
& $h_0^*(\cdot \con a_Y,a_D)$ $(a_Y \neq a_D)$      & $\beta_{h0}(a_Y) $ & $r_{h1}^{\text{\scalebox{1.35}{$\frac{2}{1+\beta_{h0}(a_Y) \wedge 2}$}}}$ & $r_{h0}(a_Y) = r_{h1}^{ \text{\scalebox{1.35}{$\frac{ \beta_{h0}(a_Y)\wedge 2 }{1+(\beta_{h0}(a_Y) \wedge 2)}$}} } $ \\ \cline{2-5} 
& $h_0^*(\cdot \con a_D,a_D)$        & $\beta_{h0}(a_D) $         & $N^{\text{\scalebox{1.35}{$-\frac{1}{1+\beta_{h0}(a_D) \wedge 2}$}}}$      & $r_{h0}(a_D) = N^{ \text{\scalebox{1.35}{$- \frac{ \beta_{h0}(a_D)\wedge 2 }{2+2 (\beta_{h0}(a_D) \wedge 2)}$}} } $     \\ \cline{2-5} 
& $h_2^*(\cdot \con a_D)$         & $\beta_{h2} $ &                $N^{\text{\scalebox{1.35}{$-\frac{1}{1+\beta_{h2} \wedge 2}$}}}$           & $r_{h2} = N^{ \text{\scalebox{1.35}{$- \frac{ \beta_{h2}\wedge 2 }{2+2 (\beta_{h2} \wedge 2)}$}} } $                \\ \cline{2-5} 
& $q_0^*(\cdot \con a_D)$          & $\beta_{q0} $                   & $N^{\text{\scalebox{1.35}{$-\frac{1}{1+\beta_{q0} \wedge 2}$}}}$           & $r_{q0} = N^{ \text{\scalebox{1.35}{$- \frac{ \beta_{q0}\wedge 2 }{2+2 (\beta_{q0} \wedge 2)}$}} } $               \\ \cline{2-5} 
& $q_1^*(\cdot \con a_Y,a_D)$ $(a_Y \neq a_D)$       & $\beta_{q1} $ & $r_{q0}^{\text{\scalebox{1.35}{$\frac{2}{1+\beta_{q1} \wedge 2}$}}}$     & $r_{q1} = r_{q0}^{ \text{\scalebox{1.35}{$\frac{ \beta_{q1}\wedge 2 }{1+ (\beta_{q1} \wedge 2)}$}} } $ \\ \hline
\multirow{2}{*}{Experimental}  & $\xi_1^*(\cdot \con a_Y)$             & $\beta_{\xi 1} $              & $N^{\text{\scalebox{1.35}{$-\frac{1}{1+\beta_{\xi 1} \wedge 2}$}}}$         & $r_{\xi 1} = N^{ \text{\scalebox{1.35}{$- \frac{ \beta_{\xi 1}\wedge 2 }{2+2 (\beta_{\xi 1} \wedge 2)}$}} } $             \\ \cline{2-5} 
& $p_1^*(\cdot \con a_Y,a_D)$ $(a_Y \neq a_D)$       & $\beta_{p1} $  & $N^{\text{\scalebox{1.35}{$-\frac{1}{1+\beta_{p1} \wedge 2}$}}}$           & $r_{p1} = N^{ \text{\scalebox{1.35}{$- \frac{ \beta_{p1}\wedge 2 }{2+2 (\beta_{p1} \wedge 2)}$}} } $               \\ \hline
\end{tabular}
\end{table}

In particular, if all $\beta$ parameters are greater than 2 and the regularization parameters are properly chosen, the PMMR estimators achieve the following convergence rates:
\begin{align*}
    &
    \big\| \widehat{h}_1 \LSS - h_1^* \big\|_{P,2} = O_P(N^{-1/3}) \ , 
    &&
    \big\| \widehat{h}_0 \LSS (\cdot \con a_Y) - h_0^*(\cdot \con a_Y) \big\|_{P,2} = O_P(N^{-2/9}) 
     \ , 
     \\
    &
    \big\| \widehat{h}_0 \LSS (\cdot \con a_D) - h_0^*(\cdot \con a_D) \big\|_{P,2} = O_P(N^{-1/3}) 
     \ , 
    &&
    \big\| \widehat{h}_2 \LSS - h_2^* \big\|_{P,2} = O_P(N^{-1/3}) 
     \ , 
     \\
    &
    \big\| \widehat{q}_0 \LSS - q_0^* \big\|_{P,2} = O_P(N^{-1/3}) 
    \ , 
    &&
    \big\| \widehat{q}_1 \LSS - q_1^* \big\|_{P,2} = O_P(N^{-2/9}) \ ,
    \\
    &
    \big\| \widehat{\xi}_1 \LSS - \xi_1^* \big\|_{P,2} = O_P(N^{-1/3}) 
    \ , 
    &&
    \big\| \widehat{p}_1 \LSS - p_1^* \big\|_{P,2} = O_P(N^{-1/3}) \ .
\end{align*}
Note that conditions \HL{A9}-\HL{A11} and \HL{A9'}-\HL{A10'} are satisfied with these convergence rates.

 \subsection{Alternative Estimation Approach} \label{sec:supp:minmax}

 As mentioned in Section \ref{sec:estimator} of the main paper, other methods can be employed to construct confounding bridge function estimators. As a concrete example, we present details of the minimax kernel machine learning approach proposed by \citet{Ghassami2022} adapted to our setting. Minimax estimator of the confounding bridge functions are obtained as solutions to the following regularized minimax optimization problems: 
 \begin{align*}
 & 
 \widehat{h}_{1}\LSS (\cdot \con a_Y)
 \\
 &
 =
 \argmin_{h \in \mathcal{H}(W,\obX_{\TIME})}
 \left[
     \begin{array}{l}
     \displaystyle{ \max_{q \in \mathcal{H}(Z,\obX_{\TIME})}  }
     \Bigg[
     \begin{array}{l}		
         \AVER\LSS \big[
         q (Z,\obX_{\TIME})
         \ind(A=a_Y,D_{\TIME+1}=0)
         \big\{
             Y - h(W,\obX_{\TIME})
         \big\} - q^2(Z,\obX_{\TIME}) \big]
         \\
         -
         \lambda_{q} \big\| q \big\|_{\mathcal{H}(Z,\obX_{\TIME})}^2
     \end{array}
     \Bigg]
     \\
     +
     \lambda_{h} \big\| h \big\|_{\mathcal{H}(W,\obX_{\TIME})}^2
     \end{array}
 \right] 
 \\
 &
 \widehat{h}_{0}\LSS  (\cdot \con a_Y, a_D)
 \\
 &
 =
 \argmin_{h \in \mathcal{H}(\bW,\bX_{0})}
 \left[
     \begin{array}{l}
     \displaystyle{ \max_{q \in \mathcal{H}(\bZ,\bX_{0})}  }
     \Bigg[
     \begin{array}{l}		
         \AVER\LSS \big[
         q (Z,\bX_{0})
         \ind(A=a_D)
         \big\{
             (1-D_{\TIME+1})
             \widehat{h}_1 \LSS(W,\obX_{\TIME} \con a_Y)
             - h(W,\bX_{0})
         \big\} - q^2(Z,\bX_{0}) \big]
         \\
         -
         \lambda_{q} \big\| q \big\|_{\mathcal{H}(Z,\bX_{0})}^2
     \end{array}
     \Bigg]
     \\
     +
     \lambda_{h} \big\| h \big\|_{\mathcal{H}(W,\bX_{0})}^2
     \end{array}
 \right] 
 \\
 &
 \widehat{h}_{0}\LSS  (\cdot \con a_D, a_D)
 \\
 &
 =
 \argmin_{h \in \mathcal{H}(\bW,\bX_{0})}
 \left[
     \begin{array}{l}
     \displaystyle{ \max_{q \in \mathcal{H}(\bZ,\bX_{0})}  }
     \Bigg[
     \begin{array}{l}		
         \AVER\LSS \big[
         q (Z,\bX_{0})
         \ind(A=a_D)
         \big\{
            Y (1-D_{\TIME+1})
             - h(W,\bX_{0})
         \big\} - q^2(Z,\bX_{0}) \big]
         \\
         -
         \lambda_{q} \big\| q \big\|_{\mathcal{H}(Z,\bX_{0})}^2
     \end{array}
     \Bigg]
     \\
     +
     \lambda_{h} \big\| h \big\|_{\mathcal{H}(W,\bX_{0})}^2
     \end{array}
 \right] 
 \\
 &
 \widehat{h}_{2}\LSS  (\cdot \con a_D)
 \\
 &
 =
 \argmin_{h \in \mathcal{H}(\bW,\bX_{0})}
 \left[
     \begin{array}{l}
     \displaystyle{ \max_{q \in \mathcal{H}(\bZ,\bX_{0})}  }
     \Bigg[
     \begin{array}{l}		
         \AVER\LSS \big[
         q (Z,\bX_{0})
         \ind(A=a_D)
         \big\{
            (1-D_{\TIME+1})
             - h(W,\bX_{0})
         \big\} - q^2(Z,\bX_{0}) \big]
         \\
         -
         \lambda_{q} \big\| q \big\|_{\mathcal{H}(Z,\bX_{0})}^2
     \end{array}
     \Bigg]
     \\
     +
     \lambda_{h} \big\| h \big\|_{\mathcal{H}(W,\bX_{0})}^2
     \end{array}
 \right] 
 \\
 &
 \widehat{q}_{0}\LSS  (\cdot \con a_D)
 \\
 &
 =
 \argmin_{q \in \mathcal{H}(Z,\bX_{0})}
 \left[
     \begin{array}{l}
     \displaystyle{ \max_{h \in \mathcal{H}(W,\bX_{0})}  }
     \Bigg[
     \begin{array}{l}		
         \AVER\LSS \big[
         h (W,\bX_{0}) 
         \big\{
             1
             - \ind(A=a_D) q(Z,\bX_{0})
         \big\} - h^2(W,\bX_{0}) \big]
         \\
         -
         \lambda_{h} \big\| h \big\|_{\mathcal{H}(W,\bX_{0})}^2
     \end{array}
     \Bigg]
     \\
     +
     \lambda_{q} \big\| q \big\|_{\mathcal{H}(Z,\bX_{0})}^2
     \end{array}
 \right] 
 \\
 &
 \widehat{q}_{1}\LSS  (\cdot \con a_Y, a_D)
 \\
 &
 =
 \argmin_{q \in \mathcal{H}(Z,\obX_{\TIME})}
 \left[
     \begin{array}{l}
     \displaystyle{ \max_{h \in \mathcal{H}(W,\obX_{\TIME})}  }
     \left[
     \begin{array}{l}		
         \AVER\LSS \left[
         h (W,\obX_{\TIME}) 
         (1-D_{\TIME+1})
         \bigg\{
         \begin{array}{l}			
             \ind(A=a_D) \widehat{q}_0\LSS (Z,\bX_{0} \con a_D)
             \\
             - \ind(A=a_Y) q(Z,\obX_{\TIME})
         \end{array}
         \bigg\} - h^2(W,\obX_{\TIME}) \right]
         \\
         -
         \lambda_{h} \big\| h \big\|_{\mathcal{H}(W,\obX_{\TIME})}^2
     \end{array}
     \right]
     \\
     +
     \lambda_{q} \big\| q \big\|_{\mathcal{H}(Z,\obX_{\TIME})}^2
     \end{array}
 \right] 
 \end{align*} 
 where $\AVER\LSS(V) = \big| \mathcal{I}_k^c \big|^{-1} \sum_{i \in \mathcal{I}_k^c} V_i$. Note that all estimated functions are estimated based on a form
 \begin{align} \label{eq-Ghassami}
 & 
 \widehat{f} \LSS (v_f) 
 =
 \argmin_{f \in \mathcal{H}(V_f)}
 \Big[
 \max_{g \in \mathcal{H}(V_g)}
 \Big[
 \AVER\LSS \big[   g (V_g) \cdot \big\{ S - T \cdot f (V_f) \big) \big\} - g^2 (V_g) \big] - \lambda_g \big\| g \big\|_{\mathcal{H}_g}^2
 \Big]
 + \lambda_f \big\| f \big\|_{\mathcal{H}_f}^2
 \Big] \ .
 \end{align}
 A closed-form representation of the solution to \eqref{eq-Ghassami} is  $\widehat{f} \LSS (v_f) = \sum_{i \in \mathcal{I}_k^c} \widehat{\gamma}_i \mathcal{K} (V_{f,i}, v_{f})$ where $\widehat{\gamma} = \big( \widehat{\gamma}_i \big)_{i \in \mathcal{I}_k^c}$ is equal to $\widehat{\gamma}
 = 
 \big(
 \mathcal{F} \mathcal{T} \mathcal{G}_{\lambda_g} \mathcal{T}   \mathcal{F}
 + M_k^2 \lambda_f  \mathcal{F} \big)^{\dagger}
 \mathcal{F} \mathcal{T} \mathcal{G}_{\lambda_g} \mathcal{S}$ 
 where 
 \begin{align*}
 & 
 \mathcal{G}_{\lambda_g} = 0.25 \mathcal{G} \big\{  M_k^{-1}   \mathcal{G} + \lambda_g I_{M_k \times M_k} \big\}^{-1}
   \in \R^{M_k \times M_k}
 \\
 & \mathcal{F}
 =
 \Big[
     \mathcal{K} \big( V_{f,i}, V_{f,j} \big)
 \Big]_{i,j \in \mathcal{I}_k^c}  
   \in \R^{M_k \times M_k}
 \quad , 
 &&
 \mathcal{G}
 =
 \Big[
     \mathcal{K} \big( V_{g,i}, V_{g,j} \big)
 \Big]_{i,j \in \mathcal{I}_k^c}  
   \in \R^{M_k \times M_k}
 \\
 &
 \mathcal{T} = \text{diag} \Big[ 
     T_i
 \Big]_{i \in \mathcal{I}_k^c}  \in \R^{M_k \times M_k}
 \quad , 
 &&
 \mathcal{S} = \Big[ S_i \Big]_{i \in \mathcal{I}_k^c}  \in \R^{M_k} \ .
 \end{align*}
 Therefore, minimax estimators of the confounding bridge functions are readily available by using $(V_f,V_g,S,T)$ as specified in Table \ref{tab:variables}. 

 Of note, the minimax estimation requires two regularization parameters $\lambda_g$ and $\lambda_f$ whereas the PMMR estimation only requires one regularization parameter. Therefore, the PMMR estimation is easier to implement as it requires fewer hyperparameters than the minimax estimation; see Section \ref{sec:supp:hyperparameter} for details on how to choose hyperparameters.

\subsection{Practical Considerations: Hyperparameter Tuning, Median Adjustment, and Multiplier Bootstrap} \label{sec:supp:practical}

In this section, we provide details on hyperparameter tuning, median adjustment for cross-fitting estimators, and the multiplier bootstrap procedure for inference.

\subsubsection{Hyperparameter Tuning via Repeated Cross-validation}	\label{sec:supp:hyperparameter}
 
The PMMR estimation requires hyperparameters, specifically the bandwidth parameters of the gram matrix and the regularization parameter. However, as discussed in Section 5 of \citet{PMMR2021}, developing a systematic hyperparameter tuning procedure remains an open question. While this is an interesting direction for future research, it is beyond the scope of this paper. Instead, we employ a cross-validation procedure for hyperparameter selection, similar to the approach proposed by \citet{Ghassami2022}, and assess its validity by evaluating the estimator's performance through a simulation study.

First, we provide details of the cross-validation procedure. Briefly, we use repeated cross-validation so that the empirical risk over the held-out validation set is minimized. We denote the number of the cross-validation repetitions by $L$ and the number of cross-validation folds by $J$. For each repetition indexed by $\ell = 1,\ldots,L$, let $\mathcal{J}_{\ell 1},\ldots,\mathcal{J}_{\ell J}$ be the $J$ cross-validation folds that partition the estimation fold $\mathcal{I}_k^c$. Let $f^{\ell,(-j)}(\kappa_{f},\kappa_{g},\lambda)$ $(\ell=1,\ldots,L, j=1,\ldots,J)$ be the regularized empirical risk minimizer of \eqref{eq-EmpRisk} where $\mathcal{I}_k^c \setminus \mathcal{J}_{\ell j} = \bigcup_{j' \neq j} \mathcal{J}_{\ell j'}$ are used and $(\kappa_{f}, \kappa_{g}, \lambda)$ are used as the bandwidth parameters of the Gaussian kernel functions of $\mathcal{H}(V_f)$, $\mathcal{H}(V_g)$, and the regularization parameter, respectively.  Recall that the Gaussian kernel function is $\mathcal{K} (\bv, \bv' \con \kappa )
 	=
 	\exp
 	\big\{ \kappa^{-1} \big\| \bv - \bv' \big\|_2^2
 	\big\}$ where $\kappa$ is the bandwidth parameter.  Then, the risk of $\widehat{f}_{1}^{\ell,(-j)}(\kappa_{f}, \kappa_{g}, \lambda)$ evaluated over $\mathcal{J}_{\ell j}$ is given as
 \begin{align*}
 &
 \widehat{R}^{\ell,(j)} ( \widehat{f}^{\ell,(-j)}( \kappa_{f}, \kappa_{g}, \lambda ) ) 
 =
 \frac{1}{ | \mathcal{J}_{\ell,j} |^2 }
 \sum_{ (i,i') \in \mathcal{J}_{\ell,j} }
 \Big[ 
  \big\{ 
  S_i - T_i \cdot f(V_{f,i})
  \big\}
  \big\{ 
  S_{i'} - T_{i'} \cdot f(V_{f,{i'}})
  \big\}
  \mathcal{K} (V_{g,i},V_{g,{i'}})
  \Big]
  \ .
 \end{align*}
 This risk $\widehat{R}^{\ell,(j)}(\widehat{f}^{\ell,(-j)}(\kappa_{f}, \kappa_{g}, \lambda) )$ can be understood as a V-statistic. The aggregated cross-validation risk is
 \begin{align*}
 \widehat{R} (\kappa_{f}, \kappa_{g}, \lambda)
 := 
 \frac{1}{L J} 
 \sum_{\ell=1}^{L}
 \sum_{j=1}^{J}
 \widehat{R}^{\ell,(j)}(\widehat{f}^{\ell,(-j)}(\kappa_{f}, \kappa_{g}, \lambda) ) \ .
 \end{align*}
 The hyperparameters are selected as the minimizer of $\widehat{R} (\kappa_{f}, \kappa_{g}, \lambda)$, i.e., $\argmin_{
 (\kappa_{f}, \kappa_{g}, \lambda) \in \Omega }
 \widehat{R} (\kappa_{f}, \kappa_{g}, \lambda)$ 
 where $\Omega$ is a user-specified space of the hyperparameters. 
 
Of note, one can select the range of the bandwidth parameter based on the median heuristic \citep{Garreau2018}. Specifically, the median pairwise distance $\kappa_{f}^{(\text{med})} := \median \big\{ \| V_{f,i} -  V_{f,i'} \|_2^2 \cond i, i' \in \mathcal{I}_k^c, i \neq i' \big\}$ and $\kappa_{g}^{(\text{med})} := \median \big\{ \| V_{g,i} - V_{g,i'} \|_{2}^2 \cond i, i' \in \mathcal{I}_k^c, i \neq i' \big\}$ are calculated and these values are used to set the scale of the bandwidth parameter values. In the simulation study and the data analysis, we set $L=J=5$, and define $\Omega$ as $\Omega = \Omega_{f} \otimes \Omega_{g} \otimes \Omega_{\lambda}$ where $
 	\Omega_{f}  = \kappa_f^{(\text{med})} \times \big[ 1 , e^{4} \big]$, 
 	$
 	\Omega_{g} = \kappa_{g}^{(\text{med})}$, and $
 	\Omega_{\lambda} = [ e^{-8}, e^{-2} ]$. 
Notably, we recommend fixing the bandwidth $\kappa_g$, as varying it can change the scale of the risk function $R$ in Section \ref{sec:supp:risk}, so lower risk values may not necessarily indicate better performance.

In Section \ref{sec:supp:Simulation CV}, we examine the relationship between sample size and estimation error of the PMMR estimator when hyperparameters are selected using the proposed repeated cross-validation procedure. The resulting estimation error decreases as the sample size increases, and the rate of error appears to satisfy Assumptions \HL{A9}-\HL{A11}. This suggests that the use of cross-validation is reasonable, at least within the context of the simulation study.

\subsubsection{Median Adjustment for Cross-fitting Estimators} \label{sec:supp:median}

 Cross-fitting estimators depend on a specific sample split and, therefore, may produce outlying estimates if some split samples do not represent the entire data. To resolve the issue, \citet{Victor2018} proposed to use median adjustment from multiple cross-fitting estimates.  First, let $\widehat{\tau}_{s}$ $(s \in \{1,\ldots,S\})$ be the $s$th cross-fitting estimate with the corresponding variance estimate $\widehat{\sigma}_{s}^2$. Then, the median-adjusted cross-fitting estimate and its variance estimate are defined as follows: 
 \begin{align*}
 & \widehat{\tau}_{\median}
 :=
 \median_{s=1,\ldots,S} \widehat{\tau}_{s}
 \ , \quad \widehat{\sigma}_{\median}^2
 :=
 \median_{s=1,\ldots,S} \big\{ \sigma_s^2 + (\widehat{\tau}_s - \widehat{\tau}_{\median} )^2 \big\} \ .
 \end{align*}
 These estimates are more robust to the particular realization of sample partition.

\subsubsection{Multiplier Bootstrap Confidence Intervals}							\label{sec:supp:MB}

We provide details on constructing a multiplier bootstrap–based variance estimator and the corresponding confidence intervals. The algorithm below outlines these steps for the observational setting; for an experimental setting, the procedure can be derived analogously using the estimated influence functions from Theorem \ref{thm:AN rand}.

     \begin{algorithm}[!htb]
 \begin{algorithmic}[1]
     \REQUIRE Number of bootstrap estimates $B$
     \STATE For $i \in \{ 1 ,\ldots, N \}$, obtain the estimated influence functions
     \begin{align*}
     \widehat{\cInfFt}_{\CSE} (\bO_i \con a_D)
     &
     =
     \frac{ 
     \widehat{\InfFt}_{\NUMER}\LSS(\bO_i \con 1, a_D)
     -
     \widehat{\InfFt}_{\NUMER}\LSS(\bO_i \con 0, a_D)
     -
     \widehat{\tau}_{\CSE}(a_D)
     \widehat{\InfFt}_{\DENOM}\LSS(\bO_i \con a_D)
      }{\widehat{\psi}_{\DENOM} (a_D)}
     \end{align*}
     where $k$ is the index that satisfies $ i \in \mathcal{I}_k$
     \FOR{$b \in \{ 1,\ldots,B \}$}
     \STATE Generate i.i.d. random variables $\epsilon_{i}^{(b)} \sim N(0,1)$ for $i \in \{1,\ldots,N\}$
     \STATE Calculate $\widehat{e}^{(b)} = N^{-1} \sum_{i=1}^{N} \epsilon_{i}^{(b)} \widehat{\cInfFt}_{\CSE} (\bO_i \con a_D)$
     \ENDFOR
     \STATE Let $\widehat{\sigma}_{\text{boot}}^{2}$ be the empirical variance of $\big\{ \widehat{e}^{(b)} \cond b \in \{ 1,\ldots,B\} \big\}$
     \STATE Let $\widehat{q}_{\text{boot},\alpha}$ be the $100\alpha$-th percentile of $\big\{ \widehat{e}^{(b)} \cond b \in \{ 1,\ldots,B\} \big\}$
     \RETURN Variance estimate $\widehat{\sigma}_{\text{boot}}^{2}$; $100(1-\alpha)$\% confidence interval $[\widehat{q}_{\text{boot},\alpha/2}, \widehat{q}_{\text{boot},1-\alpha/2}]$
 \end{algorithmic}
 \caption{Multiplier Bootstrap Procedure}
 \label{alg:MB}
 \end{algorithm}

\subsection{Extension: An Experimental Setting} \label{sec: exp setting main}

We discuss the important setting in which the treatment in the two-arm study is randomized. By virtue of randomization, there is no confounding of the association between $A$ and $(Y,D_{\stime})$. However, it is still possible that unmeasured confounders exist in the association between $D_{\stime}$ and $Y$, rendering the approach in \citet{Stensrud2022CSE} infeasible even in this experimental setting. To address such cases, we extend the proposed approach to experimental settings to establish identification and estimation of the CSE. The remainder of the section discusses details of the extension. 

We begin by making \HL{A2'} instead of \HL{A2} to formally state that $A$ is randomized:
\begin{itemize}
\item[\HT{A2'}] (\textit{Randomization}) $A  \indep (\potY{0}, \potY{1},\oD_{\TIME+1}^{(0)}, \oD_{\TIME+1}^{(1)},\obX_{\TIME}^{(0)},\obX_{\TIME}^{(1)},\bW,\bZ,\bU)$ where $\Pr(A=1) \in (0,1)$ is known.
\end{itemize}
A key to the extension to an experimental setting is to allow that $\bZ$ can be associated with $\potDt{a}{\TIME+1}$ conditional on $(A,\obX_{\TIME},\bU)$; in other words, we relax \HL{A6}-(ii) and allow that $\potDt{a}{\TIME+1}$ can be affected by $\bZ$. Therefore, in experimental settings, $\bZ$ can be viewed as a proxy variable for $\potDt{a}{\TIME+1}$. Figure \ref{fig: SWIG2} illustrates a Single World Intervention Graph \citep{RobinsRichardson2011} that is compatible with the setup of the randomization trial discussed in this Section.

\begin{figure}[!htp]
\centering 
\scalebox{0.6}{
\begin{tikzpicture}
\tikzset{line width=1pt, outer sep=0.5pt,
ell/.style={draw,fill=white, inner sep=3pt,
line width=1pt},
swig vsplit={gap=2.5pt, 
inner line width right=0.5pt,
line width right=1.5pt}};
\node[name=Ay,shape=swig vsplit] at (0,-1.5){  \nodepart{left}{$A_Y$} \nodepart{right}{$a_Y$} };
\node[name=Ad, shape=swig vsplit] at (0,0) { \nodepart{left}{$A_D$} \nodepart{right}{$a_D$} };
\node[name=A,ell,  shape=ellipse] at (-4,0) {$A$}  ;
\node[name=D1,ell,  shape=ellipse] at (4,0) {$D_1^{(a_D)}$}  ;
\node[name=D2,ell,  shape=ellipse] at (8,0) {$D_2^{(a_D)}$}  ;
\node[name=Y,ell,  shape=ellipse] (Y) at (12,0) {$Y^{(a_Y,a_D)}$};
\node[name=U,ell,  shape=ellipse] at (4,5.5) {$U$};
\node[name=X0,ell,  shape=ellipse] at (1.75,2) {$X_0$};
\node[name=X1,ell,  shape=ellipse] at (5.5,3.2) {$X_1^{(a_D)}$};
\node[name=Z,ell,  shape=ellipse] at (-4,2.5) {$Z$};
\node[name=W,ell,  shape=ellipse] at (12,2.5) {$W$};
\begin{scope}[>={Stealth[black]},
      every edge/.style={draw=black,line width=0.5pt}]
\path [->] (Ad) edge (D1);
\path [->] (Ad) edge (X1);
\path [->] (Ad) edge[bend right=20] (D2);
\path [->] (Ay) edge[bend right=10] (Y);
\path [->] (W) edge (Y);    
\path [->] (D1) edge (D2);
\path [->] (D1) edge (X1);
\path [->] (D2) edge (Y);
\path [->] (U) edge (Y);
\path [->] (U) edge (D1);
\path [->] (U) edge (Z);
\path [->] (U) edge (X0);
\path [->] (U) edge (X1);
\path [->] (X0) edge (X1);
\path [->] (X0) edge (D1);
\path [->] (X0) edge (Y);
\path [->] (X1) edge (Y);
\path [->] (X1) edge (D2);
\path [->] (X0) edge (Z);
\path [->] (X0) edge (W);
\path [->] (Z) edge (D1);
\path [->] (A) edge (Ad);
\path [->] (A) edge (Ay);
\path [->] (U) edge (W);
\end{scope}
 
\end{tikzpicture} 
}
\caption{\footnotesize Single World Intervention Graph in the Experimental Setting}
\label{fig: SWIG2} 
\end{figure} 

One can establish identification of the CSE in the experimental setting by using certain confounding bridge functions which we introduce below. Let $\xi_0^*$ and $p_1^*$ be the solution to the following integral equations, which are assumed to exist:
\begin{align}	
&
\EXP \big\{ Y  - \xi_1^* (\bW,\obX_{\TIME} \con a_Y) \cond \bZ, A=a_Y, D_{\TIME+1}=0 , \obX_{\TIME} \big\} 
=
0 \ ,
\label{eq-defxi1}
\\
&
\EXP \big\{ 
\ind (A=a_D) p_0^*(a_D) 
 - \ind(A=a_Y) p_1^* (\bZ,\obX_{\TIME} \con a_Y, a_D)  \cond \bW, D_{\TIME+1}=0 , \obX_{\TIME} \big\}  
\text{ if $a_Y  \neq a_D$} \ ,
\label{eq-defp1}
\end{align} 
where $p_0^*(a_D) = 1/\Pr(A=a_D)$. The first function $\xi_1^*(\cdot \con a_Y)$ has the same definition as the outcome confounding bridge function $h_1^*$ solving  \eqref{eq-bridgeft1 obs}. The second function $p_1^*(\cdot \con a_Y)$ has the same definition as the treatment confounding bridge function $q_1^*$ solving  \eqref{eq-bridgeft6 obs} except that $q_0^*$ is replaced with a scalar $p_0^*(a_D)=1/\Pr(A=a_D)$; we remark that $p_0^*(a_D)$ has the same role as the treatment confounding bridge function $q_0^*(\cdot \con a_D)$ solving \eqref{eq-bridgeft5 obs}. 

Using these nuisance functions, we can establish identification of the CSE in the experimental setting, as stated in the following Theorem.  
\begin{theorem} \label{thm:identifcation rand} 
Suppose that Assumptions \HL{A1}, \HL{A2'}, \HL{A3}-(i), \HL{A4}, \HL{A5}, \HL{A6}-(i),(iii),(iv), \HL{A7}-(i),(iii) are satisfied, and that there exist the confounding bridge functions $\xi_1^*$ and $p_1^*$ satisfying \eqref{eq-defxi1} and \eqref{eq-defp1}, respectively. Then, we have 
    \begin{align*}
        \tau_{\CSE}^*
        &
        =
        \frac{\EXP \big[ \big\{ \xi_1^*(\bW,\obX_{\TIME} \con a_Y) -  Y \big\} \ind(A=a_D)(1-D_{\TIME+1}) / \Pr(A=a_D) \big]  }{ \EXP \big\{ \ind(A=a_D)(1-D_{\TIME+1})/ \Pr(A=a_D) \big\} } 
        \\
        &
        =
        \frac{\EXP \big[ (1-D_{\TIME+1})Y \big\{ \ind(A=a_Y) p_1^*(\bZ,\obX_{\TIME+1})- \ind(A=a_D)/ \Pr(A=a_D) \big\} \big] }{ \EXP \big\{ \ind(A=a_D)(1-D)/ \Pr(A=a_D) \big\} }
        \ .
    \end{align*}  
\end{theorem}
We also consider the following surjectivity condition:
\begin{itemize} 
\item[\HT{S2}] Let $\mathcal{T}_1:\mathcal{L}_2(\bW,\obX_{\TIME}) \rightarrow \mathcal{L}_2(\bZ,A=1,D_{\TIME+1}=0,\obX_{\TIME})$ be the operator given by \\$\mathcal{T}_1(g)  =  \EXP \big\{ g(\bW,\obX_{\TIME}) \cond \bZ,A=a_Y,D_{\TIME+1}=0,\obX_{\TIME} \big\}$. Then, $\mathcal{T}_{1}$ is surjective. 
\end{itemize}

In parallel with the results presented in Section \ref{sec:SEB}, we characterize the semiparametric local efficiency bound for $\tau_{\CSE}^*$ under a certain semiparametric model for the experimental setting; see the Theorem below for details.

\begin{theorem} \label{thm:IF Rand Supp} Suppose that Assumptions required in Theorem \ref{thm:identifcation rand} are satisfied, and that there exists a confounding bridge function $p_1^*$ satisfying \eqref{eq-defp1} at the true data-generating law $P^*$. Let a regular semiparametric model $\M_{\RD}$ be $\M_{\RD} = \big\{
    P
    \cond
    \text{$A$ is randomized, and there exists $\xi_1^*$ 
            satisfying \eqref{eq-defxi1}}
\big\} $.
\begin{itemize}[leftmargin=0.25cm]
\item[(i)]
Then, an influence function for $\tau_{\CSE}^*$ under $\M_{\RD}$ is given by
\begin{align}	
&
\InfFt_{\RD,\CSE}^* (\bO)
=
\frac{ \InfFt_{\RD,\NUMER}^*(\bO) - \big\{ Y + \tau_{\CSE}^* \big\}\ind(A=a_D)(1-D_{\TIME+1})/ \Pr(A=a_D) }{ \EXP \big\{ \ind(A=a_D)(1-D_{\TIME+1}) / \Pr(A=a_D) \big\}  } \ , 
\nonumber
\\
& 
\InfFt_{\RD,\NUMER}^*(\bO)
=
\left[ 
\!
\mymatrixTwo{l}{ (1-D_{\TIME+1}) \ind(A=a_Y) p_1^*(Z,\obX_{\TIME}) 
\big\{ Y - \xi_1^* (\bW,\obX_{\TIME}) \big\} 
}{+
\ind(A=a_D) p_0^*(a_D) 
        (1-D_{\TIME+1})  
        \xi_1^*(\bW,\obX_{\TIME})}
\!\!
\right] \ .
\label{eq-IF_rand_N1}
\end{align}  

\item[(ii)]  

Further suppose that the true data-generating law $P^*$ belongs to $\M_{\RD, \text{sub}}$ where 
\begin{align*}
\M_{\RD,\text{sub}}
=
\big\{ P \in \M \cond \text{\text{\HL{S2} holds, and  there exist unique $\xi_1^*, p_1^*$ satisfying \eqref{eq-defxi1}, \eqref{eq-defp1}}}
\big\} \ . 
\end{align*}
Then, $\InfFt_{\RD,\CSE}^*(\bO)$ is the efficient influence function for $\tau_{\RD,\CSE}^*$ under model $\M_{\RD}$ at $P^*$. Therefore, the corresponding semiparametric local efficiency bound for $\tau_{\RD,\CSE}^*$ is $\VAR \big\{ \InfFt_{\RD,\CSE}^* (\bO) \big\}$. 

\end{itemize} 

\end{theorem}

Lastly, we construct a semiparametric estimator for $\tau_{\CSE}^*$. Specifically, let $\widehat{\xi}_{1}\LSS$ and $\widehat{p}_{1}\LSS$ be the estimators for $\xi_1^*$ and $p_1^*$ obtained from the PMMR approach by following the approach in Section \ref{sec:estimator}. By averaging the estimated influence functions over the evaluation folds, one obtains the estimator of the CSE, given by  
\begin{align*}
&
\widehat{\tau}_{\RD,\CSE}
=
\frac{
N^{-1}
\sum_{k=1}^{K}
\sum_{i \in \mathcal{I}_k}
\big\{ \widehat{\InfFt}_{\RD,\NUMER} \LSS (\bO_i)
-
\ind(A_i=a_D)(1-D_{\TIME+1,i})Y_i / \Pr(A=a_D)
\big\}
}{
N^{-1}
\sum_{k=1}^{K}
\sum_{i \in \mathcal{I}_k}
\ind(A_i=a_D)(1-D_{\TIME+1,i}) / \Pr(A=0)
} \ .
\end{align*}
Here, $\widehat{\InfFt}_{\RD,\NUMER} \LSS$ is the estimated influence function where the true confounding bridge functions in \eqref{eq-IF_rand_N1} are substituted by their estimated counterparts.

The estimator $\widehat{\tau}_{\RD,\CSE}$ is asymptotically normal under some regularity conditions. Consider the following assumptions for the estimated nuisance functions:
\begin{itemize}
\item[\HT{A8'}] (\textit{Boundedness}) For $a_Y,a_D=0,1$, there exists a finite constant $C > 0$ such that 
$\big\| \EXP \big( Y^2 \cond \bZ,A=a_Y,D_{\TIME+1}=0,\bX_{\TIME+1} \big) \big\|_{\infty}$, $\big\| \EXP \big( Y^2 \cond A=a_D,D_{\TIME+1}=0 \big) \big\|_{\infty}$, 
$\big\| \xi_1^*(\cdot \con a_Y) \big\|_{\infty} $, 
$1/\Pr(A=a_D) $, and
$\big\| p_1^*(\cdot \con a_Y,a_D) \big\|_{\infty} $ are bounded above by $C$ almost surely.
Additionally, for all $k \in \{1,\ldots,K\}$ and $a_Y,a_D \in \{0,1\}^{\otimes 2}$, 
$\big\| \widehat{\xi}\LSS_1(\cdot \con a_Y) \big\|_{\infty} $ and $\big\| \widehat{p}\LSS_1(\cdot \con a_Y,a_D) \big\|_{\infty} $ are bounded above by $C$ almost surely.

\item[\HT{A9'}] (\textit{Consistency}) For all $k \in \{1,\ldots,K\}$ and $(a_Y,a_D) \in \{0,1\}^{\otimes 2}$, 
$\big\| \widehat{\xi}\LSS_1(\cdot \con a_Y) - \xi_1^*(\cdot \con a_Y) \big\|_{P,2}$ and
$\big\| \widehat{p}\LSS_1(\cdot \con a_Y,a_D) - p_1^*(\cdot \con a_Y,a_D) \big\|_{P,2}$ are $o_P(1)$.

\item[\HT{A10'}] (\textit{Cross-product Rates for the Numerator})  $k \in \{1,\ldots,K\}$ and $(a_Y,a_D) \in \{0,1\}^{\otimes 2}$,
$\big\| \widehat{\xi}\LSS_1 (\cdot \con a_Y) - \xi_1^* (\cdot \con a_Y) \big\|_{P,2}
\times 
\big\| \widehat{p}\LSS_1 (\cdot \con a_Y,a_D) - p_1^* (\cdot \con a_Y,a_D) \big\|_{P,2}$ is $o_P(N^{-1/2})$. 
\end{itemize}
\HL{A9'} and \HL{A10'} are similar to \HL{A8}-\HL{A11}, but there are notable differences. First, \HL{A10'} does not include the cross-product rate condition involving $1/\Pr(A=a_D)$, which is present in \HL{A10} and \HL{A11} in terms of $q_0^*$. This omission is due to the fact that $1/\Pr(A=a_D)$ is known under \HL{A2'}, thus satisfying the required cross-product rate conditions under \HL{A9'}. 

Theorem \ref{thm:AN rand} below establishes the asymptotic normality of $\widehat{\tau}_{\RD,\CSE}$, enabling inference to be conducted in a manner analogous to that used in the observational setting.

\begin{theorem} \label{thm:AN rand}
Suppose that the confounding bridge functions $\xi_1^*$ and $p_1^*$ satisfying \eqref{eq-defxi1} and \eqref{eq-defp1} exist, and that Assumptions required in Theorem \ref{thm:identifcation rand} are satisfied. Then, we have that $\sqrt{N}
\big(
\widehat{\tau}_{\RD,\CSE} 
-
\tau_{\CSE}^*  
\big)
\stackrel{D}{\rightarrow}
N \big( 0 , \sigma_{\RD,\CSE}^{*2}   \big)$ 
where $\sigma_{\RD,\CSE}^{*2} = \VAR \big\{ \InfFt_{\RD, \CSE}^*(\bO) \big\}$. Moreover, a consistent estimator of the asymptotic variance of $\widehat{\tau}_{\RD,\CSE}$ is given by
\begin{align*}
& 
\widehat{\sigma}_{\RD,\CSE}^2
=
\frac{1}{N}
\sum_{k=1}^{K}
\sum_{i \in \mathcal{I}_k}
\bigg[
\frac{
\widehat{\InfFt}_{\RD,\NUMER}\LSS(\bO_i)
-
\big\{
Y_i
+
\widehat{\tau}_{\RD,\CSE} 
\big\}
\ind(A_i=a_D)(1-D_{\TIME+1,i})
/
\Pr(A=a_D)
\big\}
}{ 
N^{-1}
\sum_{k=1}^{K}
\sum_{i \in \mathcal{I}_k}
\big\{ \ind(A_i=a_D)(1-D_{\TIME+1,i})
/
\Pr(A=a_D)
\big\}
} 
\bigg]^2 \ .
\end{align*}
\end{theorem}

 
\subsection{Comparison to Previous Work} \label{sec:supp:others}

In this Section, we review results from previous studies and highlight how our framework compares to them.

\subsubsection{Comparison  between Assumptions \protect\AssumptionMixBiasNumer, \protect\AssumptionMixBiasDenom\ and Conditions in \citet{Cui2023, Dukes2023_ProxMed}} \label{sec:supp:Dukes}

First, we compare the conditions on cross-product convergence rates ensuring asymptotic normality of the ultimate treatment effect estimators in our framework versus those in \citet{Cui2023, Dukes2023_ProxMed}. Let the parameter of interest be $\psi_{\NUMER}^*(a_Y,a_D)$, and all nuisance functions are parametrically estimated. The estimator of $\psi_{\NUMER}^*(a_Y,a_D)$ proposed by \citet{Dukes2023_ProxMed} is consistent and asymptotically normal in the union model $\M_1 \cup \M_2 \cup \M_3$ where
 \begin{align*}
 &
 \M_1: \text{$h_1(\cdot \con a_Y)$ and $h_0(\cdot \con a_Y,a_D)$ are correctly specified} \ ,
 \\
 &
 \M_2: \text{$h_1(\cdot \con a_Y)$ and $q_0(\cdot \con a_D)$ are correctly specified} \ ,
 \\
 &
 \M_3: \text{$q_1(\cdot \con a_Y,a_D)$ and $q_0(\cdot \con a_D)$ are correctly specified} \ .
 \end{align*}
 We restate Assumptions \AssumptionConsistency\ and \AssumptionMixBiasNumer\ for readability:
 \begin{itemize}

     \item[\AssumptionConsistency] (\textit{Consistency}) For all $k \in \{1,\ldots,K\}$ and $a_Y  \in \{0,1\}$, 
$\big\| \widehat{h}\LSS_0(\cdot \con a_Y) - h_0^*(\cdot \con a_Y) \big\|_{P,2}$,
$\big\| \widehat{h}\LSS_1 - h_1^* \big\|_{P,2}$,
$\big\| \widehat{h}\LSS_2 - h_2^* \big\|_{P,2}$,
$\big\| \widehat{q}\LSS_0 - q_0^* \big\|_{P,2}$,  
$\big\| \widehat{q}\LSS_1 - q_1^* \big\|_{P,2}$ are $o_P(1)$.

     \item[\AssumptionMixBiasNumer] (\textit{Cross-product Rates for the Numerator}) For all $k \in \{1,\ldots,K\}$ and $a_Y \in \{0,1\}$, 
$\big\| \widehat{h}\LSS_1 - h_1^* \big\|_{P,2}
\times
\big\| \widehat{q}\LSS_1 - q_1^* \big\|_{P,2}$, 
$\big\| \widehat{h}\LSS_1 - h_1^*  \big\|_{P,2}
\times
\big\| \widehat{q}\LSS_0 -  q_0^* \big\|_{P,2}$,  
$\big\| \widehat{h}\LSS_0(\cdot \con a_Y) - h_0^*(\cdot \con a_Y) \big\|_{P,2}
\times
\big\| \widehat{q}\LSS_0 - q_0^* \big\|_{P,2}$ are $o_P(N^{-1/2})$.
 \end{itemize}

 Note that model $\M_1$ implies that $\big\| \widehat{h}\LSS_1 - h_1^* \big\|_{P,2}$ and $\big\| \widehat{h}\LSS_0 - h_0^* \big\|_{P,2}$ are $O_P(N^{-1/2})$. Likewise, model $\M_2$ implies that $\big\| \widehat{h}\LSS_1 - h_1^* \big\|_{P,2}$ and $\big\| \widehat{q}\LSS_0 - q_0^* \big\|_{P,2}$ are $O_P(N^{-1/2})$.   Lastly, model $\M_3$ implies that $\big\| \widehat{q}\LSS_1 - q_1^* \big\|_{P,2}$ and $\big\| \widehat{q}\LSS_0 - q_0^* \big\|_{P,2}$ are $O_P(N^{-1/2})$. Therefore, under the union model $\M_1 \cup \M_2 \cup \M_3$, Assumption \AssumptionMixBiasNumer\ is satisfied as long as Assumption \AssumptionConsistency\ is satisfied. 

 Next, let the parameter of interest be $\psi_{\DENOM}^*(a_D)$, and all nuisance functions are parametrically estimated. The estimator of $\psi_{\DENOM}^*(a_D)$ proposed by \citet{Cui2023} is consistent and asymptotically normal under the union model $\M_1' \cup \M_2'$ where
 \begin{align*}
     &
     \M_1': \text{$h_2(\cdot \con a_D)$ is correctly specified} \ ,
     &&
     \M_2': \text{$q_0(\cdot \con a_D)$ is correctly specified}
     \ .
 \end{align*}
 We restate Assumption \AssumptionMixBiasDenom\ for readability:
 \begin{itemize}
 \item[\AssumptionMixBiasDenom] (\textit{Cross-product Rates for the Denominator}) For all $k \in \{1,\ldots,K\}$,  $\big\| \widehat{h}\LSS_2 - h_2^* \big\|_{P,2}
\times
\big\| \widehat{q}\LSS_0 - q_0^* \big\|_{P,2}$ is $o_P(N^{-1/2})$.
 \end{itemize}

 Note that model $\M_1'$ implies that $\big\| \widehat{h}\LSS_2 - h_2^* \big\|_{P,2} = O_P(N^{-1/2})$. Likewise, model $\M_2'$ implies $\big\| \widehat{q}\LSS_0 - q_0^* \big\|_{P,2} = O_P(N^{-1/2})$. Therefore, under the union model $\M_1' \cup \M_2'$, Assumption \AssumptionMixBiasDenom\ is satisfied as long as Assumption \AssumptionConsistency\ is satisfied.

 \subsubsection{Efficient Influence Function for the CSE under No Unmeasured Confounding} \label{sec:EIF_NoU}

 We present the efficient influence functions of the CSE under both observational and experimental settings. We omit the proofs of the results, as they are direct consequences of \citet{Hahn1998}, \citet{TTS2012}, and \citet[Appendix C]{Stensrud2022CSE}. 

 Let $\bL_{0}$ and $\obL_{\TIME}$ be a complete collection of measured confounders up to time 0 and $\TIME$, respectively. To accommodate no unmeasured confounding, we modify Assumptions \HL{A2}, \HL{A3}, and \AssumptionDismissible\ as follows:
 \begin{itemize}
 \item[\HT{A2''}] (\textit{Ignorability})  For $a \in \{0,1\}$, $(\potY{a},\underline{D}_{1}^{(a)}, \underline{L}_{1}^{(a)} ) \indep A \cond \bL_{0}$.
 \item[\HT{A3'}] (\textit{Positivity}) For $a \in \{0,1\}$, (i) $\Pr(D_{\TIME+1}=0, \underline{L}_1 \cond A=a, \bL_{0}) >0  $ a.s. and (ii) $\Pr(A=a \cond \bL_{0}) >0  $ a.s.
 \item[\HT{A5'}] (\textit{Dismissible Condition}) In the four-arm trial, we have 
 \begin{itemize}[leftmargin=0cm, itemsep=0cm, topsep=0cm, partopsep=0cm,  parsep=0cm]
     \item[(i)] $Y(G) \indep A_D(G) \cond (A_Y(G), D_{\TIME+1}(G) = 0, \overline{\bL}_{\TIME}(G) ) $; 
     \item[(ii)] $D_{t+1}(G) \indep A_Y(G) \cond (A_D(G), D_{t}(G)=0, \overline{\bL}_{t}(G))$ for all $t \in \{0,\ldots,\TIME\}$; 
     \item[(iii)] $\bL_{t+1}(G) \indep A_Y(G) \cond (A_D(G), D_{t+1}(G)=0, \overline{\bL}_{t}(G))$ for all $t \in \{0,\ldots,\TIME\}$. 
 \end{itemize}
 \end{itemize}

 We denote a regular nonparametric model by $\M^\dagger$. The efficient influence functions for $\psi_{\NUMER}^*$ and $\psi_{\DENOM}^*$ in $\M^\dagger$ are given as follows.
 \begin{theorem} \label{eq-numerator no U}
 Suppose Assumptions \HL{A1}, \HL{A2''}, \HL{A3'}, \HL{A4}, \HL{A5'} are satisfied. Then, the following functions are the efficient influence functions for $\psi_{\NUMER}^*(a_Y,a_D)$, $\psi_{\NUMER}^*(a_D)$, and $\psi_{\DENOM}^*(a_D)$ in model $\M^\dagger$, respectively:
 \begin{align*}
     & \cInfFt_{\NUMER}^{\dagger} (\bO \con a_Y,a_D)
     \\
     &
     = \frac{\ind(A=a_Y)}{\Pr(A=a_D \cond \bL_{0})}  \frac{\Pr(A=a_D \cond D_{\TIME+1}=0, \obL_{\TIME})}{\Pr(A=a_Y \cond D_{\TIME+1}=0, \obL_{\TIME})} (1-D_{\TIME+1}) \big\{ Y - \EXP \big( Y \cond A=a_Y,D_{\TIME+1}=0,\obL_{\TIME} \big) \big\}
     \\
     & \quad
     +
     \frac{\ind(A=a_D) }{\Pr(A=a_D\cond \bL_{0})}
     \Bigg[
     \begin{array}{l}
     (1-D_{\TIME+1}) \EXP \big( Y \cond A=a_Y,D_{\TIME+1}=0,\obL_{\TIME} \big) \\
     -
     \EXP \big\{
     (1-D_{\TIME+1}) \EXP \big( Y \cond A=a_Y,D_{\TIME+1}=0,\obL_{\TIME} \big)
     \cond A=a_D, \bL_{0}
     \big\}
     \end{array}
      \Bigg]
     \\
     & \quad
     +
     \EXP \big\{
     (1-D_{\TIME+1}) \EXP \big( Y \cond A=a_Y,D_{\TIME+1}=0,\obL_{\TIME} \big)
     \cond A=a_D, \bL_{0}
     \big\} - \psi_{\NUMER}^*(a_Y,a_D) 
     \quad \quad \text{ if $a_Y \neq a_D$}
     \\
     &
      \cInfFt_{\NUMER}^\dagger(\bO \con a_D, a_D) \\
      &
      = \frac{\ind (A=a_D)}{\Pr(A=a_D \cond \bL_{0})} \big[ (1-D_{\TIME+1}) Y - 
     \EXP \big\{ (1-D_{\TIME+1}) Y \cond A=a_D, \bL_{0} \big\} \big]
     \\
     &
     \quad
     +
     \EXP \big\{ (1-D_{\TIME+1}) Y \cond A=a_D, \bL_{0} \big\} - \psi_{\NUMER}^*(a_D,a_D) \ ,
 \\
 & \cInfFt_{\DENOM}^\dagger(\bO \con a_D) \\
 & 
      = \frac{\ind (A=a_D)}{\Pr(A=a_D \cond \bL)} \big\{ (1-D_{\TIME+1})  -  \Pr(D_{\TIME+1}=0 \cond A=a_D,\bL_{0})  \big\}
     +
     \Pr(D=0 \cond A=a_D,\bL_{0}) - \psi_{\DENOM}^*(a_D) \ .
 \end{align*}
 \end{theorem}
 Therefore, the efficient influence function for $\tau_{\CSE}^*(a_D)$   is
 \begin{align*}
 \cInfFt_{\CSE}^\dagger(\bO \con a_D) 
 =
 \frac{ 	\cInfFt_{\NUMER}^\dagger(\bO \con a_Y=1, a_D) - \cInfFt_{\NUMER}^\dagger(\bO \con a_Y=0, a_D) - \tau_{\CSE}^*(a_D) \cInfFt_{\DENOM}^\dagger(\bO \con a_D)  }{ \psi_{\DENOM}^*(a_D) } \ .
 \end{align*}

 Next, we consider a regular model $\M_{\RD}^\dagger = \big\{ P \cond \text{$A$ is randomized with known $\Pr(A)$} \big\} $. 
The efficient influence functions for $\psi_{\RD,\NUMER}^*$ and $\psi_{\RD,\DENOM}^*$ in $\M_{\RD}^\dagger$ are given as follows.
 \begin{theorem} \label{eq-numerator no U rand}
 Suppose Assumptions \HL{A1}, \HL{A2'}, \HL{A3'}, \HL{A4}, \HL{A5'} are satisfied. Then, the following functions are the efficient influence functions for $\psi_{\RD,\NUMER}^*(a_Y,a_D)$, $\psi_{\RD,\NUMER}^*(a_D,a_D)$, and $\psi_{\RD,\DENOM}^*(a_D)$ in model $\M_{\RD}^\dagger$, respectively:
 \begin{align*}
     & \cInfFt_{\RD,\NUMER}^{\dagger} (\bO \con a_Y,a_D)
     \\
     &
     = \frac{\ind(A=a_Y)}{\Pr(A=a_D)}  \frac{\Pr(A=a_D \cond D_{\TIME+1}=0, \obL_{\TIME})}{\Pr(A=a_Y \cond D_{\TIME+1}=0, \obL_{\TIME})} (1-D_{\TIME+1}) \big\{ Y - \EXP \big( Y \cond A=a_Y,D_{\TIME+1}=0,\obL_{\TIME} \big) \big\}
     \\
     & \quad
     +
     \frac{\ind(A=a_D) }{\Pr(A=a_D)}
     (1-D_{\TIME+1}) \EXP \big( Y \cond A=a_Y,D_{\TIME+1}=0,\obL_{\TIME} \big) 
     - \psi_{\RD,\NUMER}^*(a_Y,a_D) 
     \quad \quad \text{ if $a_Y \neq a_D$} 
     \\
     &
      \cInfFt_{\RD,\NUMER}^\dagger(\bO \con a_D, a_D) = p_0^*(a_D) \ind (A=a_D)(1-D_{\TIME+1}) Y - \psi_{\RD,\NUMER}^*(a_D,a_D) \ ,
 \\
 & \cInfFt_{\RD,\DENOM}^\dagger(\bO \con a_D) = p_0^*(a_D) \ind (A=a_D)(1-D_{\TIME+1}) - \psi_{\RD,\DENOM}^*(a_D) \ .
 \end{align*}
 \end{theorem}
 Therefore, the efficient influence function for $\psi_{\RD}^*(a_Y,a_D)$  is
 \begin{align}
 & 
 \cInfFt_{\RD,\CSE}^\dagger(\bO \con a_D) 
 \nonumber
 \\
 & =
 \frac{ 	\cInfFt_{\RD,\NUMER}^\dagger(\bO \con a_Y=1, a_D) - \cInfFt_{\RD,\NUMER}^\dagger(\bO \con a_Y=0, a_D) - \tau_{\CSE}^*(a_D) \cInfFt_{\RD,\DENOM}^\dagger(\bO \con a_D)  }{ \psi_{\RD,\DENOM}^*(a_D) } 
 \ .
 \label{eq-EIF-noU-random}
 \end{align}
 Note that the form of $\cInfFt_{\RD,\NUMER}^{\dagger} $ differs from the efficient influence function of \citet{Stensrud2022CSE}, denoted by $\alpha_1(p_0)$ on page 18, because in our setting, $\Pr(A)$ is known, whereas it is assumed to be unknown in their work. The same distinction applies to the other two efficient influence functions. In case of unknown $\Pr(A)$, the efficient influence functions changes to
 \begin{align*}
     & \cInfFt_{\RD,\NUMER}^{\dagger} (\bO \con a_Y,a_D)
     \\
     &
     = \frac{\ind(A=a_Y)}{\Pr(A=a_D)}  \frac{\Pr(A=a_D \cond D_{\TIME+1}=0, \obL_{\TIME})}{\Pr(A=a_Y \cond D_{\TIME+1}=0, \obL_{\TIME})} (1-D_{\TIME+1}) \big\{ Y - \EXP \big( Y \cond A=a_Y,D_{\TIME+1}=0,\obL_{\TIME} \big) \big\}
     \\
     & \quad
     +
     \frac{\ind(A=a_D) }{\Pr(A=a_D)}
   \big\{   (1-D_{\TIME+1}) \EXP \big( Y \cond A=a_Y,D_{\TIME+1}=0,\obL_{\TIME} \big) 
     - \psi_{\RD,\NUMER}^*(a_Y,a_D) 
     \big\}
     \quad \quad \text{ if $a_Y \neq a_D$} 
     \\
     &
      \cInfFt_{\RD,\NUMER}^\dagger(\bO \con a_D, a_D) = p_0^*(a_D) \ind (A=a_D) \big\{ (1-D_{\TIME+1}) Y - \psi_{\RD,\NUMER}^*(a_D,a_D) 
      \big\}
      \ ,
 \\
 & \cInfFt_{\RD,\DENOM}^\dagger(\bO \con a_D) = p_0^*(a_D) \ind (A=a_D)
 \big\{ (1-D_{\TIME+1}) - \psi_{\RD,\DENOM}^*(a_D)
 \big\}
  \ .
 \end{align*}

 Using the efficient influence function in \eqref{eq-EIF-noU-random} as a basis, one may obtain a semiparametric estimator of $\tau_{\CSE}(a_D)$, which is asymptotically normal under regularity conditions. Specifically, we define $\mu^*(a, \overline{\bL}_{\TIME}) = \EXP \big( Y \cond A=a,D_{\TIME+1}=0,\overline{\bL}_{\TIME} \big)$ and $e^*(a,\overline{\bL}_{\TIME}) = \Pr(A=a \cond D_{\TIME+1}=0,\overline{\bL}_{\TIME})$. We obtain corresponding estimators $\widetilde{\mu}\LSS(a,\overline{\bL}_{\TIME})$ and $\widetilde{e}\LSS(a,\overline{\bL}_{\TIME})$ by only using $\mathcal{I}_k^c$ based on nonparametric machine learning approaches. In particular, we employ an ensemble learner that integrates multiple machine learning methods using the Super Learner algorithm \citep{SL2007}. For both simulation studies and data analysis, we construct the Super Learner library in \texttt{R} with the following machine learning methods: linear regression via \texttt{glm}, lasso/elastic net via \texttt{glmnet} \citep{glmnet}, splines via \texttt{earth} \citep{earth} and \texttt{polspline} \citep{polspline}, generalized additive models via \texttt{gam} \citep{gam}, boosting via \texttt{xgboost} \citep{xgboost} and \texttt{gbm} \citep{gbm}, random forests via \texttt{ranger} \citep{ranger}, and neural networks via \texttt{RSNNS} \citep{RSNNS}.

 Let $\widetilde{\psi}_{\RD}(a_Y,a_D) = \widetilde{\psi}_{\RD,\NUMER}(a_Y,a_D)/\widetilde{\psi}_{\RD,\DENOM}(a_D)$, where
 \begin{align*}
 &
 \widetilde{\psi}_{\RD,\NUMER}(a_Y,a_D)
 =
 \frac{1}{N} \sum_{k=1}^{K} \sum_{i \in \mathcal{I}_k}
 \widetilde{\InfFt}_{\RD,\NUMER}\LSS (\bO_i \con a_Y, a_D)
 \ , \quad 
 \widetilde{\psi}_{\RD,\DENOM}(a_D)
 =
 \frac{1}{N} \sum_{i=1}^{N}
 \widetilde{\InfFt}_{\RD,\DENOM}\LSS (\bO_i \con  a_D)
 \\
 &
 \widetilde{\InfFt}_{\RD,\NUMER}\LSS (\bO_i \con a_Y, a_D)
 =
 \left[
 \begin{array}{l}
 \frac{ \ind(A_i=a_Y) }{\Pr(A=a_D)} (1-D_{\TIME+1,i}) 
 \frac{\widetilde{e}\LSS(a_D,\overline{\bL}_{\TIME,i})}{\widetilde{e}\LSS(a_Y,\overline{\bL}_{\TIME,i})}
  \big\{ Y_i - \widetilde{\mu}\LSS (a_Y,\overline{\bL}_{\TIME,i}) \big\}
 \\
 + \frac{ \ind(A_i=a_D)}{\Pr(A=a_D)} (1-D_{\TIME+1,i}) \widetilde{\mu}\LSS (a_Y,\overline{\bL}_{\TIME,i})
 \end{array}
 \right] 
 \quad 
 \text{ if $a_Y \neq a_D$}
 \\
 &
 \widetilde{\InfFt}_{\RD,\NUMER}\LSS (\bO_i \con a_D, a_D)
 =
 \ind (A_i=a_D) p_0^*(a_D) (1-D_{\TIME+1,i}) Y_i
 \\
 &
 \widetilde{\InfFt}_{\RD,\DENOM}\LSS (\bO_i \con a_D)
 =
 \ind (A_i=a_D) p_0^*(a_D) (1-D_{\TIME+1,i})
 \ .
 \end{align*}
 An estimator of the CSE is given by $\widetilde{\tau}_{\RD,\CSE}(a_D) = \widetilde{\psi}_{\RD}(a_Y=1,a_D)-\widetilde{\psi}_{\RD}(a_Y=0,a_D)$. The following theorem establishes the asymptotic normality of the estimator:
 \begin{theorem}
 Consider the following assumptions:
 \begin{itemize}

 \item[\HT{A8''}] (\textit{Boundedness}) There exists a finite constant $C > 0$ such that 
 $\big\| \EXP \big( Y^2 \cond A=a_Y,D_{\TIME+1}=0,\overline{\bL}_{\TIME} \big) \big\|_{\infty}$,  $\big\| \EXP \big( Y^2 \cond A=a_D,D_{\TIME+1}=0,\overline{\bL}_{\TIME} \big) \big\|_{\infty}$,  and
 $\big\| e^*(a_D,\overline{\bL}_{\TIME}) / e^*(a_Y,\overline{\bL}_{\TIME}) \big\|_{\infty} $ for $(a_Y,a_D) \in \{0,1\}^{\otimes 2}$ are bounded above by $C$ almost surely.\\
 Additionally, for all $k  \in \{1,\ldots,K\}$, 
 $\big\| \widetilde{\mu}\LSS(a,\overline{\bL}_{\TIME}) \big\|_{\infty} $ and
 $\big\| \widetilde{e}\LSS(a_D,\overline{\bL}_{\TIME})/\widetilde{e}\LSS(a_Y,\overline{\bL}_{\TIME}) \big\|_{\infty} $ for $a \in \{0,1\}$ are bounded above by $C$ almost surely.

\item[\HT{A9''}] (\textit{Consistency}) For all $k \in \{1,\ldots,K\}$ and $a \in \{0,1\}$, 
 $\big\| \widetilde{\mu}\LSS(a,\overline{\bL}_{\TIME}) - \mu^*(a,\overline{\bL}_{\TIME}) \big\|_{P,2}$ and
 $\big\| \widetilde{e}\LSS(a,\overline{\bL}_{\TIME}) - e^*(a,\overline{\bL}_{\TIME}) \big\|_{P,2}$ are $o_P(1)$.

 \item[\HT{A10''}] (\textit{Cross-product Rates for the Numerator}) For all $k \in \{1,\ldots,K\}$ and $a \in \{0,1\}$,  
 $ \big\| \widetilde{\mu}\LSS(a,\overline{\bL}_{\TIME}) - \mu^*(a,\overline{\bL}_{\TIME}) \big\|_{P,2}
\times 
 \big\| \widetilde{e}\LSS(a,\overline{\bL}_{\TIME}) - e^*(a,\overline{\bL}_{\TIME}) \big\|_{P,2}$ is $o_P(N^{-1/2})$. 	
 \end{itemize}
 Then, we have
 \begin{align*}
     \sqrt{N}
     \big\{
         \widetilde{\tau}_{\RD,\CSE}(a_D) - \tau_{\CSE}^*(a_D)
     \big\}
     \stackrel{D}{\rightarrow}
     N \big( 0 , \VAR \big\{ \cInfFt_{\RD,\CSE}^\dagger(\bO \con a_D) \big\} \big) \ ,
 \end{align*}
 where $\cInfFt_{\RD,\CSE}^\dagger$ is given in \eqref{eq-EIF-noU-random}. A consistent estimator for the variance is
 \begin{align*}
     & 
     \widetilde{\sigma}_{\RD,\CSE}^2 (a_D)
     \\
     &
     =
     \frac{1}{N}
     \sum_{k=1}^{K}
     \sum_{i \in \mathcal{I}_k}
     \bigg\{
     \frac{\widetilde{\InfFt}_{\RD,\NUMER}\LSS(\bO_i \con a_Y=1, a_D)
     -
     \widetilde{\InfFt}_{\RD,\NUMER}\LSS(\bO_i \con a_Y=0, a_D)
     -
     \widetilde{\tau}_{\RD,\CSE}(a_D)
     \widetilde{\InfFt}_{\RD,\DENOM}\LSS(\bO_i \con  a_D)
     }{ \widetilde{\psi}_{\RD,\DENOM} (a_D) } 
     \bigg\}^2 \ .
 \end{align*}
 \end{theorem}
 The proof of the theorem is omitted because it is similar to those of Theorems \ref{thm:AN} and \ref{thm:AN rand}.

  \subsubsection{Comparison to the Survivor Average Causal Effect} \label{sec:Monotonicity}

 We provide the details on the relationship between the survivor average causal effect \citep{Robins1986, FR2002} and the conditional separable effect. Suppose that (i) Assumption \HL{A4} holds, (ii) \hyperlink{$A_D$-Partial Isolation}{($A_D$-Partial Isolation)} holds; (iii) there is determinism $A=A_Y=A_D$ in the current two-arm study; (iv) and the following monotonicity is satisfied:
 \begin{align*}
     \text{Monotonicity:} \quad \potDt{a=1}{\TIME+1} \leq \potDt{a=0}{\TIME+1} \text{ a.s.}
 \end{align*}
 Then, following Proposition 1 of \citet{Stensrud2022CSE}, we establish that
 \begin{align*}
     &
     \EXP \big\{ \potY{a=1} - \potY{a=0} \cond \potDt{a=1}{\TIME+1}=\potDt{a=0}{\TIME+1} = 0 \big\}
     \\
     &
     =
     \EXP \big\{ \potY{a=1} - \potY{a=0} \cond \potDt{a=0}{\TIME+1} = 0 \big\}
     && \Leftarrow \quad \text{Monotonicity}
     \\ 
     &
     =
     \EXP \big\{ \potY{a_Y=1,a_D=1} - \potY{a_Y=0,a_D=0} \cond \potDt{a_Y=0,a_D=0}{\TIME+1} = 0 \big\}
     && \Leftarrow \quad \text{$A=A_Y=A_D$ in the current two-arm study}
     \\ 
     &
     =
     \EXP \big\{ \potY{a_Y=1,a_D=0} - \potY{a_Y=0,a_D=0} \cond \potDt{a_Y=0,a_D=0}{\TIME+1} = 0 \big\}
     && \Leftarrow \quad \text{$A_D$-partial isolation}
     \\ 
     &
     =
     \EXP \big\{ \potY{a_Y=1,a_D=0} - \potY{a_Y=0,a_D=0} \cond \potDt{a_D=0}{\TIME+1} = 0 \big\} \ .
     && \Leftarrow \quad \text{Assumption \HL{A4}: $A_Y$-partial isolation}
 \end{align*}
 Therefore, the survivor average causal effect is equal to the conditional separable effect, i.e., 
 \begin{align*}
     \EXP \big\{ \potY{a=1} - \potY{a=0} \cond \potDt{a=1}{\TIME+1}=\potDt{a=0}{\TIME+1} = 0 \big\}
     =
     \EXP \big\{ \potY{a_Y=1,a_D=0} - \potY{a_Y=0,a_D=0} \cond \potDt{a_D=0}{\TIME+1} = 0 \big\} \ .
 \end{align*}
 Consequently, the results in the main paper can be extended to infer the survivor average causal effect under the monotonicity assumption, along with other relevant assumptions outlined therein.

\subsection{Details of the Simulation Studies} \label{sec:supp:Simulation CV}

This Section details the simulation studies by describing the data-generating process and reporting results from additional simulations.

\subsubsection{Data-generating Process}

We present the details of the data-generating process for the simulation study in the main paper. For both settings, we generated 5-dimensional $\bX_{0} = (X_{01},\ldots,X_{05})\T$ and 1-dimensional $U$ and $W$ from the following data-generating process:
 \begin{align*}
 &
 (U,X_{0}\T)\T
 \sim 
 \text{MVN}_6
 \left(
 0
 , 
 I_6
 \right) 
 \ , \
 &
 \textstyle{
 W \cond (\bX_{0},U) \sim N \big( 0.1 \sum_{j=1}^{5} X_{0j} + 2 U , 1 \big) 
 }
 \ .
 \end{align*}
 Then, the time-varying covariate $X_{\TIME}^{(a_D)} = (X_{\TIME 1}^{(a_D)},\ldots,X_{\TIME 5}^{(a_D)})\T$ is generated as follows:
 \begin{align*}
 &
 \textstyle{
 \bX_{\TIME j}^{(a_D)}
 \cond 
 (\bX_{0j},U)
 \sim 
 N
 \big(
 0.015 \ind (a_D=1) - 0.015 \ind (a_D=0) + 0.02 \bX_{0j} + 0.05 U 
 , 1
 \big)
 } \ ,  \quad
 j \in \{1,\ldots, 5\} \ .
 \end{align*} 
 For the observational setting, we then generated $(A,Z, \potDt{a_D}{\TIME+1})$ as follows:
 \begin{align*}
 &
 \textstyle{ 
 A \cond (\bX_{0},U) \sim \text{Ber} \big( \text{expit} ( 0.02 \sum_{j=1}^{5} X_{0j} + 0.1 U  )  \big)
 }
 \ , \\
 &
 \textstyle{ 
 Z \cond (A,\bX_{0},U) \sim N \big( -0.5 A + 0.1 \sum_{j=1}^{5} X_{0j} - 1.5 U , 1 \big)
 }
 \ , \\
 &
 \textstyle{ 
 \potDt{a_D=0}{\TIME+1} \cond (\obX_{\TIME}^{(a_D)},U) \sim \text{Ber} \big( 1-\exp (  -0.5 + 0.01 \sum_{j=1}^{5} \big\{ X_{0j} + 2 X_{\TIME j}^{(a_D)} \big\} + 0.025 U )   \big)
 }
 \ , \\
 &
 \textstyle{ 
 \potDt{a_D=1}{\TIME+1} \cond (\obX_{\TIME}^{(a_D)},U) \sim \text{Ber} \big( 1-\exp (  -0.5 + 0.01 \sum_{j=1}^{5} \big\{ X_{0j} - 2 X_{\TIME j}^{(a_D)} \big\} - 0.125 U )   \big)
 }
 \ .
 \end{align*}
 For the experimental setting, $(A,Z,\potDt{a_D}{2})$ were generated as follows:
 \begin{align*}
 &
 \textstyle{ 
 A \sim \text{Ber}(0.5) 
 }
 \quad , 
 \quad
 \textstyle{ 
 Z \cond (\bX_{0},U) \sim N \big( 0.1 \sum_{j=1}^{5} X_{0j} - 1.5 U , 1 \big)
 }
 \ , \\
 &
 \textstyle{ 
 \potDt{a_D=0}{\TIME+1} \cond (Z,\obX_{\TIME}^{(a_D)},U) \sim \text{Ber} \big( 1-\exp (  -0.5 + 0.075Z + 0.01 \sum_{j=1}^{5} \big\{ X_{0j} + 2 X_{\TIME j}^{(a_D)} \big\} + 0.1 U )   \big)
 }
 \ , \\
 &
 \textstyle{ 
 \potDt{a_D=1}{\TIME+1} \cond (Z,\obX_{\TIME}^{(a_D)},U) \sim \text{Ber} \big( 1-\exp (  -0.5 - 0.15Z + 0.01 \sum_{j=1}^{5} \big\{ X_{0j} - 2 X_{\TIME j}^{(a_D)} \big\} - 0.2 U )   \big)
 }
 \ .
 \end{align*}
 Lastly, for both observational and experimental settings, $\potY{a_Y,a_D}$ was generated from 
 \begin{align*}
 &
 \textstyle{ 
 \potY{a_Y=0, a_D} \cond (\potD{a_D}=0,W,\obX_{\TIME}^{(a_D)},U) \sim N \big( 2 + 0.1 W + 0.01 \sum_{j=1}^{5} \big\{ 5X_{0j} + 2 X_{\TIME j}^{(a_D)} \big\} + 1.5 U , 1 \big)
 }
 \ , \\
 &
 \textstyle{ 
 \potY{a_Y=1, a_D} \cond (\potD{a_D}=0,W,\obX_{\TIME}^{(a_D)},U) \sim N \big( 4 + 0.1 W + 0.01 \sum_{j=1}^{5} \big\{ 5X_{0j} + 2 X_{\TIME j}^{(a_D)} \big\} + 1.5 U , 1 \big)
 }
 \ .
 \end{align*}
 The CSE under $a_D=0$ is $\tau_{\CSE}^*=2$. Under the data-generating process, the confounding bridge functions are uniquely defined and available in closed-form; see Section \ref{sec:supp:CBF}  for details. 

\subsubsection{Additional Simulation Studies}

We conducted an additional simulation study under no unmeasured confounding by setting the unmeasured confounder to zero in the data-generating process, i.e., $U=0$. The empirical distribution and numerical summaries of the estimators are presented in Figure \ref{fig:Simulation no U}. Since $\widetilde{\tau}_{\CSE}$ is developed under no unmeasured confounding, it is not surprising that it performs well, exhibiting negligible bias and attaining nominal coverage. However, our proposed estimator, $\widehat{\tau}_{\CSE}$, performs competitively, also exhibiting negligible bias and attaining nominal coverage. In terms of standard error, $\widetilde{\tau}_{\CSE}$ shows a smaller standard error than $\widehat{\tau}_{\CSE}$, which is expected as the former is a semiparametric efficient estimator in this case. However, the $\widehat{\tau}_{\CSE}$ comes at the cost of its inability to address unmeasured confounding, as shown in Figure \ref{fig:Simulation}. 

 \begin{figure}[!htb]
 \centering 
 \includegraphics[width=1\textwidth]{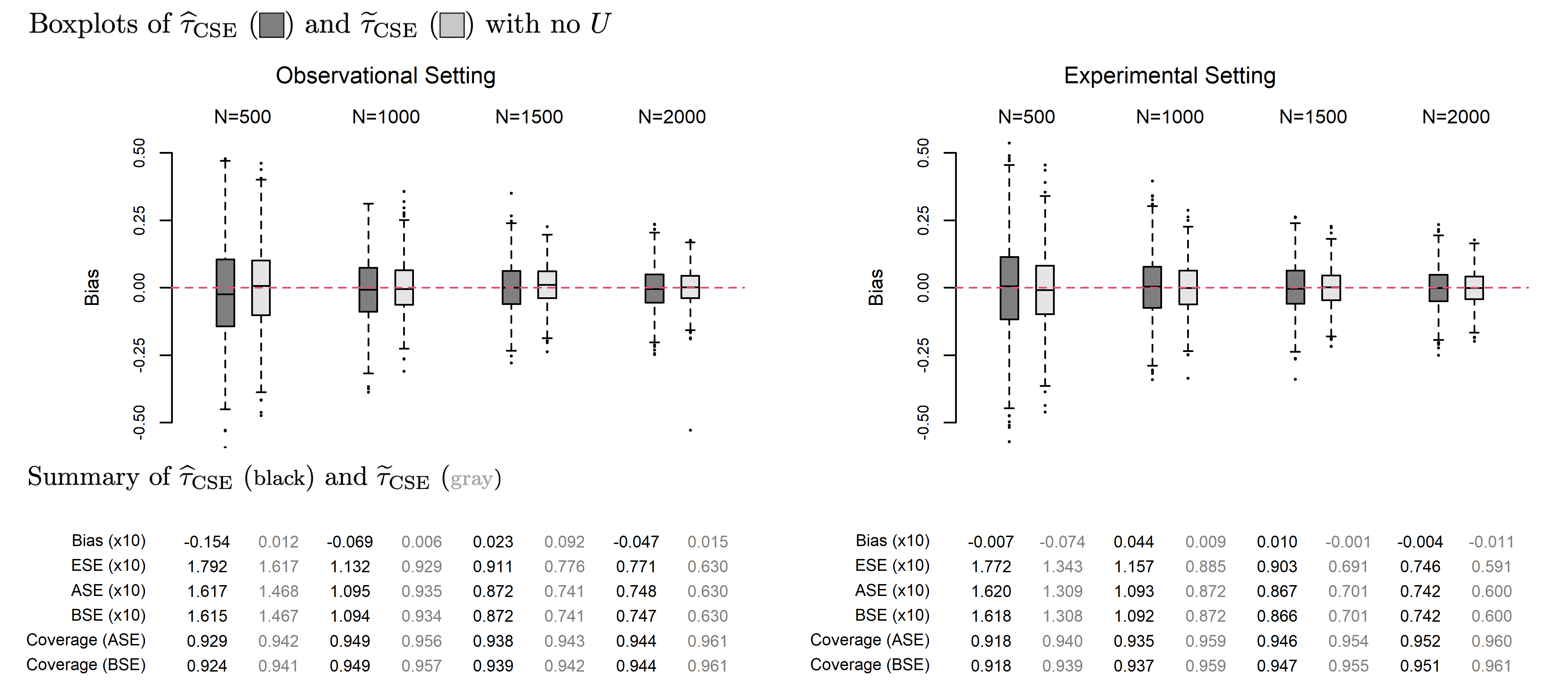} 
 \caption{\footnotesize A Graphical Summary of the Simulation Results. The left and right panels show results for observational and experimental settings, respectively. In the top panel, each column gives boxplots of biases of $\widehat{\tau}_{\CSE}$ and $\widetilde{\tau}_{\CSE}$ for  $N\in \{500,1000,1500,2000\}$, respectively. The bottom panel provides numerical summaries informing the performance of $\widehat{\tau}_{\CSE}$. Each row represents the empirical bias, empirical standard error (ESE), asymptotic standard error based on the proposed variance estimator (ASE), bootstrap standard error (BSE), empirical coverage rates of 95\% confidence intervals based on the ASE and BSE. The bias and standard errors are scaled by factors of 10.} 
 \label{fig:Simulation no U} 
 \end{figure}

Lastly, we examined estimation error of the six nuisance functions in \eqref{eq-bridgeft1 obs}-\eqref{eq-bridgeft6 obs}. Specifically, we considered the number of study units $N \in \{ 400 \times 2^{k/2} \cond k \in \{0, 1, \ldots,8\} \}$. We then generated $N$ observations from the specified observational setting, and estimated the nuisance functions, denoted by $\widehat{h}_1$, $\widehat{h}_0(a=1)$, $\widehat{h}_0(a=0)$, $\widehat{h}_2$, $\widehat{q}_0$, and $\widehat{q}_1$. We then computed the root mean squared error (RMSE) of the estimated nuisance functions using $M = 10^{4}$ observations that were independently generated from the previous $N$ observations. For example, the RMSE of $\widehat{h}_1$ was calculated by
\begin{align*}
	\text{RMSE}(f)
	 =
	 \bigg[ \frac{1}{M} \sum_{i=1}^{M} \big\{ \widehat{h}_1(\bW_i,\obX_{\TIME,i}) - h_1^*(\bW_i,\obX_{\TIME,i}) \big\}^2
	 \bigg]^{1/2} \ ,
\end{align*}
and the RMSEs of other estimated nuisance functions were similarly calculated.

Figure \ref{fig:Simulation Convergence} shows the relationship between $N$ and the RMSE of the estimated nuisance functions. Based on the simulation results, the rate of estimation error appears to be
\begin{align*}
	&
	\big\| \widehat{h}_1 - h_1^* \big\|_{P,2} = O_P(N^{-0.352})
	\ , \\
	&
	\big\| \widehat{h}_0(a=1) - h_0^*(a=1) \big\|_{P,2} = O_P(N^{-0.308})
	\ , \\
	&
	\big\| \widehat{h}_0(a=0) - h_0^*(a=0) \big\|_{P,2} = O_P(N^{-0.259})
	\ , \\
	&
	\big\| \widehat{h}_2 - h_2^* \big\|_{P,2} = O_P(N^{-0.358})
	\ , \\
	&
	\big\| \widehat{q}_0 - q_0^* \big\|_{P,2} = O_P(N^{-0.347})
	\ , \\
	&
	\big\| \widehat{q}_1 - q_1^* \big\|_{P,2} = O_P(N^{-0.315}) \ .
\end{align*}
Consequently, we find Assumptions \HL{A9}-\HL{A11} are satisfied, which demonstrates that the hyperparameter selection procedure proposed in Section \ref{sec:supp:hyperparameter} appears to be valid.

 \begin{figure}[!htb]
 \centering 
 \includegraphics[width=1\textwidth]{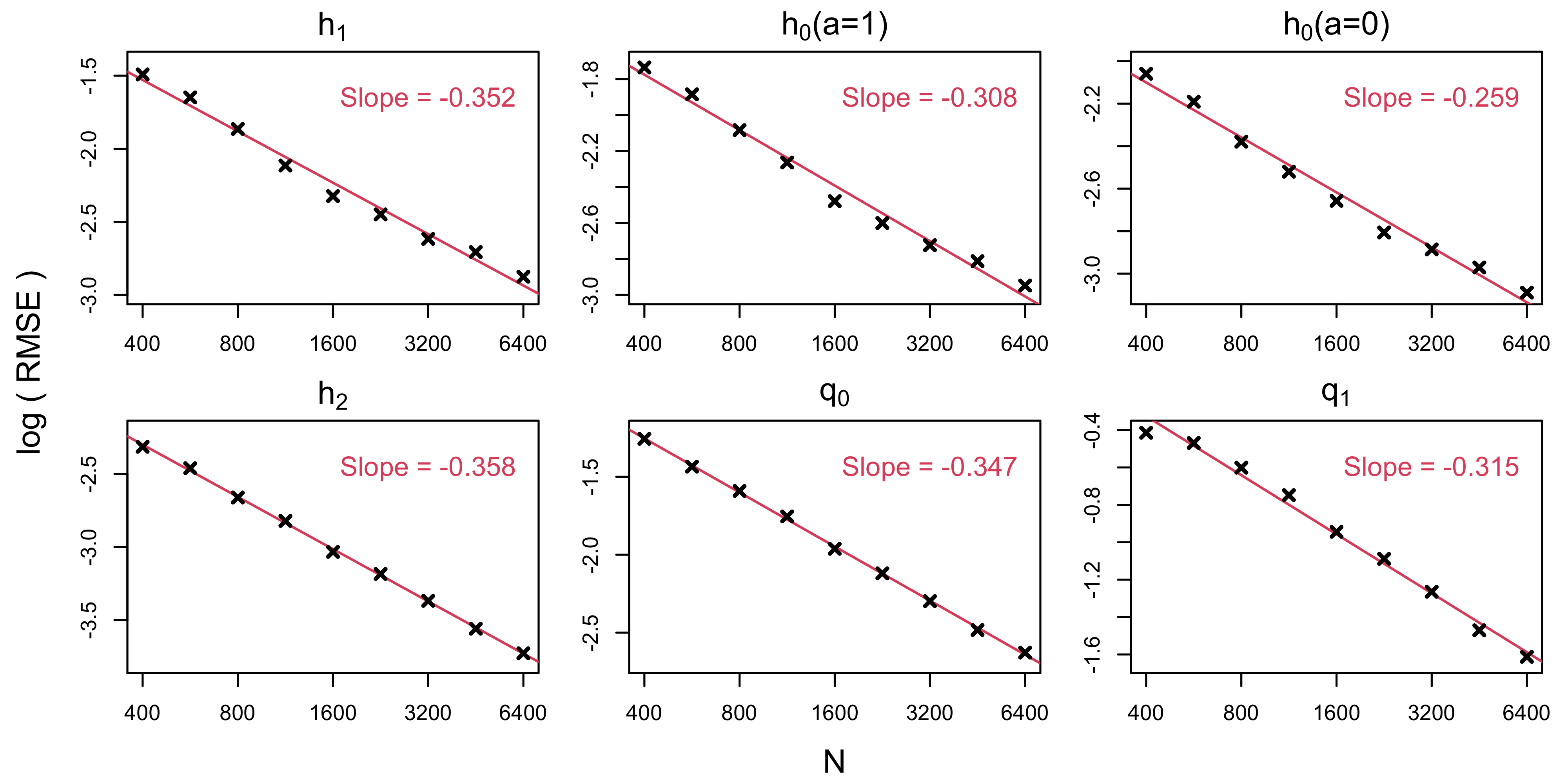} 
 \caption{\footnotesize A Graphical Summary of Estimation Error. Each plot shows the relationship between $N$ and the RMSE of the estimated nuisance functions. The red line represents the best linear fit between $\log(N)$ and $\log(\text{RMSE})$, and the number shown in each figure indicates the slope of this line.}
 \vspace*{-0.1cm}
 \label{fig:Simulation Convergence}
 \end{figure}

\subsubsection{Closed-form Representations of Confounding Bridge Functions}	\label{sec:supp:CBF}

 We present closed-form representations of confounding bridge functions in the simulation studies in Section \ref{sec:Simulation}. Our results are derived under the observational setting; however, a similar result can be established for the experimental setting.

 Recall that the data-generating process of the observational setting has the following form:
 \begin{itemize} 
 \item $p=\dim(X_0)=\dim(X_1)$
 \item $(U, X_{0}) \sim MVN(0,\Sigma_{ux_0})$
 \item $\Pr(A=1-a_D \cond X_{0},U) = \text{expit} ( \beta_{a0} + \beta_{ax_0} (1_p\T X_{0} ) + \beta_{au} U) $	\\
 $\Leftrightarrow$ \quad	$\Pr(A=a_D \cond X_{0},U) = \{ 1+ \exp( \beta_{a0} + \beta_{ax_0} (1_p\T X_{0} )+ \beta_{au} U) \}^{-1} $
 \item $Z \cond (A,X_{0},U) = \beta_{z0} + \beta_{za} A + \beta_{zx_0} \beta_{ax_0} (1_p\T X_{0} )  + \beta_{zu} U + \epsilon_z$ where $\epsilon_z \sim N(0,\sigma_z^2)$
 \item $W \cond (X_{0},U) = \beta_{w0} + \beta_{wx_0} \beta_{ax_0} (1_p\T X_{0} ) + \beta_{wu} U + \epsilon_w$ where $\epsilon_w \sim N(0,\sigma_w^2)$
\item $X_{1j}^{(a_D)} \cond (X_0,U) =  \beta_{x_1 0}^{(a_D)}+\beta_{x_1 x_0}^{(a_D)} X_{0j} + \beta_{x_1u}^{(a_D)}U  + \epsilon_{x_1 j}$ for $j=1,\ldots,p$ where $\epsilon_{x_1 j} \sim N(0,\sigma_{x_1}^2)$
\item $\Pr\{ D_{2}^{(a_D)}=0 \cond X_{1}^{(a_D)},X_0, U \} 
 = \exp \big\{ 
     \beta_{d0}^{(a_D)} + \beta_{dx_1}^{(a_D)} (1_p\T X_{1}^{(a_D)} ) + \beta_{dx_0}^{(a_D)} (1_p\T X_{0} ) + \beta_{du}^{(a_D)} U 
 \big\} $ 
\item $Y^{(a_Y,a_D)} \cond (D_{2}^{(a_D)}=0,W,X_{1}^{(a_D)},X_0,U)$\\ 
$=  \beta_{y0}^{(a_Y)} + \beta_{yx_1}^{(a_Y)} (1_p\T X_{1}^{(a_D)} ) + \beta_{yx_0}^{(a_Y)} (1_p\T X_{0} ) +  \beta_{yw}^{(a_Y)} W + \beta_{yu}^{(a_Y)} U + \epsilon_y$ where $\epsilon_y \sim N(0,\sigma_y^{2})$ 

 \end{itemize}
 In addition, the confounding bridge functions are defined as functions that solve:
 In what follows, we investigate the closed-form representation of each confounding bridge function:

 \begin{itemize}[leftmargin=0.25cm]
 \item[(i)] $h_1^*$\\[0.25cm]
 From Lemma \ref{thm:integral equation}-(i), we have
 \begin{align*}
 &	\EXP (Y \cond A=a_Y,D_{2}=0,Z,\obX_{1}) = \EXP \{ h_{1}^*(W,\obX_{1} \con a_Y) \cond A=a_Y,D_{2}=0,Z,\obX_{1} \}
 \\
 \Rightarrow \quad
 &	\EXP (Y \cond A=a_Y,D_{2}=0,\obX_{1},U) = \EXP \{ h_{1}^*(W,\obX_{1} \con a_Y) \cond A=a_Y,D_{2}=0,\obX_{1},U \} \ .
 \end{align*}	
 We further establish
 \begin{align*}
 &
 \EXP (Y \cond A=a_Y,D_{2}=0,\obX_{1},U)
 \\
 &
 =
 \EXP \big\{ \potY{a_Y,a_D=a_Y} \cond A=a_Y,\potDt{a_Y}{2}=0,X_1^{(a_Y)},X_0,U)
 \\
 &
 =
 \EXP \big\{  \beta_{y0}^{(a_Y)} + \beta_{yx_1}^{(a_Y)} (1_p\T X_{1}^{(a_Y)} ) + \beta_{yx_0} (1_p\T X_{0} ) +  \beta_{yw}^{(a_Y)} W + \beta_{yu}^{(a_Y)} U  \COND A=a_Y, \potDt{a_Y}{2}=0, X_1^{(a_Y)},X_0, U \big\}
 \\
 &
 =
 \EXP \Bigg[
 \begin{array}{l}
\beta_{y0}^{(a_Y)} + \beta_{yx_1}^{(a_Y) } (1_p\T X_{1}^{(a_Y)} ) + \beta_{yx_0} (1_p\T X_{0}) +  \beta_{yw}^{(a_Y)} W \\
+ \beta_{yu}^{(a_Y)}
 \big\{ \frac{W - \beta_{w0} - \beta_{wx_0} (1_p\T  X_{0} ) }{\beta_{wu}} \big\}
 \end{array}
  \COND A=a_Y, \potDt{a_Y}{2}=0, X_1^{(a_Y)},X_0, U \Bigg]
 \\
 &
 =
 \EXP \Bigg[
 \begin{array}{l} 
\beta_{y0}^{(a_Y)} + \beta_{yx_1}^{(a_Y)} (1_p\T X_{1}) + \beta_{yx_0} (1_p\T X_{0}) +  \beta_{yw}^{(a_Y)} W \\
+ \beta_{yu}^{(a_Y)}
 \big\{ \frac{W - \beta_{w0} - \beta_{wx_0} (1_p\T  X_{0} ) }{\beta_{wu}} \big\}
 \end{array}\COND A=a_Y, D_2=0,X_1,X_0,U \Bigg]
 \ .
 \end{align*}
 The first identity is from the consistency assumption \HL{A1}. The second and third identities are from the data-generating process. The fourth identity is again from the consistency assumption \HL{A1}. Therefore, this implies
 \begin{align*}
 h_{1}^*(W,\obX_{1} \con a_Y)
 =
 \bigg\{
 \beta_{y0}^{(a_Y)} - \frac{\beta_{yu}^{(a_Y)} \beta_{w0}}{\beta_{wu}}
 \bigg\}
 +
 \beta_{yx_1}^{(a_Y)} (1_p\T X_{1})
 +
 \bigg\{ \beta_{yx_0}^{(a_Y)} -\frac{\beta_{yu}^{(a_Y)} \beta_{wx_0} }{\beta_{wu}} \bigg\} (1_p\T X_{0})
 +
 \bigg\{ \beta_{yw}^{(a_Y)} + \frac{\beta_{yu}^{(a_Y)}}{\beta_{wu}} \bigg\} W \ .
 \end{align*}

 \item[(ii)] $h_0^*(\cdot \con a_Y,a_D)$ for $a_Y \neq a_D$\\[0.25cm]
 From Lemma \ref{thm:integral equation}-(ii), we have
 \begin{align*}
 &
 \EXP \big\{ h_{1}^* (W,\obX_{1} \con a_Y) (1-D_{2}) \cond A=a_D,Z,X_{0} \big\} 
 = 
 \EXP \big\{ h_{0}^* (W,X_{0}  \con a_Y, a_D) \cond A=a_D,Z,X_{0} \big\} 
 \\
 \Rightarrow \quad
 &
 \EXP \big\{ h_{1}^* (W,\obX_{1} \con a_Y) (1-D_{2}) \cond A=a_D,X_{0},U \big\} 
 = 
 \EXP \big\{ h_{0}^* (W,X_{0} \con a_Y, a_D) \cond A=a_D,X_{0},U \big\} 
 \ .
 \end{align*}
 We further establish
 \begin{align*}
 & 
 \Pr(D_{2}=0 \cond A=a_D,X_0,U)
 \EXP \big\{ h_{1}^*(W,\obX_{1} \con a_Y) \cond A=a_D,D_{2}=0,X_{0},U \big\} 
 \\
 &
 =	
 \EXP \big\{ h_{1}^* (W,\obX_{1} \con a_Y) (1-D_{2}) \cond A=a_D,X_{0},U \big\} 
 \\
 &
 = 
 \EXP \big\{ h_{0}^* (W,X_{0} \con a_Y, a_D) \cond A=a_D,X_{0},U \big\}   \ .
 \end{align*}
 The first identity is from the law of iterated expectation. 
 The second identity is from the above result.
 The third identity is from the law of iterated expectation. 

In addition, we establish
\begin{align*}
&
\Pr \big\{ \potDt{a_D}{2}=0 \cond A=a_D,X_{0},U \big\}
\\
&
=
\EXP
\big[
\Pr \big\{ \potDt{a_D}{2}=0 \cond A=a_D,X_{1}^{(a_D)},X_{0},U \big\}
\cond A=a_D,X_{0}, U
\big]
\\
&
=
\EXP
\big[ 
\exp \big\{ 
     \beta_{d0}^{(a_D)} + \beta_{dx_1}^{(a_D)} (1_p\T X_{1}^{(a_D)}) + \beta_{dx_0}^{(a_D)} (1_p\T X_{0}) + \beta_{du}^{(a_D)} U 
 \big\} 
 \cond 
 A=a_D,X_{0}, U
 \big] 
\\
&
=
\exp \big\{ 
     \beta_{d0}^{(a_D)} + \beta_{dx_0}^{(a_D)} (1_p\T X_{0}) + \beta_{du}^{(a_D)} U 
 \big\} 
\EXP
\big[ 
\exp \big\{   \beta_{dx_1}^{(a_D)} (1_p\T X_{1}^{(a_D)} )
 \big\} 
 \cond 
 A=a_D,X_{0}, U
 \big] 
\\
&
=
\exp \big\{ 
     \beta_{d0}^{(a_D)} + \beta_{dx_0}^{(a_D)} (1_p\T X_{0}) + \beta_{du}^{(a_D)} U 
 \big\} 
\EXP
\big[ 
\exp \big\{  \beta_{dx_1}^{(a_D)} (1_p\T X_{1}^{(a_D)}  )
 \big\} 
 \cond 
 X_{0}, U
 \big] 
\\
&
=
\exp \big\{ 
     \beta_{d0}^{(a_D)} + \beta_{dx_0}^{(a_D)} (1_p\T X_{0}) + \beta_{du}^{(a_D)} U 
 \big\} 
\exp \big[
\beta_{dx_1}^{(a_D)}
\{ p
 \beta_{x_1 0}^{(a_D)} + \beta_{x_1 x_0}^{(a_D)} (1_p\T X_{0}) + p \beta_{x_1 u}^{(a_D)} U\}
+
0.5 p \beta_{dx_1}^{(a_D) 2} \sigma_{x_1}^2
\big]
\\
&
=
 \exp 
 \left[
 \begin{array}{l} 
 \beta_{d0}^{(a_D)}
 +
p \beta_{dx_1}^{(a_D)}
\beta_{x_1 0}^{(a_D)}
+
0.5 p \beta_{dx_1}^{(a_D) 2} \sigma_{x_1}^2
\\
+
\{\beta_{dx_0}^{(a_D) } +  \beta_{dx_1}^{(a_D)} \beta_{x_1 x_0}^{(a_D)} \}  (1_p\T X_0)  \\
+ 
\{ \beta_{du}^{(a_D)} + 
p \beta_{dx_1}^{(a_D)} \beta_{x_1 u}^{(a_D)} 
\}
U
 \end{array}
\right]
\numeq \label{eq-PrD0A}
 \ .
\end{align*}
The first four lines are trivial. The fifth line holds from $X_1^{(a_D)} \indep A \cond X_{0},U$. The last line holds from the fact that 
 \begin{align}	\label{eq-MGF}
&
 V \sim N(\mu,\sigma^2) \quad
 \Rightarrow 
 \quad \EXP \big( e^{aV} \big) = \exp (\mu a + 0.5 a^2 \sigma^2 )
 \ , \quad 
 \EXP \big( V e^{aV} \big) = (\mu+ a \sigma^2) \exp( \mu a + 0.5 a^2 \sigma^2 ) \ .
 \end{align}
 
We also establish 
\begin{align*}
&
\EXP \big\{ 1_p\T X_{1}^{(a_D)} \cond  A=a_D,\potDt{a_D}{2}=0,X_{0}, U \big\}
\\
&
=
\int x
f \big\{ 1_p\T X_{1}^{(a_D)} = x \cond  A=a_D,\potDt{a_D}{2}=0,X_{0}, U \big\} \, dx
\\
&
=
\int x
\frac{ 
f \big\{ 1_p\T X_{1}^{(a_D)} = x, \potDt{a_D}{2}=0 \cond  A=a_D,X_{0}, U \big\}
}{ 
\Pr \big\{ \potDt{a_D}{2}=0 \cond  A=a_D,X_{0}, U \big\}
}
\, dx
\\
&
=
\frac{\int x
\Pr \big\{ \potDt{a_D}{2}=0 \cond 1_p\T X_{1}^{(a_D)} = x,  A=a_D,X_{0}, U \big\}
f \big\{ 1_p\T X_{1}^{(a_D)} = x \cond  A=a_D,X_{0}, U \big\}
\, dx
}{\Pr \big\{ \potDt{a_D}{2}=0 \cond  A=a_D,X_{0}, U \big\}}
\\
&
=
\frac{
\EXP \big[
(1_p\T X_{1}^{(a_D)}) 
\exp \big\{ 
     \beta_{d0}^{(a_D)} + \beta_{dx_1}^{(a_D)} (1_p\T X_{1}^{(a_D)}) + \beta_{dx_0}^{(a_D)} (1_p\T X_{0}) + \beta_{du}^{(a_D)} U 
 \big\}
 \cond A=a_D, X_{0}, U \big] 
}{\Pr \big\{ \potDt{a_D}{2}=0 \cond  A=a_D,X_{0}, U \big\}}
\\ 
&
=
\frac{
\exp \big\{ 
     \beta_{d0}^{(a_D)} + \beta_{dx_0}^{(a_D)} (1_p\T X_{0}) + \beta_{du}^{(a_D)} U 
 \big\}
}{\Pr \big\{ \potDt{a_D}{2}=0 \cond  A=a_D,X_{0}, U \big\}}
\EXP \big[ 
1_p\T X_{1}^{(a_D)}  \exp\{ \beta_{dx_1}^{(a_D)} 1_p\T X_{1}^{(a_D)}  \}
 \cond A=a_D, X_{0}, U \big] \ .
\end{align*} 
Note that $1_p\T X_{1}^{(a_D)}  \cond A=a_D, X_{0}, U \sim N (  p \beta_{x_1 0}^{(a_D)} + \beta_{x_1 x_0}^{(a_D)} (1_p\T X_{0}) + p \beta_{x_1 u}^{(a_D)} U , p \sigma_{x_1}^2)$. Therefore, 
\begin{align*}
&
\EXP \big[ 
1_p\T X_{1}^{(a_D)}  \exp\{ \beta_{dx_1}^{(a_D)} 1_p\T X_{1}^{(a_D)}  \}
 \cond A=a_D, X_{0}, U \big]
 \\
 &
 =
\big\{ p \beta_{x_1 0}^{(a_D)} + \beta_{x_1 x_0}^{(a_D)} (1_p\T X_{0}) + p \beta_{x_1 u}^{(a_D)} U + p \beta_{dx_1}^{(a_D)}  \sigma_{x_1}^2 \big\}
\\
&
\quad \times 
\exp \big\{ 
p  \beta_{dx_1}^{(a_D)} \beta_{x_1 0}^{(a_D)} +  \beta_{dx_1}^{(a_D)} \beta_{x_1 x_0}^{(a_D)} (1_p\T X_{0}) + p  \beta_{dx_1}^{(a_D)}\beta_{x_1 u}^{(a_D)} U
+
0.5 p \beta_{dx_1}^{(a_D)2} \sigma_{x_1}^2
\big\} \ ,
\end{align*} 
which results in
\begin{align*}
&
\EXP \big\{ 1_p\T X_{1}^{(a_D)} \cond  A=a_D,\potDt{a_D}{2}=0,X_{0}, U \big\}
\\
& 
=
\frac{ \displaystyle{ \left[ 
	\begin{array}{l}
	 \big\{ p \beta_{x_1 0}^{(a_D)} + p \beta_{dx_1}^{(a_D)} \sigma_{x_1}^{2} \big\} \\
	 + \beta_{x_1 x_0}^{(a_D)} (1_p\T X_{0}) \\
	 + p \beta_{x_1 u}^{(a_D)} U 
	\end{array}
\right]  } 
\times
\exp 
\left[ 
\begin{array}{l}
	\big\{
     \beta_{d0}^{(a_D)} + p \beta_{dx_1}^{(a_D)} \beta_{x_1 0}^{(a_D)} + 0.5 p \beta_{dx_1}^{(a_D)2} \sigma_{x_1}^2
     \big\}
     \\
     + 
     \big\{ \beta_{dx_0}^{(a_D)} + \beta_{dx_1}^{(a_D)} \beta_{x_1 x_0}^{(a_D)} \big\}
     (1_p\T X_{0}) \\
     + \big\{ \beta_{du}^{(a_D)} + p \beta_{dx_1}^{(a_D)} \beta_{x_1 u}^{(a_D)} \big\}  U 
\end{array}
\right]  }{\Pr \big\{ \potDt{a_D}{2}=0 \cond  A=a_D,X_{0}, U \big\}} 
\ .
\end{align*}

 In addition, from the consistency assumption \HL{A1}, Lemma \ref{thm:integral equation}-(i), and the closed-form representation of $h_1^*(W,\obX_{1} \con a_Y)$, $\EXP \big\{ h_{1}^* (W,\obX_{1} \con a_Y) \cond A=a_D,D=0, X_{0},U \big\} $ is represented as
 \begin{align*}
 &
 \EXP \big\{ h_{1}^* (W,\obX_{1} \con a_Y) \cond A=a_D,D_2=0,X_{0},U \big\} 
 \\
 &
 =
 \EXP \big\{ h_{1}^* (W,X_{1}^{(a_D)},X_{0} \con a_Y) \cond A=a_D,\potDt{a_D}{2}=0,X_{0},U \big\} 
 \\
 &
 =
 \EXP \left[
 \begin{array}{l} 
 \big\{
 \beta_{y0}^{(a_Y)} - \frac{\beta_{yu}^{(a_Y)} \beta_{w0}}{\beta_{wu}}
 \big\}
 +
 \beta_{yx_1}^{(a_Y)} (1_p\T X_{1}^{(a_D)})
 \\
 +
 \big\{ \beta_{yx_0}^{(a_Y)} -\frac{\beta_{yu}^{(a_Y)} \beta_{wx_0} }{\beta_{wu}} \big\} (1_p\T  X_{0})
 +
 \big\{ \beta_{yw}^{(a_Y)} + \frac{\beta_{yu}^{(a_Y)}}{\beta_{wu}} \big\} W
 \end{array}
	\COND
 A=a_D,\potDt{a_D}{2}=0,X_{0}, U
\right]
 \\
 &
 =
 \bigg\{
 \beta_{y0}^{(a_Y)} - \frac{\beta_{yu}^{(a_Y)} \beta_{w0}}{\beta_{wu}}
 \bigg\}  
 +
 \bigg\{ \beta_{yx_0}^{(a_Y)} -\frac{\beta_{yu}^{(a_Y)} \beta_{wx_0} }{\beta_{wu}} \bigg\} (1_p\T X_0)
 +
 \bigg\{ \beta_{yw}^{(a_Y)} + \frac{\beta_{yu}^{(a_Y)}}{\beta_{wu}} \bigg\} 
 \big\{ \beta_{w0} + \beta_{wx_0} (1_p\T X_0) + \beta_{wu} U \big\}
 \\
 &
 \quad 
 +
 \beta_{y x_1}^{(a_Y)}
 \EXP \big\{ 1_p\T X_{1}^{(a_D)} \cond  A=a_D,\potDt{a_D}{2}=0,X_{0}, U \big\}
 \\
 &
 =
 \big\{ \beta_{y0}^{(a_Y)}
  + \beta_{yw}^{(a_Y)}  \beta_{w0} 
  \big\}
 +
 \big\{ \beta_{yx_0}^{(a_Y)} + \beta_{yw}^{(a_Y)} \beta_{wx_0} \big\} (1_p\T  X_0) 
 +
 \big\{ \beta_{yu}^{(a_Y)}   + \beta_{yw}^{(a_Y)} \beta_{wu} \big\} U 
 \\
 &
 \quad + 
 \beta_{y x_1}^{(a_Y)}
 \EXP \big\{ 1_p\T X_{1}^{(a_D)} \cond  A=a_D,\potDt{a_D}{2}=0,X_{0}, U \big\}
 \ . 
 \numeq \label{eq-h11expectation}
 \end{align*}

Combining \eqref{eq-PrD0A} and \eqref{eq-h11expectation}, we get
 \begin{align*}
&
  \EXP \big\{ h_{0}^* (W,X_{0} \con a_Y, a_D) \cond A=a_D,X_{0},U \big\}
  \\
 &
 = 
 \Pr(D_{2}=0 \cond A=a_D,X_0,U)
 \EXP \big\{ h_{1}^*(W,\obX_{1} \con a_Y) \cond A=a_D,D_{2}=0,X_{0},U \big\}  
 \\
 &
 = 
\left[ 
\begin{array}{l} 
 \big\{ \beta_{y0}^{(a_Y)}
  + \beta_{yw}^{(a_Y)}  \beta_{w0} 
  \big\}
  \\
 +
 \big\{ \beta_{yx}^{(a_Y)} + \beta_{yw}^{(a_Y)} \beta_{wx} \big\} (1_p\T X_0)
 \\
 +
 \big\{ \beta_{yu}^{(a_Y)} + \beta_{yw}^{(a_Y)} \beta_{wu} \big\} U 
\end{array}
\right]  
\times 
 \exp 
 \left[
 \begin{array}{l} 
\big\{
     \beta_{d0}^{(a_D)} + p \beta_{dx_1}^{(a_D)} \beta_{x_1 0}^{(a_D)} + 0.5 p \beta_{dx_1}^{(a_D)2} \sigma_{x_1}^2
     \big\}
     \\
+
\{\beta_{dx_0}^{(a_D)} + \beta_{dx_1}^{(a_D)}  \beta_{x_1 x_0}^{(a_D)} \} (1_p\T X_0) \\
+ 
\{ \beta_{du}^{(a_D)} + 
p
\beta_{dx_1}^{(a_D)} \beta_{x_1 u}^{(a_D)} 
\}
U
 \end{array}
\right]
\\
&
\quad 
+
\beta_{y x_1}^{(a_Y)}
\left[ 
	\begin{array}{l}
	 \big\{ p \beta_{x_1 0}^{(a_D)} + p \beta_{dx_1}^{(a_D)} \sigma_{x_1}^{2} \big\} \\
	 + \beta_{x_1 x_0}^{(a_D)} (1_p\T X_{0}) \\
	 + p \beta_{x_1 u}^{(a_D)} U 
	\end{array}
\right]
\times  
\exp 
\left[ 
\begin{array}{l}
	\big\{
     \beta_{d0}^{(a_D)} + p \beta_{dx_1}^{(a_D)} \beta_{x_1 0}^{(a_D)} + 0.5 p \beta_{dx_1}^{(a_D)2} \sigma_{x_1}^2
     \big\}
     \\
     + 
     \big\{ \beta_{dx_0}^{(a_D)} + \beta_{dx_1}^{(a_D)} \beta_{x_1 x_0}^{(a_D)} \big\}
     (1_p\T X_{0}) \\
     + \big\{ \beta_{du}^{(a_D)} + p \beta_{dx_1}^{(a_D)} \beta_{x_1 u}^{(a_D)} \big\}  U 
\end{array}
\right]
\\
&
=
\left[ 
	\begin{array}{l}
	 \big\{ 
	 \beta_{y0}^{(a_Y)} + \beta_{yw}^{(a_Y)} \beta_{w0}
	 +
	 p \beta_{yx_1}^{(a_Y)} \beta_{x_1 0}^{(a_D)} + p \beta_{yx_1}^{(a_Y)}\beta_{dx_1}^{(a_D)} \sigma_{x_1}^{2} \big\} \\
	 + 
	 \big\{ \beta_{yx}^{(a_Y)} + \beta_{yw}^{(a_Y)} \beta_{wx} + \beta_{yx_1}^{(a_Y)}\beta_{x_1 x_0}^{(a_D)} 
	 \big\}
	 (1_p\T X_{0}) \\
	 + 
	 \big\{
	  \beta_{yu}^{(a_Y)} + \beta_{yw}^{(a_Y)} \beta_{wu}
	  +
	 p \beta_{yx_1}^{(a_Y)} \beta_{x_1 u}^{(a_D)} 
	 \big\}
	 U 
	\end{array}
\right] 
\times
 \exp 
 \left[
 \begin{array}{l} 
 \beta_{d0}^{(a_D)}
 +
 p
\beta_{dx_1}^{(a_D)}
\beta_{x_1 0}^{(a_D)}
+
0.5 p \beta_{dx_1}^{(a_D) 2} \sigma_{x_1}^2
\\
+
\{\beta_{dx_0}^{(a_D)} + \beta_{dx_1}^{(a_D)}  \beta_{x_1 x_0}^{(a_D)} \} (1_p\T X_0) \\
+ 
\{ \beta_{du}^{(a_D)} + 
p
\beta_{dx_1}^{(a_D)} \beta_{x_1 u}^{(a_D)} 
\}
U
 \end{array}
\right] \ .
 \end{align*}

Using \eqref{eq-MGF}, we can show that $h_0^*(W,X \con a_Y, a_D)$ has the following form with a few lines of algebra:
 \begin{align*}
 & 
 h_0^*(W,X_{0} \con a_Y, a_D) 
 =
 \big\{ \gamma_0 + \gamma_{x_0} (1_p\T X_{0}) + \gamma_w W \big\} \exp \big\{ \delta_0 + \delta_{x_0} (1_p\T X_{0}) + \delta_w W \big\} 
 \end{align*}
 where
 \begin{align*}
 & \gamma_w = \frac{ 
 \beta_{yu}^{(a_Y)}  + \beta_{yw}^{(a_Y)} \beta_{wu} + p \beta_{y x_1}^{(a_Y)} \beta_{x_1 u}^{(a_D)}
 }{\beta_{wu}}
 \ ,
 &&
 \delta_w = \frac{ \beta_{du}^{(a_D)} + p 
\beta_{dx_1}^{(a_D)} \beta_{x_1 u}^{(a_D)}  }{\beta_{wu}}
 \\
 &
 \gamma_x =  \beta_{yx}^{(a_Y)} + \beta_{yw}^{(a_Y)} \beta_{wx0} + \beta_{yx_1}^{(a_Y)} \beta_{x_1 x_0}^{(a_D)}
 - \gamma_w \beta_{wx_0}
 \ ,
 &&
 \delta_x = \beta_{dx_0}^{(a_D)} + \beta_{dx_1}^{(a_D)}  \beta_{x_1 x_0}^{(a_D)} - \delta_{w} \beta_{wx_0}
 \\
 &	
 \gamma_0 = 
 \left\{
 \begin{array}{l}
 \beta_{y0}^{(a_Y)} 
  + \beta_{yw}^{(a_Y)}  \beta_{w0} 
    \\ 
  + p \beta_{y x_1}^{(a_Y)}\beta_{x_1 0}^{(a_D)} 
  + p \beta_{y x_1}^{(a_Y)} \beta_{d x_1}^{(a_D)} \sigma_{x_1}^2
  \\ 
  - \gamma_w \delta_w \sigma_w^2	 - \gamma_w \beta_{w0}
 \end{array}
 \right\}
 \ , 
 &&
 \delta_0
 =
\left\{
 \begin{array}{l}
 \beta_{d0}^{(a_D)}
 +
 p 
\beta_{dx_1}^{(a_D)}
\beta_{x_1 0}^{(a_D)}
    \\ 
+
0.5 p \beta_{dx_1}^{(a_D) 2} \sigma_{x_1}^2 
\\ 
- \delta_w \beta_{w0} - 0.5 \delta_w^2 \sigma_w^2
 \end{array}
\right\}
 \ .
 \end{align*} 

\item[(iii)] $h_0^*(\cdot \con a_D,a_D)$\\[0.25cm] 
From Lemma \ref{thm:integral equation}-(ii), we have
 \begin{align*}
 &
 \EXP \big\{ Y (1-D_{2}) \cond A=a_D,Z,X_{0} \big\} 
 = 
 \EXP \big\{ h_{0}^* (W,X_{0}  \con a_D, a_D) \cond A=a_D,Z,X_{0} \big\} 
 \\
 \Rightarrow \quad
 &
 \EXP \big\{ Y (1-D_{2}) \cond A=a_D,X_{0},U \big\} 
 = 
 \EXP \big\{ h_{0}^* (W,X_{0} \con a_D, a_D) \cond A=a_D,X_{0},U \big\} \ . 
 \end{align*} 
 
 Based on the similar calculation in the previous section, we find
  \begin{align*}
 & 
 h_0^*(W,X_{0} \con a_D, a_D) 
 =
 \big\{ \gamma_0 + \gamma_{x_0} (1_p\T X_{0}) + \gamma_w W \big\} \exp \big\{ \delta_0 + \delta_{x_0} (1_p\T X_{0}) + \delta_w W \big\} 
 \end{align*}
 where
 \begin{align*}
 & \gamma_w = \frac{ 
 \beta_{yu}^{(a_D)}  + \beta_{yw}^{(a_D)} \beta_{wu} + p \beta_{y x_1}^{(a_D)} \beta_{x_1 u}^{(a_D)}
 }{\beta_{wu}}
 \ ,
 &&
 \delta_w = \frac{ \beta_{du}^{(a_D)} + p 
\beta_{dx_1}^{(a_D)} \beta_{x_1 u}^{(a_D)}  }{\beta_{wu}}
 \\
 &
 \gamma_x =  \beta_{yx}^{(a_D)} + \beta_{yw}^{(a_D)} \beta_{wx0} + \beta_{yx_1}^{(a_D)} \beta_{x_1 x_0}^{(a_D)}
 - \gamma_w \beta_{wx_0}
 \ ,
 &&
 \delta_x = \beta_{dx_0}^{(a_D)} + \beta_{dx_1}^{(a_D)}  \beta_{x_1 x_0}^{(a_D)} - \delta_{w} \beta_{wx_0}
 \\
 &	
 \gamma_0 = 
 \left\{
 \begin{array}{l}
 \beta_{y0}^{(a_D)} 
  + \beta_{yw}^{(a_D)}  \beta_{w0} 
    \\[-0.1cm]
  + p \beta_{y x_1}^{(a_D)}\beta_{x_1 0}^{(a_D)} 
  + p \beta_{y x_1}^{(a_D)} \beta_{d x_1}^{(a_D)} \sigma_{x_1}^2
  \\
  - \gamma_w \delta_w \sigma_w^2	 - \gamma_w \beta_{w0}
 \end{array}
 \right\}
 \ , 
 &&
 \delta_0
 =
\left\{
 \begin{array}{l}
 \beta_{d0}^{(a_D)}
 +
 p 
\beta_{dx_1}^{(a_D)}
\beta_{x_1 0}^{(a_D)}
    \\
+
0.5 p \beta_{dx_1}^{(a_D) 2} \sigma_{x_1}^2 
\\
- \delta_w \beta_{w0} - 0.5 \delta_w^2 \sigma_w^2
 \end{array}
\right\}
 \ .
 \end{align*}

 \item[(iv)] $h_2^*$\\[0.25cm]
 From Lemma \ref{thm:integral equation}-(ii), we have
 \begin{align*}
 &
 \EXP \big\{ 1-D_2 \cond A=a_D,Z,X_{0} \big\} 
 = 
 \EXP \big\{ h_{2}^* (W,X_{0} \con a_D) \cond A=a_D,Z,X_{0} \big\} 
 \\
 \Rightarrow \quad
 &
 \EXP \big\{ 1-D_{2} \cond A=a_D,X_{0},U \big\} 
 = 
 \EXP \big\{ h_{2}^* (W,X_{0} \con a_D) \cond A=a_D,X_{0},U \big\}  \ .
 \end{align*}

From \eqref{eq-PrD0A}, we get
 \begin{align*}
  \EXP \big\{ h_{2}^* (W,X_{0} \con a_D, a_D) \cond A=a_D,X_{0},U \big\}
 &
 = 
 \Pr(D_{2}=0 \cond A=a_D,X_0,U)
 \\
 &
 =
 \exp 
 \left[
 \begin{array}{l} 
 \beta_{d0}^{(a_D)}
 +
p \beta_{dx_1}^{(a_D)}
\beta_{x_1 0}^{(a_D)}
+
0.5 p \beta_{dx_1}^{(a_D) 2} \sigma_{x_1}^2
\\
+
\{\beta_{dx_0}^{(a_D) } +  \beta_{dx_1}^{(a_D)} \beta_{x_1 x_0}^{(a_D)} \}  (1_p\T X_0)  \\
+ 
\{ \beta_{du}^{(a_D)} + 
p \beta_{dx_1}^{(a_D)} \beta_{x_1 u}^{(a_D)} 
\}
U
 \end{array}
\right]
 \ .
 \end{align*} 

 We can then show that $h_2^*(W,X_{0} \con a_D)$ has the following form with a few lines of algebra using \eqref{eq-MGF}:
 \begin{align*}
 & 
 h_2^*(W,X \con a_D)
 =
 \exp \big\{ \delta_0 + p \delta_x (1_p\T X_{0}) + \delta_w W \big\}
 \ ,
 \\
 &
 \delta_w = \frac{ \beta_{du}^{(a_D)} + 
\beta_{dx_1}^{(a_D)} \beta_{x_1 u}^{(a_D)}  }{\beta_{wu}}
\ , \quad
 \delta_x = \beta_{dx_0}^{(a_D)} + \beta_{dx_1}^{(a_D)}  \beta_{x_1 x_0}^{(a_D)} - \delta_{w} \beta_{wx_0}
\ , \\
&
 \delta_0
 =
 \beta_{d0}^{(a_D)}
 +
p \beta_{dx_1}^{(a_D)}
\beta_{x_1 0}^{(a_D)}
+
0.5 p \beta_{dx_1}^{(a_D) 2} \sigma_{x_1}^2 
- \delta_w \beta_{w0} - 0.5 \delta_w^2 \sigma_w^2
  \ .
 \end{align*}

 \item[(v)] $q_0^*$\\[0.25cm]
 From Lemma \ref{thm:integral equation}-(iii), we have
  \begin{align*}
 \EXP \big\{ q_0^* (Z,X_{0} \con a_D) \cond A=a_D, X_{0}, U \big\} 	
 &
 =
 \frac{1}{\Pr(A=a_D \cond X_{0},U)} 
 =
 1 + \exp \big\{ \beta_{a0} + \beta_{ax_0} (1_p\T  X_{0}) + \beta_{au} U \big\} \ . 
  \numeq \label{eq-q0closeform}
  \end{align*}
  From \eqref{eq-MGF}, one can show that $q_0^*(Z,X_{0} \con a_D)$ is represented as
 \begin{align*}
 &
 q_0^*(Z,X_{0} \con a_D)
 =
 1+ 
  \exp\{ \gamma_0 + \gamma_{x_0} (1_p\T X_{0}) + \gamma_z Z \} 
 \\
 &
 \text{where}
 \quad 
 \gamma_z = \frac{\beta_{au}}{\beta_{zu}}
 \ , \
 \gamma_{x_0} = \beta_{ax_{0}} - \gamma_{z} \beta_{zx_{0}}
 \ , \
 \gamma_0 = \beta_{a0}   - 0.5 \gamma_z^2 \sigma_z^2 - \gamma_z \beta_{z0} \ .
 \end{align*}

 \item[(vi)] $q_1^*$\\[0.25cm]
 From Lemma \ref{thm:integral equation}-(iv), we have
 \begin{align*}
 &
 \EXP \big\{ q_0^* (Z,X_{0} \con a_D) \cond A=a_D, D_{2}=0, \obX_{1}, U \big\}
 \frac{ \Pr(A=a_D\cond D_{2}=0, \obX_{1}, U)}{ \Pr(A=a_Y\cond D_{2}=0, \obX_{1}, U) }
 \\
 &
 =
 \EXP \big\{ q_{1}^* (Z,\obX_{1} \con a_Y, a_D) \cond A=a_Y, D=0, \obX_{1}, U \big\} \ .
 \end{align*}
 If $a_Y=a_D$, $q_1^* = q_0^*$. Therefore suffices to consider $a_Y \neq a_D$.

Let $\delta_{\star} = \beta_{\star}^{(a_D)} - \beta_{\star}^{(a_Y)}$.  For simplicity, we assume $\beta_{x_1 x_0} = \beta_{x_1 x_0}^{(a_Y)}=\beta_{x_1 x_0}^{(a_D)}$ and $\beta_{x_1 u} = \beta_{x_1 u}^{(a_Y)}=\beta_{x_1 u}^{(a_D)}$, meaning that $\delta_{x_1 x_0}\delta_{x_1 u}=0$. Then, we find 
 \begin{align*}
 &
 \frac{ \Pr(A=a_D\cond D_{2}=0, \obX_{1}, U)}{ \Pr(A=a_Y\cond D_{2}=0, \obX_{1}, U) }
 \\
 & =
 \frac{\Pr(D_{2}=0 \cond A=a_D ,\obX_{1}, U)}{\Pr(D_{2}=0 \cond A=a_Y , \obX_{1}, U)}
 \frac{f(X_1 \cond A=a_D , X_0, U)}{f(X_1 \cond A=a_Y , X_0, U)}
 \frac{ \Pr(A=a_D \cond  X_0, U)}{ \Pr(A=a_Y \cond X_0, U) } 
 \\
 &
 =
 \exp \big\{ \delta_{d0} +  \delta_{dx_1} (1_p\T X_1)  +  \delta_{dx_0} (1_p\T X_0) + \delta_{du} U \big\}
 \\
 &
 \quad
 \times
 \exp \bigg[ \frac{ \delta_{x_1 0} (1_p\T X_1)  }{  \sigma_{x_1}^2}
 + 
 \frac{
	p \{ \beta_{x_10}^{(a_Y) 2} - \beta_{x_10}^{(a_D) 2} \} - 2 \beta_{x_1 x_0} \delta_{x_1 0} (1_p\T X_0) - 2 p \beta_{x_1 u} \delta_{x_1 0} U
  }{ 2\sigma_{x_1}^2 }
  \bigg]
 \\
 &
 \quad
 \times 
 \exp \big\{ - \beta_{a0} - \beta_{ax_0} (1_p\T X_0) - \beta_{au} U \big\}
  \ .
 \end{align*}

Therefore, we get
 \begin{align*}
 &
 \frac{ \Pr(A=a_D\cond D_{2}=0, \obX_{1}, U)}{ \Pr(A=a_Y\cond D_{2}=0, \obX_{1}, U) }  
 =
 \exp 
 \left[
 \begin{array}{l} 
 \displaystyle{
 \bigg[
 \frac{ 
 p \big\{
 \beta_{x_10}^{(a_Y) 2} - \beta_{x_10}^{(a_D) 2}
 \big\}
  }{ 2\sigma_{x_1}^2 } +  \delta_{d0} \bigg]
  + \bigg( \delta_{dx_1}+ \frac{  \delta_{x_1 0} }{  \sigma_{x_1}^2}  \bigg) (1_p\T X_1)  } 
  \\
  \displaystyle{
  +  
\bigg( \delta_{dx_0}
 -
\frac{  \beta_{x_1 x_0} \delta_{x_1 0} }{ \sigma_{x_1}^2}
 \bigg)
  (1_p\T X_0) 
  + 
\bigg( 
 \delta_{du} 
 -
\frac{ p \beta_{x_1 u} \delta_{x_1 0} }{ \sigma_{x_1}^2}
\bigg)
  U 
  }
  \\
  - \beta_{a0} - \beta_{ax_0} (1_p\T X_0) - \beta_{au} U
 \end{array}
  \right] 
  \ .
\numeq   \label{eq-densratio}
 \end{align*}

Combining \eqref{eq-q0closeform} and \eqref{eq-densratio}, $q_1^*(Z,\obX_{1})$ satisfies
 \begin{align*}
     & 
     \EXP \big\{ q_{1}^* (Z,\obX_{1} \con a_Y, a_D) \cond A=a_Y, D_{2}=0, \obX_{1}, U \big\}
     \\
     &
     =
      \exp 
 \left[
 \begin{array}{l} 
 \displaystyle{
 \bigg[
 \frac{ 
 p \big\{
 \beta_{x_10}^{(a_Y) 2} - \beta_{x_10}^{(a_D) 2}
 \big\}
  }{ 2\sigma_{x_1}^2 } +  \delta_{d0} \bigg]
  + \bigg( \delta_{dx_1}+ \frac{  \delta_{x_1 0} }{  \sigma_{x_1}^2}  \bigg) (1_p\T X_1)   
  }
  \\
  \displaystyle{
  +  
\bigg( \delta_{dx_0}
 -
\frac{  \beta_{x_1 x_0} \delta_{x_1 0} }{ \sigma_{x_1}^2}
 \bigg)
  (1_p\T X_0) 
  + 
\bigg( 
 \delta_{du} 
 -
\frac{ p \beta_{x_1 u} \delta_{x_1 0} }{ \sigma_{x_1}^2}
\bigg)
  U 
  }
 \end{array}
  \right]
     \\
     &
     \quad \times
 \exp \big\{ - \beta_{a0} - \beta_{ax_0} (1_p\T X_0) - \beta_{au} U \big\}
 \big[ 1 + \exp \big\{ \beta_{a0} + \beta_{ax_0} (1_p\T X_0) + \beta_{au} U \big\}  \big]  
 \\
     &
     =
 \exp \left[
 \begin{array}{l}
 \displaystyle{
 \bigg[
 \frac{
 p \big\{ 
 \beta_{x_10}^{(a_Y) 2} - \beta_{x_10}^{(a_D) 2}
 \big\}
  }{ 2\sigma_{x_1}^2 } +  \delta_{d0} - \beta_{a0} \bigg]
  + \bigg( \delta_{dx_1}+ \frac{  \delta_{x_1 0} }{  \sigma_{x_1}^2}  \bigg) (1_p\T X_1) }
   \\
\displaystyle{  +  
\bigg( \delta_{dx_0}
 -
\frac{  \beta_{x_1 x_0} \delta_{x_1 0} }{ \sigma_{x_1}^2}
- \beta_{a x_0}
 \bigg)
  (1_p\T X_0) + 
\bigg( 
 \delta_{du} 
 -
\frac{  p \beta_{x_1 u} \delta_{x_1 0} }{ \sigma_{x_1}^2}
-
\beta_{au}
\bigg)
  U }
 \end{array}
\right]
     \\
     &
     \quad 
     +
      \exp 
 \left[
 \begin{array}{l} 
 \displaystyle{
 \bigg[
 \frac{ 
 p \big\{
 \beta_{x_10}^{(a_Y) 2} - \beta_{x_10}^{(a_D) 2}
 \big\}
  }{ 2\sigma_{x_1}^2 } +  \delta_{d0} \bigg]
  + \bigg( \delta_{dx_1}+ \frac{  \delta_{x_1 0} }{  \sigma_{x_1}^2}  \bigg) (1_p\T X_1)   
  }
  \\
  \displaystyle{
  +  
\bigg( \delta_{dx_0}
 -
\frac{  \beta_{x_1 x_0} \delta_{x_1 0} }{ \sigma_{x_1}^2}
 \bigg)
  (1_p\T X_0) 
  + 
\bigg( 
 \delta_{du} 
 -
\frac{ p \beta_{x_1 u} \delta_{x_1 0} }{ \sigma_{x_1}^2}
\bigg)
  U 
  }
 \end{array}
  \right]
 \ .
 \end{align*}
 From \eqref{eq-MGF}, one can show that $q_1^*(Z,\obX_{1})
     =
 \exp\{ \gamma_0 + \gamma_{x_1} (1_p\T X_1) + \gamma_{x_0} (1_p\T X_0) + \gamma_z Z \} 
 +
 \exp\{ \gamma_0' + \gamma_{x_1} (1_p\T X_1) + \gamma_{x_0}' (1_p\T X_0) + \gamma_z' Z \}$ where
 \begin{align*}
 &
 \gamma_z
 =
 \frac{ 
  \delta_{du} 
 -
\sigma_{x_1}^{-2} p \beta_{x_1 u} \delta_{x_1 0}
-
\beta_{au}
 }{\beta_{zu}}
 \ , \
 && 
 \gamma_z'
 =
 \frac{
   \delta_{du} 
 -
\sigma_{x_1}^{-2} p \beta_{x_1 u} \delta_{x_1 0}
 }{\beta_{zu}}
  \ , \
 \\
 &
  \gamma_{x_1} = \delta_{dx_1} + \sigma_{x_1}^{-2} \delta_{x_1 0}
  \ , \
  \\
 &
 \gamma_{x_0}
 =
\delta_{dx_0}
 -
\sigma_{x_1}^{-2} \beta_{x_1 x_0} \delta_{x_1 0}
- \beta_{a x_0}
- \gamma_{z} \beta_{z x_0}
 \ , \
 &&
  \gamma_{x_0}'
 =
\delta_{dx_0}
 -
\sigma_{x_1}^{-2} \beta_{x_1 x_0} \delta_{x_1 0}
- \gamma_{z}' \beta_{z x_0}
 \ , \
 \\
 &
 \gamma_{0}
 =
 \bigg[ 
 	\begin{array}{l}
 \frac{ p \big\{ \beta_{x_10}^{(a_Y) 2} - \beta_{x_10}^{(a_D) 2} \big\} }{2 \sigma_{x_1}^{2}}  +  \delta_{d0} - \beta_{a0}
 \\
 - 0.5 \gamma_z^2 \sigma_z^2 - \gamma_z \beta_{z0}
 	\end{array}
 \bigg]
 \ , \
 &&
 \gamma_{0}'
 =
\bigg[ 
	\begin{array}{l} 
 \frac{ p \big\{ \beta_{x_10}^{(a_Y) 2} - \beta_{x_10}^{(a_D) 2} \big\} }{2 \sigma_{x_1}^{2}} 
 +  \delta_{d0}
 \\
 - 0.5 \gamma_z^2 \sigma_z^2 - \gamma_z' \beta_{z0}
	\end{array}
\bigg]
\ .
 \end{align*}

 \end{itemize}

 \subsection{Details of the Data Analysis}	\label{sec:supp:Data}
 
Lastly, we provide details of the data analysis, including assessment of the plausibility of unmeasured confounding in the application, estimation of the CSE under no unmeasured confounding, and additional analyses using complete data.

 \subsubsection{Partial Correlation Coefficients of $W$ and $Z$ Given $X_{0}$} \label{sec:supp:Data:partial corr}

 We first report partial correlation coefficients of $W$ and $Z$ given $\bX_{0}$. Specifically, we first fit three linear regression models of $W$ on $\bX_{0}$, which are (i) the main terms of $\bX_{0}$ are used; (ii) the main and second-order interaction terms of $\bX_{0}$ are used; and (iii) the main, second-order, and third-order interactions terms $\bX_{0}$ are used. Likewise, we fit three linear regression models of $Z$ on $\bX_{0}$, where we used the same three specifications of $\bX_{0}$. We then conduct hypothesis tests for correlation between residuals from the regression models of $W$ and $Z$ on $\bX_{0}$. Table \ref{tab:supp:Correlation} summarizes the p-values of the correlation tests, which suggests that $W$ and $Z$ appear to be correlated conditional on $\bX_{0}$. Therefore, this indicates that there may be a latent variable $\bU$ that is a common cause of $W$ and $Z$.

 \begin{table}[!htp]
 \footnotesize
 \renewcommand{\arraystretch}{1.1} \centering
 \begin{tabular}{|cc|ccc|}
 \hline
 \multicolumn{2}{|c|}{\multirow{2}{*}{}}                                                  & \multicolumn{3}{c|}{Correlation Coefficient}                           \\ \cline{3-5} 
 \multicolumn{2}{|c|}{}                                                                   & \multicolumn{1}{c|}{Pearson} & \multicolumn{1}{c|}{Kendall} & Spearman \\ \hline
 \multicolumn{1}{|c|}{\multirow{3}{*}{Specification of $\bX$}} & Main term                & \multicolumn{1}{c|}{0.088}   & \multicolumn{1}{c|}{0.075}   & 0.075    \\ \cline{2-5} 
 \multicolumn{1}{|c|}{}                                        & Second-order interaction & \multicolumn{1}{c|}{0.075}   & \multicolumn{1}{c|}{0.038}   & 0.040    \\ \cline{2-5} 
 \multicolumn{1}{|c|}{}                                        & Third-order interaction  & \multicolumn{1}{c|}{0.073}   & \multicolumn{1}{c|}{0.036}   & 0.038    \\ \hline
 \end{tabular}
 \caption{\footnotesize P-values of the Correlation Tests}
 \label{tab:supp:Correlation}
  \vspace*{-0.5cm}
 \end{table}

 \subsubsection{Estimation of $\tau_{\CSE}^*(\text{Docetaxel})$ Assuming No Unmeasured Confounding} \label{sec:supp:Data:no U}

Assuming there is no unmeasured common cause of $D_{\TIME+1}$ and $Y$, we may consider age ($Z$), baseline quality of life ($W$), four pre-treatment covariates ($\bX_{0}$), and cancer progression $(\bX_{\TIME})$ as a complete collection of measured confounders, denoted by $\bL_{0} = (Z,W,\bX_{0})$ and $\overline{\bL}_{\TIME} = (Z,W,\bX_{0},\bX_{\TIME})$. We then construct the estimator proposed in Section \ref{sec:EIF_NoU}.

 \subsubsection{Additional Result Using Observations with Complete Data} \label{sec:supp:Data:complete data}

We restricted the analysis to 335 patients with available one-year follow-up outcome data. Table \ref{tab:supp:Data Complete} below summarizes the results of this analysis, which are consistent with those presented in the main paper. However, the confidence intervals are wider compared to those in the main paper. Notably, the confidence intervals for our proposed proximal approach are narrower than those for the approach developed under the assumption of no $U$.
    
     \begin{table}[!htp]
 \footnotesize
 \renewcommand{\arraystretch}{1.2} \centering
 \setlength{\tabcolsep}{2.5pt}
 \begin{tabular}{|c|c|c|c|c|}
 \hline
 \multirow{2}{*}{Allow for $\bU$} & \multirow{2}{*}{Statistic} & \multicolumn{3}{c|}{Estimand}                      \\ \cline{3-5} 
 & & \multicolumn{1}{c|}{$\psi^*(\text{Prednisone},\text{Docetaxel} )$} & \multicolumn{1}{c|}{$\psi^*(\text{Estramustine},\text{Docetaxel} )$} & $\tau_{\CSE}^* (\text{Docetaxel} )$ \\ \hline
  \multirow{3}{*}{Yes}  &  Estimate  &           -6.75  &           -4.43  &           -2.32 \\ \cline{2-5} 
                         &       ASE  &            3.01  &            2.06  &            3.51 \\ \cline{2-5} 
                         &  95\% CI  &  (-12.66, -0.84)  &  (-8.46, -0.39)  &  (-9.20, 4.56)       \\ \hline
   \multirow{3}{*}{No}  &  Estimate  &           -6.27  &           -4.43  &           -1.84 \\ \cline{2-5} 
                         &       ASE  &            3.78  &            2.06  &            4.17 \\ \cline{2-5} 
                         &  95\% CI  &  (-13.68, 1.13)  &  (-8.46, -0.39)  &   (-10.01, 6.32)       \\ \hline
 \end{tabular}
 \caption{\footnotesize Summary Statistics of the Estimation of the Conditional Separable Effect $\tau_{\CSE}^*(\text{Docetaxel})$. Statistics in the top three rows (Allow for $\bU$=Yes) and the bottom three rows (Allow for $\bU$=No) are obtained from the proposed proximal approach in Section \ref{sec: exp setting main} and from the approach in Section \ref{sec:supp:Data:no U} developed under assuming no $\bU$, respectively. ASE and 95\% CI stand for the asymptotic standard error obtained from the proposed consistent variance estimator and the corresponding 95\% confidence intervals, respectively.}
 \label{tab:supp:Data Complete}
  \vspace*{-0.5cm}
 \end{table}

\newpage

\newpage 
\section{Proof}	\label{sec:supp:proof}

In this Section, we provide proofs of the Theorems in the main paper and Section \ref{sec:supp:detail} of this document. Let $\mathcal{L}_{2,0}(V) = \{ f(V) \cond \EXP\{ f(V) \} =0, \EXP \{ f^2(V) \} < \infty \} \subseteq  \mathcal{L}_2(V)$ be a Hilbert space of functions of a random variable $V$ with mean zero and finite variance equipped with the inner product $\langle f_1, f_2 \rangle = \EXP \{ f_1(V) f_2(V) \}$. 

\subsection{Proof of Theorem \ref{thm:identification}}

To facilitate the proof, we first consider the following lemma:

\begin{lemma}	\label{thm:integral equation} 
The following results hold:
\begin{itemize}[leftmargin=1cm, itemsep=0cm, topsep=0cm, partopsep=0cm,  parsep=0cm]
    \item[(i)] Suppose that Assumptions \HL{A1}-\AssumptionProxy, \AssumptionCompleteness-(i) are satisfied, and that there exists a confounding bridge function $h_1^*$ satisfying \eqref{eq-bridgeft1 obs}. Then, we have the following result almost surely:
    \begin{align*}
        & \EXP \big( Y \cond A=a_Y, D_{\TIME+1}=0, \obX_{\TIME}, \bU \big) 
        = \EXP \big\{ h_1^*(\bW,\obX_{\TIME+1} \con a_Y) \cond A=a_Y,D_{\TIME+1}=0, \obX_{\TIME}, \bU \big\} \ .
    \end{align*}
    
    \item[(ii)] Suppose that Assumptions \HL{A1}-\AssumptionProxy, \AssumptionCompleteness-(ii) are satisfied, and that there exist the confounding bridge functions $h_0^*$, $h_1^*$, and $h_2^*$ satisfying \eqref{eq-bridgeft1 obs}-\eqref{eq-bridgeft4 obs}. Then, we have  the following result almost surely:
    \begin{align*}
& \EXP \big\{ h_1^*(\bW,\obX_{\TIME} \con a_Y) (1-D_{\TIME+1}) \cond A=a_D, \bX_{0}, \bU \big\} 
= \EXP \big\{ h_0^*(\bW,\bX_{0} \con a_Y, a_D) \cond A=a_D, \bX_{0}, \bU \big\}
\nonumber
\\
& \EXP \big\{ Y (1-D_{\TIME+1}) \cond A=a_D, \bX_{0}, \bU \big\} 
= \EXP \big\{ h_0^*(\bW,\bX_{0} \con a_D, a_D) \cond A=a_D, \bX_{0}, \bU \big\}
\nonumber
\\
&
\EXP \big( 1-D_{\TIME+1} \cond A=a_D, \bX_{0}, \bU \big)
= \EXP \big\{ h_2^*(\bW,\bX_{0} \con a_D) \cond A=a_D, \bX_{0}, \bU \big\} \ .
    \end{align*}
    
    \item[(iii)] Suppose that Assumptions \HL{A1}-\AssumptionProxy, \AssumptionCompleteness-(iv) are satisfied, and that there exist the confounding bridge functions $q_0^*$ satisfying \eqref{eq-bridgeft5 obs}. Then, we have the following result almost surely:
    \begin{align*}
        &  \frac{1}{\Pr(A=a_D \cond \bX_{0},\bU)}
=
\EXP \big\{ q_0^* (\bZ,\bX_{0} \con a_D) \cond A=a_D, \bX_{0}, \bU \big\} \ .
    \end{align*}
    
    \item[(iv)] Suppose that Assumptions \HL{A1}-\AssumptionProxy, \AssumptionCompleteness-(iii) are satisfied, and that there exist the confounding bridge functions $q_0^*$ and $q_1^*$ satisfying \eqref{eq-bridgeft5 obs} and \eqref{eq-bridgeft6 obs}. Then, we have the following result almost surely:
    \begin{align*}
        & 
        \EXP \big\{ \ind (A=a_D) q_0^* (\bZ,\bX_{0} \con a_D) \cond D_{\TIME+1}=0 , \obX_{\TIME} , \bU \big\} 
= \EXP \big\{ \ind(A=a_Y) q_1^* (\bZ,\obX_{\TIME} \con a_Y, a_D) \cond D_{\TIME+1}=0 , \obX_{\TIME} , \bU \big\} \ .
    \end{align*}

\end{itemize}
\end{lemma}

We begin by proving Lemma \ref{thm:integral equation}:

\subsubsection{Proof of Lemma \ref{thm:integral equation}-(i)} \label{sec:supp:proof:thm:211}

The conditional expectation of $Y$ given $(\bZ,A=a_Y,D_{\TIME+1}=0,\obX_{\TIME})$ is represented as
\begin{align}
&
\EXP \big( Y \cond \bZ, A=a_Y, D_{\TIME+1}=0 , \obX_{\TIME}  \big)
\nonumber
\\
& =
\EXP \big\{ \EXP \big( Y \cond \bZ, A=a_Y, D_{\TIME+1}=0, \obX_{\TIME}, \bU \big) \cond \bZ,A=a_Y,D_{\TIME+1}=0 , \obX_{\TIME}  \big\}
\nonumber
\\
& 
=
\EXP \big\{ \EXP \big( Y \cond A=a_Y, D_{\TIME+1}=0, \obX_{\TIME}, \bU \big) \cond \bZ, A=a_Y, D_{\TIME+1}=0, \obX_{\TIME}  \big\}
\nonumber
\\
& 
=
\EXP \big\{ h_1^*(\bW,\obX_{\TIME} \con a_Y) \cond \bZ, A=a_Y, D_{\TIME+1}=0 , \obX_{\TIME}  \big\}
\nonumber
\\
& =
\EXP \big[
    \EXP \big\{ h_1^*(\bW,\obX_{\TIME} \con a_Y) \cond \bZ, A=a_Y, D_{\TIME+1}=0, \obX_{\TIME}, \bU \big\}
    \cond \bZ, A=a_Y, D_{\TIME+1}=0 , \obX_{\TIME}
\big]
\nonumber
\\
& 
=
\EXP \big[
    \EXP \big\{ h_1^* (\bW,\obX_{\TIME} \con a_Y) \cond A=a_Y, D_{\TIME+1}=0, \obX_{\TIME}, \bU \big\}
    \cond \bZ, A=a_Y, D_{\TIME+1}=0 , \obX_{\TIME}
\big] \ .
\nonumber
\end{align}
The first identity is from the law of iterated expectation. 
The second identity is from \HL{CI1}, \CIone. 
The third identity is from the definition of $h_1^*$.
The fourth identity is again from the law of iterated expectation.
The fifth identity is from \HL{CI5}, \CIfive. 
Therefore, from the completeness assumption \AssumptionCompleteness-(i), we get
\begin{align*}
 \EXP \big( Y \cond A=a_Y, D_{\TIME+1}=0, \obX_{\TIME}, \bU \big) 
 = \EXP \big\{ h_1^* (\bW,\obX_{\TIME} \con a_Y) \cond A=a_Y, D_{\TIME+1}=0, \obX_{\TIME}, \bU \big\} \ .
\end{align*}

\subsubsection{Proof of Lemma \ref{thm:integral equation}-(ii)}

We establish
\begin{align}
& \EXP \big\{ (1-D_{\TIME+1}) h_1^*(\bW,\obX_{\TIME} \con a_Y) \cond \bZ,A=a_D, \bX_{0} \big\} 
\nonumber
\\
&
= \EXP \big[ \EXP \big\{ (1-D_{\TIME+1}) h_1^*(\bW,\obX_{\TIME} \con a_Y) \cond \bZ, A=a_D, \bX_{0} , \bU \big\} \cond \bZ, A=a_D, \bX_{0} \big]
\nonumber
\\
& 
= \EXP \big[ \EXP \big\{ (1-D_{\TIME+1}) h_1^*(\bW,\obX_{\TIME} \con a_Y) \cond  A=a_D, \bX_{0} , \bU \big\} \cond \bZ, A=a_D, \bX_{0} \big]
\nonumber
\\
&
=
\EXP \big\{ h_0^*(\bW,\bX_{0} \con a_Y, a_D) \cond \bZ, A=a_D, \bX_{0} \big\}
\nonumber
    \\
& 
=
\EXP \big[
    \EXP \big\{ h_0^*(\bW,\bX_{0} \con a_Y, a_D) \cond A=a_D, \bX_{0}, \bU \big\}
    \cond \bZ, A=a_D, \bX_{0}
\big] \ .
\nonumber 
\end{align}
The first identity is from the law of iterated expectation.  
The second identity is from \HL{CI2}: \CItwo\ and \HL{CI6}: \CIsix. 
The third identity is from the definition of $h_0^*$. 
The last identity is from \HL{CI6}: \CIsix. 
From the completeness assumption \AssumptionCompleteness-(ii), we get
\begin{align*}
\EXP \big\{ (1-D_{\TIME+1}) h_1^*(\bW,\obX_{\TIME} \con a_Y) \cond A=a_D, \bX_{0}, \bU \big\}
 = 
 \EXP \big\{ h_0^* (\bW,\bX_{0} \con a_Y, a_D) \cond A=a_D, \bX_{0}, \bU \big\} \ .
\end{align*}

Likewise, we have
\begin{align}
&
\EXP \big\{ Y (1-D_{\TIME+1}) \cond \bZ, A=a_D,  \bX_{0}  \big\}
\nonumber
\\
& =
\EXP \big[ \EXP \big\{ Y (1-D_{\TIME+1}) \cond \bZ, A=a_D,  \bX_{0}, \bU  \big\} \cond \bZ,A=a_D, \bX_{0}  \big]
\nonumber
\\
& 
=
\EXP \big[ \EXP \big\{ Y (1-D_{\TIME+1}) \cond A=a_D, \bX_{0}, \bU \big\} \cond \bZ, A=a_D, \bX_{0}  \big]
\nonumber
\\
& 
=
\EXP \big\{ h_0^*(\bW,\bX_{0} \con a_D, a_D) \cond \bZ, A=a_D , \bX_{0}  \big\}
\nonumber
\\
& =
\EXP \big[
    \EXP \big\{ h_0^*(\bW,\bX_{0} \con a_D, a_D) \cond \bZ, A=a_D, \bX_{0}, \bU \big\}
    \cond \bZ, A=a_D , \bX_{0} 
\big]
\nonumber
\\
& 
=
\EXP \big[
    \EXP \big\{ h_0^* (\bW,\bX_{0} \con a_D, a_D) \cond A=a_D, \bX_{0}, \bU \big\}
    \cond \bZ, A=a_D , \bX_{0} 
\big] \ .
\nonumber
\end{align}
The first identity is from the law of iterated expectation. 
The second identity is from \HL{CI1} and \HL{CI2}, \CIone\ and \CItwo, which implies $(Y,\uD_{1}) \indep \bZ \cond (A,\bX_{0},\bU)$. 
The third identity is from the definition of $h_0^*$.
The fourth identity is again from the law of iterated expectation.
The fifth identity is from \HL{CI6}, \CIsix.  
Therefore, from the completeness assumption \AssumptionCompleteness-(ii), we get
\begin{align*}
 \EXP \big\{ Y (1-D_{\TIME+1}) \cond A=a_D, \bX_{0}, \bU \big\}
 = \EXP \big\{ h_0^* (\bW,\bX_{0} \con a_D, a_D) \cond A=a_D, \bX_{0}, \bU \big\} \ .
\end{align*}

Lastly, we have
\begin{align}
\EXP \big( 1-D_{\TIME+1} \cond \bZ, A=a_D, \bX_{0} \big)
&
= \EXP \big\{ \EXP \big( 1-D_{\TIME+1} \cond \bZ, A=a_D, \bX_{0}, \bU \big) \cond \bZ, A=a_D, \bX_{0} \big\}
\nonumber
\\
& 
= \EXP \big\{ \Pr(D_{\TIME+1}=0 \cond A=a_D, \bX_{0},\bU) \cond \bZ, A=a_D, \bX_{0} \big\}
\nonumber
\\
&
=
\EXP \big\{ h_2^*(\bW,\bX_{0} \con a_D) \cond \bZ, A=a_D, \bX_{0} \big\}
\nonumber
\\
&
=
\EXP \big[ \EXP \big\{ h_2^* (\bW,\bX_{0} \con a_D) \cond \bZ,A=a_D, \bX_{0},\bU \big\} \cond \bZ, A=a_D, \bX_{0} \big]
\nonumber
    \\
& 
=
\EXP \big[
    \EXP \big\{ h_2^*(\bW,\bX_{0} \con a_D) \cond A=a_D, \bX_{0}, \bU \big\}
    \cond \bZ, A=a_D, \bX_{0}
\big] \ .
\nonumber 
\end{align}
The first identity is from the law of iterated expectation.
The second identity is from \HL{CI2}: \CItwo. 
The third identity is from the definition of $h_2^*$. 
The fourth identity is from the law of iterated expectation.
The last identity is from \HL{CI6}: \CIsix.
From the completeness assumption \AssumptionCompleteness-(ii), we have
\begin{align*}
\EXP \big( 1-D_{\TIME+1} \cond A=a_D, \bX_{0}, \bU \big)
= \EXP \big\{ h_2^*(\bW,\bX_{\TIME+1} \con a_D) \cond A=a_D, \bX_{0}, \bU \big\} \ .
\end{align*}

\subsubsection{Proof of Lemma \ref{thm:integral equation}-(iii)}

The reciprocal value of $\Pr(A=a_D \cond \bX_{0}, \bU)$ is represented as
\begin{align*}
& 
\EXP \bigg\{ \frac{1}{\Pr (A=a_D \cond \bX_{0},\bU)} \, \bigg| \, \bW,A=a_D,\bX_{0} \bigg\}
\\
& =
\int \frac{1}{\Pr (A=a_D \cond \bW, \bX_{0} , \bU=\bu)} \, f^* (\bU = \bu \cond \bW,A=a_D,\bX_{0}) \, d \bu
\\
& =
\int \frac{f^* (\bW,\bX_{0},\bU=\bu)}{f^*(\bW,A=a_D,\bX_{0},\bU=\bu)} 
\frac{f^*(\bW,A=a_D,\bX_{0},\bU=\bu)}{f^*(\bW,A=a_D,\bX_{0})}
\, d\bu
\\
& =
\frac{1}{\Pr (A=a_D \cond \bW , \bX_{0}) }
\\
& =
\EXP \big\{ q_0^*(\bZ,\bX_{0} \con a_D) \cond \bW,A=a_D, \bX_{0}  \big\} 
\\
& =
\EXP \big[
\EXP \big\{ q_0^*(\bZ,\bX_{0} \con a_D) \cond \bW,A=a_D, \bX_{0}, \bU \big\} 
\cond \bW,A=a_D, \bX \big]
\\
& =
\EXP \big[
\EXP \big\{ q_0^*(\bZ,\bX_{0} \con a_D) \cond A=a_D, \bX_{0}, \bU \big\} 
\cond \bW,A=a_D, \bX_{0} \big] \ .
\end{align*}
The first identity is from \HL{CI4}: \CIfour.
The second and third identities are trivial. 
The fourth identity is from the definition of $q_0^*$. 
The fifth identity is from the law of iterated expectation.
The last identity is from \HL{CI6}: \CIsix.
From the completeness assumption \AssumptionCompleteness-(iv), we get
\begin{align*}
\frac{1}{\Pr (A=a_D \cond \bX_{0} ,\bU)}  
=
\EXP \big\{ q_0^* (\bZ,\bX_{0} \con a_D) \cond A=a_D, \bX_{0}, \bU \big\} \ .
\end{align*}

\subsubsection{Proof of Lemma \ref{thm:integral equation}-(iv)} \label{sec:supp:proof:thm:214}

We remark that the condition of Lemma \ref{thm:integral equation}-(iii) is equivalent to 
\begin{align}	\label{eq-defq1-alternative}
& \EXP \big\{ q_0^*(\bZ,\bX_{0} \con a_D) \cond \bW, A=a_D, D_{\TIME+1}=0 , \obX_{\TIME} \big\} 
\frac{ \Pr(A=a_D \cond \bW, D_{\TIME+1}=0 , \obX_{\TIME} ) }{ \Pr(A=a_Y \cond \bW, D_{\TIME+1}=0 , \obX_{\TIME} ) } 
\\
&
= \EXP \big\{ q_1^*(\bZ,\obX_{\TIME} \con a_Y, a_D) \cond \bW, A=a_Y, D_{\TIME+1}=0 , \obX_{\TIME} \big\} \ .
\nonumber
\end{align}
The left hand side of \eqref{eq-defq1-alternative} is represented as 
{\small
\begin{align}
&  
\frac{ \EXP \big\{ q_0^*(\bZ,\bX_{0} \con a_D) \cond \bW, A=a_D, D_{\TIME+1}=0 , \obX_{\TIME} \big\} 
\Pr(A=a_D \cond \bW, D_{\TIME+1}=0 , \obX_{\TIME} ) }{ \Pr(A=a_Y \cond \bW, D_{\TIME+1}=0 , \obX_{\TIME} ) } 
\nonumber
\\
& = 
\frac{ \EXP \big\{ q_0^*(\bZ,\bX_{0} \con a_D) \ind(A=a_D)  \cond \bW, D_{\TIME+1}=0 , \obX_{\TIME} \big\}  }{ \Pr (A=a_Y\cond \bW, D_{\TIME+1}=0 , \obX_{\TIME}) }  
\nonumber
\\
& =
\frac{ \EXP \big[
\EXP \big\{
q_0^* (\bZ,\bX_{0} \con a_D) \ind(A=a_D) \cond \bW, D_{\TIME+1}=0 , \obX_{\TIME}, \bU \big\} 
\cond \bW, D_{\TIME+1}=0 , \obX_{\TIME}
\big] }{ \Pr (A=a_Y \cond \bW, D_{\TIME+1}=0 , \obX_{\TIME}) }  
\nonumber
\\
& = 
\int 
\begin{bmatrix}
\EXP \big\{
q_0^* (\bZ,\bX_{0} \con a_D)  \cond \bW, A=a_D, D_{\TIME+1}=0 , \obX_{\TIME}, \bU \big\}  
\\
\times 
\Pr (A=a_D \cond \bW, D_{\TIME+1}=0 , \obX_{\TIME}, \bU) 
\end{bmatrix}
    \frac{ f^*(\bU \cond W, D_{\TIME+1}=0 , \obX_{\TIME}) }{ \Pr (A=a_Y \cond \bW, D_{\TIME+1}=0 , \obX_{\TIME})} \, d\bU
    \nonumber
\\
& 
=
\int
\begin{bmatrix}
\EXP \big\{
q_0^* (\bZ,\bX_{0} \con a_D)  \cond \bW, A=a_D, D_{\TIME+1}=0 , \obX_{\TIME}, \bU \big\}  
\\
\times 
\Pr (A=a_D \cond \bW, D_{\TIME+1}=0 , \obX_{\TIME}, \bU) 
\end{bmatrix}
\frac{ f^* (\bU \cond \bW,A=a_Y, D_{\TIME+1}=0 , \obX_{\TIME}) }{	\Pr (A=a_Y \cond \bW, D_{\TIME+1}=0 , \obX_{\TIME}, \bU)}	 
 \, d\bU
 \nonumber
\\
& = 
\int 
\begin{bmatrix}
\EXP \big\{
q_0^* (\bZ,\bX_{0} \con a_D)  \cond A=a_D, D_{\TIME+1}=0 , \obX_{\TIME}, \bU \big\}  
\\
\times 
\Pr (A=a_D \cond D_{\TIME+1}=0 , \obX_{\TIME}, \bU) 
\end{bmatrix}
\frac{ f^* (\bU \cond \bW,A=a_Y, D_{\TIME+1}=0 , \obX_{\TIME}) }{ \Pr (A=a_Y \cond D_{\TIME+1}=0 , \obX_{\TIME}, \bU)}
\, d\bU 
 \label{eq-proof1}
\\
& =
\EXP \big\{ q_1^*(\bZ,\obX_{\TIME} \con a_Y, a_D) \cond \bW, A=a_Y, D_{\TIME+1}=0 , \obX_{\TIME} \big\}	
\nonumber
\\
& =  
\EXP
\big[
\EXP \big\{
q_1^*(\bZ,\obX_{\TIME}  \con a_Y, a_D)
\cond \bW, A=a_Y, D_{\TIME+1}=0 , \obX_{\TIME}, \bU
\big\}
\cond \bW, A=a_Y, D_{\TIME+1}=0 , \obX_{\TIME} \big]
\nonumber
\\
& 
= 
\int 
\EXP \big\{
q_1^*(\bZ,\obX_{\TIME} \con a_Y, a_D)
\cond A=a_Y, D_{\TIME+1}=0 , \obX_{\TIME}, \bU
\big\}
 f^* (\bU \cond \bW,A=a_Y, D_{\TIME+1}=0 , \obX_{\TIME})
 \, d\bU \ .
\label{eq-proof2}
\end{align}}%
The first two identities are trivial from the law of iterated expectation. The third identity is trivial from the definition of the expectation. The fourth identity follows from
\begin{align*}
    \frac{f^*(\bU \cond \bW, D_{\TIME+1}=0 , \obX_{\TIME})}{\Pr(A=a_Y \cond \bW, D_{\TIME+1}=0 , \obX_{\TIME})}
    =
    \frac{f^* (\bU , \bW, D_{\TIME+1}=0 , \obX_{\TIME})}{f^*(A=a_Y , \bW, D_{\TIME+1}=0 , \obX_{\TIME})}
    =
    \frac{f^*(\bU \cond \bW, A=a_Y ,D_{\TIME+1}=0 , \obX_{\TIME})}{\Pr(A=a_Y \cond \bW,D_{\TIME+1}=0 , \obX_{\TIME},\bU)} \ .
\end{align*}
The fifth identity is from \HL{CI3}: \CIthree\ and \HL{CI5}: \CIfive. 
The sixth identity is from the definition of $q_1^*$. The last two identities are trivial from the law of iterated expectation. 
From the completeness assumption \AssumptionCompleteness-(iii), \eqref{eq-proof1} and \eqref{eq-proof2} imply
\begin{align*}
&	\EXP \big\{q_0^*(\bZ,\bX_{0} \con a_D)  \cond A=a_D, D_{\TIME+1}=0, \obX_{\TIME}, \bU \big\} 
\frac{ \Pr (A=a_D \cond D_{\TIME+1}=0,\obX_{\TIME},\bU)}{	\Pr (A=a_Y \cond D_{\TIME+1}=0,\obX_{\TIME},\bU)}  
\\
&
=
\EXP \big\{
q_1^* (\bZ,\obX_{\TIME} \con a_Y, a_D)
\cond A=a_Y,D_{\TIME+1}=0,\obX_{\TIME},\bU
\big\} \ .
\end{align*}

\subsubsection{Proof of Theorem \ref{thm:identification}-(i)}

We establish the claim for $\psi_{\NUMER}^*$ when $a_Y \neq a_D$:
\begin{align*}
    &  \EXP \big\{ h_0^* (\bW,\bX_{0} \con a_Y,a_D) \big\}
\\
& = \EXP \big[ 
\EXP \big\{  h_0^* (\bW,\bX_{0} \con a_Y,a_D) \, \big| \,  \bX_{0}, \bU \big\} \big] 
\\
& = \EXP \big[  
\EXP \big\{ h_0^* (\bW,\bX_{0} \con a_Y, a_D) 
\, \big| \, A=a_D, \bX_{0}, \bU \big\} 
\big] 
\\
&
=
\EXP \big[ 
\EXP \big\{ (1-D_{\TIME+1}) h_1^*(\bW,\obX_{\TIME} \con a_Y) \cond A=a_D, \bX_{0}, \bU \big\}
\big] 
\\
&
=
\EXP \bigg[ 
\frac{ 
\EXP \big\{ \ind(A=a_D)(1-D_{\TIME+1})h_1^*(\bW,\obX_{\TIME} \con a_Y) \cond \bX_{0}, \bU \big\} }{ \Pr(A=a_D \cond  \bX_{0},\bU) }
\bigg] 
\\
&
=
\EXP \bigg[ 
\EXP \big\{ h_1^*(\bW,\obX_{\TIME} \con a_Y) \cond A=a_D, D_{\TIME+1}=0, \obX_{\TIME}, \bU \big\}
\frac{ \Pr(A=a_D,D_{\TIME+1}=0 \cond \obX_{\TIME},\bU) }{ \Pr(A=a_D \cond  \bX_{0},\bU) }
\bigg] 
\\
&
=
\EXP \bigg[ 
\EXP \big\{ h_1^*(\bW,\obX_{\TIME} \con a_Y) \cond A=a_Y, D_{\TIME+1}=0, \obX_{\TIME}, \bU \big\}
\frac{ \Pr(A=a_D,D_{\TIME+1}=0 \cond  \obX_{\TIME},\bU) }{ \Pr(A=a_D \cond  \bX_{0},\bU) }
\bigg] 
\\
&
=
\EXP \bigg\{ 
\EXP \big( Y \cond A=a_Y, D_{\TIME+1}=0, \obX_{\TIME}, \bU \big) 
\frac{ \Pr(A=a_D,D_{\TIME+1}=0 \cond \obX_{\TIME},\bU) }{ \Pr(A=a_D \cond   \bX_{0},\bU) }
\bigg\}
\\
&
=
\psi_{\NUMER}^*(a_Y,a_D)
\ .
\end{align*}
The second line is trivial from the law of iterated expectation. 
The third line is from \HL{CI4}: \CIfour.
The fourth line is from Lemma \ref{thm:integral equation}-(ii). 
The fifth line is from the law of iterated expectation. 
The sixth line is from
\begin{align*}
&
\EXP \bigg[ 
\frac{ 
\EXP \big\{ \ind(A=a_D)(1-D_{\TIME+1})h_1^*(\bW,\obX_{\TIME} \con a_Y) \cond \bX_{0}, \bU \big\} }{ \Pr(A=a_D \cond  \bX_{0},\bU) }
\bigg] 
\\
&
=
\EXP \bigg[ 
\frac{ 
\EXP \big[ 
\EXP \big\{ \ind(A=a_D)(1-D_{\TIME+1})h_1^*(\bW,\obX_{\TIME} \con a_Y) \cond \ubX_{1},\bX_{0},\bU \big\}  \cond \bX_{0}, \bU \big]
}{ \Pr(A=a_D \cond  \bX_{0},\bU) }
\bigg] 
\\
&
=
\int 
\frac{ 
\int 
\EXP \big\{ \ind(A=a_D)(1-D_{\TIME+1})h_1^*(\bW,\obX_{\TIME} \con a_Y) \cond \ubX_{1}, \bX_{0}, \bU \big\}
\, dP (\ubX_{1} \cond \bX_{0},U)
}
{
\Pr(A=a_D \cond  \bX_{0},\bU)
}
\, dP (\bX_{0},\bU)
\\
&
=
\int 
\frac{  
\EXP \big\{ h_1^*(\bW,\obX_{\TIME} \con a_Y) \cond A=a_D, D_{\TIME+1}=0, \obX_{\TIME}, \bU \big\}
\Pr(A=a_D,D_{\TIME+1}=0 \cond \obX_{\TIME},\bU)
}
{
\Pr(A=a_D \cond  \bX_{0},\bU)
}
\, dP (\obX_{\TIME},\bU) 
\\
&
=
\EXP \bigg[ 
\EXP \big\{ h_1^*(\bW,\obX_{\TIME} \con a_Y) \cond A=a_D, D_{\TIME+1}=0, \obX_{\TIME}, \bU \big\}
\frac{ \Pr(A=a_D,D_{\TIME+1}=0 \cond \bZ,\obX_{\TIME},\bU) }{ \Pr(A=a_D \cond  \bX_{0},\bU) }
\bigg]  \ .
\numeq \label{eq-id h0 middle}
\end{align*}
The seventh line is from \HL{CI3}: \CIthree.
The last two lines are from Lemma \ref{thm:integral equation}-(i) and the form of $\psi_{\NUMER}^*$ in \eqref{eq-psi in supp}.

When $a_Y = a_D$, we have
\begin{align*}
&  \EXP \big\{ h_0^* (\bW,\bX_{0} \con a_D,a_D) \big\}
\\
& = \EXP \big[ \EXP \big\{ h_0^* (\bW,\bX_{0} \con a_D,a_D) \cond \bX_{0}, \bU \big\} \big]
\\
& = \EXP \big[ \EXP \big\{ h_0^* (\bW,\bX_{0} \con a_D,a_D) \cond A=a_D, \bX_{0}, \bU \big\} \big] 
\\
& = \EXP \big[ \EXP \big\{ (1-D_{\TIME+1}) Y \cond A=a_D, \bX_{0}, \bU \big\} \big]  
 \\
& = 
\EXP \big\{
\EXP ( Y \cond A=a_D , D_{\TIME+1} = 0 , \bX_{0} , \bU )
\Pr(D_{\TIME+1}=0 \cond A=a_D, \bX_{0}, \bU) \big\} 
\\
&
=
\psi_{\NUMER}^*(a_D,a_D)
\ .
\end{align*}
The second line is trivial from the law of iterated expectation. 
The third line is from \HL{CI4}: \CIfour.
The fourth line is from Lemma \ref{thm:integral equation}-(ii). 
The fifth line is from the law of iterated expectation. 
The last line is from the form of $\psi_{\NUMER}^*$ in \eqref{eq-psi in supp}.

We establish the claim for $\psi_{\DENOM}^*$:
\begin{align*}
&
\EXP \big\{ h_2^* (\bW,\bX_{0} \con a_D) \big\}
\\
& 
= 
\EXP \big[ \EXP \big\{ h_2^* (\bW,\bX_{0} \con a_D) \cond \bX_{0}, \bU \big\} \big]
\\
&
=
\EXP \big[ \EXP \big\{ h_2^* (\bW,\bX_{0} \con a_D) \cond A=a_D, \bX_{0} , \bU \big\} \big]
\\
&
=
\EXP \big\{ \Pr \big( D_{\TIME+1}=0 \cond A=a_D, \bX_{0} , \bU \big) \big\}
\\
& =
\psi_{\DENOM}^*(a_D)
\ .
\end{align*}
The second line is trivial from the law of iterated expectation. 
The third line is from \HL{CI4}: \CIfour.
The last two lines are from Lemma \ref{thm:integral equation}-(ii) and the form of $\psi_{\DENOM}^*$ in \eqref{eq-psi in supp}.

\subsubsection{Proof of Theorem \ref{thm:identification}-(ii)}

We establish the claim for $\psi_{\NUMER}^*$:
\begin{align*}
& \EXP \big\{ \ind (A=a_D,D_{\TIME+1}=0) q_0^*(\bZ,\bX_{0} \con a_D) h_1^*(\bW,\obX_{\TIME} \con a_Y) \big\}
\\
& =
\EXP \big[ \EXP \big\{ \ind (A=a_D,D_{\TIME+1}=0) q_0^*(\bZ,\bX_{0} \con a_D) h_1^*(\bW,\obX_{\TIME} \con a_Y) \cond \obX_{\TIME}, \bU \big\} \big]
\\
& =
\EXP \big[ 
\Pr \big( A=a_D,D_{\TIME+1}=0 \cond \obX_{\TIME} , \bU \big) 
\EXP \big\{ q_0^*(\bZ,\bX_{0} \con a_D) h_1^*(\bW,\obX_{\TIME} \con a_Y) \cond A=a_D,D_{\TIME+1}=0,\obX_{\TIME},\bU \big\} \big]
\\
& 
=
\EXP \left[
\begin{array}{l}
\Pr \big( A=a_D,D_{\TIME+1}=0 \cond \obX_{\TIME} , \bU \big) 
\EXP \big\{ q_0^*(\bZ,\bX_{0} \con a_D) \cond A=a_D,D_{\TIME+1}=0, \obX_{\TIME} , \bU \big\} 
\\
\quad
\times
\EXP \big\{ h_1^*(\bW,\obX_{\TIME} \con a_Y) \cond A=a_D,D_{\TIME+1}=0, \obX_{\TIME} , \bU \big\} 
\end{array}	
\right]
\\
& 
=
\EXP \left[
\begin{array}{l}
\Pr \big( A=a_D,D_{\TIME+1}=0 \cond \obX_{\TIME} , \bU \big) 
\EXP \big\{ q_0^*(\bZ,\bX_{0} \con a_D) \cond A=a_D,D_{\TIME+1}=0, \obX_{\TIME} , \bU \big\} 
\\
\quad
\times
\EXP \big\{ h_1^*(\bW,\obX_{\TIME} \con a_Y) \cond A=a_Y,D_{\TIME+1}=0, \obX_{\TIME} , \bU \big\} 
\end{array}	
\right]
    \\
& 
=
\EXP \left[
\begin{array}{l}
\Pr \big( A=a_D,D_{\TIME+1}=0 \cond \obX_{\TIME} , \bU \big)
\EXP \big\{ q_0^*(\bZ,\bX_{0} \con a_D) \cond A=a_D, \bX_{0} , \bU \big\} 
\\
\quad
\times
\EXP \big( Y \cond A=a_Y, D_{\TIME+1}=0, \obX_{\TIME} , U \big) 
\end{array}	
\right]
\\
& 
=
\EXP \bigg\{
\frac{\Pr \big( A=a_D,D_{\TIME+1}=0 \cond \obX_{\TIME} , \bU \big)}{ \Pr(A=a_D \cond \bX_{0} ,\bU) }
\EXP \big( Y \cond A=a_Y, D_{\TIME+1}=0, \obX_{\TIME} ,\bU \big) 
\bigg\} 
\\
& 
=
\psi_{\NUMER}^*(a_Y,a_D) \ .
\end{align*}
The second and third lines are from the law of iterated expectation. 
The fourth line holds from \HL{CI5}: \CIfive.
The fifth line holds from \HL{CI3}: \CIthree.
The sixth line holds from \HL{CI2}: \CItwo\ and Lemma \ref{thm:integral equation}-(i). 
The seventh line holds from Lemma \ref{thm:integral equation}-(iii). 
The last line holds from the form of $\psi_{\NUMER}^*$ in \eqref{eq-psi in supp}.

We establish the claim for $\psi_{\DENOM}^*$:
\begin{align*}
& \EXP \big\{ \ind (A=a_D,D_{\TIME+1}=0) q_0^*(\bZ,\bX_{0} \con a_D) \big\}
\\
& =
\EXP \big[ \EXP \big\{ \ind (A=a_D,D_{\TIME+1}=0) q_0^*(\bZ,\bX_{0} \con a_D) \cond \bX_{0}, \bU \big\} \big]
\\
& 
=
\EXP  \big[ 
\Pr \big( A=a_D,D_{\TIME+1}=0 \cond \bX_{0} , \bU \big) 
\EXP \big\{ q_0^*(\bZ,\bX_{0} \con a_D) \cond A=a_D,D_{\TIME+1}=0, \bX_{0} , \bU \big\} 
\big]
\\
& 
=
\EXP  \big[ 
\Pr \big( A=a_D,D_{\TIME+1}=0 \cond \bX_{0} , \bU \big) 
\EXP \big\{ q_0^*(\bZ,\bX_{0} \con a_D) \cond A=a_D,\bX_{0} , \bU \big\} 
\big]
\\
& 
=
\EXP \bigg\{
\frac{\Pr \big( A=a_D,D_{\TIME+1}=0 \cond \bX_{0} ,\bU \big)}{ \Pr(A=a_D \cond \bX_{0} ,\bU) }
\bigg\}
\\
& 
=
\EXP \big\{
\Pr \big(  D_{\TIME+1}=0 \cond A=a_D, \bX_{0} ,\bU \big) \big\}
\\
& 
=
\psi_{\DENOM}^*(a_D) \ .
\end{align*}
The second and third lines are from the law of iterated expectation. 
The fourth line holds from \HL{CI2}: \CItwo.
The fifth line holds from Lemma \ref{thm:integral equation}-(iii). 
The last line holds from the form of $\psi_{\DENOM}^*$ in \eqref{eq-psi in supp}.

\subsubsection{Proof of Theorem \ref{thm:identification}-(iii)}

We establish the claim for $\psi_{\NUMER}^*$:
 \begin{align*}
 &
\EXP \big\{ \ind (A=a_Y, D_{\TIME+1}=0) q_1^*(\bZ,\obX_{\TIME} \con a_Y, a_D) Y \big\}
\\
& =
\EXP \big[ \EXP \big\{ \ind (A=a_Y, D_{\TIME+1}=0) 
q_1^* (\bZ,\obX_{\TIME} \con a_Y, a_D) Y  \cond \obX_{\TIME} , U \big\}  \big]
\\
& =
\EXP \big[ \Pr(A=a_Y, D_{\TIME+1}=0 \cond \obX_{\TIME} , \bU) 
\EXP \big\{  q_1^* (\bZ,\obX_{\TIME} \con a_Y, a_D) Y  \cond A=a_Y, D_{\TIME+1}=0, \obX_{\TIME}, \bU \big\} \big]
\\
& 
=
\EXP \Bigg[
\bigg[
\begin{array}{l}
\Pr(A=a_Y, D_{\TIME+1}=0 \cond \obX_{\TIME}, \bU) 	
\EXP \big\{  q_1^*(\bZ,\obX_{\TIME} \con a_Y, a_D) \cond A=a_Y,D_{\TIME+1}=0, \obX_{\TIME}, \bU \big\}
\\
\times 
\EXP \big( Y \cond A=a_Y,D_{\TIME+1}=0, \obX_{\TIME}, \bU \big)
\end{array}
\bigg]
\Bigg]
\\
& 
=
\EXP \Bigg[
\bigg[
\begin{array}{l}
\Pr(A=a_D, D_{\TIME+1}=0 \cond \obX_{\TIME}, \bU) 	
\EXP \big\{  q_0^*(\bZ,\bX_{0} \con a_Y, a_D) \cond A=a_D,D_{\TIME+1}=0, \obX_{\TIME}, \bU \big\}
\\
\times 
\EXP \big( Y \cond A=a_Y,D_{\TIME+1}=0, \obX_{\TIME}, \bU \big)
\end{array}
\bigg]
\Bigg]
\\
& 
=
\EXP \big[ \Pr(A=a_D, D_{\TIME+1}=0 \cond \obX_{\TIME}, \bU)
\EXP \big\{  q_0^*(\bZ,\bX_{0} \con a_D) \cond A=a_D, \bX_{0}, \bU \big\}
\EXP \big( Y \cond A=a_Y,D_{\TIME+1}=0, \obX_{\TIME}, \bU \big) \big] 
\\
& 
=
\EXP \bigg\{
\frac{ \Pr(A=a_D, D_{\TIME+1}=0 \cond \obX_{\TIME}, \bU) }{ \Pr(A=a_D \cond \bX_{0} ,\bU) }
\EXP \big( Y \cond A=a_Y,D_{\TIME+1}=0, \obX_{\TIME}, \bU \big)
\bigg\} 
\\
& 
=
\psi_{\NUMER}^*(a_Y,a_D) \ .
 \end{align*}
The second and third lines are from the law of iterated expectation. 
The fourth line holds from \HL{CI1}: \CIone.
The fifth line holds from Lemma \ref{thm:integral equation}-(iv). 
The sixth line holds from \HL{CI2}: \CItwo.
The seventh line holds from Lemma \ref{thm:integral equation}-(iii). 
The last line holds from the form of $\psi_{\NUMER}^*$ in \eqref{eq-psi in supp}.

We establish the claim for $\psi_{\DENOM}^*$: 
 \begin{align*}
 &
\EXP \big\{ \ind (A=a_Y, D_{\TIME+1}=0) q_1^*(\bZ,\obX_{\TIME} \con a_Y, a_D)  \big\}
\\
& =
\EXP \big[ \EXP \big\{ \ind (A=a_Y, D_{\TIME+1}=0) 
q_1^* (\bZ,\obX_{\TIME} \con a_Y, a_D)  \cond \obX_{\TIME} , U \big\}  \big]
\\
& =
\EXP \big[ \Pr(A=a_Y, D_{\TIME+1}=0 \cond \obX_{\TIME} , \bU) 
\EXP \big\{  q_1^* (\bZ,\obX_{\TIME} \con a_Y, a_D) \cond A=a_Y, D=0, \obX_{\TIME}, \bU \big\} \big] 
\\
& 
=
\EXP \big[ \Pr(A=a_D, D_{\TIME+1}=0 \cond \obX_{\TIME}, \bU) 	
\EXP \big\{  q_0^*(\bZ,\obX_{\TIME} \con a_D) \cond A=a_D,D=0, \obX_{\TIME}, \bU \big\} 
\big]
\\
& 
=
\EXP \big[ \Pr(A=a_D, D_{\TIME+1}=0 \cond \obX_{\TIME}, \bU) 	
\EXP \big\{  q_0^*(\bZ,\bX_{0} \con a_D) \cond A=a_D, \bX_{0} \bU \big\}  \big] 
\\
& 
=
\EXP \bigg\{
\frac{\Pr \big( A=a_D,D_{\TIME+1}=0 \cond \obX_{\TIME} ,\bU \big)}{ \Pr(A=a_D \cond \bX_{0} ,\bU) } 
\bigg\}
\\
& 
=
\EXP \big\{ 
\Pr \big(  D_{\TIME+1}=0 \cond A=a_D, \bX_{0} ,\bU \big) \big\}
\\
& 
=
\psi_{\DENOM}^*(a_D) \ .
 \end{align*}
The second and third lines are from the law of iterated expectation. 
The fourth line holds from Lemma \ref{thm:integral equation}-(iv). 
The fifth line holds from \HL{CI2}: \CItwo.
The sixth line holds from Lemma \ref{thm:integral equation}-(iii). 
The last line holds from the form of $\psi_{\DENOM}^*$ in \eqref{eq-psi in supp}.

\subsection{Proof of Theorem \ref{thm:IF}}

\subsubsection{Proof of Theorem \ref{thm:IF}-(i)} 	\label{sec:supp:thm:IF}

It suffices to derive a centered influence function for $\psi^*(a_Y,a_D)$, denoted by $\cInfFt^*(O \con a_Y,a_D)$, because an influence function for $\tau_{\CSE}^*(a_D)$ is given by $\cInfFt^*(O \con a_Y=1,a_D) - \cInfFt^*(O \con a_Y=0,a_D)$. We suppress $(a_Y,a_D)$ in confounding bridge functions and influence functions for notational brevity unless necessary. Recall that model $\M$ is defined as the regular model of the form
\begin{align*}
\mathcal{M}
=
\left\{
    P
    \left|
        \begin{array}{l}
        \text{There exist $h_0^*$, $h_1^*$, $h_2^*$ satisfying }
        \\
        \EXP \big( Y \cond \bZ, A=a_Y, D_{\TIME+1}=0 , \obX_{\TIME} \big) = \EXP \big\{ h_1^*(\bW,\obX_{\TIME}) \cond \bZ,A=a_Y,D_{\TIME+1}=0, \obX_{\TIME} \big\}
        \\
        \EXP \big\{ h_1^*(\bW,\obX_{\TIME}) (1-D_{\TIME+1}) \cond \bZ,A=a_D , \bX_{0} \big\} = \EXP \big\{ h_0^* (\bW,\bX_{0}) \cond \bZ, A=a_D , \bX_{0} \big\}
        \\
        \EXP \big( 1-D_{\TIME} \cond \bZ,A=a_D , \bX_{0} \big) = \EXP \big\{ h_2^* (\bW,\bX_{0}) \cond \bZ , A=a_D , \bX_{0} \big\}
        \end{array}
    \right.
\right\} \ .
\end{align*}
Let $\M(\eta)$ be the one-dimensional parametric submodel of $\M$ and $f(\bO \con \eta) \in \M(\eta)$ be the density at $\eta$. We use $\EXP\ETA \big\{ g(\bO) \big\}$ to denote the expectation operator at $f(\bO \con \eta)$. Let $h_0(\cdot \con \eta)$, $h_1(\cdot  \con \eta)$, and $h_2(\cdot \con \eta)$ be the confounding bridge functions satisfying
\begin{align*}
&
\EXP\ETA \big\{ Y - h_1(\bW,\obX_{\TIME} \con \eta) \cond \bZ, A=a_Y, D_{\TIME+1}=0 , \obX_{\TIME} \big\}
= 
0 \ ,
\\
&
\EXP\ETA \big\{ (1-D_{\TIME+1}) h_1(\bW,\obX_{\TIME} \con \eta) - h_0(\bW,\bX_{0} \con \eta) \cond \bZ, A=a_D,\bX_{0} \big\}
= 
0 \ ,
\\
&
\EXP\ETA \big\{ (1-D_{\TIME+1}) - h_2(\bW,\bX_{0} \con \eta) \cond \bZ, A=a_D,\bX_{0} \big\}
= 
0 \ .
\end{align*}
We suppose that the true quantities are recovered at $\eta^*$, e.g., $f^*(\bO) := f(\bO \con \eta^*)$ and $h_{1}^* = h_{1}(\cdot \con \eta^*)$.

We also make certain differentiability conditions for the submodel $\mathcal{M}(\eta)$. Specifically, we assume that $f(\cdot \con \eta)$ is pointwise differentiable with respect to $\eta$. While, in principle, this condition could be relaxed to differentiability in quadratic mean \citep[Chapter 7.2]{Vaart1998}, we opt for pointwise differentiability to avoid unnecessary technical complexities.  Let $s(\cdot \cond \cdot \con \eta): = \partial \log f(\cdot \cond \cdot \con \eta) / \partial \eta$ be the score function evaluated at $\eta$, and let $s^*(\cdot \cond \cdot)$ be the score function evaluated at $\eta^*$. We also assume that the confounding bridge functions $h_0(\cdot \con \eta)$, $h_1(\cdot \con \eta)$, and $h_2(\cdot \con \eta)$ are differentiable with respect to $\eta$ under the submodel $\mathcal{M}(\eta)$, with their derivatives denoted by $\nabla_{\eta} h (\cdot \con \eta)$. We note that similar differentiability conditions have been assumed in previous literature, including \citet[Equations (67), (68)]{Kallus2021}, \citet[Proof of Theorem 3.1]{Cui2023}, \citet[Equation (B.17)]{Ying2023}, and \citet[Proof of Theorem 3]{Dukes2023_ProxMed}.

We obtain the following three restrictions by differentiating the above restrictions with respect to $\eta$ and evaluating at $\eta^*$:
{\small
\begin{align}
    &
    \EXP \Bigg[
        \begin{array}{l}
        \big\{ Y - h_1^* (\bW,\obX_{\TIME}) \big\} s^* (Y,\bW \cond \bZ, A=a_Y, D_{\TIME+1}=0, \obX_{\TIME}) \\
        - \nabla_\eta h_1(\bW,\obX_{\TIME} \con \eta^*)	
        \end{array}
        \bigg| \, \bZ, A=a_Y, D_{\TIME+1}=0, \obX_{\TIME} \Bigg] = 0
    \label{eq:tangent-1}
    \\
    &
    \EXP \Bigg[
        \begin{array}{l}
        \big\{ h_1^* (\bW,\obX_{\TIME}) (1-D_{\TIME+1}) - h_0^* (\bW,\bX_{0}) \big\} s^* (\bW,D_{\TIME+1},\ubX_{1} \cond \bZ, A=a_D, \bX_{0}) 
        \\
        + \nabla_\eta h_1(\bW,\obX_{\TIME} \con \eta^*) (1-D_{\TIME+1})
        - \nabla_\eta h_0(\bW,\bX_{0} \con \eta^*)
        \end{array}
        \bigg| \, \bZ, A=a_D, \bX_{0} \Bigg] = 0
     \label{eq:tangent-2}
     \\
    &
    \EXP \Bigg[
        \begin{array}{l}			
        \big\{ 
        (1-D_{\TIME+1}) - h_2^* (\bW,\bX_{0}) \big\} s^* (\bW,D_{\TIME+1}  \cond \bZ,A=a_D,\bX_{0})
        \\
        - \nabla_\eta h_2(\bW,\bX_{0} \con \eta^*)
        \end{array}
     \, \bigg| \, \bZ, A=a_D, \bX_{0} \Bigg] = 0 \ .
     \label{eq:tangent-3}
\end{align}}%

The target estimand at $\eta$ is defined as $\psi(a_Y,a_D \con \eta)
= \psi_{\NUMER}(a_Y,a_D \con  \eta) / \psi_{\DENOM}(a_D \con \eta)  $ where
\begin{align*}
\psi_{\NUMER}(a_Y,a_D \con \eta)
=
\EXP\ETA \big\{ h_0 (\bW,\bX_{0} \con \eta) \big\}  
\ , \
\psi_{\DENOM}(a_D \con \eta)
=
\EXP\ETA \big\{ h_2 (\bW,\bX_{0} \con \eta) \big\}  \ .
\end{align*}
The pathwise derivative of $\psi(a_Y,a_D \con \eta)$ is
\begin{align*}
\frac{\partial \psi (a_Y,a_D \con \eta) }{\partial \eta}
& =
\frac{  \frac{\partial}{\partial \eta} \psi_{\NUMER} (a_Y,a_D \con \eta) }{\psi_{\DENOM}(a_D \con \eta)}
-
\frac{ \psi_{\NUMER}(a_Y,a_D \con \eta) \frac{\partial}{\partial \eta} \psi_{\DENOM} ( a_D \con \eta)}{ \psi_{\DENOM}^2 ( a_D \con \eta) }
\\
&
=
\frac{1}{ \psi_{\DENOM} ( a_D \con \eta) }
\bigg\{
\frac{\partial \psi_{\NUMER} (a_Y,a_D \con \eta) }{\partial \eta} 
- \psi(a_Y,a_D \con \eta) \cdot \frac{\partial \psi_{\DENOM} ( a_D \con \eta) }{\partial \eta} 
\bigg\}
\ .
\end{align*}
Let $\cInfFt_{\NUMER}^* (\bO)$ and $\cInfFt_{\DENOM}^* (\bO)$ be centered influence functions for $\psi_{\NUMER}^*(a_Y,a_D)$ and $\psi_{\DENOM}^*(a_D)$, i.e., 
\begin{align*}
&
\frac{\partial \psi_{\NUMER} (a_Y,a_D \con \eta) }{\partial \eta} 
\bigg|_{\eta=\eta^*}
=
\EXP \big\{ \cInfFt_{\NUMER}^*(\bO) \times s^*(\bO) \big\}
\ ,
&&
\frac{\partial \psi_{\DENOM} (a_D \con \eta) }{\partial \eta} 
\bigg|_{\eta=\eta^*}
=
\EXP \big\{ \cInfFt_{\DENOM}^*(\bO) \times s^*(\bO) \big\} \ .
\end{align*}
We define $\cInfFt^*(\bO) = \big\{ \cInfFt_{\NUMER}^*(\bO) - \psi^*(a_Y,a_D) \cInfFt_{\DENOM}^*(\bO) \big\} / \psi_{\DENOM}^*(a_D)$. Then, we find
\begin{align} 
\EXP \big\{ \cInfFt^*(\bO) \times s^*(\bO) \big\}
& =
\frac{ 
    \EXP \big[   \big\{ \cInfFt_{\NUMER}^*(\bO) - \psi^* (a_Y,a_D) \cdot \cInfFt_{\DENOM}^*(\bO) \big\} 
\times s^*(\bO) \big]
 }{ \psi_{\DENOM}^*(a_D) } 
 \nonumber
\\
& =
\frac{1}{ \psi_{\DENOM}(a_D \con \eta^*) }
\bigg\{ 
    \frac{\partial \psi_{\NUMER} (a_Y,a_D \con \eta) }{\partial \eta} 
\bigg|_{\eta=\eta^*}
    -
    \psi(a_Y,a_D \con \eta^*)
    \frac{\partial \psi_{\DENOM} (a_D \con \eta) }{\partial \eta} 
\bigg|_{\eta=\eta^*}
\bigg\}
\nonumber
\\
& =
\frac{\partial \psi (a_Y,a_D \con \eta) }{\partial \eta} \bigg|_{\eta=\eta^*}
\ .
\label{eq-IF numer denom}
\end{align}
Therefore, it suffices to find centered influence functions for $\psi_{\NUMER}^*(a_Y,a_D)$ and $\psi_{\DENOM}^*(a_D)$, respectively. In what follows, we establish
\begin{align*}
&
\cInfFt_{\NUMER}^*(\bO)
=
\Bigg[
\begin{array}{l}
    \ind(A=a_D) q_0^* (\bZ,\bX_{0}) 
    \big\{ h_1^* (\bW,\obX_{\TIME}) (1-D_{\TIME+1}) - h_0^* (\bW,\bX_{0}) \big\}
    \\
    + \ind(A=a_Y)  (1-D_{\TIME+1}) q_1^* (\bZ,\obX_{\TIME})
    \big\{ Y - h_1^* (\bW,\obX_{\TIME}) \big\}
    + h_0^* (\bW,\bX_{0}) - \psi_{\NUMER}^* (a_Y,a_D)
\end{array}
\Bigg] 
\\
&
\cInfFt_{\DENOM}^*(\bO)
=
\ind(A=a_D) q_0^* (\bZ,\bX_{0})  \big\{ (1-D_{\TIME+1}) - h_2^* (\bW,\bX_{0}) \big\}   + \big\{ h_2^*(\bW,\bX_{0}) - \psi_{\DENOM}^*(a_D) \big\}
\end{align*}
which reduces to the influence function presented in Theorem \ref{thm:IF}:
\begin{align*}
& 
\cInfFt^*(\bO)  
\\
&
= 
\frac{ \cInfFt_{\NUMER}^*(\bO) - \psi^*(a_Y,a_D) \cInfFt_{\DENOM}^*(\bO)  }{ \psi_{\DENOM}^*(a_D) }
 \\
 &
 =	 
\frac{1}{ \psi_{\DENOM}^* (a_D) }
\left[
\begin{array}{l}
\ind(A=a_Y)  (1-D_{\TIME+1}) q_1^* (\bZ,\obX_{\TIME}) \big\{ Y - h_1^* (\bW,\obX_{\TIME}) \big\}
\\
\quad + \ind(A=a_D) q_0^* (\bZ,\bX_{0}) \big\{ h_1^* (\bW,\obX_{\TIME}) (1-D_{\TIME+1}) - h_0^* (\bW,\bX_{0}) \big\}
+
h_0^* (\bW,\bX_{0})
\\
\quad -
\psi^*(a_Y,a_D)
\big[  
         \ind(A=a_D) q_0^* (\bZ,\bX_{0})  \big\{ (1-D_{\TIME+1}) - h_2^*(\bW,\bX_{0}) \big\}  
         +  h_2^* (\bW,\bX_{0}) 
 \big]
 \end{array}
\right] \ . 
\end{align*}
The rest of the proof is to establish the form of $\cInfFt_{\NUMER}^*(\bO)$ and $\cInfFt_{\DENOM}^*(\bO)$.

\subsubsection*{Establish $\cInfFt_{\NUMER}^* (\bO)$}

We find the numerator has the pathwise derivative as
\begin{align}			\label{eq:pathwise-Numer}
\frac{\partial \psi_{\NUMER} (a_Y,a_D \con \eta) }{\partial \eta} \bigg|_{\eta=\eta^*}
& =
\frac{\partial \EXP\ETA \big\{ h_0(\bW,\bX_{0} \con \eta) \big\}  }{\partial \eta}  \bigg|_{\eta=\eta^*}
\nonumber
\\
& =
\EXP \big\{ \nabla_\eta h_0 (\bW,\bX_{0} \con \eta^* ) + h_0^* (\bW,\bX_{0}) s^*(\bW,\bX_{0}) \big\}
\nonumber
\\
& =
\EXP \big[ \nabla_\eta h_0 (\bW,\bX_{0} \con \eta^* ) + \big\{ h_0^* (\bW,\bX_{0}) - \psi_{\NUMER}^*(a_Y,a_D) \big\} s^*(\bO) \big] \ .
\end{align}
The first term is 
\begin{align}	
& 
\EXP \big\{ \nabla_\eta h_0 (\bW,\bX_{0} \con \eta^* )  \big\}
\nonumber
\\
& 
= 
\EXP \bigg\{ \frac{\ind(A=a_D)}{\Pr(A=a_D \cond \bW,\bX_{0}) }  \nabla_\eta h_0 (\bW,\bX_{0} \con \eta^* )  \bigg\}
\nonumber
\\
& 
=
\EXP \Big[  \ind(A=a_D) \EXP \big\{ q_0^* (\bZ,\bX_{0}) \cond \bW,A=a_D,\bX_{0} \big\}   \nabla_\eta h_0(\bW,\bX_{0} \con \eta^* )  \Big]
\nonumber
\\
& 
= 
\EXP \Big\{  \ind(A=a_D) q_0^* (\bZ,\bX_{0})  \nabla_\eta h_0(\bW,\bX_{0} \con \eta^* )  \Big\}
\nonumber
\\
& 
= 
\EXP \Big[  \ind(A=a_D) q_0^* (\bZ,\bX_{0}) \EXP \big\{ \nabla_\eta h_0(\bW,\bX_{0} \con \eta^* ) \cond \bZ, A=a_D, \bX_{0} \big\} \Big]
\nonumber
\\
& 
=
\EXP \left[
\begin{array}{l}
\ind(A=a_D) q_0^* (\bZ,\bX_{0}) 
\\
\times 
\EXP \left[
        \left.
        \begin{array}{l}
        \nabla_\eta h_1(\bW,\obX_{\TIME} \con \eta^*) (1-D_{\TIME+1})
        \\
        +
        \big\{ h_1^* (\bW,\obX_{\TIME}) (1-D_{\TIME+1}) - h_0^* (\bW,\bX_{0}) \big\} 
        \\
        \quad \times s^*(\bW,D_{\TIME+1},\ubX_{1} \cond \bZ, A=a_D, \bX_{0})  
        \end{array}
        \right| \, \bZ, A=a_D, \bX_{0} \right]
\end{array}
     \right]
     \ .
     \label{eq-grad-h0}
\end{align}
The second line is trivial from the law of iterated expectation. 
The third line holds from \eqref{eq-bridgeft5 obs}. 
The fourth and fifth lines are trivial from the law of iterated expectation. 
The last line holds from \eqref{eq:tangent-2}.  

Note that $\EXP \big[ \ind(A=a_D) q_0^* (\bZ,\bX_{0}) \EXP \big\{ \nabla_\eta h_1 (\bW,\obX_{\TIME} \con \eta^*) (1-D_{\TIME+1}) \cond \bZ,A=a_D,\bX_{0} \big\} \big]$ is represented as
\begin{align}
& \EXP \big[ \ind(A=a_D) q_0^* (\bZ,\bX_{0}) 
\EXP \big\{ \nabla_\eta h_1(\bW,\obX_{\TIME} \con \eta^*) (1-D_{\TIME+1}) \cond \bZ,A=a_D,\bX_{0} \big\} \big]
\nonumber
\\
& =
\EXP \big\{ \ind(A=a_D,D_{\TIME+1}=0) q_0^* (\bZ,\bX_{0}) 
\nabla_\eta h_1(\bW,\obX_{\TIME} \con \eta^*)  \big\}
\nonumber
\\
& =
\EXP \bigg[ 
\begin{array}{l}
(1-D_{\TIME+1}) \Pr(A=a_D \cond \bW,D_{\TIME+1}=0,\obX_{\TIME}) 
\\
\times
\EXP \big\{ q_0^* (\bZ,\bX_{0}) \cond \bW,A=a_D,D_{\TIME+1}=0,\obX_{\TIME} \big\} 
\nabla_\eta h_1 (\bW,\obX_{\TIME} \con \eta^*) 
\end{array}
\bigg]
\nonumber
\\
& =
\EXP 
\left[
    \begin{array}{l}
     \ind(A=a_Y) (1-D_{\TIME})  
     \Pr(A=a_D \, | \, \bW,D_{\TIME+1}=0,\obX_{\TIME}) 
     /
     \Pr(A=a_Y \, | \, \bW,D_{\TIME+1}=0,\bX_{\TIME})
\\
\quad
\times
\EXP \big\{ q_0^* (\bZ,\bX_{0}) \cond \bW,A=a_D,D_{\TIME+1}=0,\obX_{\TIME} \big\} 
\nabla_\eta h_1(\bW,\obX_{\TIME} \con \eta^*) 
    \end{array} 
\right]
\nonumber
\\
& 
=
\EXP \big[ \ind(A=a_Y) (1-D_{\TIME}) 
\EXP \big\{ q_1^* (\bZ,\obX_{\TIME}) \cond \bW,A=a_Y,D_{\TIME+1}=0,\bX_{\TIME} \big\} 
\nabla_\eta h_1(\bW,\obX_{\TIME} \con \eta^*) \big]
\nonumber
\\
& =
\EXP \big[ \ind(A=a_Y)  (1-D_{\TIME+1}) 
q_1^* (\bZ,\obX_{\TIME}) \nabla_\eta h_1(\bW,\obX_{\TIME} \con \eta^*) \big]
\nonumber
\\
& =
\EXP \big[ \ind(A=a_Y)  (1-D_{\TIME+1}) 
q_1^* (\bZ,\obX_{\TIME}) \EXP \big\{ \nabla_\eta h_1(\bW,\obX_{\TIME} \con \eta^*) \cond \bZ,A=a_Y,D_{\TIME+1}=0,\obX_{\TIME} \big\} \big]
\nonumber
\\
& 
=
\EXP 
\left[
    \begin{array}{l}
        \ind(A=a_Y)  (1-D_{\TIME+1}) 
q_1^* (\bZ,\obX_{\TIME})
\\
\times 
\EXP \bigg[ 
\begin{array}{l}
\big\{ Y - h_1^*(\bW,\obX_{\TIME}) \big\}
\\
\times 
s^*(Y,\bW \cond \bZ , A=a_Y, D_{\TIME+1}=0, \obX_{\TIME})
\end{array}
\, \bigg| \, \bZ,A=a_Y,D_{\TIME+1}=0,\obX_{\TIME} 
\bigg]
    \end{array}
\right]
\ .
 \label{eq-grad-h0-2}
\end{align}
The second, third, and fourth lines hold from the law of iterated expectation. 
The fifth line holds from \eqref{eq-bridgeft5 obs}. 
The sixth and seventh lines hold from the law of iterated expectation. 
The last line holds from \eqref{eq:tangent-1}.

Combining the results in \eqref{eq-grad-h0} and \eqref{eq-grad-h0-2}, we find
\begin{align*}
& 
\EXP \big\{ \nabla_\eta h_0(\bW,\bX_{0} \con \eta^* )  \big\}
\\
& =
\EXP \left[
\ind(A=a_D) q_0^* (\bZ,\bX_{0}) 
\EXP \left[ 
\left.
    \begin{array}{l}
\big\{ h_1^* (\bW,\obX_{\TIME}) (1-D_{\TIME+1}) - h_0^*(\bW,\bX_{0}) \big\} 	
\\
\times 
s^*(\bW,D_{\TIME+1},\ubX_{1} \cond \bZ, A=a_D, \bX_{0}) 
\end{array} \right| \,
\bZ, A=a_D, \bX_{0} \right]
\right] 	 
    \\
    & 
    \quad
    +\EXP \left[
    \begin{array}{l}
	\ind(A=a_Y,D_{\TIME+1}=0) q_1^* (\bZ,\obX_{\TIME}) 
	\\
	\times
\EXP \left[ 
\left.
\begin{array}{l}
\big\{ Y - h_1^*(\bW,\obX_{\TIME}) \big\}
\\
\times 
s^*(Y,\bW \cond \bZ , A=a_Y, D_{\TIME+1}=0, \obX_{\TIME})
\end{array} \right| \,
\bZ, A=a_Y, D_{\TIME+1}=0, \obX_{\TIME} \right]
    \end{array}
\right] 	  
\\
& =
\EXP \big[ 
\ind(A=a_D) q_0^* (\bZ,\bX_{0}) 
\big\{ h_1^* (\bW,\obX_{\TIME}) (1-D_{\TIME+1}) - h_0^*(\bW,\bX_{0}) \big\} 
s^*(\bW,D_{\TIME+1},\ubX_{1} \cond \bZ, A=a_D, \bX_{0})  \big]
    \\
    & 
    \quad
    +
    \EXP \big[ \ind(A=a_Y,D_{\TIME+1}=0) q_1^* (\bZ,\obX_{\TIME}) 
    \big\{ Y - h_1^* (\bW,\obX_{\TIME}) \big\}
    s^*(Y,\bW \cond \bZ , A=a_Y, D_{\TIME+1}=0, \obX_{\TIME})  \big]
    \\
& =
\EXP \left[ 
    \begin{array}{l}
    \ind(A=a_D) q_0^* (\bZ,\bX_{0}) 
    \big\{ h_1^* (\bW,\obX_{\TIME}) (1-D_{\TIME+1}) - h_0^* (\bW,\bX_{0}) \big\} 
    \\
    \quad \times
    \big[ 
    \EXP \big\{ s^*(\bO) \cond \bW,\bZ,A,D_{\TIME+1},\obX_{\TIME} \big\} 
    - \EXP \big\{ s^*(\bO) \cond \bZ,A,\bX_{0} \big\}
    \big]  
    \end{array} 	 
\right]
    \\
    & \quad +
    \EXP \left[
        \begin{array}{l}
            \ind(A=a_Y, D_{\TIME+1}=0) q_1^* (\bZ,\obX_{\TIME}) 
            \big\{ Y - h_1^* (\bW,\obX_{\TIME}) \big\} 
            \\
            \quad \times
            \big[ 
            \EXP \big\{ s^*(\bO) \cond Y,\bW,\bZ,A,D_{\TIME+1},\obX_{\TIME} \big\} 
            - \EXP \big\{ s^*(\bO) \cond \bZ,A,D_{\TIME+1},\obX_{\TIME} \big\}
            \big]
        \end{array}
    \right] 
\\
& =
\EXP \big[ 
\ind(A=a_D) q_0^* (\bZ,\bX_{0}) 
\big\{ h_1^* (\bW,\obX_{\TIME}) (1-D_{\TIME+1}) 
- h_0^* (\bW,\bX_{0}) \big\} 
s^*(\bO)
\big]
\\
& \quad 
-
\EXP \bigg[ 
\begin{array}{l}
\ind(A=a_D) q_0^* (\bZ,\bX_{0}) 
\EXP \big\{ s^*(\bO) \cond \bZ,A,\bX_{0} \big\} 
\\
\times 
\underbrace{ \EXP \big\{ h_1^* (\bW,\obX_{\TIME}) (1-D_{\TIME+1}) 
- h_0^* (\bW,\bX_{0}) \cond \bZ,A=a_D,\bX_{0} \big\} }_{=0}
\\[-0.5cm]
\end{array}
\bigg]
\\[0.5cm]
& \quad
+
\EXP \big[ \ind(A=a_Y, D_{\TIME+1}=0) q_1^* (\bZ,\obX_{\TIME}) 
\big\{ Y - h_1^* (\bW,\obX_{\TIME}) \big\}  s^*(\bO) \big]
\\
& \quad 
-
\EXP \bigg[ 
\begin{array}{l}
\ind(A=a_Y, D_{\TIME+1}=0)  q_1^* (\bZ,\obX_{\TIME}) 
\EXP \big\{ s^*(\bO) \cond \bZ,A,D_{\TIME+1},\obX_{\TIME} \big\} 
\\
\times 
\underbrace{ \EXP \big\{ Y - h_1^* (\bW,\obX_{\TIME}) \cond \bZ,A=a_Y,D_{\TIME+1}=0,\obX_{\TIME} \big\} }_{=0}  
\\[-0.5cm]
\end{array}
\bigg]
\\[0.5cm]
& 
=
\EXP \bigg[ 
\bigg[
	\begin{array}{l}
	\ind(A=a_D) q_0^* (\bZ,\bX_{0}) 
\big\{ h_1^* (\bW,\obX_{\TIME}) (1-D_{\TIME+1}) 
- h_0^* (\bW,\bX_{0}) \big\} 
\\
+
\ind(A=a_Y, D_{\TIME+1}=0) q_1^* (\bZ,\obX_{\TIME}) 
\big\{ Y - h_1^* (\bW,\obX_{\TIME}) \big\}
	\end{array}
\bigg]
s^*(\bO)
\bigg] \ .
\end{align*}

Returning back to \eqref{eq:pathwise-Numer}, we find
\begin{align*}
&
\frac{\partial \psi_{\NUMER} (a_Y,a_D \con \eta) }{\partial \eta} \bigg|_{\eta=\eta^*}
\\
& =
\EXP \big[ \nabla_\eta h_0 (\bW,\bX_{0} \con \eta^* ) + \big\{ h_0^* (\bW,\bX_{0}) - \psi_{\NUMER}^* (a_Y,a_D) \big\} s^*(\bO) \big]
\\
& = 
\EXP \Bigg[
\underbrace{
\Bigg[
\begin{array}{l}
    \ind(A=a_D) q_0^* (\bZ,\bX_{0}) 
    \big\{ h_1^* (\bW,\obX_{\TIME}) (1-D_{\TIME+1}) - h_0^* (\bW,\bX_{0}) \big\}
    \\
    + \ind(A=a_Y)  (1-D_{\TIME+1}) q_1^* (\bZ,\obX_{\TIME+1})
    \big\{ Y - h_1^* (\bW,\obX_{\TIME}) \big\}
    + h_0^* (\bW,\bX_{0}) - \psi_{\NUMER}^* (a_Y,a_D)
\end{array}
\Bigg] }_{=: \cInfFt_{\NUMER}^* }
\times s^*(\bO) 
\Bigg]
\\
& = \EXP \big\{ \cInfFt_{\NUMER}^*(\bO) \cdot s^*(\bO) \big\} \ .
\end{align*}

When $a_Y=a_D$, $\cInfFt_{\NUMER}^*(\bO)$ reduces to
\begin{align*}
&
\cInfFt_{\NUMER}^*(\bO)
=
\Bigg[
\begin{array}{l}
    \ind(A=a_D) q_0^* (\bZ,\bX_{0}) 
    \big\{ (1-D_{\TIME+1})Y - h_0^* (\bW,\bX_{0} \con a_D,a_D) \big\} 
    \\
    + h_0^* (\bW,\bX_{0}) - \psi_{\NUMER}^* (a_D,a_D)
\end{array}
\Bigg] \ .
\end{align*}

\subsubsection*{Establish $\cInfFt_{\DENOM}^* (\bO)$}

We find the denominator has the pathwise derivative as
\begin{align}			\label{eq:pathwise-Denom}
\frac{\partial \psi_{\DENOM} (a_D \con \eta) }{\partial \eta} \bigg|_{\eta=\eta^*}
& =
\frac{\partial \EXP \ETA \big\{ h_2 (\bW, \bX_{0} \con \eta) \big\}  }{\partial \eta}  \bigg|_{\eta=\eta^*}
\nonumber
\\
& =
\EXP \big\{ \nabla_\eta h_2 (\bW,\bX_{0} \con \eta^* ) + 
h_2^* (\bW,\bX_{0}) s^*(\bW,\bX_{0}) \big\}
\nonumber
\\
& =
\EXP \big[ \nabla_\eta h_2 (\bW,\bX \con \eta^* ) + \big\{ h_2^* (\bW,\bX) - \psi_{\DENOM}^*(a_D) \big\} s^*(\bO) \big] \ .
\end{align}
The first term is
\begin{align}
& 
\EXP \big\{ \nabla_\eta h_2(\bW,\bX_{0} \con \eta^* )  \big\}
\nonumber
\\
& = 
\EXP \bigg\{ \frac{\ind(A=a_D)}{\Pr(A=a_D \cond \bW,\bX_{0}) } 
 \nabla_\eta h_2(\bW,\bX_{0} \con \eta^* )  \bigg\}
 \nonumber
\\
& =
\EXP \Big[  \ind(A=a_D) 
\EXP \big\{ q_0^* (\bZ,\bX_{0}) \cond \bW,A=a_D,\bX_{0} \big\}  
\nabla_\eta h_2(\bW,\bX_{0} \con \eta^* )  \Big]
\nonumber
 \\
& = 
\EXP \Big\{  \ind(A=a_D) q_0^* (\bZ,\bX_{0})  
\nabla_\eta h_2(\bW,\bX_{0} \con \eta^* )  \Big\}
\nonumber
 \\
& = 
\EXP \Big[  \ind(A=a_D) q_0^* (\bZ,\bX_{0}) 
\EXP \big\{ \nabla_\eta h_2(\bW,\bX_{0} \con \eta^* ) \cond \bZ, A=a_D, \bX_{0} \big\} \Big]
\nonumber
 \\
& =
\EXP \Big[ 
\ind(A=a_D) q_0^* (\bZ,\bX_{0}) 
\EXP \big[
\big\{ (1-D_{\TIME+1}) - h_2^*(\bW,\bX_{0}) \big\} 
s^*(\bW,D_{\TIME+1} \cond \bZ, A=a_D, \bX_{0}) 
     \, \big| \, \bZ, A=a_D, \bX_{0} \big] \Big]
    \nonumber
 \\
     & =
     \EXP \Big[ 
\ind(A=a_D) q_0^* (\bZ,\bX_{0}) 
\big\{ (1-D_{\TIME+1}) - h_2^* (\bW,\bX_{0}) \big\} 
s^*(\bW,D_{\TIME+1} \cond \bZ, A=a_D, \bX_{0})  \Big]
\nonumber
 \\
     & =
     \EXP \Bigg[ 
     \begin{array}{l}
\ind(A=a_D) q_0^* (\bZ,\bX_{0}) 
\big\{ (1-D_{\TIME+1}) - h_2^* (\bW,\bX_{0}) \big\} 
\\
\times 
\big[ 
    \EXP \big\{ s^*(\bO) \cond \bW,\bZ,A,D_{\TIME+1},\bX_{0} \big\} 
    - \EXP \big\{ s^*(\bO) \cond \bZ,A,\bX_{0} \big\}
\big]  
     \end{array}
 \Bigg]
\nonumber
 \\
     & =
     \EXP \Big[ 
\ind(A=a_D) q_0^* (\bZ,\bX_{0}) 
\big\{ (1-D_{\TIME+1}) - h_2^* (\bW,\bX_{0}) \big\} 
s^*(\bO)
\Big]
\nonumber
 \\
& 
\quad - 
 \EXP \Big[ 
\ind(A=a_D) q_0^* (\bZ,\bX_{0}) 
\EXP \big\{ s^*(\bO) \cond \bZ,A,\bX_{0} \big\}
\underbrace{
\EXP \big\{ (1-D_{\TIME+1}) - h_2^* (\bW,\bX_{0}) \cond \bZ,A=a_D,\bX_{0} \big\} 
}_{=0}
\Big]
\nonumber
 \\
     & =
     \EXP \Big[ 
\ind(A=a_D) q_0^* (\bZ,\bX_{0}) 
\big\{ (1-D_{\TIME+1}) - h_2^* (\bW,\bX_{0}) \big\} 
s^*(\bO)
\Big]
\ .
 \label{eq-grad-h2}
\end{align}
The second line is trivial from the law of iterated expectation. 
The third line holds from \eqref{eq-bridgeft5 obs}. 
The fourth and fifth lines are trivial from the law of iterated expectation. 
The sixth line is from \eqref{eq:tangent-3}.
The seventh line is from the law of iterated expectation. 
The eighth line is from the property of the score functions.
The last two lines are from the law of iterated expectation. 

Combining \eqref{eq:pathwise-Denom} and \eqref{eq-grad-h2}, we get
\begin{align*}
&
\frac{\partial \psi_{\DENOM} (a_D \con \eta) }{\partial \eta} \bigg|_{\eta=\eta^*}
\\
& =
\EXP \big[ \nabla_\eta h_2(\bW,\bX_{0} \con \eta^* ) 
+ \big\{ h_2^* (\bW,\bX_{0}) - \psi_{\DENOM}^* (a_D) \big\} s^*(\bO) \big]
\\
& =
\EXP \Big[ 
    \underbrace{ \big[ \ind(A=a_D) q_0^* (\bZ,\bX_{0})  \big\{ (1-D_{\TIME+1}) - h_2^* (\bW,\bX_{0}) \big\}   + \big\{ h_2^*(\bW,\bX_{0}) - \psi_{\DENOM}^*(a_D) \big\} \big] }_{=: \cInfFt_{\DENOM}^* }
\cdot s^*(\bO) \Big]
\\
& = \EXP \big\{ \cInfFt_{\DENOM}^* (\bO) \cdot s^*(\bO) \big\} \ .
\end{align*}

\subsubsection{Proof of Theorem \ref{thm:IF}-(ii)}

It suffices to derive the efficient influence function for $\psi^*(a_Y,a_D)$, denoted by $\cInfFt^*(O \con a_Y,a_D)$, because the efficient influence function for $\tau_{\CSE}^*(a_D)$ is given by $\cInfFt^*(O \con a_Y=1,a_D) - \cInfFt^*(O \con a_Y=0,a_D)$. We suppress $(a_Y,a_D)$ in confounding bridge functions and influence functions for notational brevity unless necessary. To show that the influence function $\cInfFt^*$ achieves the semiparametric local efficiency bound, we need to establish that $\cInfFt^*$ belongs to the tangent space of model $\M$ when the true data-generating law belongs to $\M_{\text{sub}}$.  Note that the model imposes restrictions \eqref{eq:tangent-1}-\eqref{eq:tangent-3} on the score functions. These restrictions imply that the tangent space of model $\M$ consists of the functions $S(\bO) \in \mathcal{L}_{2,0}(\bO)$ satisfying
\begin{align}	\label{eq-restriction}
& 
\EXP \bigg[
\begin{array}{l}
    \big\{ Y - h_1^* (\bW,\obX_{\TIME} \con a_Y) \big\} 
    \\
    \times 
    S(Y,\bW \cond \bZ, A=a_Y, D_{\TIME+1}=0, \obX_{\TIME}) 
\end{array}
	\COND
     \bZ, A=a_Y, D_{\TIME+1}=0, \obX_{\TIME} \bigg]
     \nonumber
     \\
     &
     =
     \EXP \big\{ 
     \nabla_\eta h_1(\bW,\obX_{\TIME} \con a_Y, \eta^*)	
     \cond \bZ, A=a_Y, D_{\TIME+1}=0, \obX_{\TIME}
     \big\}
    &&  
    \in  \text{Range}(\mathcal{T}_1)
    \nonumber
    \\
    & 
    \EXP \bigg[ 
    \begin{array}{l}
    \big\{ h_1^* (\bW,\obX_{\TIME} \con a_Y) (1-D_{\TIME+1}) - h_0^* (\bW,\bX_{0} \con a_Y, a_D) \big\} 
    \\
    \times 
    S (\bW,D_{\TIME+1} \cond \bZ, A=a_D, \bX_{0}) 
    \end{array}
    \COND \bZ, A=a_D, \bX_{0} \bigg]
    \nonumber 
    \\
    & 
    =
    \EXP \big\{
    \nabla_{\eta} h_{0} (W, \bX_{0} \con a_Y,a_D , \eta^*)
    \cond \bZ, A=a_D, \bX_{0} \big\}
    &&
    \in  \text{Range}(\mathcal{T}_0)
    \nonumber
    \\
    & 
    \EXP \bigg[ 
    \begin{array}{l}
    \big\{ (1-D_{\TIME+1})Y - h_0^* (\bW,\bX_{0} \con a_D, a_D) \big\} 
    \\
    \times 
    S (Y,\bW,D_{\TIME+1} \cond \bZ, A=a_D, \bX_{0}) 
    \end{array}
    \COND \bZ, A=a_D, \bX_{0} \bigg]
    \nonumber 
    \\
    & 
    =
    \EXP \big\{
    \nabla_{\eta} h_{0} (W, \bX_{0} \con a_D,a_D , \eta^*)
    \cond \bZ, A=a_D, \bX_{0} \big\}
    &&
    \in  \text{Range}(\mathcal{T}_0)
    \nonumber 
    \\
    & 
    \EXP \bigg[
    \begin{array}{l}
        \big\{ 
        (1-D_{\TIME+1}) - h_2^* (\bW,\bX_{0}) \big\} 
        \\
        \times
        S(\bW,D_{\TIME+1} \cond \bZ,A=a_D,\bX_{0}) 
    \end{array}
        \COND \bZ, A=a_D, \bX_{0} \bigg] \nonumber 
    \\
    & 
    =
    \EXP \big\{
    \nabla_{\eta} h_{2} (W, \bX_{0} \con a_D , \eta^*)
    \cond \bZ, A=a_D, \bX_{0} \big\}
    && 
     \in  \text{Range}(\mathcal{T}_0)
     \ .
\end{align}
Under the surjectivity condition \HL{S1}, we have $\text{Range}(\mathcal{T}_0) = \mathcal{L}_{2} (\bZ,A=a_D,\bX_{0})$ and $\text{Range}(\mathcal{T}_1) = \mathcal{L}_{2} ( \bZ,A=a_Y,D_{\TIME+1}=0,\obX_{\TIME})$. 
In other words, when the true data-generating law belongs to submodel $\M_{\text{sub}}$, any $S(\bO) \in \mathcal{L}_{2,0}(\bO)$ satisfies the three restrictions in \eqref{eq-restriction}. Therefore, the influence function $\cInfFt^*$ belongs to the tangent space of $\M$ when the true data-generating law belongs to submodel $\M_{\text{sub}}$, which proves the claim of Theorem \ref{thm:IF}-(ii).

\subsection{Proof of Theorem \ref{thm:AN}} \label{sec:supp:thm:AN}

We set $a_D=0,$ as the proof for $a_D=1$ case can be shown in a similar manner. We also suppress $(a_Y,a_D)$ in confounding bridge functions and influence functions for notational brevity unless necessary. To simplify notation and reduce clutter,  we denote $D=D_{\TIME+1}$. Let $N_k = N/K= |\mathcal{I}_k|$, $\AVERk(V) = \sum_{i \in \mathcal{I}_k} V_i / N_k$, and $\EMP_k(V) = |N_k|^{-1/2} \sum_{i \in \mathcal{I}_k } \{ V_i - \EXP (V_i) \} $. Additionally, let $\EXP\LSS \{ \widehat{g}\LSS (\bO) \} = \int \widehat{g}\LSS (\bo) f^*(\bo) \, d \bo$ and $\VAR\LSS \{ \widehat{g}\LSS (\bO) \} = \int [ \widehat{g}\LSS (\bo) - \EXP\LSS \{ \widehat{g}\LSS (\bO) \}  ]^2 f^*(\bo) \, d \bo$. Let $\InfFt_{\NUMER}^*$ and $\InfFt_{\DENOM}^*$ be the uncentered influence functions, i.e., 
\begin{align*}
& 
\InfFt_{\NUMER}^*(\bO \con a_Y=1)
=
\left[ 
\!
\mymatrixThree{l}{ A  (1-D) q_1^* (\bZ,\obX_{\TIME} ) \big\{ Y - h_1^* (\bW,\obX_{\TIME}) \big\}}{+ (1-A) q_0^* (\bZ,\bX_{0} ) 
\big\{ h_1^* (\bW,\obX_{\TIME} ) (1-D) - h_0^* (\bW,\bX_{0}  \con a_Y=1) \big\}}{+
h_0^* (\bW,\bX_{0} \con a_Y=1) }
\!\!
\right]  \ ,
\\
&
\InfFt_{\NUMER}^* (\bO \con a_Y=0)
=
(1-A) q_0^* (\bZ,\bX_{0}) \big\{ (1-D) Y - h_0^* (\bW,\bX_{0} \con a_Y=0) \big\} + h_0^* (\bW,\bX_{0} \con a_Y=0)   \ ,
\\
& 
\InfFt_{\DENOM}^*(\bO)
=
(1-A) q_0^* (\bZ,\bX_{0})  \big\{ (1-D) - h_2^*(\bW,\bX_{0}) \big\}  
+  h_2^* (\bW,\bX_{0})  \ .
\end{align*} 
Let $\widehat{\psi}_{\NUMER} (a_Y) = K^{-1} \sum_{k=1}^{K} \widehat{\psi}_{\NUMER}\SSS (a_Y)$ and $\widehat{\psi}_{\DENOM} = K^{-1} \sum_{k=1}^{K} \widehat{\psi}_{\DENOM}\SSS $ where
 \begin{align*}
    \widehat{\psi}_{\NUMER}\SSS (a_Y=1)
    &
    =
    \AVERk \big\{ \widehat{\InfFt}_{\NUMER} \LSS(\bO \con a_Y=1) \big\}
    \\
    &
    =
    \AVERk 
    \left[
    \begin{array}{l}
    A(1-D)
    \EqOone 
    \big\{ Y - \EhOone \big\}
    \\
    + 
    (1-A) \EqOzero \big\{ (1-D) \EhOone - \EhOzone \big\}
    \\
    +
    \EhOzone
    \end{array}
    \right]  \ ,
    \\
    \widehat{\psi}_{\NUMER}\SSS (a_Y=0)
    &
    =
    \AVERk \big\{ \widehat{\InfFt}_{\NUMER} \LSS(\bO \con a_Y=0) \big\}
    \\
    &
    =
    \AVERk 
    \left[
    (1-A) \EqOzero \big\{ (1-D) Y - \EhOzzero \big\} + \EhOzzero  
    \right]
    \\
    \widehat{\psi}_{\DENOM}\SSS
    &
    =
    \AVERk \big\{ \widehat{\InfFt}_{\DENOM} \LSS(\bO) \big\}
    \\
    &
    =
    \AVERk
    \left[ 
    (1-A) \EqOzero \big\{ (1-D) - \EhOtwo \big\}
    +
    \EhOtwo 
    \right]
    \ .
 \end{align*}
We will establish 
\begin{align}	\label{eq-AsymptoticNormality}
\sqrt{N} 
\left\{
    \begin{pmatrix}
        \widehat{\psi}_{\NUMER} (a_Y=1) \\
        \widehat{\psi}_{\NUMER} (a_Y=0) \\
        \widehat{\psi}_{\DENOM} (a_D)
    \end{pmatrix}
    -
    \begin{pmatrix}
        \psi_{\NUMER}^*(a_Y=1) \\ 
        \psi_{\NUMER}^*(a_Y=0) \\ 
        \psi_{\DENOM}^* (a_D)
    \end{pmatrix}
\right\}	
& =
\frac{1}{\sqrt{N}}
\sum_{i=1}^{N}
\begin{pmatrix}
    \cInfFt_{\NUMER}^* (\bO_i \con a_Y=1) \\ 
    \cInfFt_{\NUMER}^* (\bO_i \con a_Y=0) \\
    \cInfFt_{\DENOM}^* (\bO_i)
\end{pmatrix}
+o_P(1)  
\end{align}
which is asymptotically normal as $N
\big( 0 , \Sigma^* \big)$ where
\begin{align*}
\Sigma^*
=
\VAR \left\{ \begin{pmatrix}
    \cInfFt_{\NUMER}^*(\bO \con a_Y=1) \\
    \cInfFt_{\NUMER}^*(\bO \con a_Y=0) \\
    \cInfFt_{\DENOM}^*(\bO)
\end{pmatrix} \right\} \ ,
\end{align*}
of which a consistent variance estimator is
\begin{align}	\label{eq-ConsistencyVar}
&
\widehat{\Sigma}
=
\frac{1}{K} \sum_{k=1}^{K} \widehat{\Sigma}\SSS
\ , \
\widehat{\Sigma}\SSS
=
\AVERk
\left\{
    \begin{pmatrix}
        \widehat{\cInfFt}_{\NUMER}\LSS(\bO \con a_Y=1)
        \\
        \widehat{\cInfFt}_{\NUMER}\LSS(\bO \con a_Y=0)
        \\
        \widehat{\cInfFt}_{\DENOM}\LSS(\bO)
    \end{pmatrix}^{\otimes 2}
\right\}
\\
&
\widehat{\cInfFt}_{\NUMER}\LSS(\bO \con a_Y=1)
\nonumber
=
\left[
    \begin{array}{l}
    A(1-D) \EqOone 
    \big\{ Y - \EhOone \big\}
    \\
    + 
    (1-A) \EqOzero \big\{ (1-D) \EhOone - \EhOzone \big\}
    \\
    +
    \EhOzone - \widehat{\psi}_{\NUMER}(a_Y=1)
    \end{array}
    \right]
    \nonumber
\\
&
\widehat{\cInfFt}_{\NUMER}\LSS(\bO \con a_Y=0)
=
\left[
    \begin{array}{l}
    (1-A) \EqOzero \big\{ (1-D) Y - \EhOzzero \big\}
    \\
    +
    \EhOzzero - \widehat{\psi}_{\NUMER}(a_Y=0)
    \end{array}
    \right]
    \nonumber
    \\
 &
\widehat{\cInfFt}_{\DENOM}\LSS(\bO)
= 
    (1-A) \EqOzero \big\{ (1-D) - \EhOtwo \big\}
    +
    \EhOtwo 
    -
    \widehat{\psi}_{\DENOM} \ .
        \nonumber
\end{align} 
Since $\widehat{\tau}_{\CSE} = \big\{ \widehat{\psi}_{\NUMER}(1) - \widehat{\psi}_{\NUMER}(0) \big\}/ \widehat{\psi}_{\DENOM}$, we have
\begin{align*}
&
\sqrt{N} 
\Big\{ \widehat{\tau}_{\CSE} - \tau_{\CSE}^* \Big\}
\\
 &
 =
 \sqrt{N}
 \frac{	\big\{ \widehat{\psi}_{\NUMER}(1) - \widehat{\psi}_{\NUMER}(0) \big\} \psi_{\DENOM}^* - \tau_{\CSE}^* \widehat{\psi}_{\DENOM} }{\widehat{\psi}_{\DENOM} \psi_{\DENOM}^* } 
 \\
 &
 =
 \sqrt{N}
 \frac{ 1 }{2 \widehat{\psi}_{\DENOM} \psi_{\DENOM}^* }
 \left[
    \begin{array}{l} 		
 \big[
 \big\{
    \widehat{\psi}_{\NUMER}(1) - \widehat{\psi}_{\NUMER}(0) \big\} - 
    \big\{ \psi_{\NUMER}^* (1) - \psi_{\NUMER}^* (0) \big\}
 \big]
 \big\{
    \widehat{\psi}_{\DENOM} + \psi_{\DENOM}^*
 \big\}
 \\
 -
 \big[
 \big\{
    \widehat{\psi}_{\NUMER}(1) - \widehat{\psi}_{\NUMER}(0) \big\} + 
    \big\{ \psi_{\NUMER}^* (1) - \psi_{\NUMER}^* (0) \big\}
 \big]
 \big\{
 \widehat{\psi}_{\DENOM} - \psi_{\DENOM}^* 
   \big\}
    \end{array}
 \right]
  \\
 &
 =
 \sqrt{N}
 \frac{ 1 + o_P(1) }{2 \{ \psi_{\DENOM}^{*}  \}^2 } 
 \left[
    \begin{array}{l}
        \big[
 \big\{
    \widehat{\psi}_{\NUMER}(1) - \widehat{\psi}_{\NUMER}(0) \big\} - 
    \big\{ \psi_{\NUMER}^* (1) - \psi_{\NUMER}^* (0) \big\}
 \big]
        \big\{ 2 \psi_{\DENOM}^* + o_P(1) \big\}
        \\
        -
        \big[ 2 \big\{ \psi_{\NUMER}^*(1) - \psi_{\NUMER}^*(0) \big\} + o_P(1) \big]
        \big\{ \widehat{\psi}_{\DENOM} - \psi_{\DENOM}^* \big\}
    \end{array}
 \right] 
  \\
 &
 =
 \frac{1}{\sqrt{N}}
\sum_{i=1}^{N}
 \Bigg[
    \frac{\InfFt_{\NUMER}^* (\bO_i \con a_Y=1) - \InfFt_{\NUMER}^* (\bO_i \con a_Y=0)}{\psi_{\DENOM}^* } 
    -
    \frac{\psi_{\NUMER}^*(1) - \psi_{\NUMER}^*(0)}{\{ \psi_{\DENOM}^*   \}^2} \InfFt_{\DENOM}^* (\bO_i)
 \Bigg]
 +
 o_P(1)
  \\
 &
 =
 \frac{1}{\sqrt{N}}
\sum_{i=1}^{N}
 \Bigg[
    \frac{\cInfFt_{\NUMER}^* (\bO_i \con a_Y=1) - \cInfFt_{\NUMER}^* (\bO_i \con a_Y=0)}{\psi_{\DENOM}^* } 
    -
    \frac{\psi_{\NUMER}^*(1) - \psi_{\NUMER}^*(0)}{\{ \psi_{\DENOM}^*   \}^2} \cInfFt_{\DENOM}^* (\bO_i)
 \Bigg]
 +
 o_P(1)
   \\
 &
 =
 \frac{1}{\sqrt{N}}
\sum_{i=1}^{N}
 \cInfFt_{\CSE}^* (\bO_i)
 +
 o_P(1) \ ,
\end{align*} 
which is asymptotically normal with the limiting distribution $N \big( 0 ,  \VAR \big\{ \cInfFt_{\CSE}^*(\bO) \big\} \big)$. Additionally, the proposed variance estimator is consistent:
\begin{align*}
\widehat{\sigma}_{\CSE}^2
    & 
    =
    \frac{1}{N}
    \sum_{k=1}^{K}
    \sum_{i \in \mathcal{I}_k}
    \bigg\{
    \frac{\widehat{\InfFt}_{\NUMER}\LSS(\bO_i \con a_Y=1) - \widehat{\InfFt}_{\NUMER}\LSS(\bO_i \con a_Y=0)
    -
    \widehat{\tau}_{\CSE}
    \widehat{\InfFt}_{\DENOM}\LSS(\bO_i)}{ \widehat{\psi}_{\DENOM} } 
    \bigg\}^2
    \\
    &
    =
    \begin{bmatrix}
    1/\widehat{\psi}_{\DENOM} 
    \ & \
    -
    1/\widehat{\psi}_{\DENOM} 
        \ & \
        -
        {\widehat{\tau}_{\CSE}}/{\widehat{\psi}_{\DENOM} }
    \end{bmatrix}
    \widehat{\Sigma}
    \begin{bmatrix}
    1/\widehat{\psi}_{\DENOM} 
    \\
    -
    1/\widehat{\psi}_{\DENOM} 
        \\
        -
        {\widehat{\tau}_{\CSE}}/{\widehat{\psi}_{\DENOM} }
    \end{bmatrix}
    \\
    &
    =
    \begin{bmatrix}
        1/\psi_{\DENOM}^* 
        \ & \ 
        - 1/\psi_{\DENOM}^* 
        \ & \
        -
        \tau_{\CSE}^*/\psi_{\DENOM}^* 
    \end{bmatrix}
    \Sigma^*
    \begin{bmatrix}
        1/\psi_{\DENOM}^* 
        \\
        - 1/\psi_{\DENOM}^* 
        \\
        -
        \tau_{\CSE}^*/\psi_{\DENOM}^* 
    \end{bmatrix}
    +
    o_P(1)
    \\
    &
    =
    \EXP \big[ \big\{ \cInfFt_{\CSE}^*(\bO) \big\}^2 \big]
    +
    o_P(1) \ .
\end{align*}
This yields the desired result.

\subsubsection{Proof of the Asymptotic Normality in \eqref{eq-AsymptoticNormality}}

\subsubsection*{Numerator Part Related to $a_Y=1$}

We use shorthands $\cInfFt^*(O)=\cInfFt^*(O \con a_Y=1)$, $\InfFt^*(O)=\InfFt^*(O \con a_Y=1)$, $\psi_{\NUMER}^* = \psi_{\NUMER}^*(a_Y=1)$, and $\widehat{\psi}_{\NUMER} = \widehat{\psi}_{\NUMER} (a_Y=1)$, and we drop variables in the confounding bridge functions, e.g., $h_0^* = h_0^*(W,X_{0})$ and $\widehat{h}_0\LSS = h_0\LSS(W,X_{0})$.

Recall that the centered influence function of the numerator part related to $a_Y=1$ is
 \begin{align}
    \cInfFt_{\NUMER}^* (\bO) 
    &
    =
    \left[
        \begin{array}{l}	 		
    A(1-D) \TqOone \big\{ Y - \ThOone \big\}  
    \\
    +
    (1-A) \TqOzero  \big\{ (1-D) \ThOone - \ThOzero \big\} 
    +  \ThOzero - \psi_{\NUMER}^*
        \end{array}
    \right]
    \nonumber
    \\
    &
    =
    A(1-D) \Tq{1} \big\{ Y - \Th{1} \big\}  
    +
    (1-A) \Tq{0}  \big\{ (1-D) \Th{1} - \Th{0} \big\} 
    +  \Th{1} - \psi_{\NUMER}^*
    \nonumber
 \end{align}
 
Let $\widehat{\psi}_{\NUMER} = K^{-1} \sum_{k=1}^{K} \widehat{\psi}_{\NUMER}\SSS$ where
 \begin{align*}
    \widehat{\psi}_{\NUMER}\SSS
    &
    =
    \AVERk
    \left[
    \begin{array}{l}
    A(1-D) \EqOone 
    \big\{ Y - \EhOone \big\}
    \\
    + 
    (1-A) \EqOzero \big\{ (1-D) \EhOone - \EhOzero \big\}
    \\
    +
    \EhOzero
    \end{array}
    \right]
    \ .
 \end{align*}

 The empirical process of $\widehat{\psi}_{\NUMER}\SSS$ is 
 {\small
 \begin{align}
    &
    \sqrt{N_k} \big\{ \widehat{\psi}_{\NUMER}\SSS - {\psi}_{\NUMER}^* \big\}
    \nonumber
    \\
    & =
    \sqrt{N_k}
    \Bigg[
        \begin{array}{l}
        \AVERk
            \big[
    A(1-D) \Eq{1} 
    \big\{ Y - \Eh{1} \big\}
    + 
    (1-A) \Eq{0} \big\{ (1-D) \Eh{1} - \Eh{0} \big\}
    +
    \Eh{0}
    \big]
    \nonumber
    \\
    -
    \AVERk
            \big[
    A(1-D) \Tq{1} 
    \big\{ Y - \Th{1} \big\}
    + 
    (1-A) \Tq{0} \big\{ (1-D) \Th{1} - \Th{0} \big\}
    +
    \Th{0}
    \big]
        \end{array}
    \Bigg]
    \nonumber
    \\
    & 
    \hspace*{0.5cm}
    + 
    \sqrt{N_k}
    \AVERk
            \big[
    A(1-D) \Tq{1} 
    \big\{ Y - \Th{1} \big\}
    + 
    (1-A) \Tq{0} \big\{ (1-D) \Th{1} - \Th{0} \big\}
    +
    \Th{0}
    - \psi_{\NUMER}^*
    \big]
    \nonumber
    \\
    & =
    \sqrt{N_k}
    \left[
        \begin{array}{l}
        \AVERk
        \big[
    A(1-D) \Eq{1} 
    \big\{ Y - \Eh{1} \big\}
    + 
    (1-A) \Eq{0} \big\{ (1-D) \Eh{1} - \Eh{0} \big\}
    +
    \Eh{0}
    \big]
    \nonumber
    \\
    -
    \EXPk \big[
    A(1-D) \Eq{1} 
    \big\{ Y - \Eh{1} \big\}
    + 
    (1-A) \Eq{0} \big\{ (1-D) \Eh{1} - \Eh{0} \big\}
    +
    \Eh{0}
    \big]
    \nonumber
    \\
    -
    \AVERk
    \big[
    A(1-D) \Tq{1} 
    \big\{ Y - \Th{1} \big\}
    + 
    (1-A) \Tq{0} \big\{ (1-D) \Th{1} - \Th{0} \big\}
    +
    \Th{0}
    \big]
    \nonumber
    \\
    +
    \EXPk \big[
    A(1-D) \Tq{1} 
    \big\{ Y - \Th{1} \big\}
    + 
    (1-A) \Tq{0} \big\{ (1-D) \Th{1} - \Th{0} \big\}
    +
    \Th{0}
    \big]
        \end{array}
    \right]
    \nonumber
    \\
    &
    \hspace*{0.5cm} 
    +
    \sqrt{N_k}
    \left[
        \begin{array}{l}
    \EXPk \big[
    A(1-D) \Eq{1} 
    \big\{ Y - \Eh{1} \big\}
    + 
    (1-A) \Eq{0} \big\{ (1-D) \Eh{1} - \Eh{0} \big\}
    +
    \Eh{0}
    \big]
    \nonumber
    \\
    -
    \EXPk \big[
    A(1-D) \Tq{1} 
    \big\{ Y - \Th{1} \big\}
    + 
    (1-A) \Tq{0} \big\{ (1-D) \Th{1} - \Th{0} \big\}
    +
    \Th{0}
    \big]
        \end{array}
    \right]
    \nonumber
    \\
    & 
    \hspace*{0.5cm}
    + 
    \sqrt{N_k}
    \AVERk
            \big[
    A(1-D) \Tq{1} 
    \big\{ Y - \Th{1} \big\}
    + 
    (1-A) \Tq{0} \big\{ (1-D) \Th{1} - \Th{0} \big\}
    +
    \Th{0}
    - \psi_{\NUMER}^*
    \big]
    \nonumber
    \\
    &
    =
    \EMPk
    \left[
        \begin{array}{l}
        \big[ A(1-D) \Eq{1} 
    \big\{ Y - \Eh{1} \big\}
    + 
    (1-A) \Eq{0} \big\{ (1-D) \Eh{1} - \Eh{0} \big\}
    +
    \Eh{0}
    \big]
    \\
    -
    \big[
    A(1-D) \Tq{1} 
    \big\{ Y - \Th{1} \big\}
    + 
    (1-A) \Tq{0} \big\{ (1-D) \Th{1} - \Th{0} \big\}
    +
    \Th{0}
    \big]
        \end{array}
    \right]
    \label{eq-Bias-Numer-T1}
    \\
    & 
    \hspace*{0.5cm}
    +
    \sqrt{N_k}
    \left[
        \begin{array}{l}
    \EXPk \big[
    A(1-D) \Eq{1} 
    \big\{ Y - \Eh{1} \big\}
    + 
    (1-A) \Eq{0} \big\{ (1-D) \Eh{1} - \Eh{0} \big\}
    +
    \Eh{0}
    \big]
    \\
    -
    \EXPk \big[
    A(1-D) \Tq{1} 
    \big\{ Y - \Th{1} \big\}
    + 
    (1-A) \Tq{0} \big\{ (1-D) \Th{1} - \Th{0} \big\}
    +
    \Th{0}
    \big]
        \end{array}
    \right]
    \label{eq-Bias-Numer-T2}
    \\
    & 
    \hspace*{0.5cm}
    + 
    \EMPk
            \big[
    A(1-D) \Tq{1} 
    \big\{ Y - \Th{1} \big\}
    + 
    (1-A) \Tq{0} \big\{ (1-D) \Th{1} - \Th{0} \big\}
    +
    \Th{0}
    \big] \ .
    \label{eq-Bias-Numer-T3}
 \end{align}}%

The last term \eqref{eq-Bias-Numer-T3} is equal to 
\begin{align*}
    & \EMPk
    \big[
    A(1-D) \Tq{1} 
    \big\{ Y - \Th{1} \big\}
    + 
    (1-A) \Tq{0} \big\{ (1-D) \Th{1} - \Th{0} \big\}
    +
    \Th{0}
    \big]
    =
    \sqrt{\frac{K}{N}}
    \sum_{i \in \mathcal{I}_k} \cInfFt_{\NUMER}^* (\bO_i) \ .
\end{align*}
If \eqref{eq-Bias-Numer-T1} and \eqref{eq-Bias-Numer-T2} are $o_P(1)$, then we obtain
 \begin{align*}
    \sqrt{N}\Big( \widehat{\psi}_{\NUMER} - \psi_{\NUMER}^* \Big)
    & =
    \frac{ \sqrt{N_k}\sum_{k=1}^{K} \big\{ \widehat{\psi}_{\NUMER}\SSS - \psi_{\NUMER}^* \big\} }{\sqrt{K}}
    =
    \frac{\sum_{i = 1}^{N} \cInfFt_{\NUMER}^* (\bO_i)}{\sqrt{N}}  + o_P(1)  \ . 
 \end{align*}

Term \eqref{eq-Bias-Numer-T2} is decomposed as
\begin{align*}
& 
\eqref{eq-Bias-Numer-T2}
\\
& =
\sqrt{N_k}
    \left[
        \begin{array}{l}
    \EXPk \big[
    A(1-D) \Eq{1} 
    \big\{ Y - \Eh{1} \big\}
    + 
    (1-A) \Eq{0} \big\{ (1-D) \Eh{1} - \Eh{0} \big\}
    +
    \Eh{0}
    \big]
    \\
    -
    \EXPk \big[
    A(1-D) \Tq{1} 
    \big\{ Y - \Th{1} \big\}
    + 
    (1-A) \Tq{0} \big\{ (1-D) \Th{1} - \Th{0} \big\}
    +
    \Th{0}
    \big]
        \end{array}
    \right]
    \\
    &
    =
    \sqrt{N_k}
    \EXPk 
    \left[
        \begin{array}{l}
            \big[ A(1-D) \big\{ \EqOone Y  - \TqOone Y \big\}  \big]
            \\
            -
            \big[ A(1-D) \big\{ \EqOone \EhOone  - \TqOone \ThOone \big\} \big]
            \\
            +  
            \big[ (1-A)(1-D) \big\{ \EqOzero \EhOone - \TqOzero \ThOone \big\} \big]
            \\
            -
            \big[ (1-A) \big\{ \EqOzero \EhOzero - \TqOzero \ThOzero \big\} \big]
            \\
            +
            \big[ \EhOzero - \ThOzero \big]
        \end{array}
    \right]
    \quad 
    \begin{array}{l}
        \text{\HT{Numer-T1}}
        \\
        \text{\HT{Numer-T2}}
        \\
        \text{\HT{Numer-T3}}
        \\
        \text{\HT{Numer-T4}}
        \\
        \text{\HT{Numer-T5}}
    \end{array}
    \\
    &
    \stackrel{(*)}{=}
    \sqrt{N_k}
    \EXPk
    \left[
        \begin{array}{l}
            A (1-D) \big\{ \EqOone - \TqOone \big\} \big\{  \ThOone  - \EhOone \big\}
            \\
            +
            (1-A)(1-D) \big\{ \EqOzero - \TqOzero \big\} \big\{ \EhOone - \ThOone \big\}
            \\
            + 
            (1-A) \big\{ \EqOzero - \TqOzero \big\} \big\{ \ThOzero - \EhOzero \big\}
        \end{array}
    \right]
    \\
    &
    \leq
    \sqrt{N_k}
    \left[
        \begin{array}{l}
             \big\| \EqOone - \TqOone \big\|_{P,2}  \big\| \EhOone - \ThOone \big\|_{P,2}
             \\
             +
             \big\| \EqOzero - \TqOzero \big\|_{P,2}  \big\| \EhOone - \ThOone \big\|_{P,2}
             \\
             +
             \big\| \EqOzero - \TqOzero \big\|_{P,2}  \big\| \EhOzero - \ThOzero \big\|_{P,2}
        \end{array}
    \right]
    \\
    &
    =
    o_P(1) \ .
\end{align*}
Identity $(*)$ is established below in (a)-(c).  The inequality is from the Cauchy-Schwarz inequality, and the convergence rate holds from Assumption \AssumptionMixBiasNumer. 
\begin{itemize}[leftmargin=0cm,itemsep=0cm]
\item[(a)] \HL{Numer-T1}+\HL{Numer-T2}
\begin{align*}
&
\text{\HL{Numer-T1}}
\\
&
=
\EXPk \big[ A(1-D) \big\{ \EqOone Y  - \TqOone Y \big\} \big]
\\
&
=
\EXPk \big[ \Pr(A=1,D=0 \cond Z, \obX_{\TIME}) \big\{ \EqOone  - \TqOone \big\} \EXP \big( Y \cond Z,A=1,D=0, \obX_{\TIME} \big) \big]
\\
&
\stackrel{(*)}{=}
\EXPk \big[ \Pr(A=1,D=0 \cond Z, \obX_{\TIME}) \big\{ \EqOone  - \TqOone \big\} \EXP \big\{ \ThOone \cond Z,A=1,D=0,\obX_{\TIME} \big\} \big]
\\
&
=
\EXPk \big[ A (1-D) \big\{ \EqOone \ThOone  - \TqOone \ThOone \big\} \big] \ .
\end{align*}
Equality $(*)$ holds from \eqref{eq-bridgeft1 obs}. Therefore, we find
\begin{align*}
& 
\text{\HL{Numer-T1}} + \text{\HL{Numer-T2}}
\\
&
=
\EXPk \Bigg[ 
\begin{array}{l}
    \big[ A (1-D) \big\{ \EqOone \ThOone  - \TqOone \ThOone \big\}  \big]
    \\
    -
    \big[ A (1-D) \big\{ \EqOone \EhOone  - \TqOone \ThOone \big\} \big]
\end{array}
\Bigg]
\\
&
=
\EXPk \big[ A (1-D) \EqOone \big\{  \ThOone  - \EhOone \big\}  \big]
\\
&
=
\EXPk \big[ A (1-D) \big\{ \EqOone - \TqOone \big\} \big\{  \ThOone  - \EhOone \big\}  \big]
\\
& 
\hspace*{2cm}
+
\EXPk \big[ A (1-D) \TqOone \big\{  \ThOone  - \EhOone \big\}  \big]
\\
&
\stackrel{(*)}{=}
\EXPk \big[ A (1-D) \big\{ \EqOone - \TqOone \big\} \big\{  \ThOone  - \EhOone \big\}  \big]
\\
& 
\hspace*{2cm}
+
\EXPk \big[ (1-A) (1-D) \TqOzero \big\{  \ThOone  - \EhOone \big\}  \big] \ .
\end{align*}
Equality $(*)$ holds from \eqref{eq-bridgeft6 obs}.

\item[(b)] \HL{Numer-T3} 
\begin{align*}
&
\text{\HL{Numer-T3}}
\\
&
=
\EXPk \big[ (1-A)(1-D) \big\{ \EqOzero \EhOone - \TqOzero \ThOone \big\} \big]
\\
&
=
\EXPk \big[ (1-A)(1-D) \big\{ \EqOzero \EhOone - \TqOzero \EhOone \big\} \big]
\\
& 
\hspace*{2cm}
-
\EXPk \big[ (1-A) (1-D) \TqOzero \big\{  \ThOone  - \EhOone \big\}  \big]
\\
&
=
\EXPk \big[ (1-A)(1-D) \big\{ \EqOzero - \TqOzero \big\} \big\{ \EhOone - \ThOone \big\} \big]
\\
& 
\hspace*{2cm}
+
\EXPk \big[ (1-A)(1-D) \big\{ \EqOzero - \TqOzero \big\} \ThOone \big]
\\
& 
\hspace*{2cm}
-
\EXPk \big[ (1-A) (1-D) \TqOzero \big\{  \ThOone  - \EhOone \big\}  \big]
\\
&
\stackrel{(*)}{=}
\EXPk \big[ (1-A)(1-D) \big\{ \EqOzero - \TqOzero \big\} \big\{ \EhOone - \ThOone \big\} \big]
\\
& 
\hspace*{2cm}
+
\EXPk \big[ (1-A) \big\{ \EqOzero - \TqOzero \big\} \ThOzero \big]
\\
& 
\hspace*{2cm}
-
\EXPk \big[ (1-A) (1-D) \TqOzero \big\{  \ThOone  - \EhOone \big\}  \big] \ .
\end{align*}
Equality $(*)$ holds from \eqref{eq-bridgeft1 obs}.

\item[(c)] \HL{Numer-T4} 
\begin{align*}
&
\text{\HL{Numer-T4}}
\\
&=
\EXPk \big[ (1-A) \big\{ \TqOzero \ThOzero - \EqOzero \EhOzero \big\} \big]
\\
&
=
\EXPk \big[ (1-A) \big\{ \EqOzero \ThOzero - \EqOzero \EhOzero \big\} \big]
\\
&
\hspace*{2cm}
-
\EXPk \big[ (1-A) \big\{ \EqOzero - \TqOzero \big\} \ThOzero \big]
\\
&
=
\EXPk \big[ (1-A) \big\{ \EqOzero - \TqOzero \big\} \big\{ \ThOzero - \EhOzero \big\} \big]
\\
&
\hspace*{2cm}
+
\EXPk \big[ (1-A) \TqOzero  \big\{ \ThOzero - \EhOzero \big\} \big]
\\
&
\hspace*{2cm}
-
\EXPk \big[ (1-A) \big\{ \EqOzero - \TqOzero \big\} \ThOzero \big]
\\
&
\stackrel{(*)}{=}
\EXPk \big[ (1-A) \big\{ \EqOzero - \TqOzero \big\} \big\{ \ThOzero - \EhOzero \big\} \big]
\\
&
\hspace*{2cm}
+
\EXPk \big\{ \ThOzero - \EhOzero \big\}
\\
&
\hspace*{2cm}
-
\EXPk \big[ (1-A) \big\{ \EqOzero - \TqOzero \big\} \ThOzero \big]
\\
&
=
\EXPk \big[ (1-A) \big\{ \EqOzero - \TqOzero \big\} \big\{ \ThOzero - \EhOzero \big\} \big]
\\
&
\hspace*{2cm}
- \text{\HL{Numer-T5}}
\\
&
\hspace*{2cm}
-
\EXPk \big[ (1-A) \big\{ \EqOzero - \TqOzero \big\} \ThOzero \big] \ .
\end{align*}
Equality $(*)$ holds from \eqref{eq-bridgeft5 obs}.

\end{itemize}

Combining (a), (b), and (c), we find Term \eqref{eq-Bias-Numer-T2} is $o_P(1)$. 

Next, we establish that Term \eqref{eq-Bias-Numer-T1} is $o_P(1)$.  Suppose $g(\bO)$ is a mean-zero function. Then, $\EXPk \big\{ \EMPk(g) \big\} = 0$. Therefore, it suffices to show that $\VARk \big\{ \EMPk(g) \big\} = o_P(1)$, i.e., 
\begin{align*}
\VARk \big\{ \EMPk(g) \big\}
=
\VARk \bigg\{
\frac{1}{\sqrt{N_k}} \sum_{i \in \mathcal{I}_k} g(\bO_i)
\bigg\}
=
\VARk \big\{ g(\bO) \big\}
=
\EXPk \big\{ g(\bO)^2 \big\}
=
o_P(1) \ .
\end{align*}
The variance of \eqref{eq-Bias-Numer-T1} is
\begin{align*}
& 
\VARk \big\{
\eqref{eq-Bias-Numer-T1}
\big\}
\\
& =
\VARk \left[
\EMPk
    \left[
        \begin{array}{l}
        \big[ A(1-D) \Eq{1} 
    \big\{ Y - \Eh{1} \big\}
    + 
    (1-A) \Eq{0} \big\{ (1-D) \Eh{1} - \Eh{0} \big\}
    +
    \Eh{0}
    \big]
    \\
    -
    \big[
    A(1-D) \Tq{1} 
    \big\{ Y - \Th{1} \big\}
    + 
    (1-A) \Tq{0} \big\{ (1-D) \Th{1} - \Th{0} \big\}
    +
    \Th{0}
    \big]
        \end{array}
    \right]
    \right]
    \\
    &
    =
    \EXPk 
    \left[
    \left[
        \begin{array}{l}
            \big[ A(1-D) \big\{ \EqOone Y  - \TqOone Y \big\}  \big]
            \\
            -
            \big[ A(1-D) \big\{ \EqOone \EhOone  - \TqOone \ThOone \big\} \big]
            \\
            +  
            \big[ (1-A)(1-D) \big\{ \EqOzero \EhOone - \TqOzero \ThOone \big\} \big]
            \\
            -
            \big[ (1-A) \big\{ \EqOzero \EhOzero - \TqOzero \ThOzero \big\} \big]
            \\
            +
            \big[ \EhOzero - \ThOzero \big]
        \end{array}
    \right]^2
    \right]
    \\
    &
    \stackrel{(*)}{\precsim }
    \EXPk 
    \left[
        \begin{array}{l}
            \big[ A(1-D) \big\{ \EqOone Y  - \TqOone Y \big\}  \big]^2
            \\
            +
            \big[ A(1-D) \big\{ \EqOone \EhOone  - \TqOone \ThOone \big\} \big]^2
            \\
            +  
            \big[ (1-A)(1-D) \big\{ \EqOzero \EhOone - \TqOzero \ThOone \big\} \big]^2
            \\
            +
            \big[ (1-A) \big\{ \EqOzero \EhOzero - \TqOzero \ThOzero \big\} \big]^2
            \\
            +
            \big[ \EhOzero - \ThOzero \big]^2
        \end{array}
    \right]
    \quad 
    \begin{array}{l}
        \text{\HT{Numer-Q1}}
        \\
        \text{\HT{Numer-Q2}}
        \\
        \text{\HT{Numer-Q3}}
        \\
        \text{\HT{Numer-Q4}}
        \\
        \text{\HT{Numer-Q5}}
    \end{array}
    \\
    &
    \stackrel{(\dagger)}{\precsim}
    \big\| \Eq{0}  - \Tq{0} \big\|_{P,2}^2
    +
    \big\| \Eq{1}  - \Tq{1} \big\|_{P,2}^2
    +
    \big\| \Eh{0}  - \Th{0} \big\|_{P,2}^2
    +
    \big\| \Eh{1}  - \Th{1} \big\|_{P,2}^2 
    \\
    &
    =
    o_P(1) \ .
\end{align*}
Inequality $(*)$ holds from the Cauchy-Schwarz inequality $(\sum_{j=1}^{M} a_j)^2 \leq M (\sum_{j=1}^{M} a_j^2)$. Inequality $(\dagger)$ is established below in (a)-(e). The last convergence rate is obtained  from Assumption \AssumptionConsistency. 
\begin{itemize}[leftmargin=0cm,itemsep=0cm]
\item[(a)] \HL{Numer-Q1}
\begin{align*}
&
\text{\HL{Numer-Q1}}
\\
&
=
\EXPk \big[ \big[ A(1-D) \big\{ \EqOone Y  - \TqOone Y \big\}  \big]^2 \big]
\\
&
=
\EXPk \big[ A(1-D) \big\{ \EqOone  - \TqOone \big\}^2 Y^2 \big]
\\
&
=
\EXPk \big[ \Pr(A=1,D=0 \cond Z, \obX_{\TIME}) \big\{ \EqOone  - \TqOone \big\}^2 \EXP \big(Y^2 \cond Z,A=1,D=0,\obX_{\TIME} \big) \big]
\\
&
\stackrel{(*)}{\precsim}
\EXPk \big[ \big\{ \EqOone  - \TqOone \big\}^2 \big]
\\
& =
\big\| \Eq{1}  - \Tq{1} \big\|_{P,2}^2 \ .
\end{align*}
Inequality $(*)$ holds from Assumption \AssumptionStrongOv: $\EXP (Y^2 \cond Z,A=1,D=0,\obX_{\TIME}) \leq C$ and $\Pr(A=1,D=0 \cond Z, \obX_{\TIME}) \leq 1$.

\item[(b)] \HL{Numer-Q2} 
\begin{align*}
&
\text{\HL{Numer-Q2}}
\\
&
=
\EXPk \big[ \big[ A(1-D) \big\{ \EqOone \EhOone  - \TqOone \ThOone \big\} \big]^2 \big]
\\
&
\stackrel{(*)}{\leq}
\frac{1}{2}
\EXPk
\Bigg[ 
    \begin{array}{l}
        \big\{ \EqOone - \TqOone \big\}^2 \big\{ \EhOone + \ThOone \big\}^2
        \\
        +
        \big\{ \EqOone + \TqOone \big\}^2 \big\{ \EhOone - \ThOone \big\}^2
    \end{array}
\Bigg]
\\
&
\stackrel{(\dagger)}{\leq }
2 C_{h1}^2
\EXPk
\big[ \big\{ \EqOone - \TqOone \big\}^2 \big]
+
2 C_{q1}^2
\EXPk
\big[ \big\{ \EhOone - \ThOone \big\}^2 \big]
\\
&
\precsim
\big\| \Eq{1} - \Tq{1} \big\|_{P,2}^2 + \big\| \Eh{1} - \Th{1} \big\|_{P,2}^2 \ .
\end{align*}
Inequality $(*)$ holds from $A(1-D) \leq 1$ and 
\begin{align*}
(ab-cd)^2 
= 
\frac{\{ (a-c)(b+d) + (a+c)(b-d) \}^2}{4}
\leq \frac{ (a-c)^2(b+d)^2 + (a+c)^2(b-d)^2 }{2} \ .
\end{align*}
Inequality $(\dagger)$ holds from $\big| \Eh{1} + \Th{1} \big| \leq \big| \Eh{1} \big| + \big| \Th{1} \big| \leq 2 C$ and 
$\big| \Eq{1} + \Tq{1} \big| \leq \big| \Eq{1} \big| + \big| \Tq{1} \big| \leq 2 C$ where the upper bound constant $C$ can be chosen from Assumption \AssumptionStrongOv.

\item[(c)] \HL{Numer-Q3} 
Based on analogous algebra, we find
\begin{align*}
&
\text{\HL{Numer-Q3}}
\\
&
=
\EXPk \big[ \big[ (1-A)(1-D) \big\{ \EqOzero \EhOone - \TqOzero \ThOone \big\} \big]^2 \big]
\\
&
\stackrel{(*)}{\leq}
\frac{1}{2}
\EXPk
\Bigg[ 
    \begin{array}{l}
        \big\{ \EqOzero - \TqOzero \big\}^2 \big\{ \EhOone + \ThOone \big\}^2
        \\
        +
        \big\{ \EqOzero + \TqOzero \big\}^2 \big\{ \EhOone - \ThOone \big\}^2
    \end{array}
\Bigg]
\\
&
\stackrel{(\dagger)}{\leq }
2 C_{h1}^2
\EXPk
\big[ \big\{ \EqOzero - \TqOzero \big\}^2 \big]
+
2 C_{q0}^2
\EXPk
\big[ \big\{ \EhOone - \ThOone \big\}^2 \big]
\\
&
\precsim
\big\| \Eq{0} - \Tq{0} \big\|_{P,2}^2 + \big\| \Eh{1} - \Th{1} \big\|_{P,2}^2 \ .
\end{align*}
Inequality $(*)$ holds from $(1-A)(1-D) \leq 1$ and 
\begin{align*}
(ab-cd)^2 
= 
\frac{\{ (a-c)(b+d) + (a+c)(b-d) \}^2}{4}
\leq \frac{ (a-c)^2(b+d)^2 + (a+c)^2(b-d)^2 }{2} \ .
\end{align*}
Inequality $(\dagger)$ holds from $\big| \Eh{1} + \Th{1} \big| \leq \big| \Eh{1} \big| + \big| \Th{1} \big| \leq 2 C$ and 
$\big| \Eq{0} + \Tq{0} \big| \leq \big| \Eq{0} \big| + \big| \Tq{0} \big| \leq 2 C$ where the upper bound constant $C$ can be chosen from Assumption \AssumptionStrongOv.

\item[(d)] \HL{Numer-Q4} 
Based on analogous algebra, we find
\begin{align*}
&
\text{\HL{Numer-Q4}}
\\
&
=
\EXPk \big[ \big[ (1-A) \big\{ \EqOzero \EhOzero - \TqOzero \ThOzero \big\} \big]^2 \big]
\\
&
\stackrel{(*)}{\leq}
\frac{1}{2}
\EXPk
\Bigg[ 
    \begin{array}{l}
        \big\{ \EqOzero - \TqOzero \big\}^2 \big\{ \EhOzero + \ThOzero \big\}^2
        \\
        +
        \big\{ \EqOzero + \TqOzero \big\}^2 \big\{ \EhOzero - \ThOzero \big\}^2
    \end{array}
\Bigg]
\\
&
\stackrel{(\dagger)}{\leq }
2 C_{h0}^2
\EXPk
\big[ \big\{ \EqOzero - \TqOzero \big\}^2 \big]
+
2 C_{q0}^2
\EXPk
\big[ \big\{ \EhOzero - \ThOzero \big\}^2 \big]
\\
&
\precsim
\big\| \Eq{0} - \Tq{0} \big\|_{P,2}^2 + \big\| \Eh{0} - \Th{0} \big\|_{P,2}^2 \ .
\end{align*}
Inequality $(*)$ holds from $(1-A) \leq 1$ and 
\begin{align*}
(ab-cd)^2 
= 
\frac{\{ (a-c)(b+d) + (a+c)(b-d) \}^2}{4}
\leq \frac{ (a-c)^2(b+d)^2 + (a+c)^2(b-d)^2 }{2} \ .
\end{align*}
Inequality $(\dagger)$ holds from $\big| \Eh{0} + \Th{0} \big| \leq \big| \Eh{0} \big| + \big| \Th{0} \big| \leq 2 C$ and 
$\big| \Eq{0} + \Tq{0} \big| \leq \big| \Eq{0} \big| + \big| \Tq{0} \big| \leq 2 C$ where the upper bound constant $C$ can be chosen from Assumption \AssumptionStrongOv.

\item[(e)] \HL{Numer-Q5}
\begin{align*}
\text{\HL{Numer-Q5}}
=
\EXPk \big[ \big\{ \EhOzero - \ThOzero \big\}^2 \big]
=
\big\| \Eh{0} - \Th{0} \big\|_{P,2}^2 \ .
\end{align*}

\end{itemize}

\subsubsection*{Numerator Part Related to $a_Y=0$}

The proof is omitted because it is similar to the \textbf{Denominator Part} below except that we use $(1-D) Y$ and $h_0(\bW,\bX_{0} \con a_Y=0)$ instead of $(1-D)$ and $h_2(\bW,\bX_{0})$, respectively. 

\subsubsection*{Denominator Part}
Recall that the centered  influence function of the denominator part is
 \begin{align}
    & \cInfFt_{\DENOM}^* (\bO) 
    =
    (1-A) \TqOzero  \big\{ (1-D) -\ThOtwo \big\} + \ThOtwo - \psi_{\DENOM}^* \ .
    \nonumber
 \end{align}
 
Let $\widehat{\psi}_{\DENOM} = K^{-1} \sum_{k=1}^{K} \widehat{\psi}_{\DENOM}\SSS$ where
 \begin{align*}
            \widehat{\psi}_{\DENOM}\SSS
    &
    =
    \AVERk
    \left[
    \begin{array}{l}
    (1-A) \EqOzero \big\{ (1-D) - \EhOtwo \big\}
    +
    \EhOtwo
    \end{array}
    \right]
    \ .
 \end{align*}
 
 The empirical process of $\widehat{\psi}_{\DENOM}\SSS$ is: 
 \begin{align}
    &
    \sqrt{N_k} \big\{ \widehat{\psi}_{\DENOM}\SSS - {\psi}_{\DENOM}^* \big\}
    \nonumber
    \\
    & =
    \sqrt{N_k}
    \Bigg[
        \begin{array}{l}
        \AVERk
            \big[
    (1-A) \Eq{0} \big\{ (1-D) - \Eh{2} \big\}
    +
    \Eh{2}
    \big]
    \nonumber
    \\
    -
    \AVERk
            \big[
    (1-A) \Tq{0} \big\{ (1-D) - \Th{2} \big\}
    +
    \Th{2}
    \big]
        \end{array}
    \Bigg]
    \nonumber
    \\
    &
    \hspace*{1cm}
    + 
    \sqrt{N_k}
    \AVERk
            \big[
    (1-A) \Tq{0} \big\{ (1-D) - \Th{2} \big\}
    +
    \Th{2}
    - \psi_{\DENOM}^*
    \big]
    \nonumber
    \\
    & =
    \sqrt{N_k}
    \left[
        \begin{array}{l}
        \AVERk
        \big[
    (1-A) \Eq{0} \big\{ (1-D) - \Eh{2} \big\}
    +
    \Eh{2}
    \big]
    \nonumber
    \\
    -
    \EXPk \big[
    (1-A) \Eq{0} \big\{ (1-D) - \Eh{2} \big\}
    +
    \Eh{2}
    \big]
    \nonumber
    \\
    -
    \AVERk
    \big[
    (1-A) \Tq{0} \big\{ (1-D) - \Th{2} \big\}
    +
    \Th{2}
    \big]
    \nonumber
    \\
    +
    \EXPk \big[
    (1-A) \Tq{0} \big\{ (1-D) - \Th{2} \big\}
    +
    \Th{2}
    \big]
        \end{array}
    \right]
    \nonumber
    \\
    & \hspace*{1cm}
    +
    \sqrt{N_k}
    \left[
        \begin{array}{l}
    \EXPk \big[
    (1-A) \Eq{0} \big\{ (1-D) - \Eh{2} \big\}
    +
    \Eh{2}
    \big]
    \nonumber
    \\
    -
    \EXPk \big[
    (1-A) \Tq{0} \big\{ (1-D) - \Th{2} \big\}
    +
    \Th{2}
    \big]
        \end{array}
    \right]
    \nonumber
    \\
    &
    \hspace*{1cm}
    + 
    \sqrt{N_k}
    \AVERk
            \big[
    (1-A) \Tq{0} \big\{ (1-D) - \Th{2} \big\}
    +
    \Th{2}
    - \psi_{\DENOM}^*
    \big]
    \nonumber
    \\
    &
    =
    \EMPk
    \left[
        \begin{array}{l}
        \big[ 
        (1-A) \Eq{0} \big\{ (1-D) - \Eh{2} \big\}
    +
    \Eh{2}
    \big]
    \\
    -
    \big[
    (1-A) \Tq{0} \big\{ (1-D) - \Th{2} \big\}
    +
    \Th{2}
    \big]
        \end{array}
    \right]
    \label{eq-Bias-Denom-T1}
    \\
    & \hspace*{1cm}
    +
    \sqrt{N_k}
    \left[
        \begin{array}{l}
    \EXPk \big[
    (1-A) \Eq{0} \big\{ (1-D) - \Eh{2} \big\}
    +
    \Eh{2}
    \big]
    \\
    -
    \EXPk \big[
    (1-A) \Tq{0} \big\{ (1-D) - \Th{2} \big\}
    +
    \Th{2}
    \big]
        \end{array}
    \right]
    \label{eq-Bias-Denom-T2}
    \\
    &
    \hspace*{1cm}
    + 
    \EMPk
            \big[
    (1-A) \Tq{0} \big\{ (1-D) - \Th{2} \big\}
    +
    \Th{2}
    \big] \ .
    \label{eq-Bias-Denom-T3}
 \end{align}

The last term \eqref{eq-Bias-Denom-T3} is equal to 
\begin{align*}
    & \EMPk
    \big[
        (1-A) \Tq{0} \big\{ (1-D) - \Th{2} \big\}
    +
    \Th{2}
    \big]
    =
    \sqrt{\frac{K}{N}}
    \sum_{i \in \mathcal{I}_k} \cInfFt_{\DENOM}^* (\bO_i) \ .
\end{align*}
If \eqref{eq-Bias-Denom-T1} and \eqref{eq-Bias-Denom-T2} are $o_P(1)$, then we obtain
 \begin{align*}
    \sqrt{N}\Big( \widehat{\psi}_{\DENOM} - \psi_{\DENOM}^* \Big)
    & =
    \frac{ \sqrt{N_k}\sum_{k=1}^{K} \big\{ \widehat{\psi}_{\DENOM}\SSS - \psi_{\DENOM}^* \big\} }{\sqrt{K}}
    =
    \frac{\sum_{i = 1}^{N} \cInfFt_{\DENOM}^* (\bO_i)}{\sqrt{N}}  + o_P(1)  \ . 
 \end{align*}

Term \eqref{eq-Bias-Denom-T2} is decomposed as
\begin{align*}
& 
\eqref{eq-Bias-Denom-T2}
\\
& =
\sqrt{N_k}
    \left[
        \begin{array}{l}
    \EXPk \big[
    (1-A) \Eq{0} \big\{ (1-D) - \Eh{2} \big\}
            +
            \Eh{2}
    \big]
    \\
    -
    \EXPk \big[
    (1-A) \Tq{0} \big\{ (1-D) - \Th{2} \big\}
            +
            \Th{2}
    \big]
        \end{array}
    \right]
    \\
    &
    =
    \sqrt{N_k}
    \EXPk 
    \left[
        \begin{array}{l}
            \big[ (1-A)(1-D) \big\{ \EqOzero - \TqOzero \big\} \big]
            \\
            -
            \big[ (1-A) \big\{ \EqOzero \EhOtwo - \TqOzero \ThOtwo \big\} \big]
            \\
            +
            \big[ \EhOtwo - \ThOtwo \big]
        \end{array}
    \right]
    \quad 
    \begin{array}{l}
        \text{\HT{Denom-T1}}
        \\
        \text{\HT{Denom-T2}}
        \\
        \text{\HT{Denom-T3}}
    \end{array}
    \\
    &
    \stackrel{(*)}{=}
    \sqrt{N_k}
    \EXPk \big[ (1-A) \big\{ \EqOzero - \TqOzero \big\} \big\{ \ThOtwo - \EhOtwo \big\} \big]
    \\
    &
    \leq
    \sqrt{N_k}  \big\| \EqOzero - \TqOzero \big\|_{P,2}  \big\| \EhOtwo - \ThOtwo \big\|_{P,2}
    \\
    &
    =
    o_P(1)
    \ .
\end{align*}
Identity $(*)$ is established below in (a)-(b).  The inequality is from the Cauchy-Schwarz inequality, and the convergence rate holds from Assumption \AssumptionMixBiasDenom. 
\begin{itemize}[leftmargin=0cm,itemsep=0cm]
\item[(a)] \HL{Denom-T1}	
\begin{align*}
\text{\HL{Denom-T1}}
&
=
\EXPk \big[ (1-A)(1-D)  \big\{ \EqOzero  - \TqOzero \big\} \big]
\\
&
\stackrel{(*)}{=}
\EXPk \big[ (1-A)  \big\{ \EqOzero  - \TqOzero \big\} \ThOtwo \big]
\ .
\end{align*}
Equality $(*)$ holds from \eqref{eq-bridgeft4 obs}.

\item[(b)] \HL{Denom-T2}
\begin{align*}
\text{\HL{Denom-T2}}
&=
\EXPk \big[ (1-A) \big\{ \TqOzero \ThOtwo - \EqOzero \EhOtwo \big\} \big]
\\
&=
\EXPk \big[ (1-A) \EqOzero \big\{ \ThOtwo - \EhOtwo \big\} \big]
\\
& \hspace*{2cm}
-
\EXPk \big[ (1-A)  \big\{ \EqOzero  - \TqOzero \big\} \ThOtwo \big]
\\
&=
\EXPk \big[ (1-A) \big\{ \EqOzero - \TqOzero \big\} \big\{ \ThOtwo - \EhOtwo \big\} \big]
\\
& \hspace*{2cm}
+
\EXPk \big[ (1-A) \TqOzero \big\{ \ThOtwo - \EhOtwo \big\} \big]
-
\text{\HL{Denom-T1}}
\\
& \stackrel{(*)}{=}
\EXPk \big[ (1-A) \big\{ \EqOzero - \TqOzero \big\} \big\{ \ThOtwo - \EhOtwo \big\} \big]
\\
& \hspace*{2cm}
+
\EXPk \big\{ \ThOtwo - \EhOtwo \big\} 
- \text{\HL{Denom-T1}}
\\
& \stackrel{(*)}{=}
\EXPk \big[ (1-A) \big\{ \EqOzero - \TqOzero \big\} \big\{ \ThOtwo - \EhOtwo \big\} \big]
\\
& \hspace*{2cm}
- \text{\HL{Denom-T1}}- \text{\HL{Denom-T3}}
\ .
\end{align*}
Equality $(*)$ holds from \eqref{eq-bridgeft5 obs}.

\end{itemize}

Next, we establish that Term \eqref{eq-Bias-Denom-T1} is $o_P(1)$.  Suppose $g(\bO)$ is a mean-zero function. Then, $\EXPk \big\{ \EMPk(g) \big\} = 0$. Therefore, it suffices to show that $\VARk \big\{ \EMPk(g) \big\} = o_P(1)$, i.e., 
\begin{align*}
\VARk \big\{ \EMPk(g) \big\}
=
\VARk \bigg\{
\frac{1}{\sqrt{N_k}} \sum_{i \in \mathcal{I}_k} g(\bO_i)
\bigg\}
=
\VARk \big\{ g(\bO) \big\}
=
\EXPk \big\{ g(\bO)^2 \big\}
=
o_P(1) \ .
\end{align*}
The variance of \eqref{eq-Bias-Denom-T1} is
\begin{align*}
& 
\VARk \big\{
\eqref{eq-Bias-Denom-T1}
\big\}
\\
& =
\VARk \left[
\EMPk
    \left[
        \begin{array}{l}
        \big[ 
        (1-A) \Eq{0} \big\{ (1-D) - \Eh{2} \big\}
    +
    \Eh{2}
    \big]
    \\
    -
    \big[
    (1-A) \Tq{0} \big\{ (1-D) - \Th{2} \big\}
    +
    \Th{2}
    \big]
        \end{array}
    \right]
    \right]
    \\
    &
    =
    \EXPk 
    \left[
    \left[
        \begin{array}{l}
            \big[ (1-A)(1-D) \big\{ \EqOzero - \TqOzero \big\} \big]
            \\
            -
            \big[ (1-A) \big\{ \EqOzero \EhOtwo - \TqOzero \ThOtwo \big\} \big]
            \\
            +
            \big[ \EhOtwo - \ThOtwo \big]
        \end{array}
    \right]^2
    \right]
    \\
    &
    \stackrel{(*)}{\precsim }
    \EXPk 
    \left[
        \begin{array}{l}
            \big[ (1-A)(1-D) \big\{ \EqOzero - \TqOzero \big\} \big]^2
            \\
            +
            \big[ (1-A) \big\{ \EqOzero \EhOtwo - \TqOzero \ThOtwo \big\} \big]^2
            \\
            +
            \big[ \EhOtwo - \ThOtwo \big]^2
        \end{array}
    \right]
    \quad 
    \begin{array}{l}
        \text{\HT{Denom-Q1}}
        \\
        \text{\HT{Denom-Q2}}
        \\
        \text{\HT{Denom-Q3}}
    \end{array}
    \\
    &
    \stackrel{(\dagger)}{\precsim}
    \big\| \Eq{0}  - \Tq{0} \big\|_{P,2}^2 
    +
    \big\| \Eh{2}  - \Th{2} \big\|_{P,2}^2 
    \\
    &
    =
    o_P(1) \ . 
\end{align*}
Inequality $(*)$ holds from the Cauchy-Schwarz inequality $(\sum_{j=1}^{M} a_j)^2 \leq M (\sum_{j=1}^{M} a_j^2)$. Inequality $(\dagger)$ is established below in (a)-(c). The last convergence rate is obtained  from Assumption \AssumptionConsistency. 
\begin{itemize}
\item[(a)] \HL{Denom-Q1}
Based on analogous algebra, we find
\begin{align*}
\text{\HL{Denom-Q1}}
&
=
\EXPk \big[ \big[ (1-A)(1-D) \big\{ \EqOzero - \TqOzero \big\} \big]^2 \big]
\\
&
\leq \EXPk \big[ \big\{ \EqOzero - \TqOzero \big\}^2 \big]
\\
& =
\big\| \Eq{0} - \Tq{0} \big\|_{P,2}^2 \ .
\end{align*}

\item[(b)] \HL{Denom-Q2}
\begin{align*}
&
\text{\HL{Denom-Q2}}
\\
&
=
\EXPk \big[ \big[ (1-A) \big\{ \EqOzero \EhOtwo - \TqOzero \ThOtwo \big\} \big]^2 \big]
\\
&
\stackrel{(*)}{\leq}
\frac{1}{2}
\EXPk
\Bigg[ 
    \begin{array}{l}
        \big\{ \EqOzero - \TqOzero \big\}^2 \big\{ \EhOtwo + \ThOtwo \big\}^2
        \\
        +
        \big\{ \EqOzero + \TqOzero \big\}^2 \big\{ \EhOtwo - \ThOtwo \big\}^2
    \end{array}
\Bigg]
\\
&
\stackrel{(\dagger)}{\leq }
2 C_{h2}^2
\EXPk
\big[ \big\{ \EqOzero - \TqOzero \big\}^2 \big]
+
2 C_{q0}^2
\EXPk
\big[ \big\{ \EhOtwo - \ThOtwo \big\}^2 \big]
\\
&
\precsim
\big\| \Eq{0} - \Tq{0} \big\|_{P,2}^2 + \big\| \Eh{2} - \Th{2} \big\|_{P,2}^2 \ .
\end{align*}
Inequality $(*)$ holds from $(1-A) \leq 1$ and 
\begin{align*}
(ab-cd)^2 
= 
\frac{\{ (a-c)(b+d) + (a+c)(b-d) \}^2}{4}
\leq \frac{ (a-c)^2(b+d)^2 + (a+c)^2(b-d)^2 }{2} \ .
\end{align*}
Inequality $(\dagger)$ holds from $\big| \Eh{2} + \Th{2} \big| \leq \big| \Eh{2} \big| + \big| \Th{2} \big| \leq 2 C$ and 
$\big| \Eq{0} + \Tq{0} \big| \leq \big| \Eq{0} \big| + \big| \Tq{0} \big| \leq 2 C$ where the upper bound constant $C$ can be chosen from Assumption \AssumptionStrongOv. 

\item[(c)] \HL{Denom-Q3}
\begin{align*}
\text{\HL{Denom-Q3}}
=
\EXPk \big[ \big\{ \EhOtwo - \ThOtwo \big\}^2 \big]
=
\big\| \Eh{2} - \Th{2} \big\|_{P,2}^2 \ .
\end{align*}

\end{itemize}

\subsubsection{Proof of the Consistency of $\widehat{\Sigma}$ in \eqref{eq-ConsistencyVar}}

For notational brevity, we define vectors of uncentered influence functions:
\begin{align*}
&
\bIF^* (\bO)
=
\begin{pmatrix}
        {\InfFt}_{\NUMER}^* (\bO \con a_Y=1)
        \\
        {\InfFt}_{\NUMER}^* (\bO \con a_Y=0)
        \\
        {\InfFt}_{\DENOM}^* (\bO)
    \end{pmatrix} 
    ,
    &&
\widehat{\bIF}\LSS(\bO)
=
\begin{pmatrix}
        \widehat{\InfFt}_{\NUMER}\LSS(\bO \con a_Y=1)
        \\
        \widehat{\InfFt}_{\NUMER}\LSS(\bO \con a_Y=0)
        \\
        \widehat{\InfFt}_{\DENOM}\LSS(\bO)
    \end{pmatrix}  
\end{align*}
vectors of parameters and their estimators:
\begin{align*}
&
\bm{\psi}^* 
= \begin{pmatrix}
    \psi_{\NUMER}^*(a_Y=1) \\ \psi_{\NUMER}^*(a_Y=0) \\ \psi_{\DENOM}^*(a_D)
\end{pmatrix}
, &&
\widehat{\bm{\psi}}
= \begin{pmatrix}
    \widehat{\psi}_{\NUMER} (a_Y=1) \\ \widehat{\psi}_{\NUMER} (a_Y=0) \\ \widehat{\psi}_{\DENOM} (a_D)
\end{pmatrix} \
 ,
\end{align*}
and vectors of centered influence functions:
\begin{align*}
&
\cbIF^* (\bO)
=
\bIF^* (\bO) - \bm{\psi}^*
\ , 
&&
\widehat{\cbIF}\LSS(\bO)
=
\widehat{\bIF}\LSS(\bO)-\widehat{\bm{\psi}} \ .
\end{align*}

We first present four useful results:
\begin{itemize}[leftmargin=0cm]

\item[] \HT{Result 1} $\AVERk
\big\{
    \widehat{\bIF}\LSS (\bO)
\big\}
-
\AVERk
\big\{
    {\bIF}^* (\bO)
\big\} = o_P(1)$.

\begin{proof}
        The result can be achieved by straightforward algebra:
\begin{align*}
&
\Big\|
\AVERk
\big[
    \big\{ \widehat{\bIF}\LSS (\bO)
    -
    {\bIF}^* (\bO)
\big\} \big]
\Big\|_2^2
\\
&
\stackrel{*}{=}
\Big\|
\EXPk
\big[
    \big\{ \widehat{\bIF}\LSS (\bO)
    -
    {\bIF}^* (\bO)
\big\} \big]
+
o_P(1)
\Big\|_2^2
\\
&
\precsim
\EXPk
\big[
    \big\| \widehat{\bIF}\LSS (\bO)
    -
    {\bIF}^* (\bO)
\big\|_2^2 \big]
+
o_P(1)
\\
&
=
\underbrace{
\EXPk
\big[
    \big\{ \widehat{\InfFt}_{\NUMER} \LSS (\bO \con a_Y=1)
     - \InfFt_{\NUMER}^* (\bO \con a_Y=1) \big\}^2
\big]
}_{= \VARk\{ \eqref{eq-Bias-Numer-T1} \}}
\\
&
\hspace*{1cm}
+
\underbrace{
\EXPk
\big[
    \big\{ \widehat{\InfFt}_{\NUMER} \LSS (\bO \con a_Y=0)
     - \InfFt_{\NUMER}^* (\bO \con a_Y=0) \big\}^2
\big]
}_{= \VARk\{ \eqref{eq-Bias-Denom-T1} \} \text{ with $(1-D)Y$ and $\widehat{h}_0(\cdot \con a_Y=0)$} }
\\
&
\hspace*{1cm}
+
\underbrace{
\EXPk
\big[
    \big\{ \widehat{\InfFt}_{\DENOM} \LSS (\bO) - \InfFt_{\DENOM}^* (\bO) \big\}^2
\big]
}_{= \VARk\{ \eqref{eq-Bias-Denom-T1} \}}
+
o_P(1)
\\
& = o_P(1) \ .
\end{align*}
Equality $(*)$ holds from the law of large numbers. The last line holds from the fact that \eqref{eq-Bias-Numer-T1} and \eqref{eq-Bias-Denom-T1} are $o_P(1)$.
\end{proof}

\item[] \HT{Result 2} $\AVERk \big\{ \widehat{\bIF}\LSS (\bO) \big\} \cdot \widehat{\bm{\psi}} \T
-
\AVERk \big\{ 	{\bIF}^* (\bO) \big\} \cdot \bm{\psi}\sT = o_P(1)$.

\begin{proof} 
From straightforward algebra using $ab - cd = \{(a-c)(b+d) + (a+c)(b-d) \}/2$, we have	
\begin{align*}
&
\AVERk
\big\{
    \widehat{\bIF}\LSS(\bO)
\big\}
\cdot \widehat{\bm{\psi}} \T
-
\AVERk
\big\{
    {\bIF}(\bO)
\big\}
\cdot \bm{\psi}\sT
=
\frac{1}{2}
\left[
    \begin{array}{l}
\AVERk
\big[
    \big\{ \widehat{\bIF}\LSS(\bO)
    -
    {\bIF}^* (\bO)
\big\} \big] \cdot 
\big\{ \widehat{\bm{\psi}} + \bm{\psi}^* \big\} \T
\\
+
\AVERk
\big[
    \big\{ \widehat{\bIF}\LSS(\bO)
    +
    {\bIF}^*(\bO)
\big\} \big] \cdot 
\big\{ \widehat{\bm{\psi}} - \bm{\psi}^* \big\} \T
    \end{array}
\right] \ .
\end{align*}

From \HL{Result 1}, we have $\AVERk \big[ \big\{ \widehat{\bIF}\LSS(\bO) - {\bIF}^*(\bO) \big\} \big]=o_P(1)$. Additionally, $\widehat{\bm{\psi}} - \bm{\psi}^* =o_P(1)$ from the consistency of $\widehat{\bm{\psi}}$. Since $\widehat{\bm{\psi}} + \bm{\psi}^* =O_P(1)$ and $\AVERk
\big[
    \big\{ \widehat{\bIF}\LSS(\bO)
    +
    {\bIF}^*(\bO)
\big\} \big]=O_P(1)$, we have $\AVERk
\big\{
    \widehat{\bIF}\LSS(\bO) 
\big\}
\cdot \widehat{\bm{\psi}} \T
-
\AVERk
\big\{
    {\bIF}^*(\bO)
\big\}
\cdot \bm{\psi}\sT = o_P(1)$ from the Slutsky's theorem.

\end{proof}

\item[] \HT{Result 3} $\widehat{\bm{\psi}} \cdot \widehat{\bm{\psi}} \T - \bm{\psi}^* \cdot \bm{\psi}\sT = o_P(1)$.
\begin{proof}
The result is trivial from $ \widehat{\bm{\psi}} - \bm{\psi}^* =o_P(1)$ and $\widehat{\bm{\psi}} + \bm{\psi}^*=O_P(1)$ along with the Slutsky's theorem.
    \begin{align*}
        & \widehat{\bm{\psi}} \cdot \widehat{\bm{\psi}} \T - \bm{\psi}^* \cdot \bm{\psi}\sT
        =
        \frac{1}{2}
        \left\{
            \begin{array}{l}
                \big( \widehat{\bm{\psi}} - \bm{\psi}^* \big)
            \big( \widehat{\bm{\psi}} + \bm{\psi}^* \big) \T
            \\
            +
            \big( \widehat{\bm{\psi}} + \bm{\psi}^* \big)
            \big( \widehat{\bm{\psi}} - \bm{\psi}^* \big) \T
            \end{array}
        \right\}
        =
        o_P(1) \ .
    \end{align*}
\end{proof}

\item[] \HT{Result 4} $\AVERk
\big\{
    \widehat{\bIF}\LSS (\bO) \cdot \widehat{\bIF}\tLSS (\bO)
\big\}
-
\AVERk
\big\{
    {\bIF}^*(\bO) \cdot {\bIF}\sT(\bO)
\big\} = o_P(1)$.
\begin{proof}
The result is trivial from \HL{Result 1}: $\AVERk \big[ \big\{ \widehat{\bIF}\LSS(\bO) - {\bIF}^*(\bO) \big\} \big]=o_P(1)$ and $\AVERk \big[ \big\{ \widehat{\bIF}\LSS(\bO) + {\bIF}^*(\bO) \big\} \big]=O_P(1)$ along with the Slutsky's theorem.
    \begin{align*}
        & \AVERk \big\{ \widehat{\bIF}\LSS(\bO) \cdot \widehat{\bIF}\tLSS(\bO) \big\}
-
\AVERk \big\{ {\bIF}^*(\bO) \cdot {\bIF}\sT(\bO)	\big\}
\\
&
        =
        \frac{1}{2}
        \left\{
            \begin{array}{l}
                \AVERk \big[ \big\{ \widehat{\bIF}\LSS(\bO) - {\bIF}^*(\bO) \big\} \big]
                \AVERk \big[ \big\{ \widehat{\bIF}\LSS(\bO) + {\bIF}^*(\bO) \big\} \big] \T
            \\
            +
                \AVERk \big[ \big\{ \widehat{\bIF}\LSS(\bO) + {\bIF}^*(\bO) \big\} \big]
                \AVERk \big[ \big\{ \widehat{\bIF}\LSS(\bO) - {\bIF}^*(\bO) \big\} \big] \T
            \end{array}
        \right\}
        =
        o_P(1) \ .
    \end{align*}
\end{proof}

\end{itemize}

Now we show the consistency of $\widehat{\Sigma}$, which is defined as follows:
\begin{align*}
\widehat{\Sigma}
=
\frac{1}{K} \sum_{k=1}^{K} \widehat{\Sigma}\SSS
\ , \
\widehat{\Sigma}\SSS
=
\AVERk
\left\{
    \begin{pmatrix}
        \widehat{\cInfFt}_{\NUMER}\LSS(\bO \con a_Y=1)
        \\
        \widehat{\cInfFt}_{\NUMER}\LSS(\bO \con a_Y=0)
        \\
        \widehat{\cInfFt}_{\DENOM}\LSS(\bO)
    \end{pmatrix}^{\otimes 2}
\right\}
\end{align*}
Therefore, for the consistency of $\widehat{\Sigma}$, it suffices to show that $\widehat{\Sigma}\SSS$ is consistent for $\Sigma^*$. We have
\begin{align*}
&
\widehat{\Sigma}\SSS
-
\Sigma^*
\\
& 
=
\AVERk
\big\{
    \widehat{\cbIF}\LSS(\bO) \cdot \widehat{\cbIF}\tLSS(\bO)
\big\}
-
\EXPk 
\big\{
    \cbIF^*(\bO) \cdot \cbIF\sT(\bO)
\big\}
\\
&
=
\AVERk
\big\{
    \widehat{\cbIF}\LSS(\bO) \cdot \widehat{\cbIF}\tLSS(\bO)
\big\}
-
\AVERk
\big\{
    \cbIF^* (\bO) \cdot \cbIF\sT(\bO)
\big\}
\\
& \hspace*{1cm}
+
\underbrace{
\AVERk
\big\{
    \cbIF^*(\bO) \cdot \cbIF\sT(\bO)
\big\}
-
\EXPk 
\big\{
    \cbIF^*(\bO) \cdot \cbIF\sT(\bO)
\big\}
}_{=o_P(1)}
\\
&
\stackrel{(*)}{=}
\AVERk
\big\{
    \widehat{\bIF}\LSS(\bO) \cdot \widehat{\bIF}\tLSS(\bO)
\big\}
-
\AVERk
\big\{
    {\bIF}^*(\bO) \cdot {\bIF}\sT(\bO)
\big\}
\\
& 
\hspace*{1cm}
- \big[ \AVERk \big\{ \widehat{\bIF}\LSS(\bO) \big\} \cdot \widehat{\bm{\psi}} \T
-
\AVERk \big\{ 	{\bIF}^*(\bO) \big\} \cdot \bm{\psi}\sT \big]
\\
& 
\hspace*{1cm}
- \big[ \AVERk \big\{ \widehat{\bIF}\LSS(\bO) \big\} \cdot \widehat{\bm{\psi}} \T
-
\AVERk \big\{ 	{\bIF}^*(\bO) \big\} \cdot \bm{\psi}\sT \big]\T
\\
& 
\hspace*{1cm}
+ \big( \widehat{\bm{\psi}} \cdot \widehat{\bm{\psi}} \T
-
\bm{\psi} \cdot \bm{\psi}\sT \big)
+o_P(1) \\
&	\stackrel{(\dagger)}{=}
o_P(1) \ .
\end{align*}
Equality $(*)$ holds from the law of large numbers and the relationships between centered and uncentered influence functions. Equality $(\dagger)$ holds from \HL{Result 1}-\HL{Result 4} above. This concludes the proof.

\subsection{Proof of Theorem \ref{thm:convergence:1} and Theorem \ref{lemma:convergence:1}} \label{sec:proof:lemma:convergence:1}

The proof of the lemma can be established from using intermediate results of \citet{Carrasco2007, Darolles2011,  Florens2011, PMMR2021}.

To avoid unnecessary notational clutter, we omit the subscripts $_{h1}$ in $\beta$ and $\lambda$. Let $h_1(\lambda)$ be the optimal PMMR solution in the population given a regularization parameter $\lambda \in (0,\infty)$:
\begin{align*}
	& h_1(\lambda) = \argmin_{h \in \HH_{W \obX_{\TIME}}} \Big\{ R_{h1}(h) + \lambda \big\| h \big\|_{\HH}^2 \Big\}  \ .
\end{align*}
Then, following equation (86) and (87) of \citet{PMMR2021}, we have 
\begin{align*}
	h_1(\lambda) = (\mathcal{T}^{\text{ad}}\mathcal{T} + \lambda I)^{-1} \mathcal{T}^{\text{ad}} g  \ , \quad 
	\widehat{h}_1\LSS = (\widehat{\mathcal{T}}^{\text{ad}}\widehat{\mathcal{T}} + \lambda I)^{-1} \widehat{\mathcal{T}}^{\text{ad}} \widehat{g} \ .
\end{align*}
Here, $g  = \EXP \big\{ \ind(A=a_Y,D_{\TIME+1}=0) Y \varphi(Z,\obX_{\TIME}) \big\}$ where $\varphi(z,x) = \mathcal{K}((z,x), \cdot)$ is the canonical feature map of $(Z,\obX_{\TIME})$. Also, $\widehat{g}$ and $\widehat{\mathcal{T}}$ are the empirical estimates of $g$ and $\mathcal{T}$ based on the training data $\mathcal{I}_k\LSS$, i.e.:
\begin{align*}
&
\widehat{g} = \frac{1}{|\mathcal{I}_k\LSS|} \sum_{i \in \mathcal{I}_k\LSS} \big\{ \ind(A_i=a_Y,D_{\TIME+1_i}=0) Y_i \varphi(Z_i,\obX_{\TIME,i})  \big\} \ ,
\\
&
\widehat{\mathcal{T}}(h) =  \frac{1}{|\mathcal{I}_k\LSS|} \sum_{i \in \mathcal{I}_k\LSS} \big\{ \ind(A_i=a_Y,D_{\TIME+1_i}=0) h(W_i,\obX_{\TIME,i}) \varphi(Z_i,\obX_{\TIME,i})  \big\} \ .
\end{align*} 
Likewise, the adjoint operator of $\widehat{\mathcal{T}}$ is denoted by $\widehat{\mathcal{T}}^{\text{ad}}$.

Following Lemma A.1 of \citet{Florens2011}, the proof of Theorem 2.1 of \citet{Darolles2011}, and Lemma 12 of \citet{PMMR2021}, we establish
\begin{align}
	&
    \text{Page 1563 of \citet{Darolles2011} and Lemma A.1 of \citet{Florens2011}}
    \nonumber
    \\
    &
    \Rightarrow \quad 
    \left\{
    \begin{array}{l}         
	\big\| (\widehat{\mathcal{T}}^{\text{ad}} \widehat{\mathcal{T}} + \lambda I)^{-1} \widehat{\mathcal{T}}^{\text{ad}} \big\|
	=
	\big\| \widehat{\mathcal{T}}(\widehat{\mathcal{T}}^{\text{ad}} \widehat{\mathcal{T}} + \lambda I)^{-1}  \big\| = O_P(\lambda^{-1/2})   \\         
	\big\| ({\mathcal{T}}^{\text{ad}} {\mathcal{T}} + \lambda I)^{-1} {\mathcal{T}}^{\text{ad}} \big\|
	=
	\big\| {\mathcal{T}}({\mathcal{T}}^{\text{ad}} {\mathcal{T}} + \lambda I)^{-1}  \big\| = O_P(\lambda^{-1/2})  
    \end{array}
    \right.\ ,
	\label{eq-A2-Fl}
	\\
    &
    \text{Page 1563 of \citet{Darolles2011} and Lemma A.1 of \citet{Florens2011}}
    \nonumber
    \\
	&
    \Rightarrow \quad 
	\big\| (\widehat{\mathcal{T}}^{\text{ad}} \widehat{\mathcal{T}} + \lambda I)^{-1} \big\|
	= O_P(\lambda^{-1}) \ ,
    \quad
    \big\| ({\mathcal{T}}^{\text{ad}} {\mathcal{T}} + \lambda I)^{-1} \big\|
	= O_P(\lambda^{-1})
	\label{eq-A3-Fl}
	\\
	&
    \text{Lemma 12 of \citet{PMMR2021}}
    \nonumber
    \\
	&
    \Rightarrow \quad 
	\big\| \widehat{g} - g \big\| = O_P(N^{-1/2}) \ , 
	\label{eq-PMMR-g}
	\\
	&
    \text{Lemma 12 of \citet{PMMR2021}}
    \nonumber
    \\
	&
    \Rightarrow \quad  
	\big\| \widehat{\mathcal{T}} - \mathcal{T} \big\| = O_P(N^{-1/2}) \ , 
	\label{eq-PMMR-T}
	\\
	&
    \text{Lemma 12 of \citet{PMMR2021}}
    \nonumber
    \\
	&
    \Rightarrow \quad  
	\big\| \widehat{\mathcal{T}}^{\text{ad}} - \mathcal{T}^{\text{ad}} \big\| = O_P(N^{-1/2}) \ .
	\label{eq-PMMR-Tad}
    \\
    &
    \text{Lemma 14 of \citet{PMMR2021}}
    \nonumber
    \\
	&
    \Rightarrow \quad  
    \big\|  \widehat{\mathcal{T}}^{\text{ad}} \widehat{g} -  \widehat{\mathcal{T}}^{\text{ad}} \widehat{\mathcal{T}} h_1^* \big\| = O_P(N^{-1/2})
    \label{eq-PMMR-Tg}
\end{align}

Then, we can decompose the estimation bias into three parts:
\begin{align}
	h_1^* - \widehat{h}_1\LSS 
	&
	=
	h_1^* - h_1(\lambda) + h_1(\lambda) - \widehat{h}_1\LSS
	\nonumber
	\\
	&
	=
	h_1^* - h_1(\lambda)
	+
	\big( {\mathcal{T}}^{\text{ad}} {\mathcal{T}} + \lambda I \big)^{-1}
\big( {\mathcal{T}}^{\text{ad}} {\mathcal{T}} h_1^* \big)
-
\big( \widehat{\mathcal{T}}^{\text{ad}} \widehat{\mathcal{T}} + \lambda I \big)^{-1}
\widehat{\mathcal{T}}^{\text{ad}} \widehat{g} 
\nonumber
\\
&
=
\underbrace{
\big( \widehat{\mathcal{T}}^{\text{ad}} \widehat{\mathcal{T}} + \lambda I \big)^{-1}
\big( \widehat{\mathcal{T}}^{\text{ad}} \widehat{\mathcal{T}} h_1^* - \widehat{\mathcal{T}}^{\text{ad}} \widehat{g}
\big)}_{=:(\texttt{A})}
+
\underbrace{h_1^* - h_1(\lambda)}_{=:(\texttt{B})}
\nonumber
\\
&
\quad 
+
\underbrace{ 
\big( {\mathcal{T}}^{\text{ad}} {\mathcal{T}} + \lambda I \big)^{-1}
\big( {\mathcal{T}}^{\text{ad}} {\mathcal{T}} h_1^* \big)
-
\big( \widehat{\mathcal{T}}^{\text{ad}} \widehat{\mathcal{T}} + \lambda I \big)^{-1}
\big( \widehat{\mathcal{T}}^{\text{ad}} \widehat{\mathcal{T}} h_1^*\big) 
}_{=:(\texttt{C})}
	 \ .
	\label{eq-PMMR-convergence1}
\end{align}

First, following the proof of Theorem 2.2 of \citet{Florens2011}, we find $\texttt{(A)}$ is upper bounded by:
\begin{align*}
\big\| \texttt{(A)} \big\| ^2
&
\leq
\big\| (\widehat{\mathcal{T}}^{\text{ad}} \widehat{\mathcal{T}} + \lambda I)^{-1} \widehat{\mathcal{T}}^{\text{ad}} \big\|^2
\big\| \widehat{g} - \widehat{\mathcal{T}} h_1^* \big\|^2
\\
&
=
\big\| (\widehat{\mathcal{T}}^{\text{ad}} \widehat{\mathcal{T}} + \lambda I)^{-1} \widehat{\mathcal{T}}^{\text{ad}} \big\|^2
\big\| \widehat{g} - g + \mathcal{T} h_1^*  - \widehat{\mathcal{T}} h_1^* \big\|^2
\\
&
\lesssim
\big\| (\widehat{\mathcal{T}}^{\text{ad}} \widehat{\mathcal{T}} + \lambda I)^{-1} \widehat{\mathcal{T}}^{\text{ad}} \big\|^2
\big\{ \big\|\widehat{g} - g \big\|^2 + \big\| \widehat{\mathcal{T}} - \mathcal{T} \big\|^2 \big\}
\\
&
=
O_P \bigg( \frac{1}{N \lambda } \bigg) \ .
\numeq
\label{eq-PMMR-convergence-piece1}
\end{align*}
The last line holds from \eqref{eq-A2-Fl}, \eqref{eq-PMMR-g}, \eqref{eq-PMMR-T}.  We remark that a slower convergence rate of $O_P( N^{-1} \lambda^{-2} )$ is obtained in \citet[Proposition 4.1]{Carrasco2007} and \citet[equation 100]{PMMR2021}, and is also assumed in \citet[Assumption A.4]{Darolles2011}. Note that this convergence rate is obtained by
\begin{align*}
\big\| \texttt{(A)} \big\| ^2
&
\leq
\big\| (\widehat{\mathcal{T}}^{\text{ad}} \widehat{\mathcal{T}} + \lambda I)^{-1} \big\|^2
\big\|  \widehat{\mathcal{T}}^{\text{ad}} \widehat{g} -  \widehat{\mathcal{T}}^{\text{ad}} \widehat{\mathcal{T}} h_1^* \big\|^2
=
O_P \bigg( \frac{1}{N \lambda^2 } \bigg) \ ,
\numeq
\label{eq-PMMR-convergence-piece1-slower}
\end{align*}
where the last rate is obtained from \eqref{eq-A3-Fl} and \eqref{eq-PMMR-Tg}.

Next, from the proof of Theorem 2.1 of \citet{Florens2011}, we establish
\begin{align}
\big\| \texttt{(B)} \big\|^2
=
\big\| h_1^* - h_1(\lambda) \big\| = O ( \lambda^{\min(\beta,2)} ) \ .
\label{eq-PMMR-convergence-piece2}
\end{align}
Similar results are found in Proposition 12 of \citet{PMMR2021} and Proposition 3.12 of \citet{Carrasco2007}. However, these works assume that the solution to the integral equation \eqref{eq-bridgeft1 obs} is unique. In contrast, \citet{Florens2011} accommodates the possibility of multiple solutions, defining $h_1^*$ as the minimal norm solution among them.

Lastly, from the proof of Theorem 4.1 of \cite{Darolles2011}, we obtain  
\begin{align*}
\texttt{(C)} 
&
=
\underbrace{
\lambda \big( \widehat{\mathcal{T}}^{\text{ad}}\widehat{\mathcal{T}} + \lambda I \big)^{-1} \widehat{\mathcal{T}}^{\text{ad}} \big( \widehat{\mathcal{T}} - \mathcal{T} \big) \big( \mathcal{T}^{\text{ad}} \mathcal{T} + \lambda I \big)^{-1} h_1^*
}_{=: (\texttt{D})}
\\
&
\quad \quad 
+
\underbrace{
\lambda \big( \widehat{\mathcal{T}}^{\text{ad}}\widehat{\mathcal{T}} + \lambda I \big)^{-1}  \big( \widehat{\mathcal{T}}^{\text{ad}} - \mathcal{T}^{\text{ad}} \big) \mathcal{T} \big( \mathcal{T}^{\text{ad}} \mathcal{T} + \lambda I \big)^{-1} h_1^*
}_{=: (\texttt{E})}
 \ .
\end{align*}
Note that
\begin{align*}
&
\big\| \lambda  \big( \mathcal{T}^{\text{ad}} \mathcal{T} + \lambda I \big)^{-1} h_1^* \big\|^2
=
\big\| h_1^* - h_1(\lambda) \big\|^2
=
O(\lambda^{\min(\beta,2)} )
\ ,
\\
&
\big\| \lambda \mathcal{T}  \big( \mathcal{T}^{\text{ad}} \mathcal{T} + \lambda I \big)^{-1} h_1^* \big\|^2
=
O(\lambda^{\min(\beta+1,2)}) \ .
\end{align*}
We remark that these results can be also found in \citet[Page 1563]{Darolles2011}. Therefore, combined with \eqref{eq-A2-Fl}, \eqref{eq-A3-Fl}, \eqref{eq-PMMR-T}, \eqref{eq-PMMR-Tad}, we have 
\begin{align*}
\big\| \texttt{(D)} \big\|^2
&
\leq 
\big\| 
\big( \widehat{\mathcal{T}}^{\text{ad}}\widehat{\mathcal{T}} + \lambda I \big)^{-1} \widehat{\mathcal{T}}^{\text{ad}}
\big\|^2
\big\|  \widehat{\mathcal{T}} - \mathcal{T} \big\|^2
\big\| \lambda 
\big( \mathcal{T}^{\text{ad}} \mathcal{T} + \lambda I \big)^{-1} h_1^*
 \big\|^2 
 =
 O_P \bigg(
 	\frac{ \lambda^{\min(\beta,2)} }{N \lambda} 	
 \bigg) 
=
 O_P \bigg(
 	\frac{ \lambda^{\min(\beta-1,1)} }{N } 	
 \bigg) 
 \ ,
 \\
\big\| \texttt{(E)} \big\|^2
&
\leq 
\big\| 
\big( \widehat{\mathcal{T}}^{\text{ad}}\widehat{\mathcal{T}} + \lambda I \big)^{-1} \big\|^2
\big\| \widehat{\mathcal{T}}^{\text{ad}} - \mathcal{T}^{\text{ad}} \big\|^2
\big\| \lambda \mathcal{T} \big( \mathcal{T}^{\text{ad}} \mathcal{T} + \lambda I \big)^{-1} h_1^*
 \big\|^2
 =
 O_P \bigg(
 	\frac{ \lambda^{\min(\beta+1,2)} }{N \lambda^2} 	
 \bigg)
 =
 O_P \bigg(
 	\frac{ \lambda^{\min(\beta-1,0)} }{N} 	
 \bigg)
  \ .
\end{align*}
Therefore,
\begin{align}
\big\| \texttt{(C)} \big\|^2
\lesssim 
\big\| \texttt{(D)} \big\|^2
+
\big\| \texttt{(E)} \big\|^2
=
O_P
\bigg(
	\frac{ \lambda^{\min(\beta-1,0)} }{N} 	
\bigg) \ .
\label{eq-PMMR-convergence-piece3}
\end{align}

%
%

Combining all results in \eqref{eq-PMMR-convergence1}, \eqref{eq-PMMR-convergence-piece1}, \eqref{eq-PMMR-convergence-piece2}, \eqref{eq-PMMR-convergence-piece3}, we have
\begin{align*}
\big\| h_1^* - \widehat{h}_1\LSS \big\|^2
	 =
	 O_P
	 \bigg( 	 
	 \frac{1}{N \lambda }
	 +
	 \lambda^{\min(\beta,2)}
	 +
	 \frac{ \lambda^{\min(\beta-1,0)}}{N   }
	 \bigg) 
	  \ .
\end{align*}
Therefore, if we take $\lambda = N^{-\max( 1/(1+\beta), 1/3 )}$, the rate becomes
\begin{align*} 
	 &
	\frac{1}{N \lambda }
	 +
	 \lambda^{\min(\beta,2)}
	 +
	 \frac{ \lambda^{\min(\beta-1,0)}}{N   } 
	 =
	 \left\{
	 	\begin{array}{ll}
		N^{- \frac{2\beta}{2\beta+2} }
	 		& \text{ if $\beta \in (0,2]$}
	 		\\
	 	N^{-\frac{2}{3}}
	 		& \text{ if $\beta \in (2,\infty)$}
	 	\end{array}
	 \right. \ .
\end{align*}
From the compact support and uniformly bounded density conditions, we find
\begin{align*}
	\big\| h_1^* - \widehat{h}_1\LSS \big\|_{P,2}
	 =
	 \left\{
	 	\begin{array}{ll}
	 	O_P \big( 
		N^{- \frac{\beta}{2\beta+2} }
		\big)
	 		& \text{ if $\beta \in (0,2]$}
	 		\\
	 		O_P \big( 
	 	N^{-\frac{1}{3}}
	 	\big)
	 		& \text{ if $\beta \in (2,\infty)$}
	 	\end{array}
	 \right. \ .
\end{align*} 
This agrees with the convergence rate given by \citet[Theorem 2.2]{Florens2011}.

Of note, if we use the conservative convergence rate \eqref{eq-PMMR-convergence-piece1-slower} instead of  \eqref{eq-PMMR-convergence1}, we have
\begin{align*}
\big\| h_1^* - \widehat{h}_1\LSS \big\|^2
	 =
	 O_P
	 \bigg( 	 
	 \frac{1}{N \lambda^2}
	 +
	 \lambda^{\min(\beta,2)}
	 +
	 \frac{ \lambda^{\min(\beta-1,0)}}{N   }
	 \bigg) 
	  \ .
\end{align*}
Therefore, if we take $\lambda = N^{-\max( 1/(2+\beta), 1/4 )}$, the rate becomes
\begin{align*} 
	 &
	\frac{1}{N \lambda^2 }
	 +
	 \lambda^{\min(\beta,2)}
	 +
	 \frac{ \lambda^{\min(\beta-1,0)}}{N   } 
	 =
	 \left\{
	 	\begin{array}{ll}
		N^{- \frac{\beta}{\beta+2} }
	 		& \text{ if $\beta \in (0,2]$}
	 		\\
	 	N^{-\frac{1}{2}}
	 		& \text{ if $\beta \in (2,\infty)$}
	 	\end{array}
	 \right. \ .
\end{align*}
By leveraging the compact support and uniformly bounded density conditions again, we find
\begin{align*}
	\big\| h_1^* - \widehat{h}_1\LSS \big\|_{P,2}
	 =
	 \left\{
	 	\begin{array}{ll}
	 	O_P \big( 
		N^{- \frac{\beta}{2\beta+4} }
		\big)
	 		& \text{ if $\beta \in (0,2]$}
	 		\\
	 		O_P \big( 
	 	N^{-\frac{1}{4}}
	 	\big)
	 		& \text{ if $\beta \in (2,\infty)$}
	 	\end{array}
	 \right. \ ,
\end{align*}
which coincides with the convergence rate given by \citet{Carrasco2007, Darolles2011, PMMR2021}.

\subsection{Proof of Theorem \ref{lemma:convergence:2}} \label{sec:proof:lemma:convergence:2}

      The proof is similar to that of Lemma \ref{lemma:convergence:1}. Note $g$ and $\widehat{g}$ are redefined as 
\begin{align*}
&
g = \EXP \big\{ \ind(A=a_D , D_{\TIME+1}=0) h_1^*(W,\obX_{\TIME}) \varphi(Z,\bX_{0}) \big\}
\\
&
\widehat{g} = \frac{1}{|\mathcal{I}_k\LSS|} \sum_{i \in \mathcal{I}_k\LSS} \big\{ \ind(A_i=a_D,D_{\TIME+1_i}=0) \widehat{h}_{1}\LSS (W,\obX_{\TIME})
\varphi(Z_i,\bX_{0,i})  \big\} \ . 
\end{align*} 
We also consider the following intermediate quantity:
\begin{align*}
    \widetilde{g}
    =
    \frac{1}{|\mathcal{I}_k\LSS|} \sum_{i \in \mathcal{I}_k\LSS} \big\{ \ind(A_i=a_D,D_{\TIME+1_i}=0) h_1^*(W,\obX_{\TIME})
\varphi(Z_i,\bX_{0,i})  \big\} \ .
\end{align*}
Then, we find $ g - \widehat{g}  =    g - \widetilde{g} + \widetilde{g} - \widehat{g}$. 

Following Lemma 12 of \citet{PMMR2021}, we find $	\big\| \widetilde{g} - g \big\| = O_P(N^{-1/2})$. Next, we establish the rate of $\big\| \widehat{g} - \widetilde{g} \big\|$. From Lemma \ref{lemma:convergence:1}, we get
\begin{align*}
    \big\| \widehat{h}_1\LSS(W,\obX_{\TIME}) - h_1^*(W,\obX_{\TIME}) \big\|_{P,2} = O_P(r_{h1}) = o_P(1) \ ,
\end{align*}
which implies
\begin{align*}
    \big\| \ind  (A=a_D,D_{\TIME+1}=0) \varphi(Z,\bX_{0}) 
    \big\{ \widehat{h}_1\LSS(W,\obX_{\TIME}) - h_1^* (W,\obX_{\TIME}) \big\} \big\|_{P,2} = O_P(r_{h1}) = o_P(1) \ .
\end{align*}
Let $\mathcal{F} = \{ \ind  (A=a_D,D_{\TIME+1}=0) h(W,\obX_{\TIME}) \varphi(Z,\bX_{0}) \cond h \in \mathcal{H}_c \big\}$. As we establish at the end of the proof, $\mathcal{F}$ is a Donsker class. Thus, Theorem 19.24 of \citet{Vaart1998} applies, yielding the following result:
\begin{align*} 
    &
    \mathbbm{G}
    :
    = 
    \frac{1}{
    \sqrt{ \big| \mathcal{I}_k^c \big| }
    }
    \sum_{i \in \mathcal{I}_k\LSS}
    \bigg[ 
    \begin{array}{l}         
    \ind  (A_i=a_D,D_{\TIME+1,i}=0) \varphi(Z,\obX_{0,i}) \big\{ \widehat{h}_1\LSS(W_i,\obX_{\TIME,i}) - h_1^*(W_i,\obX_{\TIME,i}) \big\} 
    \\
    - \EXP \big[ [ 
    \ind  (A=a_D,D_{\TIME+1}=0) \varphi(Z,\obX_{0}) \big\{ \widehat{h}_1\LSS(W,\obX_{\TIME}) - h_1^*(W,\obX_{\TIME}) \big\} 
    \big]
    \end{array} 
    \bigg] 
    = o_P(1) \ .
\end{align*}
This implies that
\begin{align*}
    \big\|
    \widehat{g} - \widetilde{g}
    \big\|
    &
    \leq 
    \bigg\|
    \frac{1}{| \mathcal{I}_{k}\LSS | }
    \sum_{i \in \mathcal{I}_k\LSS} 
    \ind  (A_i=a_D,D_{\TIME+1,i}=0) \varphi(Z,\obX_{0,i}) \big\{ \widehat{h}_1\LSS(W_i,\obX_{\TIME,i}) - h_1^*(W_i,\obX_{\TIME,i}) \big\} 
    \bigg\|
    \\
    &
    \leq 
    \frac{1}{\sqrt{ | \mathcal{I}_{k}\LSS | }}
    \big\| 
    \mathbbm{G}
    \big\|
    +
    \big\|
    \EXP \big[ [ 
    \ind  (A=a_D,D_{\TIME+1}=0) \varphi(Z,\obX_{0}) \big\{ \widehat{h}_1\LSS(W,\obX_{\TIME}) - h_1^*(W,\obX_{\TIME}) \big\} 
    \big]
    \big\|
    \\
    &
    \leq
    o_P(N^{-1/2})
    +
    \big\| \ind  (A=a_D,D_{\TIME+1}=0) \varphi(Z,\bX_{0}) 
    \big\{ \widehat{h}_1\LSS(W,\obX_{\TIME}) - h_1^* (W,\obX_{\TIME}) \big\} \big\|_{P,2}
    \\
    &
    \leq
    o_P(N^{-1/2})
    +
    O_P(r_{h1})
    \\
    &
    =
    O_P(r_{h1}) \ .
\end{align*}
Therefore, \eqref{eq-PMMR-g} is now substituted with $
    \big\| \widehat{g} - g \big\| = O_P(r_{h1})$. 

Following similar steps as in the proof of Lemma \ref{lemma:convergence:1}, one can obtain
\begin{align*}
\big\| h_0^*(\cdot \con a_Y) - \widehat{h}_0\LSS(\cdot \con a_Y) \big\|^2
	 =
	 O_P
	 \bigg( 	 
	 \frac{r_{h1}^2}{\lambda_{h0} }
	 +
	 \lambda_{h0}^{\min(\beta_{h0},2)}
	 +
	 \frac{ \lambda^{\min(\beta_{h0}-1,0)}}{N   }
	 \bigg) 
	  \ .
\end{align*}
Therefore, if we select $\lambda_{h0} = r_{h1}^{ \max(2/(1+\beta_{h0}),2/3)}$, we obtain
\begin{align*}
&
\big\| h_0^*(\cdot \con a_Y) - \widehat{h}_0\LSS(\cdot \con a_Y) \big\|^2
=
	 \left\{
	 	\begin{array}{ll}
	 	O_P \big( 
		r_{h1}^{2\beta_{h0}/(\beta_{h0}+1) }
		\big)
	 		& \text{ if $\beta_{h0} \in (0,2]$}
	 		\\
	 		O_P \big( 
	 	r_{h1}^{4/3}
	 	\big)
	 		& \text{ if $\beta_{h0} \in (2,\infty)$}
	 	\end{array}
	 \right.  
	  \ .
\end{align*}
Leveraging the compact support and uniformly bounded density conditions, we find
\begin{align*}
&
\big\| h_0^*(\cdot \con a_Y) - \widehat{h}_0\LSS(\cdot \con a_Y) \big\|_{P,2}
=
	 \left\{
	 	\begin{array}{ll}
	 	O_P \big( 
		r_{h1}^{\beta_{h0}/(\beta_{h0}+1) }
		\big)
	 		& \text{ if $\beta_{h0} \in (0,2]$}
	 		\\
	 		O_P \big( 
	 	r_{h1}^{2/3}
	 	\big)
	 		& \text{ if $\beta_{h0} \in (2,\infty)$}
	 	\end{array}
	 \right.  
	  \ .
\end{align*}

Now, it remains to show that $\mathcal{F}$ is a Donsker class. Example 2.10.10 of \cite{VW1996} implies that if $\mathcal{G}$ is a Donsker class and $g$ is a uniformly bounded measurable function, then $\mathcal{G} \cdot g$ is a Donsker class. Consequently, $\mathcal{F}$ is a Donsker class if $\mathcal{H}_c$ is a Donsker class, which is shown below.  The proof is based on Lemma A.9 of \citet{Hable2012}, with a similar argument presented in Appendix A.2 of \citet{BG2023}.

The Gaussian kernel $\mathcal{K}$ is infinitely differentiable. Hence, for any fixed $m \in \{1,2,\ldots\}$, the quantity $\sup_{\bv \in \mathcal{V}} (\partial^{\alpha,\alpha} \mathcal{K}(\bv,\bv) )$ is finite where $\bv = (w,\obx_{\TIME})$ and $\mathcal{V} = \text{supp}(W,\obX_{\TIME})$. Given $m$ let $\mathcal{C}^{m}(\mathcal{V})$ denote 
    \begin{align*}
        \mathcal{C}^{m}(\mathcal{V})
        =
        \bigg\{ 
            f \, \bigg| \, 
            \max_{\alpha\in \{0,1,\ldots,m\}}
        \sup_{\bv \in \mathcal{V}}
        | \partial^\alpha f(\bv)|
        \leq 1
        \bigg\} \ ,
    \end{align*}
    i.e., the set of all functions $f : \mathcal{V} \rightarrow \R$  of which partial derivatives up to order $m$ are uniformly bounded by 1. Then, we find the following result for any $f \in \mathcal{H}_c$:
    \begin{align*}
        \max_{\alpha\in \{0,1,\ldots,m\}}
        \sup_{\bv \in \mathcal{V}}
        | \partial^\alpha f(\bv)|
        &
        \leq \big\| f \big\|_{\HH} 
        \max_{\alpha\in \{0,1,\ldots,m\}}
        \sup_{\bv \in \mathcal{X}} (\partial^{\alpha,\alpha} \mathcal{K}(\bv,\bv) )
        \\
        &
        \leq c
        \max_{\alpha\in \{0,1,\ldots,m\}}
        \sup_{\bv \in \mathcal{V}} (\partial^{\alpha,\alpha} \mathcal{K}(\bv,\bv) )
        \\
        &
        =: a_c \in (0,\infty)
    \end{align*}
    The first inequality is from Corollary 4.36 of \citet{SVM2008}. The second inequality is from the definition of $\HH_c$ with the Gaussian kernel. Therefore, we find $\mathcal{H}_c / a_c \subseteq \mathcal{C}^{m}(\mathcal{V})$. Therefore, Theorem 2.7.1 of \citet{VW1996} implies that there exists a constant $r \in (0,\infty)$ such that, for every $\epsilon>0$, 
    \begin{align*}
        \log \mathcal{N}(a_c \epsilon, \mathcal{H}_c, \| \cdot \|_{\infty} )
        =
        \log  \mathcal{N} \bigg( \epsilon, \frac{\mathcal{H}_c}{a_c}, \| \cdot \|_{\infty} \bigg)
        \leq r \bigg( \frac{1}{\epsilon} \bigg)^{d/m} \ ,
    \end{align*}
    where $\mathcal{N}(\cdot, \cdot, \cdot)$ is the covering number. This implies
    \begin{align*}
        \log  \mathcal{N}(\epsilon, \mathcal{H}_c, \| \cdot \|_{\infty} )
        \leq r a_c^{d/m} \bigg( \frac{1}{\epsilon} \bigg)^{d/m} \ , \quad \forall \epsilon > 0 \ .
    \end{align*}
    We take $c' = c \| k \|_{\infty} + 1$ as the envelope of $\HH_c$, which implies
    \begin{align*}
        &
        \log  \mathcal{N}\big( \epsilon \|c'\|_{\infty}, \mathcal{H}_c, \| \cdot \|_{\infty} \big) 
        \leq r \bigg(\frac{a_c}{c'}\bigg)^{d/m} \bigg( \frac{1}{\epsilon} \bigg)^{d/m} \ , \quad && \epsilon \in (0,1) \ , 
        \\
        &
        \log  \mathcal{N}\big( \epsilon \|c'\|_{\infty}, \mathcal{H}_c, \| \cdot \|_{\infty} \big) = 0 \ , &&
        \epsilon \geq 1
    \end{align*}
    In addition, $\big\| f \big\|_{P,2} \leq \big\| f \big\|_{\infty}$. Therefore,
    \begin{align*}
        &
        \sup_{P} 
         \log  \mathcal{N} \big( \epsilon \| c' \|_{P,2}, \mathcal{H}_c, \| \cdot \|_{P,2} \big) 
        \leq r \bigg( \frac{a_c}{c'} \bigg)^{d/m} \bigg( \frac{1}{\epsilon} \bigg)^{d/m} \ , && \epsilon \in (0,1) \ , 
        \\
        &
        \sup_{P} 
         \log  \mathcal{N} \big( \epsilon \| c' \|_{P,2}, \mathcal{H}_c, \| \cdot \|_{P,2} \big) 
        = 0
        && \epsilon \geq 1 \ .
    \end{align*}
Since $m$ can be arbitrarily chosen, we can take $m>d/2$, which results in
\begin{align*}
&
\int_{0}^{\infty}
    \sqrt{ 
        \sup_{P} 
         \log  \mathcal{N} \big( \epsilon \| c' \|_{P,2}, \mathcal{H}_c, \| \cdot \|_{P,2} \big) 
    }
    \, d \epsilon
    =
    \int_{0}^{1}
    \sqrt{ 
        \sup_{P} 
         \log  \mathcal{N} \big( \epsilon \| c' \|_{P,2}, \mathcal{H}_c, \| \cdot \|_{P,2} \big) 
    }
    \, d \epsilon < \infty \ .
    \numeq \label{eq-UEN}
\end{align*}
By Theorem 2.5.2 of \citet{VW1996}, \eqref{eq-UEN} implies that $\mathcal{H}_c$ is a Donsker class. This completes the proof.

\subsection{Proof of Theorem \ref{thm:identifcation rand}}

In Section \ref{sec:supp:psi exp}, we established $\psi_{\RD,\NUMER}^*(a_D,a_D) = \EXP \big\{ p_0^*(a_D) \ind (A=a_D)(1-D_{\TIME+1})Y \big\}$ and $\psi_{\RD,\DENOM}^*(a_D) = \EXP \big\{ p_0^*(a_D) \ind (A=a_D)(1-D_{\TIME+1}) \big\}$. Therefore, it suffices to establish identification of $\psi_{\RD,\NUMER}^*(a_Y,a_D)$ when $a_Y \neq a_D$. Hereafter, we suppress $(a_Y,a_D)$ in confounding bridge functions and influence functions for notational brevity unless necessary.  We first provide a useful lemma to establish the results in Theorem \ref{thm:identifcation rand}. 

\begin{lemma} \label{lemma:identifcation rand}

\begin{itemize} Let $a_Y \neq a_D$. Then,
    \item[(i)] Suppose that Assumptions \HL{A1}, \HL{A2'}, \HL{A3}-(i), \HL{A4}, \HL{A5}, \HL{A6}-(i),(iii),(iv), \HL{A7}-(i) are satisfied, and that there exist the confounding bridge function $\xi_1^*$ satisfying \eqref{eq-bridgeft1 exp}. Then, we have 
    \begin{align*}
        & \EXP \big( Y \cond A=a_Y, D_{\TIME+1}=0, \obX_{\TIME}, \bU \big) 
        = \EXP \big\{ \xi_1^*(\bW,\bX) \cond A=a_Y, D_{\TIME+1}=0, \obX_{\TIME}, \bU \big\}
        \text{ a.s.}
    \end{align*}
    
    \item[(ii)] Suppose that Assumptions \HL{A1}, \HL{A2'}, \HL{A3}-(i), \HL{A4}, \HL{A5}, \HL{A6}-(i),(iii),(iv), \HL{A7}-(iii) are satisfied, and that there exist the confounding bridge function $p_1^*$ satisfying \eqref{eq-bridgeft5 exp}. Then, we have
    \begin{align*}
        & 
        \EXP \big\{ \ind (A=a_D) p_0^*(a_D) \cond D_{\TIME+1}=0 , \obX_{\TIME} , \bU \big\} 
= \EXP \big\{ \ind(A=a_Y) p_1^* (\bZ,\obX_{1}) \cond D_{\TIME+1}=0 , \obX_{\TIME} , \bU \big\}
\text{ a.s.}
    \end{align*}

\end{itemize}
\end{lemma}
The proof of Lemma \ref{lemma:identifcation rand}-(i) and (ii) are similar to those of Lemma \ref{thm:integral equation}-(i) and (iv); see Sections \ref{sec:supp:proof:thm:211} and \ref{sec:supp:proof:thm:214}.

 \subsubsection{First Identification Result in Theorem \ref{thm:identifcation rand}}

We establish the claim for $\psi_{\RD,\NUMER}^*$:
\begin{align*}
& \EXP \big\{ \ind (A=a_D,D_{\TIME+1}=0) p_0^*(a_D) \xi_1^*(\bW,\obX_{\TIME}) \big\}
\\
& =
\EXP \big[ \EXP \big\{ \ind (A=a_D,D_{\TIME+1}=0) p_0^*(a_D) \xi_1^*(\bW,\obX_{\TIME}) \cond \obX_{\TIME}, \bU \big\} \big]
\\
& =
\EXP \big[ 
\Pr \big( A=a_D,D_{\TIME+1}=0 \cond \obX_{\TIME} , \bU \big) p_0^*(a_D) 
\EXP \big\{ \xi_1^*(\bW,\obX_{\TIME}) \cond A=a_D,D_{\TIME+1}=0,\obX_{\TIME},\bU \big\} \big] 
\\
& 
=
\EXP \bigg\{
\frac{\Pr \big( A=a_D,D_{\TIME+1}=0 \cond \obX_{\TIME} , \bU \big)}{ \Pr(A=a_D) }
\EXP \big( Y \cond A=a_Y, D_{\TIME+1}=0, \obX_{\TIME} ,\bU \big) 
\bigg\} 
\\
& 
=
\psi_{\RD,\NUMER}^*(a_Y,a_D) \ .
\end{align*}
The second and third lines are from the law of iterated expectation.  
The fourth line holds from $p_0^*(a_D) = 1/\Pr(A=a_D)$.
The last line holds from the form of $\psi_{\RD,\NUMER}^*$ in \eqref{sec:supp:psi exp}.

\subsubsection{Third Identification Result in Theorem \ref{thm:identifcation rand}}

We establish the claim for $\psi_{\RD,\NUMER}^*$:
 \begin{align*}
 &
\EXP \big\{ \ind (A=a_Y, D_{\TIME}=0) p_1^*(\bZ,\obX_{\TIME}) Y \big\}
\\
& =
\EXP \big[ \EXP \big\{ \ind (A=a_Y, D_{\TIME}=0) 
p_1^* (\bZ,\obX_{\TIME}) Y  \cond \obX_{\TIME} , U \big\}  \big]
\\
& =
\EXP \big[ \Pr(A=a_Y, D_{\TIME+1}=0 \cond \obX_{\TIME} , \bU) 
\EXP \big\{  p_1^* (\bZ,\obX_{\TIME}) Y  \cond A=a_Y, D_{\TIME+1}=0, \obX_{\TIME}, \bU \big\} \big]
\\
& 
=
\EXP \bigg[ 
\begin{array}{l}
\Pr(A=a_Y, D_{\TIME+1}=0 \cond \obX_{\TIME}, \bU) 	
\EXP \big\{  p_1^*(\bZ,\obX_{\TIME}) \cond A=a_Y,D_{\TIME+1}=0,\obX_{\TIME}, \bU \big\}
\\
\times
\EXP \big( Y \cond A=a_Y,D_{\TIME+1}=0, \obX_{\TIME}, \bU \big)
\end{array}
 \bigg]
\\
& 
=
\EXP \big[ \Pr(A=a_D, D_{\TIME+1}=0 \cond \obX_{\TIME}, \bU) 	
p_0^*(a_D)
\EXP \big( Y \cond A=a_Y,D_{\TIME+1}=0, \obX_{\TIME}, \bU \big) \big]
\\
& 
=
\EXP \bigg\{
\frac{\Pr \big( A=a_D,D_{\TIME+1}=0 \cond \obX_{\TIME} ,\bU \big)}{ \Pr(A=a_D) }
\EXP \big( Y \cond A=a_Y, D_{\TIME+1}=0  , \obX_{\TIME} ,\bU \big) 
\bigg\} 
\\
& 
=
\psi_{\RD,\NUMER}^*(a_Y,a_D) \ .
 \end{align*}
The second and third lines are from the law of iterated expectation. 
The fourth line holds from \HL{CI1}: $\bZ \indep Y \cond A,D,\bX,\bU$. 
The fifth line holds from Lemma \ref{lemma:identifcation rand}-(ii). 
The sixth line holds from $p_0^*(a_D) = 1/\Pr(A=a_D)$.
The last line holds from the form of $\psi_{\RD,\NUMER}^*$ in \eqref{sec:supp:psi exp}. 


\subsection{Proof of Theorem \protect{\ref{thm:IF Rand Supp}}}

We suppress $(a_Y,a_D)$ in confounding bridge functions and influence functions for notational brevity unless necessary.

\subsubsection{$\cInfFt_{\RD,\CSE}^*$ Is an Influence Function}

It suffices to derive an centered influence function for $\psi_{\RD}^*(a_Y,a_D)$, denoted by $\cInfFt_{\RD}^*(O \con a_Y,a_D)$, because an influence function for $\tau_{\CSE}^*(a_D)$ is given by $\cInfFt_{\RD}^*(O \con a_Y=1,a_D) - \cInfFt_{\RD}^*(O \con a_Y=0,a_D)$. In addition, the efficient influence functions for $\psi_{\RD,\NUMER}^*(a_D,a_D)$ and $\psi_{\RD,\DENOM}^*(a_D)$ are given by
\begin{align*}
    &
    \psi_{\RD,\NUMER}^*(a_D,a_D)
    =
    p_0^*(a_D)
     \EXP \big\{ \ind (A=a_D) (1-D_{\TIME+1}) Y \big\}  
    \\
    & \Rightarrow
    \quad 
    \text{Centered EIF of $\psi_{\RD,\NUMER}^*(a_D,a_D)$}
    =
    p_0^*(a_D) \ind (A=a_D) (1-D_{\TIME+1}) Y
   -
    \psi_{\RD,\NUMER}^*(a_D,a_D)
    \\
    &
    \psi_{\RD,\DENOM}^*(a_D)
    =
    p_0^*(a_D) \EXP \big\{ \ind (A=a_D) (1-D_{\TIME+1})  \big\} 
    \\
    & \Rightarrow
    \quad 
    \text{Centered EIF of $\psi_{\RD,\DENOM}^*(a_D)$}
    =
    p_0^*(a_D) \ind (A=a_D) (1-D_{\TIME+1})
    -
     \psi_{\RD,\DENOM}^*(a_D)  \ .
\end{align*} 
This is because the expected value of a random variable, and the efficient influence function for $\mu = \EXP(V)$ is $V-\mu$, and $p_0^*(a_D)$ is known. Therefore, the efficient influence function for $\psi_{\RD}^*(a_D,a_D)$ is
\begin{align*}
    &
    \cInfFt_{\RD}^*(\bO \con a_D,a_D)
    \\    
    &
    =
    \frac{
    \bigg[ 
    	\begin{array}{l} 
     \big\{ p_0^*(a_D) \ind (A=a_D) (1-D_{\TIME+1}) Y
    -
    \psi_{\RD,\NUMER}^*(a_D,a_D)
    \big\}
    \\
    -
    \psi^*(a_D,a_D)
    \big\{ p_0^*(a_D) \ind (A=a_D) (1-D_{\TIME+1})
    -
     \psi_{\RD,\NUMER}^*(a_D) 
     \big\}
    	\end{array}
    \bigg]
    }{
        \psi_{\RD,\NUMER}^*(a_D) 
    }
    \\    
    &
    =
    \frac{ 
        p_0^*(a_D) \ind (A=a_D) (1-D_{\TIME+1}) 
        \big\{ Y - \psi_{\RD}^*(a_D,a_D)
        \big\}
    }{
        \psi_{\RD,\DENOM}^*(a_D) 
    }
    \ .
\end{align*}
Therefore, it suffices to derive an influence function for $\psi_{\RD,\NUMER}^*(a_Y, a_D)$ where $a_Y \neq a_D$ because an influence function for $\psi_{\RD}^*(a_Y,a_D)$ is obtained from \eqref{eq-IF numer denom}. In what follows, we establish
\begin{align*}
&
\cInfFt_{\RD,\NUMER}^*(\bO)
=
\bigg[
        \begin{array}{l}
             (1-D_{\TIME+1}) \ind(A=a_Y) p_1^*(Z,\obX_{\TIME}) 
\big\{ Y - \xi_1^* (\bW,\obX_{\TIME}) \big\} 
\\
+
\ind(A=a_D) p_0^*(a_D) 
        (1-D_{\TIME+1})  
        \xi_1^*(\bW,\obX_{\TIME})  - \psi_{\RD,\NUMER}^*(a_Y,a_D)
        \end{array}
\bigg] \ ,
\end{align*}
which reduces to an influence function:
\begin{align*}
& 
\cInfFt_{\RD}^*(\bO)  
\\
&
= 
\frac{ \cInfFt_{\RD,\NUMER}^*(\bO) - \psi_{\RD}^*(a_Y,a_D) \cInfFt_{\RD,\DENOM}^*(\bO)  }{ \psi_{\RD,\DENOM}^*(a_D) }
 \\
 &
 =	 
\frac{1}{ \psi_{\RD,\DENOM}^* (a_D) }
\left[
\begin{array}{l}
\ind(A=a_Y)  (1-D_{\TIME+1}) p_1^* (\bZ,\obX_{\TIME}) \big\{ Y - \xi_1^* (\bW,\obX_{\TIME}) \big\}
\\
\quad + \ind(A=a_D) p_0^* (a_D) (1-D_{\TIME+1}) \xi_1^* (\bW,\obX_{\TIME})  -
\psi_{\RD}^*(a_Y,a_D)
\\
\quad 
-
\psi_{\RD}^*(a_Y,a_D)
\big\{
         \ind(A=a_D) p_0^* (a_D) (1-D_{\TIME+1})
         -  \psi_{\RD,\DENOM}^*(a_D)
 \big\}
 \end{array}
\right] 
 \\
 &
 =	 
\frac{1}{ \psi_{\RD,\DENOM}^* (a_D) }
\left[
\begin{array}{l}
\ind(A=a_Y)  (1-D_{\TIME+1}) p_1^* (\bZ,\obX_{\TIME}) \big\{ Y - \xi_1^* (\bW,\obX_{\TIME}) \big\}
\\
\quad + \ind(A=a_D) p_0^* (a_D) (1-D_{\TIME+1}) \xi_1^* (\bW,\obX_{\TIME})  
\\
\quad 
-
\psi_{\RD}^*(a_Y,a_D)
\ind(A=a_D) p_0^* (a_D) (1-D_{\TIME+1})
 \end{array}
\right] 
\numeq \label{eq-IF rand}
\ . 
\end{align*}
The rest of the proof is to establish the form of $\cInfFt_{\RD,\NUMER}^*(\bO)$.

Recall that model $\M_{\RD}$ is defined as the regular model of the form
\begin{align*}
\mathcal{M}_{\RD}
=
\left\{
    P
    \left|
        \begin{array}{l}
        \text{$A$ is randomized with known $\Pr(A)$,}
        \\ \text{and  there exists $\xi_1^*$ satisfying } 
        \EXP \big\{ Y - \xi_1^*(\bW,\obX_{\TIME}) \cond \bZ,A=a_Y,D_{\TIME+1}=0, \obX_{\TIME} \big\}
        \end{array}
    \right.
\right\} \ .
\end{align*}
Let $\M_{\RD}(\eta)$ be the one-dimensional parametric submodel of $\M_{\RD}$ and $f(\bO \con \eta) \in \M(\eta)$ be the density at $\eta$. We suppose that the true density is recovered at $\eta^*$, i.e., $f^*(\bO) := f(\bO \con \eta^*)$. We use $\EXP\ETA \big\{ g(\bO) \big\}$ to denote the expectation operator at $f(\bO \con \eta)$. Let $\xi_1(\cdot \con \eta)$ be the confounding bridge functions satisfying
\begin{align*}
&
\EXP\ETA \big\{ Y - \xi_1(\bW,\obX_{\TIME} \con \eta) \cond \bZ, A=a_Y, D_{\TIME+1}=0 , \obX_{\TIME} \big\}
= 
0
\ .
\end{align*}

We also make certain differentiability conditions for the submodel $\mathcal{M}_{\RD}(\eta)$. Specifically, we assume that $f(\cdot \con \eta)$ is pointwise differentiable with respect to $\eta$. While, in principle, this condition could be relaxed to differentiability in the quadratic mean sense, we opt for pointwise differentiability to avoid unnecessary technical digressions.  Let $s(\cdot \cond \cdot \con \eta): = \partial \log f(\cdot \cond \cdot \con \eta) / \partial \eta$ be the score function evaluated at $\eta$, and let $s^*(\cdot \cond \cdot)$ be the score function evaluated at $\eta^*$. We also assume that the confounding bridge functions $\xi_1(\cdot \con \eta)$ is differentiable with respect to $\eta$.

Therefore, taking the derivative of the above restrictions with respect to $\eta$ and evaluating at $\eta^*$, the score function $s^*$ satisfies the following restriction:
\begin{align}
    &
    \EXP \Bigg[
        \begin{array}{l}
        \big\{ Y - \xi_1^* (\bW,\obX_{\TIME}) \big\} s^* (Y,\bW \cond \bZ, A=a_Y, D_{\TIME+1}=0, \obX_{\TIME}) \\
        - \nabla_\eta \xi_1(\bW,\obX_{\TIME} \con \eta^*)	
        \end{array}
     \, \bigg| \, \begin{array}{l}
     \bZ, A=a_Y,
     \\
      D_{\TIME+1}=0, \obX_{\TIME}
     \end{array} \Bigg] = 0
     \ .
    \label{eq:tangent-1 exp}
\end{align}

The target estimand at $\eta$ is defined as $\psi_{\RD,\NUMER}(a_Y,a_D \con  \eta) = \EXP\ETA \big\{ \xi_0 (\bW,\bX_{0} \con \eta) \big\}  $. The pathwise derivative of $\psi_{\RD,\NUMER}(a_Y,a_D \con \eta)$ is  
\begin{align}			\label{eq:pathwise-Numer exp}
&
\frac{\partial \psi_{\RD,\NUMER} (a_Y,a_D \con \eta) }{\partial \eta} \bigg|_{\eta=\eta^*}
\nonumber 
\\
& =
p_0^*(a_D)
\frac{\partial \EXP\ETA \big\{ \ind(A=a_D)  (1-D_{\TIME+1}) \xi_1(\bW,\obX_{\TIME} \con \eta)  \big\}  }{\partial \eta}  \bigg|_{\eta=\eta^*}
\nonumber
\\
& =
\EXP \big[  \ind(A=a_D) p_0^*(a_D) (1-D_{\TIME+1})
\big\{ \nabla_\eta \xi_1(\bW,\obX_{\TIME} \con \eta^*)  + 
\xi_1^*(\bW,\obX_{\TIME}) s^*(W, \obX_{\TIME})
\big\} \big]
\nonumber
\\
& =
\EXP \bigg[ 
\begin{array}{l}
 \ind(A=a_D) p_0^*(a_D) (1-D_{\TIME+1}) \nabla_\eta \xi_1(\bW,\obX_{\TIME} \con \eta^*) 
\\
+
\big\{ \ind(A=a_D) p_0^*(a_D) (1-D_{\TIME+1})
\xi_1^*(\bW,\obX_{\TIME}) 
-
\psi_{\RD,\NUMER}^*(a_Y,a_D) \big\}
s^*(O) 
\end{array}
\bigg] 
 \ .
\end{align}

Note that $\EXP \big\{  \ind(A=a_D) p_0^*(a_D) (1-D_{\TIME+1}) \nabla_\eta \xi_1(\bW,\obX_{\TIME} \con \eta^*)  \big\}$ is represented as
\begin{align}
& 
\EXP \big\{ \ind(A=a_D,D_{\TIME+1}=0) p_0^* (a_D) 
\nabla_\eta \xi_1(\bW,\obX_{\TIME} \con \eta^*)  \big\}
\nonumber
\\
& =
\EXP \big[ (1-D_{\TIME+1}) \EXP \big\{ \ind(A=a_D) p_0^*(a_D) \cond W,D_{\TIME+1}=0,\obX_{\TIME} \big\}
\nabla_\eta \xi_1 (\bW,\obX_{\TIME} \con \eta^*) \big]
\nonumber
\\
& =
\EXP \big[ (1-D_{\TIME+1}) \EXP \big\{ \ind(A=a_Y) p_1^*(\bZ,\obX_{\TIME}) \cond W,D_{\TIME+1}=0,\obX_{\TIME} \big\}
\nabla_\eta \xi_1 (\bW,\obX_{\TIME} \con \eta^*) \big]
\nonumber
\\
& =
\EXP \big[ (1-D_{\TIME+1}) \ind(A=a_Y) p_1^*(\bZ,\obX_{\TIME}) 
\nabla_\eta \xi_1 (\bW,\obX_{\TIME} \con \eta^*) \big]
\nonumber
\\
& =
\EXP \big[ (1-D_{\TIME+1}) \ind(A=a_Y) p_1^*(\bZ,\obX_{\TIME}) 
\EXP \big\{ \nabla_\eta \xi_1 (\bW,\obX_{\TIME} \con \eta^*) \cond Z, A=a_Y,D_{\TIME+1}=0,\obX_{\TIME} \big\} \big]
\nonumber
\\
& =
\EXP \Bigg[ 
\begin{array}{l}
(1-D_{\TIME+1}) \ind(A=a_Y) p_1^*(\bZ,\obX_{\TIME}) 
\\
\times 
\EXP \big[ \big\{ Y - \xi_1^* (\bW,\obX_{\TIME}) \big\} s^* (Y,\bW \cond \bZ, A=a_Y, D_{\TIME+1}=0, \obX_{\TIME}) \cond Z, A=a_Y,D_{\TIME+1}=0,\obX_{\TIME} \big]
\end{array} \Bigg]
\nonumber
\\
&
=
\EXP \bigg[ 
\begin{array}{l}
(1-D_{\TIME+1}) \ind(A=a_Y) p_1^*(\bZ,\obX_{\TIME}) 
\big\{ Y - \xi_1^* (\bW,\obX_{\TIME}) \big\} 
\\
\times 
s^* (Y,\bW \cond \bZ, A=a_Y, D_{\TIME+1}=0, \obX_{\TIME})
\end{array}
\bigg]
     \nonumber
\\
&
=
\EXP \left[ 
        \begin{array}{l}     
        (1-D_{\TIME+1}) \ind(A=a_Y) p_1^*(Z,\obX_{\TIME}) 
\big\{ Y - \xi_1^* (\bW,\obX_{\TIME}) \big\} 
\\
\times \big[ 
\EXP \big\{ s^*(O) \cond Y,W,Z,A,D_{\TIME+1},\obX_{\TIME}
\big\}
-
\EXP \big\{ s^*(O) \cond Z,A,D_{\TIME+1},\obX_{\TIME}
\big\}
\big]
        \end{array} 
     \right] 
     \nonumber
     \\
     &
     = 
\EXP \left[ 
        \begin{array}{l}     
        (1-D_{\TIME+1}) \ind(A=a_Y) p_1^*(Z,\obX_{\TIME}) 
\big\{ Y - \xi_1^* (\bW,\obX_{\TIME}) \big\} 
s^*(O)
\\
-
(1-D_{\TIME+1}) \ind(A=a_Y) p_1^*(Z,\obX_{\TIME}) 
\EXP \big\{ s^*(O) \cond Z,A,D_{\TIME+1},\obX_{\TIME}
\big\}
\\
\quad 
\times 
\underbrace{
\EXP \big\{ Y - \xi_1^* (\bW,\obX_{\TIME}) \cond Z,A=a_Y,D_{\TIME+1}=0,\obX_{\TIME} \big\} }_{=0} 
\\[-0.5cm]
        \end{array} 
     \right]
     \nonumber
     \\[0.5cm]
     &
     =
     \EXP \big[ (1-D_{\TIME+1}) \ind(A=a_Y) p_1^*(Z,\obX_{\TIME}) 
\big\{ Y - \xi_1^* (\bW,\obX_{\TIME}) \big\} 
s^*(O) \big]
\ .
 \label{eq-grad-xi0-2}
\end{align}
The second line holds from the law of iterated expectation. 
The third line holds from \eqref{eq-bridgeft5 exp}. 
The fourth and fifth lines hold from the law of iterated expectation. 
The sixth line holds from \eqref{eq:tangent-1 exp}.
The last lines hold from the law of iterated expectation. 

Combining \eqref{eq:pathwise-Numer exp} and \eqref{eq-grad-xi0-2}, we have
\begin{align*}
&
\frac{\partial \psi_{\RD,\NUMER} (a_Y,a_D \con \eta) }{\partial \eta} \bigg|_{\eta=\eta^*}
\\
& = 
\EXP
\bigg[
\underbrace{
\bigg[
        \begin{array}{l}
             (1-D_{\TIME+1}) \ind(A=a_Y) p_1^*(Z,\obX_{\TIME}) 
\big\{ Y - \xi_1^* (\bW,\obX_{\TIME}) \big\} 
\\
+
\ind(A=a_D) p_0^*(a_D) 
        (1-D_{\TIME+1})  
        \xi_1^*(\bW,\obX_{\TIME})  - \psi_{\RD,\NUMER}^*(a_Y,a_D)
        \end{array}
\bigg]}_{=:\InfFt_{\RD,\NUMER}^*(\bO \con a_Y,a_D) - \psi_{\RD,\NUMER}^*(a_Y,a_D)}
\times s^*(\bO) 
\bigg]
\\
& = \EXP \big[ \underbrace{ \big\{ \InfFt_{\RD,\NUMER}^*(\bO \con a_Y,a_D) - \psi_{\RD,\NUMER}^*(a_Y,a_D) \big\}}_{=:\cInfFt_{\RD,\NUMER}^*(\bO \con a_Y,a_D)} \cdot s^*(\bO) \big] \\
&
=
\EXP \big\{ \cInfFt_{\RD,\NUMER}^*(\bO \con a_Y,a_D) 
\cdot s^*(\bO) \big\}
 \ .
\end{align*}

Therefore, an centered influence function for $\tau_{\CSE}^*(a_D)$ is
\begin{align*}
    &
    \cInfFt_{\RD,\CSE}^*(\bO)
    \\    
    &
    =
    \frac{
    \left[
    \begin{array}{l}
    \big\{ \InfFt_{\RD,\NUMER}^*(\bO \con a_Y,a_D) - \psi_{\RD,\NUMER}^*(a_Y,a_D) \big\}
        -
         \big\{ p_0^*(a_D) \ind (A=a_D) (1-D_{\TIME+1}) Y
   -
    \psi_{\RD,\NUMER}^*(a_D,a_D)
    \big\}
    \\
    -
    \tau_{\CSE}^*(a_D) 
    \big\{ p_0^*(a_D) \ind (A=a_D) (1-D_{\TIME+1})
    -
     \psi_{\RD,\DENOM}^*(a_D) 
     \big\}
    \end{array}
    \right] 
    }{
        \psi_{\RD,\DENOM}^*(a_D) 
    }
    \\    
    &
    =
    \frac{ 
        \InfFt_{\RD,\NUMER}^*(\bO \con a_Y, a_D) 
        -
        p_0^*(a_D) \ind (A=a_D) (1-D_{\TIME+1}) 
        \big\{ Y + \tau_{\CSE}^*(a_D)
        \big\}
    }{
        \psi_{\RD,\DENOM}^*(a_D) 
    }
\end{align*}

This completes the proof.

\subsubsection{$\cInfFt_{\RD,\CSE}^*$ Is the Efficient Influence Function at Submodel $\M_{\RD,\text{sub}}$}

To show that the influence function $\cInfFt_{\RD,\CSE}^*$ achieves the semiparametric local efficiency bound, we need to establish that $\cInfFt_{\RD}^*$ in \eqref{eq-IF rand} belongs to the tangent space of model $\M_{\RD}$ when the true data-generating law belongs to submodel $\M_{\RD,\text{sub}}$. Similar to observational settings in Section \ref{sec:supp:thm:IF}, the model imposes restriction
\begin{align}
\EXP \Bigg[
        \begin{array}{l}
        \big\{ Y - \xi_1^* (\bW,\obX_{\TIME}) \big\} s^* (Y,\bW \cond \bZ, A=a_Y, D_{\TIME+1}=0, \obX_{\TIME}) \\
        - \nabla_\eta \xi_1(W,\obX_{\TIME} \con \eta^*)	
        \end{array}
     \, \bigg| \, \bZ, A=a_Y, D_{\TIME+1}=0, \obX_{\TIME} \Bigg] = 0 \ .
     \nonumber
\end{align}
This restriction implies that the tangent space of model $\M_{\RD}$ consists of the functions $S(\bO) \in \mathcal{L}_{2,0}(\bO)$ satisfying
\begin{align}	
\label{eq-restriction rand}
& 
\EXP \Bigg[
\begin{array}{l}
    \big\{ Y - \xi_1^* (\bW,\obX_{\TIME}) \big\} 
    \\
    \times
    S(Y,\bW \cond \bZ, A=a_Y, D=0, \obX_{\TIME})  
\end{array}
    \COND \bZ, A=a_Y, D_{\TIME+1}=0, \obX_{\TIME} \Bigg] 
    \nonumber
    \\
    &
    =
    \EXP \big\{
    \nabla_\eta \xi_1(W,\obX_{\TIME} \con \eta^*)	
    \cond \bZ, A=a_Y, D_{\TIME+1}=0, \obX_{\TIME} 
    \big\}
    \in  \text{Range}(\mathcal{T}_1) 
     \ , \\
& 
\label{eq-restriction rand A}
\EXP \big\{ S(\bO) \cond A \big\} = 0 \ .
\end{align}
Of note, the second restriction is from \HL{A2'}, which is derived as follows:
\begin{align*}
& \text{Density factorization: }
&&
 f(\bO) = f(Y,D_{\TIME+1},\bW,\bZ,\obX_{\TIME} \cond A) f(A) 
\\
& \text{Parametric submodel: } 
&&
f(\bO \con \eta) = f(Y,D_{\TIME+1},\bW,\bZ,\obX_{\TIME} \cond A \con \eta) f(A) 
\\
& \text{Score function: } 
&&
S(\bO) = S(Y,D_{\TIME+1},\bW,\bZ,\obX_{\TIME} \cond A)
\end{align*}

Under the surjectivity condition \HL{S1}-(ii), we have $\text{Range}(\mathcal{T}_1) = \mathcal{L}_2 ( \bZ,A=a_Y,D_{\TIME+1}=0,\obX_{\TIME})$. In other words, when the true data-generating law belongs to submodel $\M_{\RD,\text{sub}}$, any $S(\bO) \in \mathcal{L}_{2,0}(\bO)$ satisfies restriction \eqref{eq-restriction rand}. 

Next, we show that $\cInfFt_{\RD}^*$ satisfies \eqref{eq-restriction rand A}:
\begin{align*}
\EXP \big\{ \cInfFt_{\RD}^*(\bO) \cond A \big\}
&
=
\frac{1}{ \psi_{\RD,\DENOM}^* (a_D) }
\EXP  
\left.
\left[
\begin{array}{l}
\ind(A=a_Y)  (1-D_{\TIME+1}) p_1^* (\bZ,\obX_{\TIME}) \big\{ Y - \xi_1^* (\bW,\obX_{\TIME}) \big\}
\\
\quad + \ind(A=a_D) p_0^* (a_D) (1-D_{\TIME+1}) \xi_1^* (\bW,\obX_{\TIME})  
\\
\quad 
-
\psi_{\RD}^*(a_Y,a_D)
\ind(A=a_D) p_0^* (a_D) (1-D_{\TIME+1})
 \end{array}
\right|
A
\right] 
\\
&
=
\frac{1}{ \psi_{\RD,\DENOM}^* (a_D) }
\EXP
\left[
\begin{array}{l}
\EXP \big[ 
\ind(A=a_Y)  (1-D_{\TIME+1}) p_1^* (\bZ,\obX_{\TIME}) \big\{ Y - \xi_1^* (\bW,\obX_{\TIME}) \big\} \cond A \big]
\\
\quad 
+ \EXP \bigg[ 
\begin{array}{l}
\ind(A=a_D) p_0^* (a_D) (1-D_{\TIME+1}) \xi_1^* (\bW,\obX_{\TIME})  
\\
-
\psi_{\RD}^*(a_Y,a_D)
\ind (A=a_D) p_0^*(a_D) (1-D_{\TIME+1}) 
\end{array}
\COND A \bigg] 
 \end{array} 
\right]  
\\
&
=
0 \ .
\end{align*} 
The last line is from 
\begin{align*}
&
\EXP \big[
\ind(A=a_Y)  (1-D_{\TIME+1}) p_1^* (\bZ,\obX_{\TIME}) \big\{ Y - \xi_1^* (\bW,\obX_{\TIME}) \big\}
\cond A=a_D
\big] = 0  \ ,
\\
&
\EXP \big[
\ind(A=a_Y)  (1-D_{\TIME+1}) p_1^* (\bZ,\obX_{\TIME}) \big\{ Y - \xi_1^* (\bW,\obX_{\TIME}) \big\}
\cond A=a_Y
\big] 
\\
& =
\EXP \big[(1-D_{\TIME+1}) p_1^* (\bZ,\obX_{\TIME}) \big\{ Y - \xi_1^* (\bW,\obX_{\TIME}) \big\}
\cond A=a_Y
\big]
\\
&
=
\Pr(D_{\TIME+1}=0 \cond A=a_Y)
\\
&
\quad 
\times 
\EXP \big[
p_1^* (\bZ,\obX_{\TIME}) 
\underbrace{
\EXP \big\{ Y - \xi_1^* (\bW,\obX_{\TIME}) 
\cond
\bZ,
A=a_Y, D_{\TIME+1}=0, \obX_{\TIME}
\big\}
}_{=0}
\cond A=a_Y, D_{\TIME+1}=0
\big]	
\\
&
=
0 \ ,
\\  
&
\EXP \bigg[ 
\begin{array}{l}
\ind(A=a_D) p_0^* (a_D) (1-D_{\TIME+1}) \xi_1^* (\bW,\obX_{\TIME})  
\\
-
\psi_{\RD}^*(a_Y,a_D)
\ind (A=a_D) p_0^*(a_D) (1-D_{\TIME+1}) 
\end{array}
\COND A=a_Y \bigg]  = 0 \ ,
\\  
&
\EXP \bigg[ 
\begin{array}{l}
\ind(A=a_D) p_0^* (a_D) (1-D_{\TIME+1}) \xi_1^* (\bW,\obX_{\TIME})  
\\
-
\psi_{\RD}^*(a_Y,a_D)
\ind (A=a_D) p_0^*(a_D) (1-D_{\TIME+1}) 
\end{array}
\COND A=a_D \bigg]  
\\
&
=
p_0^* (a_D) 
\EXP \bigg[ 
\begin{array}{l}
\ind(A=a_D)  (1-D_{\TIME+1}) \xi_1^* (\bW,\obX_{\TIME})  
\\
-
\psi_{\RD}^*(a_Y,a_D)
\ind (A=a_D)  (1-D_{\TIME+1}) 
\end{array}
\COND A=a_D \bigg] 
\\
&
=
p_0^* (a_D) \EXP \big\{
(1-D_{\TIME+1}) \xi_1^* (\bW,\obX_{\TIME})  
-
\psi_{\RD}^*(a_Y,a_D)
(1-D_{\TIME+1}) 
\cond A = a_D \big\}
         \\
         &
         =
          \psi_{\RD,\NUMER}(a_Y,a_D)
-
         \psi_{\RD}^*(a_Y,a_D) \psi_{\RD,\DENOM}^*(a_D)
         \\
         &
         =
         0 \ .
\end{align*}
Therefore, the influence function $\cInfFt_{\RD}^*$ belongs to the tangent space of $\M_{\RD}$ when the true data-generating law belongs to submodel $\M_{\RD,\text{sub}}$, i.e., $\cInfFt_{\RD}^*$ is the efficient influence function for $\psi_{\RD}^*(a_Y,a_D)$ under model $\M_{\RD}$ when the true data-generating law belongs to submodel $\M_{\RD,\text{sub}}$. This ends the proof.

\subsection{Proof of Theorem \ref{thm:AN rand}}

The proof is similar to that of Theorem \ref{thm:AN} where the nuisance functions $(\xi_1^*,p_1^*)$ in \eqref{eq-bridgeft1 exp}-\eqref{eq-bridgeft5 exp} and the corresponding estimator $(\widehat{\xi}_1\LSS,\widehat{p}_1\LSS)$ are used instead of the confounding bridge functions $(h_1^*,q_1^*)$ in \eqref{eq-bridgeft1 obs} and \eqref{eq-bridgeft6 obs} and the corresponding estimator $(\widehat{h}_1\LSS,\widehat{q}_1\LSS)$, respectively. We refer the readers to Section \ref{sec:supp:thm:AN}.  

To complete the proof, we show that \HL{A8'}-\HL{A10'} satisfies \AssumptionStrongOv-\AssumptionMixBiasDenom\ by taking 
\begin{align*}
&
(h_0^*,h_1^*,h_2^*,q_0^*,q_1^*)
&& 
= 
&&
(0 , \xi_1^*, 0 ,p_0^*,p_1^*) 
\\
&
(\widehat{h}_0\LSS,\widehat{h}_1\LSS,\widehat{h}_2\LSS,\widehat{q}_0\LSS,\widehat{q}_1\LSS)
&&
=
&&
(0 , \widehat{\xi}_1\LSS, 0 ,p_0^*,\widehat{p}_1\LSS) \ .
\end{align*} 

First, Assumption \HL{A8'} implies $
\big\| \xi_0^* \big\|_{\infty}
=
\big\| \widehat{\xi}_0\LSS \big\|_{\infty} = 0 \leq C^2$, satisfying  Assumption \AssumptionStrongOv.

Next, Assumption \HL{A9'} implies \AssumptionConsistency\ because 
\begin{align*}
&
\big\| \widehat{\xi}_0\LSS - \xi_0^* \big\|_{P,2} = \big\| \widehat{\xi}_2\LSS - \xi_2^* \big\|_{P,2} = \big\| \widehat{p}_0\LSS - p_0^* \big\|_{P,2} = 0 \ ,
\\
&
\big\| \widehat{\xi}_1\LSS - \xi_1^* \big\|_{P,2} = o_P(1) \ , \quad 
\big\| \widehat{p}_1\LSS - p_1^* \big\|_{P,2} = o_P(1) \ .
\end{align*} 

Lastly, we show that Assumption \AssumptionMixBiasNumer\ and \AssumptionMixBiasDenom\ are satisfied under \HL{A10'}:
\begin{align*}
    &
    \text{\AssumptionMixBiasNumer}:
    &&
    \big\| \widehat{\xi}_1\LSS - \xi_1^* \big\|_{P,2}
    \big\| \widehat{p}_1\LSS - p_1^* \big\|_{P,2}
    =
    o_P(N^{-1/2}) \ ,
    \\
    &
    &&
    \big\| \widehat{\xi}_1\LSS - \xi_1^* \big\|_{P,2}
    \big\| \widehat{p}_0\LSS - p_0^* \big\|_{P,2}
    =
    0 \ ,
    \\
    &
    &&
    \big\| \widehat{\xi}_0\LSS - \xi_0^* \big\|_{P,2}
    \big\| \widehat{p}_0\LSS - p_0^* \big\|_{P,2}
    =
    0 \ ,
    \\
    &
    \text{\AssumptionMixBiasDenom}:
    &&
    \big\| \widehat{\xi}_2\LSS - \xi_2^* \big\|_{P,2}
    \big\| \widehat{p}_0\LSS - p_0^* \big\|_{P,2}
    =
    0 \ .
\end{align*}
This completes the proof.  

}

\newpage

\bibliographystyle{apa}
\bibliography{PCSE.bib}

\begin{thebibliography}{}

\bibitem[\protect\astroncite{Andrews and Didelez}{2021}]{AndrewsDidelez2021}
Andrews, R.~M. and Didelez, V. (2021).
\newblock Insights into the cross-world independence assumption of causal
  mediation analysis.
\newblock {\em Epidemiology}, 32(2).

\bibitem[\protect\astroncite{Bakoyannis}{2023}]{BG2023}
Bakoyannis, G. (2023).
\newblock Estimating optimal individualized treatment rules with multistate
  processes.
\newblock {\em Biometrics}, 79(4):2830--2842.

\bibitem[\protect\astroncite{Bennett and Kallus}{2024}]{Bennett2024}
Bennett, A. and Kallus, N. (2024).
\newblock Proximal reinforcement learning: Efficient off-policy evaluation in
  partially observed markov decision processes.
\newblock {\em Operations Research}, 72(3):1071--1086.

\bibitem[\protect\astroncite{Bennett et~al.}{2023}]{Bennett2023}
Bennett, A., Kallus, N., Mao, X., Newey, W., Syrgkanis, V., and Uehara, M.
  (2023).
\newblock Source condition double robust inference on functionals of inverse
  problems.
\newblock {\em Preprint arXiv:2307.13793}.

\bibitem[\protect\astroncite{Bergmeir and Ben\'itez}{2012}]{RSNNS}
Bergmeir, C. and Ben\'itez, J.~M. (2012).
\newblock Neural networks in {R} using the stuttgart neural network simulator:
  {RSNNS}.
\newblock {\em Journal of Statistical Software}, 46(7):1--26.

\bibitem[\protect\astroncite{Carrasco et~al.}{2007}]{Carrasco2007}
Carrasco, M., Florens, J.-P., and Renault, E. (2007).
\newblock Linear inverse problems in structural econometrics estimation based
  on spectral decomposition and regularization.
\newblock In Heckman, J.~J. and Leamer, E.~E., editors, {\em Handbook of
  Econometrics}, volume~6, pages 5633--5751. Elsevier.

\bibitem[\protect\astroncite{Chen and Guestrin}{2016}]{xgboost}
Chen, T. and Guestrin, C. (2016).
\newblock Xgboost: A scalable tree boosting system.
\newblock In {\em Proceedings of the 22nd ACM SIGKDD International Conference
  on Knowledge Discovery and Data Mining}, KDD '16, page 785–794.

\bibitem[\protect\astroncite{Chen et~al.}{2014}]{Chen2014}
Chen, X., Chernozhukov, V., Lee, S., and Newey, W.~K. (2014).
\newblock Local identification of nonparametric and semiparametric models.
\newblock {\em Econometrica}, 82(2):785--809.

\bibitem[\protect\astroncite{Chen and Christensen}{2018}]{ChenChristensen2018}
Chen, X. and Christensen, T.~M. (2018).
\newblock Optimal sup-norm rates and uniform inference on nonlinear functionals
  of nonparametric iv regression.
\newblock {\em Quantitative Economics}, 9(1):39--84.

\bibitem[\protect\astroncite{Chernozhukov et~al.}{2018}]{Victor2018}
Chernozhukov, V., Chetverikov, D., Demirer, M., Duflo, E., Hansen, C., Newey,
  W., and Robins, J. (2018).
\newblock Double/debiased machine learning for treatment and structural
  parameters.
\newblock {\em The Econometrics Journal}, 21(1):C1--C68.

\bibitem[\protect\astroncite{Cui et~al.}{2024}]{Cui2023}
Cui, Y., Pu, H., Shi, X., Miao, W., and {Tchetgen Tchetgen}, E. (2024).
\newblock Semiparametric proximal causal inference.
\newblock {\em Journal of the American Statistical Association},
  119(546):1348--1359.

\bibitem[\protect\astroncite{Darolles et~al.}{2011}]{Darolles2011}
Darolles, S., Fan, Y., Florens, J.~P., and Renault, E. (2011).
\newblock Nonparametric instrumental regression.
\newblock {\em Econometrica}, 79(5):1541--1565.

\bibitem[\protect\astroncite{Dawid}{1979}]{Dawid1979}
Dawid, A.~P. (1979).
\newblock Conditional independence in statistical theory.
\newblock {\em Journal of the Royal Statistical Society: Series B
  (Methodological)}, 41(1):1--15.

\bibitem[\protect\astroncite{Dawid and Didelez}{2012}]{DawidDidelez2012}
Dawid, P. and Didelez, V. (2012).
\newblock ``imagine a can opener''--the magic of principal stratum analysis.
\newblock {\em The International Journal of Biostatistics}, 8(1).

\bibitem[\protect\astroncite{Deaner}{2018}]{Deaner2018}
Deaner, B. (2018).
\newblock Proxy controls and panel data.
\newblock {\em Preprint arXiv:1810.00283}.

\bibitem[\protect\astroncite{Ding et~al.}{2011}]{Ding2011}
Ding, P., Geng, Z., Yan, W., and Zhou, X.-H. (2011).
\newblock Identifiability and estimation of causal effects by principal
  stratification with outcomes truncated by death.
\newblock {\em Journal of the American Statistical Association},
  106(496):1578--1591.

\bibitem[\protect\astroncite{Ding and Lu}{2016}]{Ding2016}
Ding, P. and Lu, J. (2016).
\newblock {Principal Stratification Analysis Using Principal Scores}.
\newblock {\em Journal of the Royal Statistical Society Series B: Statistical
  Methodology}, 79(3):757--777.

\bibitem[\protect\astroncite{Dukes et~al.}{2023}]{Dukes2023_ProxMed}
Dukes, O., Shpitser, I., and {Tchetgen Tchetgen}, E.~J. (2023).
\newblock {Proximal mediation analysis}.
\newblock {\em Biometrika}, 110(4):973--987.

\bibitem[\protect\astroncite{Engl et~al.}{2000}]{Engl2000}
Engl, H.~W., Hanke, M., and Neubauer, A. (2000).
\newblock {\em Regularization of inverse problems}, volume 375.
\newblock Kluwer Academic Publishers, Dordrecht, The Netherlands.

\bibitem[\protect\astroncite{Feller et~al.}{2017}]{Feller2017}
Feller, A., Mealli, F., and Miratrix, L. (2017).
\newblock Principal score methods: Assumptions, extensions, and practical
  considerations.
\newblock {\em Journal of Educational and Behavioral Statistics},
  42(6):726--758.

\bibitem[\protect\astroncite{Florens et~al.}{2011}]{Florens2011}
Florens, J.-P., Johannes, J., and {Van Bellegem}, S. (2011).
\newblock Identification and estimation by penalization in nonparametric
  instrumental regression.
\newblock {\em Econometric Theory}, 27(3):472–496.

\bibitem[\protect\astroncite{Forastiere et~al.}{2018}]{Laura2018}
Forastiere, L., Mattei, A., and Ding, P. (2018).
\newblock {Principal ignorability in mediation analysis: through and beyond
  sequential ignorability}.
\newblock {\em Biometrika}, 105(4):979--986.

\bibitem[\protect\astroncite{Frangakis and Rubin}{2002}]{FR2002}
Frangakis, C.~E. and Rubin, D.~B. (2002).
\newblock Principal stratification in causal inference.
\newblock {\em Biometrics}, 58(1):21--29.

\bibitem[\protect\astroncite{Freyberger}{2017}]{Freyberger2017}
Freyberger, J. (2017).
\newblock {Non-parametric Panel Data Models with Interactive Fixed Effects}.
\newblock {\em The Review of Economic Studies}, 85(3):1824--1851.

\bibitem[\protect\astroncite{Friedman et~al.}{2010}]{glmnet}
Friedman, J., Hastie, T., and Tibshirani, R. (2010).
\newblock Regularization paths for generalized linear models via coordinate
  descent.
\newblock {\em Journal of Statistical Software}, 33(1):1--22.

\bibitem[\protect\astroncite{Friedman}{1991}]{earth}
Friedman, J.~H. (1991).
\newblock Multivariate adaptive regression splines.
\newblock {\em The Annals of Statistics}, 19(1):1 -- 67.

\bibitem[\protect\astroncite{Garreau et~al.}{2018}]{Garreau2018}
Garreau, D., Jitkrittum, W., and Kanagawa, M. (2018).
\newblock Large sample analysis of the median heuristic.
\newblock {\em Preprint arXiv:1707.07269}.

\bibitem[\protect\astroncite{Ghassami et~al.}{2024}]{Ghassami2024}
Ghassami, A., Yang, A., Shpitser, I., and {Tchetgen Tchetgen}, E. (2024).
\newblock {Causal inference with hidden mediators}.
\newblock {\em Biometrika}, 112(1):asae037.

\bibitem[\protect\astroncite{Ghassami et~al.}{2022}]{Ghassami2022}
Ghassami, A., Ying, A., Shpitser, I., and Tchetgen~Tchetgen, E. (2022).
\newblock Minimax kernel machine learning for a class of doubly robust
  functionals with application to proximal causal inference.
\newblock In Camps-Valls, G., Ruiz, F. J.~R., and Valera, I., editors, {\em
  Proceedings of The 25th International Conference on Artificial Intelligence
  and Statistics}, volume 151 of {\em Proceedings of Machine Learning
  Research}, pages 7210--7239. PMLR.

\bibitem[\protect\astroncite{Greenwell et~al.}{2019}]{gbm}
Greenwell, B., Boehmke, B., Cunningham, J., and Developers, G. (2019).
\newblock {\em {gbm: Generalized Boosted Regression Models}}.
\newblock R package version 2.1.5.

\bibitem[\protect\astroncite{Hable}{2012}]{Hable2012}
Hable, R. (2012).
\newblock Asymptotic normality of support vector machine variants and other
  regularized kernel methods.
\newblock {\em Journal of Multivariate Analysis}, 106:92--117.

\bibitem[\protect\astroncite{Hahn}{1998}]{Hahn1998}
Hahn, J. (1998).
\newblock On the role of the propensity score in efficient semiparametric
  estimation of average treatment effects.
\newblock {\em Econometrica}, 66(2):315--331.

\bibitem[\protect\astroncite{Hastie and Tibshirani}{1986}]{gam}
Hastie, T. and Tibshirani, R. (1986).
\newblock Generalized additive models.
\newblock {\em Statistical Science}, 1(3):297 -- 310.

\bibitem[\protect\astroncite{Hu and Schennach}{2008}]{Hu2008}
Hu, Y. and Schennach, S.~M. (2008).
\newblock Instrumental variable treatment of nonclassical measurement error
  models.
\newblock {\em Econometrica}, 76(1):195--216.

\bibitem[\protect\astroncite{Jemiai et~al.}{2007}]{Jemiai2007}
Jemiai, Y., Rotnitzky, A., Shepherd, B.~E., and Gilbert, P.~B. (2007).
\newblock {Semiparametric Estimation of Treatment Effects Given Base-Line
  Covariates on an Outcome Measured After a Post-Randomization Event Occurs}.
\newblock {\em Journal of the Royal Statistical Society Series B: Statistical
  Methodology}, 69(5):879--901.

\bibitem[\protect\astroncite{Jo and Stuart}{2009}]{JoStuart2009}
Jo, B. and Stuart, E.~A. (2009).
\newblock On the use of propensity scores in principal causal effect
  estimation.
\newblock {\em Statistics in Medicine}, 28(23):2857--2875.

\bibitem[\protect\astroncite{Joffe}{2011}]{Joffe2011}
Joffe, M. (2011).
\newblock Principal stratification and attribution prohibition: Good ideas
  taken too far.
\newblock {\em The International Journal of Biostatistics},
  7(1):0000102202155746791367.

\bibitem[\protect\astroncite{Kallus et~al.}{2021}]{Kallus2021}
Kallus, N., Mao, X., and Uehara, M. (2021).
\newblock Causal inference under unmeasured confounding with negative controls:
  {A} minimax learning approach.
\newblock {\em Preprint arXiv:2103.14029}.

\bibitem[\protect\astroncite{Kimeldorf and Wahba}{1970}]{KW1970}
Kimeldorf, G.~S. and Wahba, G. (1970).
\newblock A correspondence between bayesian estimation on stochastic processes
  and smoothing by splines.
\newblock {\em Annals of Mathematical Statistics}, 41(2):495--502.

\bibitem[\protect\astroncite{Kooperberg}{2020}]{polspline}
Kooperberg, C. (2020).
\newblock {\em polspline: {Polynomial Spline Routines}}.
\newblock R package version 1.1.19.

\bibitem[\protect\astroncite{Kress}{2014}]{Kress2014}
Kress, R. (2014).
\newblock {\em Linear Integral Equations}.
\newblock Springer, 3 edition.

\bibitem[\protect\astroncite{Kuroki and Pearl}{2014}]{Kuroki2014}
Kuroki, M. and Pearl, J. (2014).
\newblock {Measurement bias and effect restoration in causal inference}.
\newblock {\em Biometrika}, 101(2):423--437.

\bibitem[\protect\astroncite{Lipsitch et~al.}{2010}]{Lipsitch2010}
Lipsitch, M., Tchetgen~Tchetgen, E., and Cohen, T. (2010).
\newblock Negative controls: {A} tool for detecting confounding and bias in
  observational studies.
\newblock {\em Epidemiology}, 21(3):383.

\bibitem[\protect\astroncite{Luo et~al.}{2024}]{LuoLiMiaoHe2024}
Luo, S., Li, W., Miao, W., and He, Y. (2024).
\newblock Identification and estimation of causal effects in the presence of
  confounded principal strata.
\newblock {\em Statistics in Medicine}, 43(22):4372--4387.

\bibitem[\protect\astroncite{Mastouri et~al.}{2021}]{PMMR2021}
Mastouri, A., Zhu, Y., Gultchin, L., Korba, A., Silva, R., Kusner, M., Gretton,
  A., and Muandet, K. (2021).
\newblock Proximal causal learning with kernels: Two-stage estimation and
  moment restriction.
\newblock In Meila, M. and Zhang, T., editors, {\em Proceedings of the 38th
  International Conference on Machine Learning}, volume 139 of {\em Proceedings
  of Machine Learning Research}, pages 7512--7523. PMLR.

\bibitem[\protect\astroncite{Meza and Singh}{2021}]{Meza2021}
Meza, I. and Singh, R. (2021).
\newblock Nested nonparametric instrumental variable regression.
\newblock {\em Preprint arXiv:2112.14249}.

\bibitem[\protect\astroncite{Miao et~al.}{2018}]{Miao2018}
Miao, W., Geng, Z., and Tchetgen~Tchetgen, E.~J. (2018).
\newblock {Identifying causal effects with proxy variables of an unmeasured
  confounder}.
\newblock {\em Biometrika}, 105(4):987--993.

\bibitem[\protect\astroncite{Miao et~al.}{2020}]{Miao2020}
Miao, W., Shi, X., and {Tchetgen Tchetgen}, E. (2020).
\newblock A confounding bridge approach for double negative control inference
  on causal effects.
\newblock {\em Preprint arXiv:1808.04945}.

\bibitem[\protect\astroncite{Muandet et~al.}{2020}]{Muandet2020}
Muandet, K., Jitkrittum, W., and K\"{u}bler, J. (2020).
\newblock Kernel conditional moment test via maximum moment restriction.
\newblock In Peters, J. and Sontag, D., editors, {\em Proceedings of the 36th
  Conference on Uncertainty in Artificial Intelligence (UAI)}, volume 124 of
  {\em Proceedings of Machine Learning Research}, pages 41--50. PMLR.

\bibitem[\protect\astroncite{Newey and Powell}{2003}]{Newey2003}
Newey, W.~K. and Powell, J.~L. (2003).
\newblock Instrumental variable estimation of nonparametric models.
\newblock {\em Econometrica}, 71(5):1565--1578.

\bibitem[\protect\astroncite{Olivas-Martinez and
  Rotnitzky}{2025}]{Rotnizky2025}
Olivas-Martinez, A. and Rotnitzky, A. (2025).
\newblock Source-condition analysis of kernel adversarial estimators.
\newblock {\em Preprint arXiv:2508.17181}.

\bibitem[\protect\astroncite{Pearl}{2001}]{Pearl2001}
Pearl, J. (2001).
\newblock Direct and indirect effects.
\newblock In {\em Proceedings of the Seventeenth Conference on Uncertainty in
  Artificial Intelligence}, UAI'01, page 411–420. Morgan Kaufmann Publishers
  Inc.

\bibitem[\protect\astroncite{Pearl}{2009}]{Pearl2009}
Pearl, J. (2009).
\newblock {\em Causality: Models, Reasoning and Inference}.
\newblock Cambridge university press, 2 edition.

\bibitem[\protect\astroncite{Petrylak et~al.}{2004}]{SWOG2004}
Petrylak, D.~P., Tangen, C.~M., Hussain, M.~H., Lara, P.~N., Jones, J.~A.,
  Taplin, M.~E., Burch, P.~A., Berry, D., Moinpour, C., Kohli, M., Benson,
  M.~C., Small, E.~J., Raghavan, D., and Crawford, E.~D. (2004).
\newblock Docetaxel and estramustine compared with mitoxantrone and prednisone
  for advanced refractory prostate cancer.
\newblock {\em New England Journal of Medicine}, 351(15):1513--1520.

\bibitem[\protect\astroncite{Plato}{2025}]{Plato2025}
Plato, R. (2025).
\newblock A class of regularization schemes for linear ill-posed problems in
  banach spaces under low order source conditions.
\newblock {\em Preprint arXiv:2509.05418}.

\bibitem[\protect\astroncite{Richardson and Robins}{2013}]{SWIG2013}
Richardson, T.~S. and Robins, J.~M. (2013).
\newblock Single world intervention graphs {(SWIGs)}: {A} unification of the
  counterfactual and graphical approaches to causality.
\newblock {\em Center for the Statistics and the Social Sciences, University of
  Washington Series. Working Paper}, 128(30):2013.

\bibitem[\protect\astroncite{Robins}{1986}]{Robins1986}
Robins, J. (1986).
\newblock A new approach to causal inference in mortality studies with a
  sustained exposure period—application to control of the healthy worker
  survivor effect.
\newblock {\em Mathematical Modelling}, 7(9):1393--1512.

\bibitem[\protect\astroncite{Robins et~al.}{2007}]{Robins2007Discussion}
Robins, J., Rotnitzky, A., and Vansteelandt, S. (2007).
\newblock Discussions on ``principal stratification designs to estimate input
  data missing due to death''.
\newblock {\em Biometrics}, 63(3):650--658.

\bibitem[\protect\astroncite{Robins and Greenland}{1992}]{RG1992}
Robins, J.~M. and Greenland, S. (1992).
\newblock Identifiability and exchangeability for direct and indirect effects.
\newblock {\em Epidemiology}, 3(2):143--155.

\bibitem[\protect\astroncite{Robins and
  Richardson}{2011}]{RobinsRichardson2011}
Robins, J.~M. and Richardson, T.~S. (2011).
\newblock {Alternative Graphical Causal Models and the Identification of Direct
  Effects}.
\newblock In {\em {Causality and Psychopathology: Finding the Determinants of
  Disorders and their Cures}}. Oxford University Press.

\bibitem[\protect\astroncite{Robins et~al.}{2022}]{RRS2022}
Robins, J.~M., Richardson, T.~S., and Shpitser, I. (2022).
\newblock {\em An Interventionist Approach to Mediation Analysis}, page
  713–764.
\newblock Association for Computing Machinery, New York, NY, USA, 1 edition.

\bibitem[\protect\astroncite{Rotnitzky et~al.}{2020}]{Rotnitzky2020}
Rotnitzky, A., Smucler, E., and Robins, J.~M. (2020).
\newblock {Characterization of parameters with a mixed bias property}.
\newblock {\em Biometrika}, 108(1):231--238.

\bibitem[\protect\astroncite{Rubin}{2006}]{Rubin2006}
Rubin, D.~B. (2006).
\newblock Causal inference through potential outcomes and principal
  stratification: application to studies with ``censoring'' due to death.
\newblock {\em Statistical Science}, pages 299--309.

\bibitem[\protect\astroncite{Schick}{1986}]{Schick1986}
Schick, A. (1986).
\newblock On asymptotically efficient estimation in semiparametric models.
\newblock {\em The Annals of Statistics}, 14(3):1139--1151.

\bibitem[\protect\astroncite{Sch{\"o}lkopf et~al.}{2001}]{SHS2001}
Sch{\"o}lkopf, B., Herbrich, R., and Smola, A.~J. (2001).
\newblock A generalized representer theorem.
\newblock In Helmbold, D. and Williamson, B., editors, {\em Computational
  Learning Theory}, pages 416--426. Springer, Berlin, Heidelberg.

\bibitem[\protect\astroncite{Serfling}{2009}]{Serfling2009}
Serfling, R.~J. (2009).
\newblock {\em Approximation Theorems of Mathematical Statistics}.
\newblock John Wiley \& Sons.

\bibitem[\protect\astroncite{Shi et~al.}{2020a}]{Shi2020}
Shi, X., Miao, W., Nelson, J.~C., and Tchetgen~Tchetgen, E.~J. (2020a).
\newblock {Multiply Robust Causal Inference with Double-Negative Control
  Adjustment for Categorical Unmeasured Confounding}.
\newblock {\em Journal of the Royal Statistical Society Series B: Statistical
  Methodology}, 82(2):521--540.

\bibitem[\protect\astroncite{Shi et~al.}{2020b}]{Shi2020_Epi}
Shi, X., Miao, W., and {Tchetgen Tchetgen}, E. (2020b).
\newblock A selective review of negative control methods in epidemiology.
\newblock {\em Current Epidemiology Reports}, 7(4):190--202.

\bibitem[\protect\astroncite{Singh}{2022}]{Singh2022}
Singh, R. (2022).
\newblock A finite sample theorem for longitudinal causal inference with
  machine learning: {L}ong term, dynamic, and mediated effects.
\newblock {\em Preprint arXiv:2112.14249}.

\bibitem[\protect\astroncite{Singh et~al.}{2019}]{Singh2019}
Singh, R., Sahani, M., and Gretton, A. (2019).
\newblock Kernel instrumental variable regression.
\newblock In Wallach, H., Larochelle, H., Beygelzimer, A., d\textquotesingle
  Alch\'{e}-Buc, F., Fox, E., and Garnett, R., editors, {\em Advances in Neural
  Information Processing Systems}, volume~32. Curran Associates, Inc.

\bibitem[\protect\astroncite{Steinwart and Christmann}{2008}]{SVM2008}
Steinwart, I. and Christmann, A. (2008).
\newblock {\em Support Vector Machines}.
\newblock Springer-Verlag, New York.

\bibitem[\protect\astroncite{Stensrud et~al.}{2021}]{Stensrud2021}
Stensrud, M.~J., Hern\'an, M.~A., {Tchetgen Tchetgen}, E.~J., Robins, J.~M.,
  Didelez, V., and Young, J.~G. (2021).
\newblock A generalized theory of separable effects in competing event
  settings.
\newblock {\em Lifetime Data Analysis}, 27(4):588--631.

\bibitem[\protect\astroncite{Stensrud et~al.}{2023}]{Stensrud2022CSE}
Stensrud, M.~J., Robins, J.~M., Sarvet, A., {Tchetgen Tchetgen}, E.~J., and
  Young, J.~G. (2023).
\newblock Conditional separable effects.
\newblock {\em Journal of the American Statistical Association},
  118(544):2671--2683.

\bibitem[\protect\astroncite{Stensrud et~al.}{2022}]{Stensrud2022}
Stensrud, M.~J., Young, J.~G., Didelez, V., Robins, J.~M., and Hern\'an, M.~A.
  (2022).
\newblock Separable effects for causal inference in the presence of competing
  events.
\newblock {\em Journal of the American Statistical Association},
  117(537):175--183.

\bibitem[\protect\astroncite{Tchetgen~Tchetgen and Shpitser}{2012}]{TTS2012}
Tchetgen~Tchetgen, E.~J. and Shpitser, I. (2012).
\newblock {Semiparametric theory for causal mediation analysis: Efficiency
  bounds, multiple robustness and sensitivity analysis}.
\newblock {\em The Annals of Statistics}, 40(3):1816 -- 1845.

\bibitem[\protect\astroncite{{Tchetgen Tchetgen}
  et~al.}{2024}]{TT2024_Proximal}
{Tchetgen Tchetgen}, E.~J., Ying, A., Cui, Y., Shi, X., and Miao, W. (2024).
\newblock An introduction to proximal causal inference.
\newblock {\em Statistical Science}, 39(3):375--390.

\bibitem[\protect\astroncite{{van Buuren} and
  Groothuis-Oudshoorn}{2011}]{MICE2011}
{van Buuren}, S. and Groothuis-Oudshoorn, K. (2011).
\newblock {mice}: Multivariate imputation by chained equations in r.
\newblock {\em Journal of Statistical Software}, 45(3):1--67.

\bibitem[\protect\astroncite{van~der Laan et~al.}{2007}]{SL2007}
van~der Laan, M.~J., Polley, E.~C., and Hubbard, A.~E. (2007).
\newblock Super learner.
\newblock {\em Statistical Applications in Genetics and Molecular Biology},
  6(1).

\bibitem[\protect\astroncite{van~der Vaart}{1998}]{Vaart1998}
van~der Vaart, A.~W. (1998).
\newblock {\em Asymptotic Statistics}.
\newblock Cambridge Series in Statistical and Probabilistic Mathematics.
  Cambridge University Press, New York.

\bibitem[\protect\astroncite{van~der Vaart and Wellner}{1996}]{VW1996}
van~der Vaart, A.~W. and Wellner, J.~A. (1996).
\newblock {\em Weak Convergence and Empirical Processes: With Applications to
  Statistics}.
\newblock Springer.

\bibitem[\protect\astroncite{VanderWeele}{2011}]{Vanderweele2011}
VanderWeele, T.~J. (2011).
\newblock Principal stratification--{U}ses and limitations.
\newblock {\em The International Journal of Biostatistics}, 7(1).

\bibitem[\protect\astroncite{Wang et~al.}{2017}]{Wang2017}
Wang, L., Zhou, X.-H., and Richardson, T.~S. (2017).
\newblock {Identification and estimation of causal effects with outcomes
  truncated by death}.
\newblock {\em Biometrika}, 104(3):597--612.

\bibitem[\protect\astroncite{Wright and Ziegler}{2017}]{ranger}
Wright, M.~N. and Ziegler, A. (2017).
\newblock {ranger}: A fast implementation of random forests for high
  dimensional data in {C++} and {R}.
\newblock {\em Journal of Statistical Software}, 77(1):1--17.

\bibitem[\protect\astroncite{Yang and Ding}{2018}]{YangDing2018}
Yang, F. and Ding, P. (2018).
\newblock Using survival information in truncation by death problems without
  the monotonicity assumption.
\newblock {\em Biometrics}, 74(4):1232--1239.

\bibitem[\protect\astroncite{Ying et~al.}{2023}]{Ying2023}
Ying, A., Miao, W., Shi, X., and Tchetgen~Tchetgen, E.~J. (2023).
\newblock {Proximal causal inference for complex longitudinal studies}.
\newblock {\em Journal of the Royal Statistical Society Series B: Statistical
  Methodology}, 85(3):684--704.

\bibitem[\protect\astroncite{Zhang et~al.}{2023}]{Zhang2023}
Zhang, R., Imaizumi, M., Schölkopf, B., and Muandet, K. (2023).
\newblock Instrumental variable regression via kernel maximum moment loss.
\newblock {\em Journal of Causal Inference}, 11(1):20220073.

\end{thebibliography}

\end{document}